    \RequirePackage[english=usenglishmax]{hyphsubst}
    \documentclass[twoside,openright,titlepage,numbers=noenddot,headinclude,footinclude=true,cleardoublepage=empty,listof=totoc,paper=a4,fontsize=11pt,australian,twoside=semi,DIV=calc]{book} 

    \newcommand{\myTitle}{Dynamics and Interaction of Solitons in the BPS Limit and Their Internal Modes}

    \newcommand{\myFirstName}{Sergio Navarro Obregón}

    \newcommand{\myFaculty}{Facultad de Ciencias}
    \newcommand{\myUni}{Universidad de Valladolid}
    
    \PassOptionsToPackage{utf8}{inputenc}
	\usepackage{inputenc}

\PassOptionsToPackage{eulerchapternumbers,listings,pdfspacing,beramono,dottedtoc}{classicthesis}

\usepackage{bbm}

\newcounter{dummy} 
\providecommand{\mLyX}{L\kern-.1667em\lower.25em\hbox{Y}\kern-.125emX\@}

\usepackage[spanish,english]{babel}

\usepackage{csquotes}
\PassOptionsToPackage{
  style=numeric-comp,
  bibencoding=ascii,
  backend=bibtex,
  sorting=none,
  maxbibnames=10
}{biblatex}
\usepackage{biblatex}

\renewbibmacro{in:}{}
\DeclareFieldFormat{pages}{#1} 
\DeclareFieldFormat{page}{#1}

  \usepackage{amsmath}

\usepackage{lettrine}
\usepackage{amssymb}
\usepackage{gensymb}
\usepackage{lipsum}
\PassOptionsToPackage{T1}{fontenc} 
    \usepackage{fontenc}
\usepackage{textcomp} 
\usepackage{scrhack} 
\usepackage{xspace} 
\usepackage{mparhack} 
\PassOptionsToPackage{printonlyused,smaller}{acronym}
  \usepackage{acronym} 

\usepackage{afterpage}
\usepackage[version=4]{mhchem}

\usepackage{epigraph}
\usepackage{tabularx} 
\usepackage{makecell}
\usepackage{multirow}
\usepackage{caption}
\usepackage{subcaption}
\usepackage{notoccite} 
\usepackage{longtable}%
\usepackage{ragged2e}
\usepackage{graphicx}
\usepackage{bigints}
\usepackage{pdfpages}
\usepackage{esvect}
\usepackage{mathtools}

\usepackage{comment}
\usepackage{pdflscape}
\usepackage[svgnames]{xcolor} 
\usepackage{colortbl}%
  
\usepackage{booktabs}

\newcolumntype{Y}{>{\centering\arraybackslash}X}

\usepackage{etoolbox}
\makeatletter
\newif\if@in@acrolist

\AtBeginEnvironment{acronym}{\@in@acrolisttrue}
\newrobustcmd{\LU}[2]{\if@in@acrolist#1\else#2\fi}

\newcommand{\ACF}[1]{{\@in@acrolisttrue\acf{#1}}}
\makeatother

\usepackage{listings}
\PassOptionsToPackage{pdftex,hyperfootnotes=false,pdfpagelabels}{hyperref}
\usepackage{hyperref}  
\PassOptionsToPackage{pdftex}{graphicx}
    \usepackage{graphicx}

\hypersetup{breaklinks=true}
\hypersetup{linktocpage=true}
\hypersetup{colorlinks=true}
\hypersetup{urlcolor=webbrown}
\hypersetup{linkcolor=RoyalBlue}
\hypersetup{citecolor=webgreen}
\hypersetup{pageanchor=true}
\hypersetup{plainpages=false}
\hypersetup{pdfstartpage=3}
\hypersetup{pdfstartview=FitV}
\hypersetup{pdfpagemode=UseNone}
\hypersetup{pageanchor=true}
\hypersetup{pdfpagemode=UseOutlines}
\hypersetup{pdftitle={\myTitle}}
\hypersetup{pdfauthor={\textcopyright\ \myFirstName, \myUni, \myFaculty}}
\hypersetup{pdfsubject={}}
\hypersetup{pdfkeywords={}}
\hypersetup{pdfhighlight=/O}

\makeatletter
\@ifpackageloaded{babel}%
{%
  \addto\extrasaustralian{%
  }%
  \addto\extrasngerman{%
  }%
}{\relax}
\makeatother

\usepackage{classicthesis}

\usepackage{colortbl}
\usepackage{setspace}
\usepackage{fp}
\usepackage{soul}

\usepackage{textcomp}

\usepackage{pgfplots}
\usepackage{pgfplotstable}
\pgfplotsset{compat=newest}
\usepackage[separate-uncertainty = true,multi-part-units=single]{siunitx}

\usepackage[shortcuts]{extdash}

\usepackage{tikz-timing}[2011/01/09]
\usetikzlibrary{arrows, backgrounds, patterns, shapes}
\usepackage{arydshln}

\usepackage{algorithm}
\usepackage{algorithmic}

\renewcommand{\lstlistlistingname}{List of Listings}

\newcolumntype{L}[1]{>{\raggedright\arraybackslash}m{#1}}
\newcolumntype{C}[1]{>{\centering\arraybackslash}m{#1}}
\newcolumntype{R}[1]{>{\raggedleft\arraybackslash}m{#1}}

\numberwithin{equation}{chapter}
\numberwithin{figure}{chapter}
\numberwithin{table}{chapter}
\numberwithin{algorithm}{chapter}
\newtheorem{definition}{Definition}[chapter]
\numberwithin{definition}{chapter}

\newtheorem{theorem}{Theorem}[chapter]
\numberwithin{theorem}{chapter}

\DeclareMathOperator{\sech}{sech}
\DeclareMathOperator{\cosech}{cosech}
\DeclareMathOperator{\csch}{csch}

\DeclareMathOperator{\arctanh}{arctanh}

\newcommand{\Phiv}{\Phi^{\raisebox{0.36ex}{$\scriptstyle v$}}}

\usepackage{amssymb} 
    \usepackage[a4paper,top=25.4mm,bottom=25.4mm,left=25mm, right=25mm,bindingoffset=6mm]{geometry} 

\begin{document}
        \frenchspacing
        \raggedbottom
        \selectlanguage{english}
      
        \pagestyle{plain}
        \pagenumbering{roman}
      

        \clearpage
        \thispagestyle{empty}
        \phantomsection
        \clearpage
        \includepdf[pages=-]{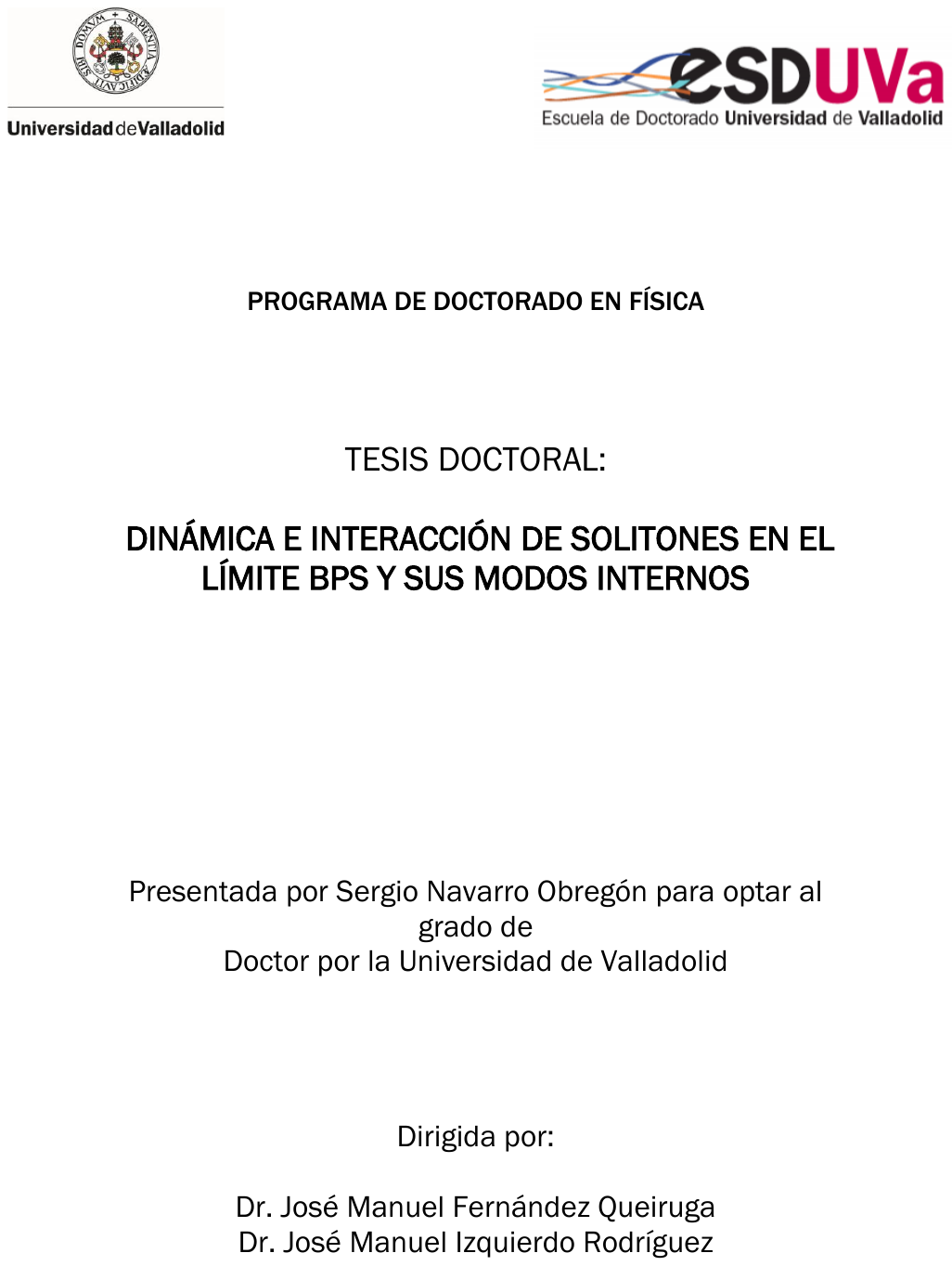}

        \clearpage
        \thispagestyle{empty}
        \newpage
        \mbox{~}
        \clearpage
        \newpage
        \thispagestyle{empty}
        \clearpage

        \clearpage
        \thispagestyle{empty}
        \phantomsection
        \clearpage
        \includepdf[pages=-]{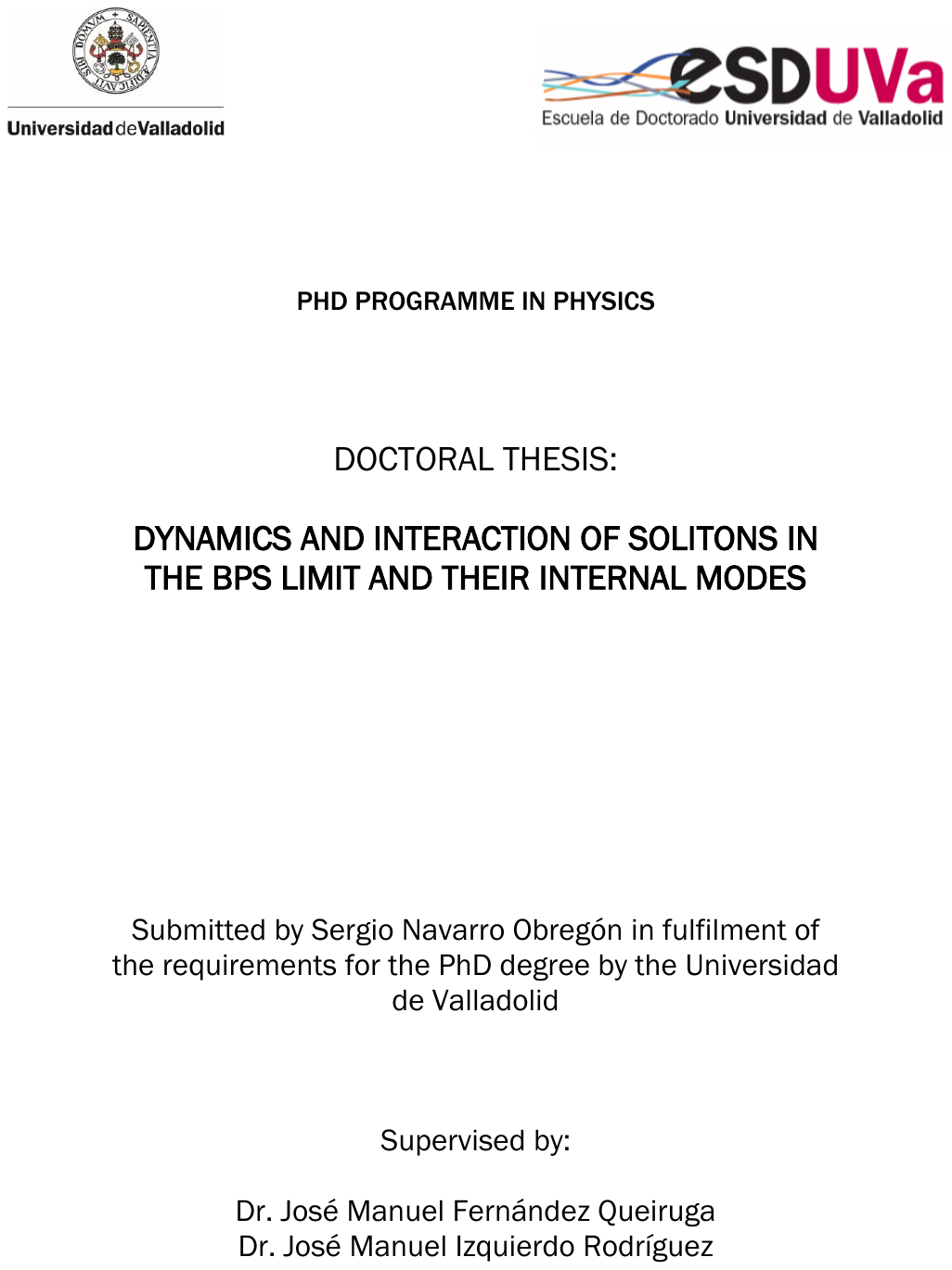}
      
        \onehalfspacing
    
    \refstepcounter{dummy}
    \pdfbookmark[1]{\contentsname}{tableofcontents}
    \setcounter{tocdepth}{2}
    \setcounter{secnumdepth}{3}
    \manualmark
    \markboth{\spacedlowsmallcaps{\contentsname}}{\spacedlowsmallcaps{\contentsname}}
    \tableofcontents
    \automark[section]{chapter}
    \renewcommand{\chaptermark}[1]{\markboth{\spacedlowsmallcaps{#1}}{\spacedlowsmallcaps{#1}}}
    \renewcommand{\sectionmark}[1]{\markright{\thesection\enspace\spacedlowsmallcaps{#1}}}
    
    \clearpage
    
    \begingroup 
    \let\clearpage\relax
    \let\cleardoublepage\relax
    \let\cleardoublepage\relax

    \newpage


    \newpage

   

    \newpage
    
\endgroup
                          

\chapter*{Acknowledgments}
\addcontentsline{toc}{chapter}{Acknowledgments}

El trabajo que aquí presento no es solo el resultado de años de esfuerzo y dedicación, sino también del constante estímulo y apoyo de quienes me han acompañado a lo largo de este camino. A continuación, deseo expresar mi agradecimiento a todas las personas que han hecho posible la realización de esta tesis doctoral.

    En primer lugar, quiero agradecer a mis padres, quienes siempre han confiado plenamente en cada uno de los objetivos que me he propuesto y me han brindado la oportunidad de alcanzarlos. Sin su cariño y sus sacrificios, nada de esto habría sido posible. También quiero dedicar esta tesis a mi hermano, que ha sido una fuente constante de inspiración y un motivo más por el que seguir esforzándome. Eres un símbolo de lucha y de superación, y estoy seguro de que conseguirás todo aquello que te propongas.
    
    Estoy profundamente agradecido a Adrián, Diego, Emmanuel, Fabio, Gonzalo, Javier, Marcos y Martín por su papel en todo este proceso. Por muy agotado que estuviese en ocasiones, regresar a Burgos siempre fue una manera de desconectar y volver a ser yo mismo entre risas y charlas con vosotros. Me siento un afortunado por teneros a mi lado. Vuestra esencia os hace únicos, no cambiéis nunca.
    
    Valoro todo el cariño que me han ofrecido Esther y Vera desde el primer día. Sois dos chicas con un corazón enorme y una actitud de admirar. Gracias por hacerme sentir siempre tan bien acompañado.
    
    Tambi\'en quiero destacar a Álvaro, que siempre ha creído en mí y me ha animado desde el primer momento. Admiro la paciencia que has tenido conmigo, y agradezco todo lo que has hecho por m\'i. Te deseo lo mejor en tu nueva etapa en Madrid. 
    
    No puedo olvidarme de Marina, ni de todas aquellas veces que ha tenido que soportar mis largos mon\'ologos. Gracias por tu paciencia y por estar cuando lo he necesitado. Siempre valorar\'e todo el cariño y la ayuda que me has proporcionado. 

    Estos años en Valladolid también me han permitido cruzar mi camino con bellísimas personas. Irene, tu simpatía y sentido del humor siempre han sido algo que te ha caracterizado. Este año ha sido complicado, pero eso no cambia el talento que tienes. Todos sabemos que eres y ser\'as una gran profesora. Itto, me siento afortunado de haberte conocido, eres un apoyo constante y una persona en la que siempre puedo confiar. Ojalá más personas tuvieran tu bondad y forma de ser. No hay nada que desee más que verte cumplir tus sueños. Muy pronto vamos a celebrar todo el esfuerzo que has puesto en ellos. Roberto, agradezco el apoyo que has supuesto. Todo lo que sé de docencia te lo debo a ti; espero, algún día, llegar a ser tan buen profesor como tú. Tu vocación por la enseñanza y tu carisma harán que seas la inspiración de muchos estudiantes. Jorge, fuiste mi primera amistad en Valladolid y, desde entonces, me has acompañado tanto en lo académico como en lo personal. Espero que sigamos compartiendo terraceos durante mucho tiempo más y que cumplas todas tus aspiraciones.
    
    Aunque no hayamos tenido mucho tiempo para conocernos, no puedo dejar de reconocer el afecto que me ha brindado Nerea. En el poco tiempo que hemos compartido piso, siempre estuviste presente y te preocupaste por mí. Por todo ello, muchas gracias. Espero que consigas todo lo que te propongas.
    
    También quiero mencionar el cobijo que ha supuesto compartir despacho con Albert, Julio, Samane, Sergio y Paz. Los largos días de trabajo y las calurosas tardes de verano se vuelven mucho más llevaderos gracias a vuestra compañía. A mayores, doy las gracias a David, con quien no solo he crecido como científico, sino también he compartido innumerables viajes a congresos y una inolvidable estancia en Cracovia.

    I would also like to thank Andrzej, Carlos, Kasia and Tom for their hospitality during the months that I stayed in Krakow. I really had an enjoyable experience learning about Polish culture and discussing science.
    
    Finalmente, quiero destacar el papel fundamental que han desempeñado mi tutor Dr. Luis Miguel Nieto Calzada y mis directores, Dr. Xose Manuel Fern\'andez Queiruga y Dr. Jos\'e Manuel Izquierdo Rodr\'iguez, durante mi desarrollo, tanto personal como investigador. Siempre me habéis guiado de la mejor manera, me habéis formado con dedicación y habéis estado presentes en cada uno de mis primeros pasos en la ciencia. Os estaré siempre agradecido.

    Agradezco la financiaci\'on proporcionada por el proyecto "New developments in mathematical modelling of quantum phenomena" (\href{https://portaldelaciencia.uva.es/proyectos/181076/detalle}{Ref: PID2020-113406GB-I00}) de la Agencia Estatal de In\-ves\-ti\-ga\-ci\'on y del Ministerio de Ciencia e Innovaci\'on, al igual que valoro la financiaci\'on otorgada por el proyecto "Q-CAYLE, Plan Complementario en Comunicaci\'on Cu\'antica" (\href{https://portaldelaciencia.uva.es/proyectos/276569/detalle}{Ref: PRTRC17.I1}) dentro del Plan de Recuperaci\'on, Transformaci\'on y Resiliencia - Next Generation UE, cofinanciado por el Ministerio de Ciencia e Innovaci\'on y la Junta de Castilla y Le\'on. 

    \begin{figure}[h]	\centerline{\includegraphics[width=0.6\textwidth]{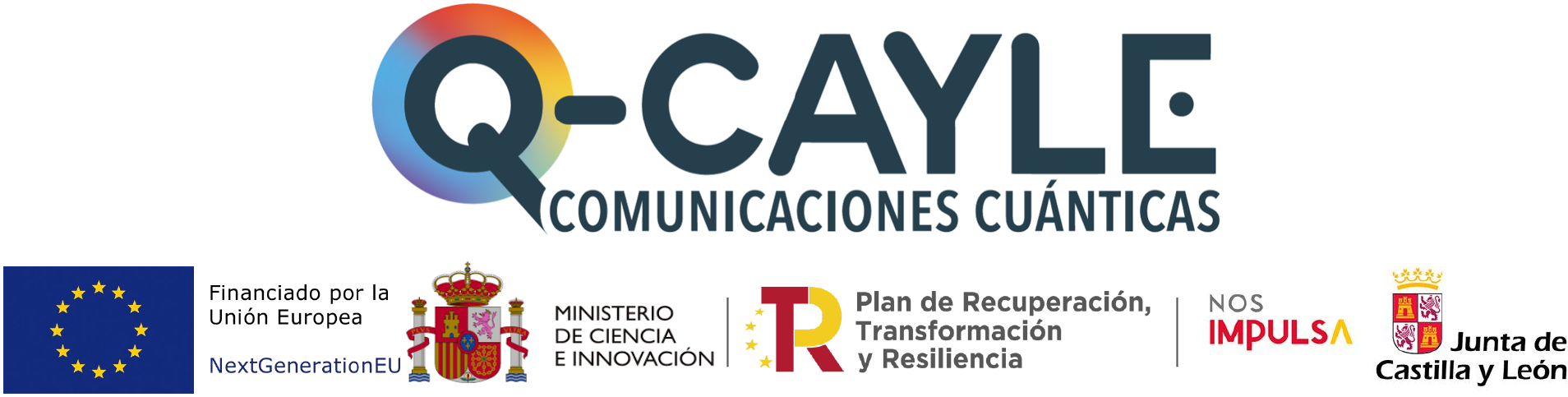}} 
    \end{figure}
    
    Igualmente, aprecio la financiaci\'on recibida del proyecto "Dynamics of Topological Defects: New Analytical and Numerical Developments with Applications" (\href{https://www.aei.gob.es/sites/default/files/convocatory_info/file/2024-07/PID2023-PRP-Texto\%2BAnexos\_fda.pdf}{Ref: PID2023-148409NB-I00}) otorgado por la Agencia Estatal de Investigaci\'on, as\'i como la ayuda recibida por el proyecto "Programas estratégicos de investigación: Laboratory of disruptive and interdisciplinary sciences" (\href{https://portaldelaciencia.uva.es/proyectos/913645/detalle}{Ref: CLU-2023-1-05}), financiado por la Junta de Castilla y Le\'on - Conserjer\'ia de Educac\'on y por el Fondo Europeo de Desarrollo Regional (FEDER). 
    \vspace{-0.7cm}
    \begin{figure}[h]	\centerline{\includegraphics[width=0.6\textwidth]{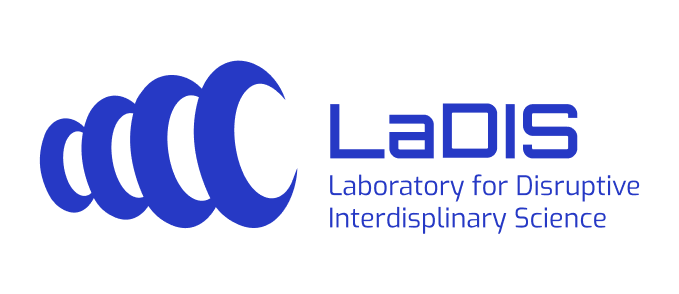}} 
    \end{figure}
    \vspace{-0.7cm}
    
    Por \'ultimo, destacar que esta tesis ha sido financiada a trav\'es de una ayuda para la contrataci\'on de personal investigador (\href{https://bocyl.jcyl.es/boletines/2022/12/28/pdf/BOCYL-D-28122022-18.pdf}{Ref: EDU/1868/2022}) cofinanciadas por el Fondo Social Europeo Plus (FSE+).  

    \begin{figure}[h]	\centerline{\includegraphics[width=0.6\textwidth]{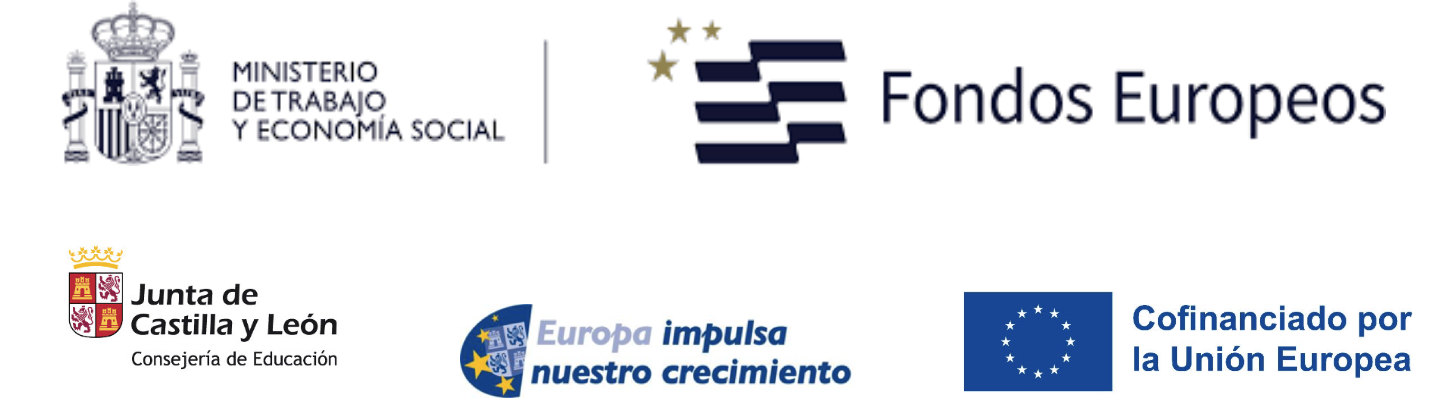}} 
    \end{figure}

        \chapter*{Abstract}
        \addcontentsline{toc}{chapter}{Abstract}

        Solitons constitute a distinguished subset within the spectrum of solutions to certain non-linear classical field theories. These solutions are characterised by exhibiting properties analogous to those of extended particle-like states, making them particularly appealing from the perspective of quantum field theory as well. Moreover, solitons have found applications in various areas of physics, ranging from condensed matter and nuclear physics to cosmology.

        These inherently non-perturbative objects can be classified as either stable or unstable solitons, depending on whether or not they exhibit dispersive behaviour. Among stable solitons, two broad families can be distinguished, based on the mechanism responsible for their stability. On one hand, there are the \textit{topological solitons}, such as \textit{kinks}, \textit{vortices}, and \textit{monopoles}, whose stability arises from the existence of topological invariants. Within the class of field theories that admit topological solitons, there is a particularly interesting subclass known as \textit{BPS theories} (Bogomolny–Prasad–Sommerfield). These theories have the remarkable property of lacking static interactions between solitons, allowing multiple configurations to coexist with the same energy. Furthermore, BPS solutions satisfy first-order differential equations, the solutions of which automatically fulfil the static Euler–Lagrange equations of the system. On the other hand, there are the \textit{non-topological solitons}, such as \textit{Q-balls}, \textit{soliton stars}, or \textit{oscillons}, which are stabilised by the conservation of charges associated with continuous global symmetries, by the integrability of the theory, or by certain adiabatic invariants. In contrast, \textit{unstable solitons} include configurations such as \textit{sphalerons}, which represent metastable states.
        
        It is important to emphasise that solitons possess a rich internal structure, which can be interpreted in terms of collective excitations around the base configuration. Depending on their nature, these excitations can induce various behaviours: from the fragmentation of the soliton into more fundamental entities, to oscillations that effectively modify its size. Such modes can be triggered during the soliton formation process following a phase transition, or may emerge during its evolution, for example, during interactions with other entities such as other solitons, impurities, or incoming radiation.
        
        The main objective of this thesis has been to analyse soliton dynamics in detail, with special attention paid to the role of possible internal modes associated with these configurations. Specifically, the research has focused on the study of one- and two-dimensional models, with the aim of developing a solid basis for understanding soliton dynamics, which can then be extended to the study of their behaviour in three-dimensional theories. Among the broad spectrum of solitons, this thesis concentrates on the study of the aforementioned kinks, oscillons, vortices, and sphalerons. Nevertheless, regardless of dimensionality, field theories constitute systems with an infinite number of degrees of freedom, which poses challenges both for obtaining analytical results and for predictive modelling. To address these challenges, this thesis employs the construction of effective models that retain the essential degrees of freedom required to capture the phenomenology observed in numerical simulations of the full theory, using the well-known \textit{collective coordinate method}. In addition, other complementary mathematical tools, such as perturbative techniques, have also been employed.
        
        Among the main achievements of this doctoral thesis, it is worth highlighting the introduction, for the first time, of genuine radiation modes within the collective coordinate framework to describe phenomena associated with radiation emission and interaction in solitonic systems. Furthermore, a generalisation of Samols’ moduli space metric for local vortices in the Abelian-Higgs model has been developed through the incorporation of vibrational degrees of freedom. This provides suitable initial conditions for modelling vortex–vortex collisions in effective models. Additionally, a new class of sphalerons that we have coined semi-BPS sphalerons has been identified and analysed; these exhibit properties analogous to those of BPS solitons. Finally, the role of oscillatory internal modes in the decay process of sphalerons has been studied in detail, leading to the proposal of a dynamic stabilisation mechanism. This mechanism has been further explained and extended to more general models, demonstrating the robustness and potential applicability of this phenomenon to physically relevant theories.


    \chapter*{Resumen}
    \addcontentsline{toc}{chapter}{Resumen}

    \indent \indent Los solitones constituyen un subconjunto distinguido dentro del espectro de soluciones de ciertas teor\'ias de campo cl\'asicas no lineales. Estas soluciones se caracterizan por exhibir propiedades an\'alogas a las de estados extendidos de tipo part\'icula, lo que les ha hecho bastante atractivas tambi\'en desde el punto de vista de la teor\'ia cu\'antica de campos. Adem\'as, han encontrado aplicaciones en diversas \'areas dentro de f\'isica, desde materia condensada o f\'isica nuclear, hasta cosmolog\'ia.

    Estos objetos, de naturaleza no perturbativa, pueden clasificarse en solitones estables e inestables, en funci\'on de si presentan o no un comportamiento dispersivo. Dentro del conjunto de solitones estables, es posible distinguir a su vez dos grandes familias, determinadas por el mecanismo que asegura su estabilidad. Por un lado, se encuentran los \textit{solitones topol\'ogicos}, como los "\textit{kinks}", \textit{v\'ortices}, o los \textit{monopolos}, cuya estabilidad se fundamenta en la existencia de invariantes topol\'ogicos. Dentro de las teor\'ias de campos que contienen solitones topol\'ogicos, existe una clase particularmente interesante conocida como \textit{teor\'ias BPS} (Bogomolny-Prasad-Sommerfield).  Estas teor\'ias cuentan con la propiedad de que no hay interacci\'on est\'atica entre los solitones, lo que implica que m\'ultiples configuraciones pueden coexistir con la misma energ\'ia.  Adem\'as, las soluciones BPS satisfacen ecuaciones diferenciales de primer orden, cuya resoluci\'on garantiza autom\'aticamente el cumplimiento de las ecuaciones est\'aticas de Euler–Lagrange del sistema. Por otro lado, se distinguen los s\textit{olitones no topol\'ogicos}, como las "\textit{Q-balls}", las \textit{estrellas de solitones} o los \textit{oscilones}, estabilizados por la conservaci\'on de cargas asociadas a simetr\'ias globales continuas, por la integrabilidad de la teor\'ia, o por ciertas invarianzas adiab\'aticas. Por otro lado, en la categor\'ia de \textit{solitones inestables} se encuentran configuraciones como los "\textit{sphalerons}", representando estados metaestables. 
    
    Es crucial destacar que los solitones cuentan con una rica estructura interna, que puede in\-ter\-pre\-tar\-se en t\'erminos de excitaciones colectivas alrededor de la configuraci\'on base. Estas excitaciones, dependiendo de su naturaleza, pueden inducir distintos comportamientos: desde la fragmentación del solit\'on en entidades m\'as fundamentales, hasta oscilaciones que modifican efect\'ivamente su tamaño. Tales modos pueden ser activados durante el proceso de formaci\'on del solit\'on tras una transici\'on de fase, o pueden emerger a lo largo de su evoluci\'on, por ejemplo durante interacciones con otras entidades tales como otros solitones, impurezas o radiaci\'on incidente. 
    
    El objetivo principal de esta tesis ha sido el de analizar en detalle la din\'amica de solitones, pres\-tan\-do especial atenci\'on al papel que desempeñan los posibles modos internos asociados a dichas configuraciones. En concreto, la investigaci\'on se ha centrado en el estudio de modelos unidimensionales y bidimensionales, con el prop\'osito de adquirir una primera intuici\'on sobre la din\'amica de los solitones, que sirva como base para una posterior extrapolaci\'on al an\'alisis de su evoluci\'on en teor\'ias tridimensionales. Dentro de la pl\'etora de solitones, esta tesis se ha centrado en el estudio de los anteriormente mencionados "kinks", oscilones, v\'ortices y "sphalerons". No obstante, independientemente de la dimensionalidad, las teor\'ias de campos constituyen sistemas con un n\'umero infinito de grados de libertad, lo que dificulta tanto la obtenci\'on de resultados anal\'iticos como la capacidad predictiva. Para lidiar con estas dificultades, en esta tesis se ha recurrido a la cons\-truc\-ci\'on de modelos efectivos que contengan los grados de libertad esenciales para describir la fenomenolog\'ia observada en las simulaciones num\'ericas de la teor\'ia completa, mediante el uso del conocido como \textit{m\'etodo de las coordenadas colectivas}. Adem\'as, se han empleado otras herramientas matem\'aticas complementarias, como t\'ecnicas perturbativas. 
    
    Entre los principales logros alcanzados en el desarrollo de esta tesis doctoral, cabe destacar la introducci\'on, por primera vez, de modos genuinos de radiaci\'on en el marco de las coordenadas colectivas para describir fen\'omenos asociados a la emisi\'on e interacci\'on de radiaci\'on en sistemas con solitones. Asimismo, se ha conseguido una generalizaci\'on de la m\'etrica del espacio de m\'odulos de Samols para v\'ortices locales en el modelo de Higgs abeliano mediante la incorporaci\'on de grados de libertad vibracionales, lo que otorga las condiciones iniciales id\'oneas para el modelado de colisiones v\'ortice-v\'ortice en modelos efectivos. Por otro lado, se ha identificado y analizado una nueva clase de "sphalerons", que hemos denominado "sphalerons" semi-BPS, y que presentan propiedades an\'alogas a las de los solitones BPS. Finalmente, se ha analizado en detalle el papel de los modos internos oscilatorios en el proceso de decaimiento de los "sphalerons", lo que nos ha permitido proponer un mecanismo de estabilizaci\'on din\'amica que ha sido explicado y posteriormente extendido a otros modeos m\'as generales, lo cual muestra la robustez y la posible aplicabilidad de este fen\'omeno a modelos de mayor inter\'es f\'isico.


    \chapter*{Publications}
    \addcontentsline{toc}{chapter}{Publications}
    
        The research presented in this thesis is based on the following publications:
    
        \begin{enumerate}
            \item S. Navarro-Obreg\'on, L.M. Nieto and J.M. Queiruga, \textit{Inclusion of radiation in the collective coordinate method approach of the $\phi^4$ model}, \href{https://doi.org/10.1103/PhysRevE.108.044216}{Phys. Rev. E \textbf{108} (2023), 044216}.
            
            \item A. Alonso-Izquierdo, S. Navarro-Obreg\'on, K. Oles, J. Queiruga, T. Romanczukiewicz and A. Wereszczynski, \textit{Semi-Bogomol'nyi-Prasad-Sommerfield sphaleron and its dynamics}, \href{https://doi.org/10.1103/PhysRevE.108.064208}{Phys. Rev. E \textbf{108} (2023), 064208}.
    
            \item S. Navarro-Obreg\'on and J. Queiruga, \textit{Impact of the internal modes on the sphaleron decay}, \href{https://doi.org/10.1140/epjc/s10052-024-13175-w}{Eur. Phys. J. C \textbf{84} (2024), 821}.

            \item D. Migu\'elez-Caballero, S. Navarro-Obreg\'on and A. Wereszczynski, \textit{Moduli space metric of the excited vortex}, \href{https://doi.org/10.1103/PhysRevD.111.105008}{Phys. Rev. D \textbf{111} (2025), 105008}.
    
            \item S. Navarro-Obreg\'on and J.M. Queiruga, \textit{Perturbed nonlinear dynamics and decay of sphalerons on circles}, \href{https://doi.org/10.1016/j.physd.2025.134805}{Physica D \textbf{481} (2025), 134805}.          
        \end{enumerate}

        \noindent Moreover, the research activity conducted during the doctoral studies has also resulted in the following publication, which is not included in the present thesis:

        \begin{enumerate}
            \setcounter{enumi}{5}
            \item A. Alonso-Izquierdo, J.J. Blanco-Pillado, D. Migu\'elez-Caballero, S. Navarro-Obreg\'on and J. Queiruga, \textit{Excited Abelian-Higgs vortices: Decay rate and radiation emission}, \href{https://doi.org/10.1103/PhysRevD.110.065009}{Phys. Rev. D \textbf{110} (2024), 065009}.        
        \end{enumerate}
              
        \cleardoublepage
        \pagestyle{scrheadings}
        \pagenumbering{arabic}
        \onehalfspacing

        \chapter{Introduction}\label{c:Introduction}

    \epigraph{"Somewhere, something incredible is waiting to be known."}{-- Carl Sagan}

    At a first glance, one could naively assume that physical systems are always governed by linear differential equations, given that Newton's laws, Maxwell’s equations in the vacuum and even the equations of quantum mechanics are linear in their fundamental formulation. However, the reality is more complex, and the majority of physical situations are non-linear phenomena. The understanding of these systems has required significant effort from mathematicians and theoretical physicists. Over the past few decades, considerable progress has been made in this field. For example, a vast number of mathematical structures have been uncovered, and new techniques have been developed to determine exact or approximate solutions of these systems.  

    One type of non-linear solution that has drawn particular attention is the \textit{soliton}, which represents the principal object of study in the present thesis. However, it is important to emphasise that the notion of a soliton has been used in somewhat different senses, depending on the framework or specific application. For that reason, a brief discussion on the evolution of this concept is insightful for properly understanding the background of this thesis.
    
    This introductory chapter is divided into four sections. First, in \autoref{s:Solitons} we explain the historical development of the concepts of \textit{solitary wave} and \textit{soliton}. Then, in \autoref{s:SolitonsField}, we specify our notion of soliton within the context of field theories. With this framework explained, \autoref{s:Objetives} outlines the main objectives of the thesis. Finally, in \autoref{s:Structure_Methodology} we describe the structure of the presented work and the methodology employed throughout the research. 

\section{From solitary waves to solitons}\label{s:Solitons}

\begin{figure}[htb]
        \centering
        \includegraphics[width=0.44\linewidth]{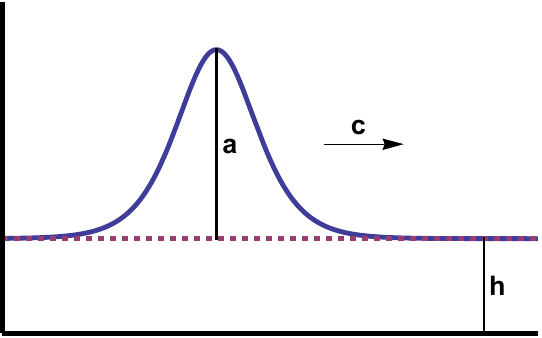}
        \caption{A scheme of a solitary wave in a shallow channel of water.}
        \label{f:Solitary}
    \end{figure}

    Our exploration of the concept of soliton -- from its origins in hydrodynamics to its modern applications in physics and mathematics -- begins in a narrow channel near Edinburgh, Scotland. In 1834, J. Scott Russell witnessed "a rounded, smooth and well-defined heap of water, which continued its course along the channel apparently without change of form or diminution of speed" \cite{Russell:1984}. This milestone could be considered as the origin of the \textit{soliton theory}. He coined the term \textit{wave of translation} to describe this large solitary elevation, although it would later become known as \textit{solitary wave}. After that, Russell, fascinated with this phenomenon, performed his own laboratory experiments, generating these waves of unchanging profile and constant velocity by dropping a weight at one end of a water channel (see \autoref{f:Solitary}). Remarkably, he was able to empirically determine an expression relating the speed of the wave $c$, its amplitude $a$, the undisturbed depth of water $h$ and the acceleration of gravity $g$
    \begin{equation}\label{e:Russell}
        c^2 = g(a + h)\,.   
    \end{equation}

    At the very beginning, the idea of a solitary wave faced substantial criticism and provoked controversy among some of the leading figures of the time; for example, Airy and Stokes claimed that any such wave would be unstable \cite{Stokes:1847}. The scientific community had to wait over thirty years until Boussinesq \cite{Boussinesq:1871} and Lord Rayleigh \cite{Rayleigh:1876} deduced from the equations of motion for an inviscid incompressible fluid the same relation (\ref{e:Russell}) as Russell. However, this possibility was not fully accepted until the seminal work by D.J. Korteweg and G. de Vries \cite{Korteweg:1895}, where these authors derived a non-linear differential equation (the so-called \textit{KdV equation}) governing the evolution of long one-dimensional, small-amplitude waves propagating in a shallow channel of water\footnote{This partial differential equation is simplified in dimensionless units to $ u_\tau + 6\, u\,u_\xi + u_{\xi\xi\xi}= 0\,.$}
    \begin{equation}\label{e:KdV}
        \dfrac{\partial\eta}{\partial t} = \dfrac{3}{2}\sqrt{\dfrac{g}{h}}\dfrac{\partial}{\partial x}\left(\dfrac{1}{2}\eta^2 + \dfrac{2}{3}\alpha\eta + \dfrac{1}{3}\sigma\dfrac{\partial^2\eta}{\partial x^2}\right)\,, \quad \sigma = \dfrac{1}{3}h^3 - \dfrac{T h}{\rho g}\,,
    \end{equation} 
    where $\eta$ is the elevation of the wave over the equilibrium depth $h$, $\alpha$ is a constant parameter related to the uniform motion of the liquid, $g$ the acceleration of gravity, $\rho$ the density and $T$ the surface tension. This partial differential equation admits solitary waves, and thanks to this breakthrough solitary waves gained theoretical foundation.  
    
    The description of these entities through the KdV equation allowed to provide a rigorous definition based on the features outlined above.
    \begin{definition}[Solitary wave: Ablowitz and Clarkson \cite{Ablowitz:1991}]
    A solitary wave is a non-singular solution $u: \mathbb{R}^2 \rightarrow \mathbb{R}$ of a non-linear differential equation $\Phi(x,t,u) = 0$, with $x,t \in \mathbb
    R$, that fulfills the following conditions:
    \begin{enumerate}
        \item The field $u(x,t)$ has a space-time dependence of the form $u(x,t) = u(x - c\,t) \equiv u(\xi)$\,.
        \item The field $u(\xi)$ approaches a constant value asymptotically when $|\xi| \rightarrow \infty$\,. 
    \end{enumerate}
    \end{definition}
    To appreciate the peculiarity of this behaviour, note that, in general, even the addition of the simplest contributions to a wave equation leads to dispersive or dissipative effects. Only in certain special cases a balance among the dispersive and the non-linear terms results in waves that preserve their shape.
    
    After the discovery of solitary waves, it took sixty years for solitons to emerge, specifically, in the context of the so-called Fermi-Pasta-Ulam (FPU) experiment \cite{Fermi:1955}. Between 1954 and 1955 on the Los Alamos MANIAC computer, Fermi, Pasta, and Ulam conducted numerical simulations of the evolution of a vibrating string with fixed ends and non-linear elastic restoring forces. In order to perform these simulations, they modeled the string of length $l$ as $N$ oscillators with equilibrium positions $p_i = ih$, $i \in \{0, \ldots, N-1\}$, where $h = l/(N-1)$ is the lattice spacing and the new positions read $X_i(t) = p_i + x_i(t)$. Then, they discretised the partial differential equation, obtaining a finite set of ordinary differential equations of the form
    \begin{equation}\label{e:FPU}
        m\,\ddot{x}_i = k(x_{i+1} + x_{i-1} - 2x_i)[1 + \beta(x_{i+1} - x_{i-1})]\,, \quad x_0(t) = x_{N-1}(t) = 0\,,    
    \end{equation}
    describing Newton's equation for each oscillator, with $\beta$ accounting for the strength of the non-linear contribution. 
    
    They expected that, due to the ergodic hypothesis -- or in more physical terms, the equipartition principle -- if the system had many degrees of freedom and was relatively close to a stable equilibrium, the non-linear contributions would thermalise the energy of the system. In other words, they sought to demonstrate that, after a sufficiently long time, the energy would be distributed among all the normal modes of the corresponding linearised system. Nevertheless, they observed that there was no equipartition of the energy among the modes. Instead, after a very long time, the system returned to a state arbitrarily close to the initial condition. This process was repeated almost periodically, serving as an example of Poincar\'e recurrence \cite{Palais:1997,Drazin:1996}. 

\begin{figure}[htb]
      \centering
      \noindent
      \begin{minipage}[t]{0.33\textwidth}
        \includegraphics[width=\linewidth]{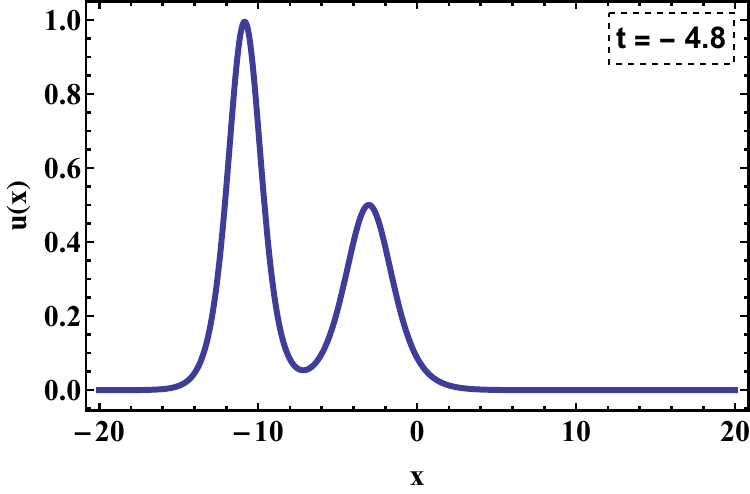}
      \end{minipage}%
      \begin{minipage}[t]{0.331\textwidth}
        \includegraphics[width=\linewidth]{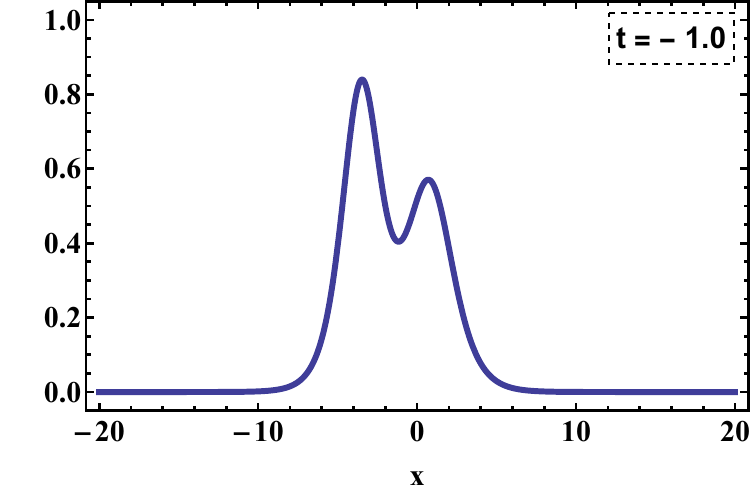}
      \end{minipage}%
      \begin{minipage}[t]{0.333\textwidth}
        \includegraphics[width=\linewidth]{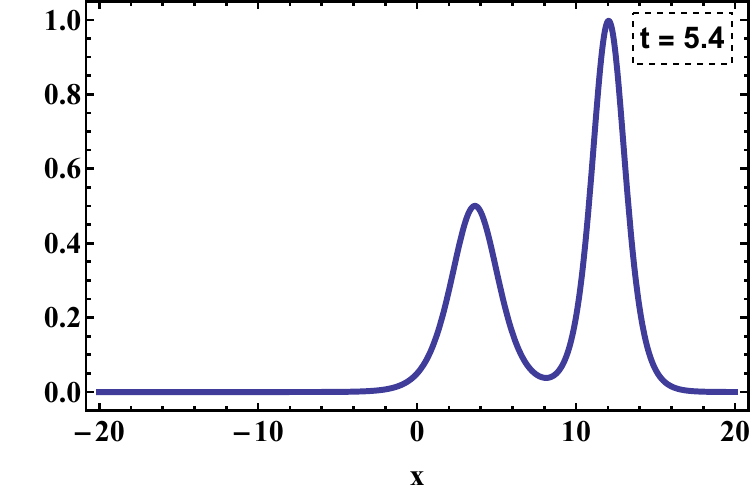}
      \end{minipage}
      
      \caption{Dynamics of a two-soliton solution of the KdV equation.}
      \label{f:KdV}
    \end{figure}

    The explanation for this surprising behaviour was developed by Kruskal and Zabusky \cite{Kruskal:1965}. They first showed that, given a solution $x_i(t)$ of the FPU lattice, one could construct a continuous function $u(x,t)$ that satisfies $u(ih,t) = x_i(t)$. For small lattice spacing $h$ and non-linearity parameter $\beta$, there would be solutions $x_i(t)$ where the corresponding $u(x,t)$ would describe a travelling wave with slowly varying shape, and those solutions would be intimately related to a KdV equation. Moreover, they discovered that the solitary wave solutions of the KdV equation interact elastically, recovering their individual velocities and shapes after the interaction (see \autoref{f:KdV} for an illustrative example). 
    During the collision, the only possible distinction after the interaction was a time delay. In the words of Kruskal and Zabusky \cite{Palais:1997}: "\textit{Here we have a nonlinear physical process in which interacting localised pulses do not scatter irreversibly.}" These features motivated Kruskal and Zabusky to coin the name "solitary wave pulse" or "soliton", to emphasise the particle-like behaviour of these solutions.
    \begin{definition}[Soliton: Drazin and Johnson \cite{Drazin:1996}]
        A soliton is a solitary wave that can interact with other solitary waves without losing its identity.
    \end{definition}

    To summarise, solitons are particle-like solutions that exhibit a non-linear superposition. It is worthwhile to remark that solitons do not only emerge from the KdV equation. They also appear in the sine-Gordon or the non-linear Schrödinger equation, which are among the most fundamental equations supporting these entities due to their multiple applications (see \cite{Cuevas:2014} and \cite{Ablowitz:2004} and references quoted therein). We will study the sine-Gordon model in more detail in the next chapter.

    \section{Solitons in field theories}\label{s:SolitonsField}

     Despite the relevance of the non-linear superposition explained before to characterise solitons, it is a highly restrictive property. The only way to recover the individual character of each soliton demands an infinite number of conserved quantities. This infinite number of conservation laws is intimately related to the integrability of the model. The rich mathematical structure of the integrable theories has provided a wide range of analytical tools, such as the inverse scattering transformation \cite{Miura:1967}, Lax pairs \cite{Lax:1968}, and Bäcklund transformations \cite{Backlund:1880}. Nevertheless, the majority of the systems that will be considered in this thesis are non-integrable. Therefore, following the standard approach in the literature, we will adopt a relaxed definition of soliton in order to include non-integrable cases, and we shall reserve the term \textit{integrable soliton} for those objects that exhibit non-linear superposition.
     
     Moreover, in the previous section, we introduced the concept of solitary wave and soliton only attending to the profile of the solution itself. Nevertheless, this thesis is framed within the context of field theories, where fields possess an associated energy density bounded from below that we can assume positive semi-definite $\epsilon(x,t) \geq 0$. In conclusion, in order to include non-integrable theories and the availability of an energy density, we introduce an alternative formulation for the notion of a soliton, which will be used throughout the present text.

    \begin{definition}[Soliton]\label{d:soliton}
        A soliton is a non-perturbative solution of a non-linear field theory that exhibits particle-like behaviour and whose energy density $\epsilon(\vec{x},t)$ is localised, that is, \[\max_{\vec{x} \in \mathbb{R}^n} \{ \epsilon(\vec{x},t) \} < \infty\,, \quad \quad \lim_{||\vec{x}|| \to \infty} \epsilon(\vec{x},t) = 0\,.\]
    \end{definition}
    This type of configuration has also been referred to by other authors as a \textit{lump} \cite{Coleman:1977,Coleman:2010}. Within this definition of soliton, one can distinguish between \textit{stable} and \textit{unstable solitons}.
    \begin{definition}[Stable and unstable soliton]\label{d:StableSoliton}
    A stable soliton corresponds to a solitonic configuration whose energy density $\epsilon(\vec{x},t)$ does not disperse. In other words, \[\lim_{t \to \infty} \max_{\vec{x} \in \mathbb{R}^n} \{ \epsilon(\vec{x},t) \} \neq 0\,.\] Otherwise, the configuration is referred to as an unstable soliton.
    \end{definition}

    When it is said that solitons are non-perturbative, it means that they cannot arise from small corrections to solutions of the corresponding linearised equations of motion and then treating the non-linear terms in a perturbative expansion. Conversely, they require the whole non-linearity to be conceived. This is particularly interesting because solitons, by capturing non-perturbative aspects of the field theory, can play a key role in processes beyond the reach of perturbative methods, thereby contributing to a more complete understanding of different phenomena. Therefore, solitons are of significant interest within the framework of Quantum Field Theory \cite{Rajamaran:1987}. However, we emphasise that this thesis does not address the correspondence between classical soliton solutions and their quantum counterparts; all discussions throughout the manuscript are restricted to the classical regime.
     
\section{Objectives of the thesis}\label{s:Objetives}

    Despite the theoretical and experimental importance of solitons, their dynamical properties are far from being fully understood. This complexity arises partly from the infinite number of degrees of freedom inherent to field theories, and partly from the non-integrability and non-linearity of the models in which these solitons emerge. In particular, the particle-like behaviour of solitons allows for mutual interaction, interaction with external fields, or with radiation. Moreover, solitons exhibit collective excitations that can be understood as internal degrees of freedom, which can store and exchange energy, thereby altering non-trivially their expected dynamics. 

    This thesis aims to further investigate the role of internal modes in soliton dynamics. These collective excitations can be triggered either during the formation of solitons or through the interaction of solitons with other entities. Therefore, gaining a deeper understanding of the effects of the internal modes may yield valuable insights into the physical/experimental realisation of solitons. In general, the complexity of the underlying theories increases significantly with spatial dimensionality. Hence, we will restrict our analysis to models in one and two spatial dimensions, with the expectation that the insights gained may eventually extend to higher dimensions.   

    Specifically, this thesis aims to address the following topics:
    \begin{itemize}
        \item To describe effectively radiation in field theories with solitons. The inclusion of radiation in an effective theory is essential to include dissipative degrees of freedom, that could account for the de-excitation of massive modes storing energy or the annihilation in scattering processes. Moreover, this approach could describe the influence of radiation on the internal degrees of freedom. Specifically, the inclusion of genuine radiation modes will be sought in the prototypical one-dimensional $\phi^4$ model.     
        \item To generalise the moduli space describing a moving Abelian-Higgs vortex by adding its single massive bound mode. This extension has been made before by other authors in the context of kinks. In that scenario, it has been unveiled that the massive bound mode is able to capture approximately the Lorentz contraction of a moving kink. It will be proved whether that is the case for Abelian-Higgs vortices. Specifically, a generalised effective model for a single moving vortex could provide the suitable initial conditions for studying the scattering of multiple vortices. 
        \item To understand the influence of the massive modes on sphaleron dynamics. These unstable solitons are believed to have played a role in processes that contributed to baryogenesis in the early universe. Despite their importance, the inherent instability of these objects means that their dynamical properties are still poorly understood. One-dimensional field theories will be sought that can facilitate the analysis of the sphaleron dynamics, drawing special attention to the role of the internal modes during the decay of these solutions. Moreover, stabilisation mechanisms that can enhance the life span of these solutions will be examined. 
    \end{itemize}

\section{Structure of the thesis and methodology employed}\label{s:Structure_Methodology}

    The present thesis is written in order to fulfil the requirements needed to obtain the degree of Doctor of Philosophy in Physics with the International Mention from the University of Va\-lla\-do\-lid. Specifically, this thesis is presented as a compendium of five peer-reviewed publications in high-impact scientific journals.    
    This thesis is intended to be self-contained. In order to do that, an extensive description of the main results that constitute the foundations of the soliton theory is provided, as well as the basic ingredients that will be used throughout the contributions achieved during the different research works. A mathematically sound discussion is combined with a physical description of the diverse concepts exposed. The aim is to provide an accessible and engaging thesis for readers with a background in physics or mathematics. However, certain technical mathematical results will only be sketched when a complete understanding is far from the scope of this thesis. 

    The main body of the thesis is divided into two parts. The first offers a comprehensive introduction to the fundamental concepts related to topological solitons, along with the contributions made during the PhD within this framework. The second part focuses on the contextualisation of the basic concepts surrounding sphalerons and, once again, on the research articles related to them. The reason behind this partition is that topological solitons are stable, whereas sphalerons are not. Therefore, the details differ and the objectives are not the same. Following that reasoning, it has been considered that they deserve their own chapters. A more detailed description of this organisation is provided below.

    \vspace{0.7cm}

    \noindent \noindent \underline{Part I: On the Role of Internal Modes in the Dynamics of Topological Solitons.} 

    In \autoref{c:Topological}, the fundamental ideas regarding the Lagrangian formalism of field theory are presented. Within that framework, it is explained how the topological solitons emerge as a consequence of topological and energetic considerations. However, although the topological structure of the theory prevents topological solitons from dissipating, it does not guarantee that the configurations are time-independent. To ensure the stability of the topological solitons, the Bogomolny arrangement and the study of the linear spectrum of perturbations are explained. Moreover, a criterion to rule out the existence of non-trivial, static, finite energy solutions in field theories is provided. Finally, some prototypical examples of topological solitons are given in one and two spatial dimensions, and a semi-analytical scheme to extract insights into the soliton dynamics is introduced.

    In \autoref{c:Radiation}, an effective theory is constructed within the frame of the collective coordinates approximation to describe the phenomena related to radiation in soliton dynamics. Specifically, the $\phi^4$ model in one spatial dimension is considered. The relevance of this novel effective theory is that, although an effective description of radiation in theories with solitons has already been performed in earlier works by other authors, this is the first time that genuine radiation modes are considered. In this chapter, it is shown that this approach is able to analytically reproduce numerous phenomena well established in the literature, such as the decay law of the shape mode held by the $\phi^4$ kink or its resonant excitation by monochromatic radiation waves of frequency twice the frequency of the shape mode. Moreover, it is unveiled that the construction of effective theories that partially reproduce relativistic invariance requires not only the inclusion of the shape mode as suggested in previous studies, but also of radiation modes, because they completely capture the first-order correction to the Lorentz contraction. Likewise, the decay of unstable oscillons is described within this framework. Finally, through a complementary model, the degrees of freedom responsible for the amplitude modulation in the dynamics of excited oscillons are also captured.
    
    In \autoref{c:Vortex}, a generalisation of Samols's moduli space for a local vortex in the Abelian-Higgs model in $2+1$ dimensions is developed by incorporating its single positive bound mode. The analysis shows that including this internal mode not only induces an effective potential, but also modifies Samols's metric through a non-trivial coupling between the shape mode amplitude and the kinetic degrees of freedom. Interestingly, a solution of the model describes a local vortex moving at constant velocity with its internal mode excited at a non-zero, constant amplitude, which mimics the Lorentz invariance of the field theory. However, the radial symmetry of the shape mode is insufficient to capture Lorentz contraction. To address this, an alternative effective model is constructed based on a Derrick mode associated with spatial rescaling along the direction of motion. While the non-radial symmetry of the Derrick mode better captures the Lorentz contraction, it fails to reproduce the shape mode's frequency. To overcome the limitations of both effective models, a new model is introduced incorporating two orthogonal Derrick modes. As will be shown, this extended model successfully captures both the frequency of the internal bound mode and the spatial asymmetry characteristic of a Lorentz boost. This last model provides optimal initial conditions for studying vortex–vortex and vortex–antivortex collisions in the presence of internal excitations when an effective description is used.

    \vspace{0.7cm}

    \noindent \noindent \underline{Part II: On the Role of Internal Modes in the Dynamics of Sphalerons.}

    In \autoref{c:Sphalerons}, the basic ideas surrounding sphalerons are explained. Specifically, it is shown that these particle-like objects can also have a relationship with the non-trivial topological structure of the configuration space. Contrary to topological solitons, they represent unstable solutions of the equations of motion. Despite this property, the potential physical relevance that these solutions can have in particle physics and in the early universe is discussed. Concretely, it is shown that there exists a sphaleron configuration in the bosonic sector of the electroweak theory. Finally, sphaleron solutions in one-dimensional models are discussed in detail.
    
    In \autoref{c:BPS_Sph}, a spatial impurity is introduced in a scalar field theory with a false vacuum resembling the $\phi^4$ model. Remarkably, it is demonstrated, for the first time, the existence of a new class of solutions that we have called semi-BPS sphalerons, characterised by the absence of static interaction with the impurity. This scenario provides an ideal framework to study the dynamical properties of sphalerons, since the lack of static interactions with the impurity allows the isolation of phenomena due to internal modes. In particular, it is unveiled that many of the typical properties of BPS solitons are partially preserved in these unstable configurations, such as geodesic flow in moduli space when only the translational mode is excited, or the appearance of the spectral wall phenomenon.
    
    In \autoref{c:Sph_False}, two one-dimensional deformations of the $\phi^6$ model are constructed, both characterised by the presence of at least one false vacuum. The fundamental difference between them lies in the spectral structure of the sphaleron configurations: while in one of the models the sphaleron lacks positive bound modes, in the other it possesses a variable, though never zero, number of such modes. The numerical analysis shows that, after the initial collapse, both sphalerons evolve into an excited oscillon, provided that the initial sphaleron has a moderate size. However, when the initial size is sufficiently large, differences emerge in the evolution of the sphaleron with positive bound modes. In particular, the sphaleron with positive internal bound modes exhibits a sequence of bounces after the collapse, due to an energy transfer mechanism between the modes. Furthermore, for sufficiently large amplitudes of the unstable mode, it is observed that the sphaleron does not collapse, but expands, suggesting the existence of a critical amplitude for which the sphaleron is stabilised. Through a perturbative development, a general expression for this critical amplitude is obtained, which shows excellent agreement with numerical results.
    
    In \autoref{c:Sph_Circle}, the decay of sphaleron configurations appearing in the sine-Gordon, $\phi^4$, and $\phi^6$ models defined on the circle is analysed. In the case of the sine-Gordon model, due to its integrable character, the decay of the sphaleron gives rise to a periodic configuration interpolating between a sphaleron and an antisphaleron. It is shown that this evolution can be described by an explicit analytical expression. By contrast, the non-integrable $\phi^4$ and $\phi^6$ models show different dynamics, influenced by the presence or absence of reflection symmetry. In the $\phi^4$ model, which possesses such symmetry, the sphaleron decays exclusively into an oscillon, and the two decay directions are equivalent. In contrast, in the $\phi^6$ model, where no reflection symmetry exists, the two decay directions are not equivalent. Depending on the decay channel, one can observe either collapse into an oscillon, or the formation of bounces due to the presence of positive internal bound modes. Moreover, in both models, the possibility of dynamic stabilisation of the sphaleron has been observed. Although in both cases the mechanism requires excitation of positive internal modes, the underlying mechanisms are not completely analogous, and the differences can be attributed to the presence or absence of reflection symmetry.

    \vspace{0.7cm}

    I would like to emphasise that the chapters mentioned previously are adapted versions of the published articles and include additional comments and further explanations. Moreover, recent references have been included to reflect developments in the field that have emerged since the original works. The purpose of these modifications is to provide a clearer connection between the chapters, to facilitate the understanding of some of the reasoning and computations presented in the articles, and to reflect a more up-to-date state of the art. To fulfil the requirements set by the University of Valladolid for submitting a thesis by compendium of publications, I have included the original articles in Appendices A–E.

    Finally, in \nameref{c:Conclusions}, the conclusions extracted during the thesis are summarised and various future research lines are proposed.

    \vspace{0.7cm}
    
    Regarding the methodology employed, the non-integrability and the non-linearity of the field theories with solitons generally complicate the extraction of analytical results. As a consequence, this thesis partly relies on numerical simulations conducted using C++ and Pyhton softwares. In particular, C++ has been primarily used to perform time evolutions and extract the physical quantities of the system, while Python has been employed to study the spectral problems numerically by using its optimised packages. Nevertheless, this thesis also combines approximate analytical techniques to reproduce and explain the phenomena observed numerically. More precisely, perturbative techniques have been employed to compute corrections around a background configuration. Moreover, the collective coordinate method has been used to construct effective theories based on the main degrees of freedom involved in each individual phenomenon. Finally, the graphical illustrations have been done within the Wolfram Language using the program Mathematica as well as the library Matplotlib in Python.

        \part{On the Role of Internal Modes in the Dynamics of Topological Solitons}

        \chapter{Introduction to topological solitons}\label{c:Topological}

    \epigraph{"There is no branch of mathematics, no matter how abstract, which may not someday be applied to phenomena of the real world."}{-- Nikolai Lobachevsky}

    In \autoref{c:Introduction}, we presented the concept of soliton as a non-perturbative, localised solution of a classical non-linear field theory. There, we anticipated that solitons might be classified into two main families: \textit{stable} and \textit{unstable solitons}. Among the stable ones, a further distinction can be made based on the underlying mechanism that ensures their stability. In this context, we identify two important categories: \textit{topological} and \textit{non-topological solitons}. 
    
    This chapter is devoted to the introduction and study of topological solitons. We begin in \mbox{\autoref{s:Lagrangian}} with a brief overview of the Lagrangian formulation of field theory. In \autoref{s:Topology}, we clarify the conditions under which topological solitons can arise. As will become evident, their existence is deeply connected to both topological and energetic considerations. However, while these conditions are necessary, they are not sufficient to ensure stability of higher-charge topological solitons. A deeper examination of the criteria for the existence of stable topological solitons is conducted in \autoref{s:Modes}. Remarkably, in certain field theories, the field equations can be reduced to first-order differential equations, significantly simplifying the search for topological solitons, and ensuring the existence of global minimisers of the energy distinct from the vacuum states. This possibility is discussed in \autoref{s:Bogomolny}. Moreover, an important theorem ruling out the presence of stable, time-independent solitons with finite energy in a given model is introduced in \autoref{s:Derrick}. Prototypical examples of topological solitons in one and two spatial dimensions are discussed in \autoref{s:1D} and \autoref{s:2D}, respectively. To conclude this chapter, a semi-analytical approach that allows to obtain insights into the dynamics of solitons is provided in \autoref{s:CCM}.
    
\section{Lagrangian field theory}\label{s:Lagrangian}

    Lagrangian field theory concerns the dynamics of one or more fields defined throughout space and evolving in time. We are specially interested in theories that exhibit Poincar\'e invariance,  as they are suitable for describing relativistic phenomena. Consequently, the base space is taken to be the $(d+1)$-dimensional Minkowski spacetime. Time and space coordinates will be denoted by $x_0 = t$ and $\vec{x} = (x_1,x_2,\ldots,x_d)$, respectively, and henceforth we adopt the metric signature $\eta_{\mu\nu} = \text{diag}(+,-,\ldots,-)$.
    
    The simplest examples of field theories are the real scalar field theories. In these theories, the main ingredient is a map $\vec{\phi}: \mathbbm{R}^{1+d} \rightarrow \mathbbm{R}^{m}$ consisting of $m \in \mathbbm{N}$ scalar fields that take values on the real line. The dynamics is determined by a functional named \textit{action}
    \begin{equation}\label{e:action}
        S[\vec{\phi}] =  \int_{\mathbbm{R}^{1+d}} \mathcal{L}[\vec{\phi}, \partial_{\mu}\vec{\phi}]\,dt\,dx_1\,\ldots\,dx_d\,,
    \end{equation}
    where $\mathcal{L}$ is the Lagrangian density defining the theory.

    Hamilton's principle states that the dynamics of a physical system is determined by a variational procedure. Hence, assuming a variation $\delta\phi^a(t,\vec{x})$ of the components of the scalar field $\vec{\phi}(t,\vec{x})$ and imposing that the variation of the action (\ref{e:action}) at first order vanishes $\delta S = 0$, we are left with the so-called  \textit{Euler-Lagrange field equations}  
    \begin{equation}\label{e:EL}
    \partial_{\mu}\dfrac{\partial \mathcal{L}}{\partial(\partial_{\mu}\phi^a)} - \dfrac{\partial \mathcal{L}}{\partial \phi^a} = 0\,,\qquad a = 1,2,\ldots,m\,,
    \end{equation}
    where Einstein's summation convention is assumed. Due to the Poincar\'e invariance of the theory, it possesses a set of conserved charges by virtue of Noether's theorem.
    \begin{theorem}[Noether's theorem \cite{Noether:1918}]\label{t:Noether} Let $\phi^a \rightarrow \phi^a + \epsilon\,\delta\phi^a$ be a continuous symmetry of a Lagrangian density, that is, a continuous transformation that changes the Lagrangian density by a divergence
    \begin{equation}
        \mathcal{L} \rightarrow \mathcal{L} + \epsilon\,\partial_{\mu}K^{\mu}\,.
    \end{equation}
    To this continuous symmetry corresponds a conserved current $J^{\mu} = (J^0,\vec{J})$ such that
    \begin{equation}
        \partial_{\mu}J^{\mu} = 0\,,  \quad \text{with} \quad J^{\mu} = \dfrac{\partial\mathcal{L}}{\partial(\partial_{\mu}\phi^a)}\delta\phi^a - K^{\mu}\,,
    \end{equation}
    and a conserved charge given by
    \begin{equation}
        Q = \int_{\mathbbm{R}^d}J^0\, \,dx_1\ldots dx_d\,.    
    \end{equation}
    \end{theorem}
    We proceed to state the symmetries due to the Poincar\'e invariance of the theory and the corresponding conserved charges:
    \begin{enumerate}
        \item \textbf{Space-time translations:} $x^{\mu} \rightarrow x^{\mu} + \epsilon^{\mu}$. \\
        The conserved current associated to this symmetry is the energy-momentum tensor
        \begin{equation}\label{e:EM-tensor}
            T^{\mu\nu} = \dfrac{\partial\mathcal{L}}{\partial(\partial_{\mu}\phi^a)}\partial^{\nu}\phi^a - \eta^{\mu\nu}\, \mathcal{L}\,,    
        \end{equation}
        and the corresponding conserved charges are the energy $E$ and the momentum $\vec{P}$
        \begin{equation}\label{e:Energy_Momentum}
            E = \int_{\mathbbm{R}^{d}} T^{00}\,dx_1\,\ldots\,dx_d\,, \qquad P^i = \int_{\mathbbm{R}^{d}} T^{0i}\,dx_1\,\ldots\,dx_d\,.     
        \end{equation}
        \item \textbf{Lorentz transformations:} $x^{\mu} \rightarrow \Lambda^{\mu}_{\phantom{\mu}\nu} x^{\nu}$. \\
        The conserved current associated to this symmetry is the angular momentum tensor
        \begin{equation}\label{e:Angular_Tot}
         M^{\mu\nu\rho} = x^{\mu}T^{\nu\rho} - x^{\nu}T^{\mu\rho}\,,
        \end{equation}
        and the conserved charges are the total angular momentum and the boosts 
        \begin{equation}\label{e:Angular}
            M^{0i} =  \int_{\mathbbm{R}^{d}}\left(x^0T^{0i} - x^{i}T^{00}\right)\,dx_1\,\ldots\,dx_d\,, \qquad M^{ij} =  \int_{\mathbbm{R}^{d}}\left(x^iT^{0j} - x^{j}T^{0i}\right)\,dx_1\,\ldots\,dx_d\,.
        \end{equation}
    \end{enumerate}
    Apart from the previously discussed symmetries, the theory may possess additional symmetries that do not act on the spacetime itself. Instead, these transformations act on the internal space, that is, directly on the fields. The associated symmetry groups may be discrete or continuous. The latter case is particularly relevant for the formulation of so-called \textit{gauge theories} when the transformation is local.  

    It follows from the preceding analysis that invariance under time translations implies the conservation of the total energy. For a Lagrangian density in the standard form
    \begin{equation}\label{e:LagrangianDensity}
    \mathcal{L} = \dfrac{1}{2}\partial_{\mu}\vec{\phi}\cdot\partial^{\mu}\vec{\phi} - U(\phi^1,\phi^2,\ldots,\phi^m)\,,
    \end{equation}
    with $U(\phi^1,\phi^2,\ldots,\phi^m)$ the potential function,
    the energy functional takes the form
    \begin{equation}\label{e:Energy}
    E[\hspace{0.02cm} \vec{\phi} \hspace{0.05cm}] = \int_{\mathbbm{R}^{d}} \left( \dfrac{1}{2}\partial_{0}\vec{\phi}\cdot\partial_{0}\vec{\phi} +  \dfrac{1}{2}\nabla \vec{\phi} \cdot \nabla \vec{\phi} + U (\phi^1,\phi^2,\ldots,\phi^m) \right) \,dx_1\,\ldots\,dx_d\,.
    \end{equation}
    Here, the symbol "$\cdot$" denotes the scalar product in the internal space $\mathbbm{R}^m$, that is, $\delta_{ij} = (+,+ ,\ldots,+)$. The energy functional (\ref{e:Energy}) can be split into a kinetic and potential term as follows
    \begin{align}\label{e:KE_PE}
    T &= \dfrac{1}{2}\int_{\mathbbm{R}^{d}} \left[\partial_{0}\vec{\phi}\cdot\partial_{0}\vec{\phi}\right]\,dx_1\,\ldots\,dx_d\,, \\
    V &= \int_{\mathbbm{R}^{d}} \left[\dfrac{1}{2} \nabla \vec{\phi} \cdot \nabla \vec{\phi} + U(\phi^1,\phi^2,\ldots,\phi^m) \right] \,dx_1\,\ldots\,dx_d\, .
    \end{align}
    In this thesis, our focus is on field configurations with finite energy. Without loss of generality, we can assume that $U(\phi^1,\phi^2,\ldots,\phi^m) \geq 0$. A natural set of configurations with finite energy are the vacua of the theory.
    \begin{definition}
        The vacuum manifold $\mathcal{V}$ is defined as the set of constant field configurations that satisfy
        \begin{equation}
            \mathcal{V} := \{\vec{\phi} \in \text{Maps}(\mathbbm{R}^{d},\mathbbm{R}^{m}): \phi^a = const.,\ U[(\phi^1,\phi^2,\ldots,\phi^m)] = 0 \}.
        \end{equation}
    \end{definition}
    More generally, a quick inspection of the energy functional (\ref{e:Energy}) unveils that, for a field configuration to have finite energy, it must fulfil 
    \begin{equation}\label{e:BC_time}
        \lim_{||\vec{x}|| \rightarrow \infty} \dfrac{\partial \phi^a}{\partial t} = 0\,,
    \end{equation}
    as well as
    \begin{equation}\label{e:BC}
        \lim_{||\vec{x}|| \rightarrow \infty} \vec{\phi} \in \mathcal{V}\,, \quad  \lim_{||\vec{x}|| \rightarrow \infty} \dfrac{\partial \phi^a}{\partial x_j} = 0\,.
    \end{equation}
    \begin{definition}
        The configuration space $\mathcal{C}$ is defined as the set of static field configurations with finite energy
        \begin{equation}
            \mathcal{C} := \{\vec{\phi}(\vec{x}) \in \text{Maps}(\mathbbm{R}^{d},\mathbbm{R}^{m}):  \ V[(\phi^1,\phi^2,\ldots,\phi^m)] < \infty\}\,.
        \end{equation}
    \end{definition}
    Importantly, the field configuration must converge sufficiently fast to the vacuum field value to ensure energy finiteness. Otherwise, the energy may diverge, as is the case for the so-called \textit{global vortices} and \textit{global monopoles} \cite{Manton:2002}.

    It is worth mentioning that, if the field $\vec{\phi}$ takes values in a compact manifold $Y$, that is, $\vec{\phi}\nobreak:\nobreak \mathbbm{R}^{1+d} \nobreak\rightarrow\nobreak Y$, the requirement of finite energy changes, and a potential function is not completely necessary in order to have non-trivial solutions. In this situation, the field should tend to the same constant value at infinity independently of the direction for the energy to be finite
    \begin{equation}\label{e:BC_nonlinear}
        \lim_{||\vec{x}|| \rightarrow \infty}\vec{\phi} = \vec{y}_0\,.
    \end{equation}
    As a result, we can identify the spatial infinity with a single point and compactify the base space $\mathbbm{R}^d\cup\{\infty\} \cong \mathbbm{S}^d$. These types of field theories are referred to as \textit{non-linear $\sigma$ models} or \textit{Skyrme models}, depending on the specific structure of the theory. Sigma models are defined by a Lagrangian density of the form
    \begin{equation}
        \mathcal{L} = \dfrac{1}{2} g_{ij}(\vec{\phi})\partial_{\mu} \phi^i\partial^{\mu} \phi^j\,,
    \end{equation}
    where $g_{ij}(\vec{\phi})$ denotes the components of the metric tensor on the internal space manifold $Y$. Therefore, the energy depends quadratically on the gradient of the field. Skyrme models, on the other hand, include additional terms of fourth order in derivatives in the Lagrangian density \cite{Skyrme:1961}. In fact, the Skyrme term represents the unique Lorentz-invariant expression of degree four in derivatives that leads to field equations which remain second order in time derivatives. This is crucial, as avoiding higher-order time derivatives prevents the emergence of Ostrogradsky instabilities, which are associated with Hamiltonians unbounded from below. Despite the relevance of these theories, they will not be addressed in this thesis.

    In the discussion that follows, we will refer to theories in which the fields are maps of the form $\vec{\phi}: \mathbb{R}^{1+d} \to \mathbb{R}^m$ as linear field theories, and those where $\vec{\phi}: \mathbb{R}^{1+d} \to Y$, with $Y$ a compact manifold, as non-linear field theories. In this context, the terms "linear" and "non-linear" classify the Lagrangian density according to whether the derivatives of the fields appear linearly or non-linearly in the equations of motion.
    
\section{Topological protection}\label{s:Topology}

    As explained in \autoref{s:SolitonsField}, we are mainly interested in the emergence of solitons in field theories. The requirement of Lorentz invariance and the high dimensionality of a theory, makes integrable field theories extremely rare. While solitons can arise in integrable systems as \textit{integrable solitons}, these are not the only scenarios in which stable bound states may exist. This consideration is precisely what motivated us to relax the definition of a soliton in \autoref{d:soliton} to include these cases. In this section, we explore how the topological structure of a field theory can give rise to configurations of this nature. 

\begin{figure}[htb]
        \centering{\includegraphics[width=0.7\linewidth]{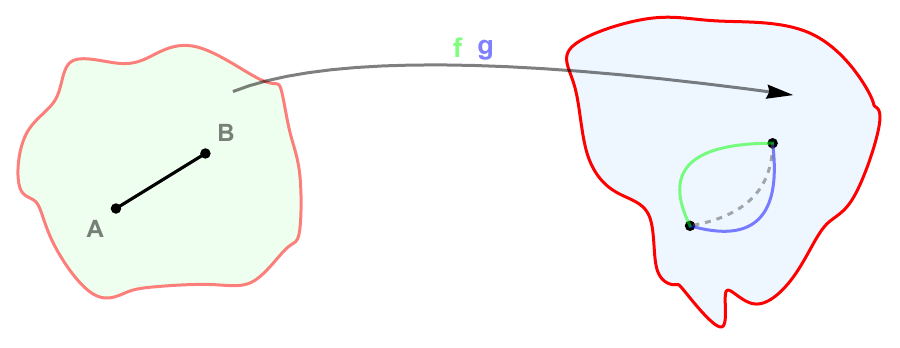}
        }
      
        \caption{The scheme illustrates two homotopic functions, $f$ and $g$, with $f(A) = g(A)$ and $f(B) = g(B)$, which map the segment $AB$ onto the green upper curve and the blue lower curve, respectively. The dashed curve represents the action of the map $F:X \times[0,1] \rightarrow Y$ for an intermediate value $c \in [0,1]$.}
        \label{f:Homotopy}
    \end{figure}

    To proceed, we first introduce the notion of homotopy, one of the central concepts in topology.
    \begin{definition}[Homotopy]
        Let $X$ and $Y$ be two topological spaces, and let $f$ and $g$ be two continuous functions from $X$ to $Y$. The functions $f$ and $g$ are said to be homotopic if and only if there exists a continuous function $F: X \times [0,1] \rightarrow Y$ such that $F(x,0) = f(x)$ and $F(x,1) = g(x)$ for all $x \in X$.
    \end{definition}
    An intuitive idea of two homotopic functions is depicted in \autoref{f:Homotopy}. 
    
    The homotopy relation defines an equivalence relation on the set of all continuous functions from $X$ to $Y$, and the equivalence classes are called \textit{homotopy classes}. Note that solutions of the Euler–Lagrange equations (\ref{e:EL}) typically belong to the class $\mathcal{C}^2(\mathbbm{R}^{1+d})$, and thus serve as examples of continuous functions. Therefore, this topological classification can be applied to fields in the configuration space.

    To determine the topological character of a field configuration $\vec{\phi}(t,\vec{x})$ in our context, it suffices to consider the homotopy class of its asymptotic behaviour $\vec{\phi}^{\infty}(\vec{x})$ \cite{Manton:2002}. This is justified because, as previously discussed, a static field configuration must satisfy specific boundary conditions to ensure finite energy and be part of the configuration space. Namely, they must fulfil the boundary conditions given by (\ref{e:BC}) or by (\ref{e:BC_nonlinear}) depending on whether the field theory is linear or non-linear, respectively, in the sense defined earlier. Accordingly, we must focus on the values taken by the field at $\partial\mathbbm{R}^d = \mathbbm{S}^{d-1}$ in the former case, or at $\mathbbm{S}^{d}$ in the latter. Therefore, the following definition is relevant in light of the previous discussion. 
    \begin{definition}[$n$-th homotopy group]
        The set of homotopy classes of the maps $f: \mathbbm{S}^{n} \rightarrow Y$ is denoted by $\pi_n(Y)$. For $n\geq 1$, the set $\pi_n(Y)$ forms a group under the composition of maps, and is referred to as the $n$-th homotopy group of $Y$.
    \end{definition}
    As a consequence, the topological classification of the field configurations is given by the homotopy class $\pi_{d-1}(\mathcal{V})$ in the case of a linear theory, or of $\pi_{d}(Y)$ in the case of a non-linear theory. Through this topological classification, we can realise the configuration space as a disjoint union of connected components
    \begin{equation}
        \mathcal{C} = \displaystyle\bigcup_{i} \mathcal{C}_i\,.
    \end{equation}
    Each of these connected components is usually referred to as \textit{topological sectors}, and the number of connected components is given by the number of elements of $\pi_0(\mathcal{C})$. In order to have multiple connected components, the relevant homotopy groups must be non-trivial. For example, in order for the vacuum manifold to possess a non-trivial homotopy group, it is necessary for the potential function to be non-linear \cite{Manton:2002}. This non-linearity allows for a degenerate set of vacuum states, although this does not guarantee a non-trivial homotopy group. 
    
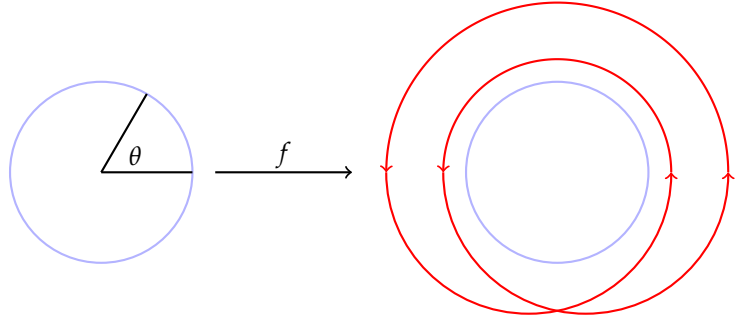
\begin{figure}[htb]
        \hspace{1.6cm}
    \begin{tikzpicture}[scale=1.5]
        \def\Rmax{1.5}
        \def\Rmin{1.0}
        \hspace{1.6cm}
        \draw[blue!30, thick] (-4,0) circle (\Rmin - 0.2);
        \draw[black, thick, ->] (-3,0) -- (-1.8,0);
        \draw[black, thick] (-4,0) -- (-3.2,0);
        \draw[black, thick] (-4,0) -- ({-4 + 0.8*cos(60)},{ 0.8*sin(60)});
        \draw[blue!30, thick] (0,0) circle (\Rmin - 0.2);
    
        \node[] at (-2.4,0.15) {$f$};
        \node[] at (-3.7,0.15) {$\theta$};
    
        \draw[red, thick, domain=0:180, samples=100, ->]
        plot ({\Rmax*cos(\x)}, {\Rmax*sin(\x)});
    
        \path coordinate (A) at ({\Rmax*cos(180)}, {\Rmax*sin(180)});
        \path coordinate (B) at ({\Rmin*cos(360)}, {\Rmin*sin(360)});
        \draw[red, thick, ->] 
        (A) arc[start angle=180, end angle=360, radius={( \Rmax + \Rmin )/2}];
    
        \draw[red, thick, domain=360:540, samples=100, ->]
        plot ({\Rmin*cos(\x)}, {\Rmin*sin(\x)});
    
        \draw[red, thick, ->] 
        ({\Rmin*cos(180)}, {\Rmin*sin(180)}) 
        arc[start angle=180, end angle=360, radius={( \Rmax + \Rmin )/2}];
    
        \end{tikzpicture}
        \caption{Example of a function $f$ mapping between two one-spheres $\mathbbm{S}^1$. As the domain variable ranges over $\theta \in [0,2\pi)$ once, the image of $f$ winds twice around the target one-sphere. In this case, deg\,$f$ = 2.}
        \label{f:Winding}
    \end{figure}

    Nevertheless, determining the homotopy groups of a topological space is in general a complex task. Hence, it is in certain circumstances easier to compute the so-called \textit{topological degree} of a map.  
    \begin{definition}[Topological degree \cite{Manton:2002}]
        Let $f: X \rightarrow Y$ be a differentiable map between two closed manifolds of the same dimension. Suppose that $X$ is connected. Then, the topological degree of the map is given by
        \begin{equation}
            \text{deg}\,f = \int_X\,f^*(\Omega)\,,
        \end{equation}
        where $\Omega$ is the normalised volume form and the symbol "$*$" denotes the pull-back.
    \end{definition}
    This topological degree, sometimes referred to as \textit{topological charge}, is a conserved integer that serves to label the equivalence classes. Intuitively, this integer characterises the number of times the domain manifold wraps around the target manifold under the action of the map (see \autoref{f:Winding}).

    While this definition may seem only applicable for non-linear theories, it is also useful in the context of linear field theories: assume that the linear field theory has a continuous symmetry group $G$, and let $\vec{\phi}_v$ be an element of the vacuum manifold $\mathcal{V}$. If $\vec{\phi}_v$ is invariant under a subgroup $H \subset G$, then the vacuum manifold can be identified with the coset space $\mathcal{V} \cong G/H$. In most cases considered in this thesis, $\mathcal{V} \cong G/H \cong \mathbbm{S}^n$. Therefore, the topological character of the fields can be classified using the degree of the asymptotic map $\vec{\phi}^{\infty}: \mathbbm{S}^{d-1} \rightarrow \mathbbm{S}^n$. The prototypical example occurs when $d-1 = n$, since $\text{deg}\,\,\vec{\phi}^{\infty} = k$, with $k \in \mathbbm{Z}$.
    
    The requirement of finite energy suggests that, in each topological sector, there must exist at least one configuration that minimises the energy functional. One such possibility is the set of vacua of the theory. Usually, the set of field configurations homotopically equivalent to a vacuum configuration is referred to as the \textit{vacuum sector}. However, another possibility is the existence of energy-minimising configurations that are not homotopic to the vacuum solutions. 
    \begin{definition}[Topological soliton]
        Let $\vec{\phi} \in \mathcal{C}$ be a non-singular configuration of a field theory. Let us assume that $\vec{\phi}$ is a local minimum of the energy functional that does not belong to the vacuum sector. We refer to this energy minimiser as a topological soliton.
    \end{definition}
    In the literature, it is usual to say that topological solitons are \textit{topologically protected}. This terminology arises from the following reasoning. Let us assume that the configuration space $\mathcal{C}$ is path-connected. Then, any two field configurations would be connected by a continuous path. In particular, the configuration space $\mathcal{C}$ would be connected. As time evolution is an example of continuous transformation, we can anticipate that any field configuration would decay into a vacuum configuration \cite{Rajamaran:1987}. On the other hand, if the configuration space $\mathcal{C}$ is not connected, it is not possible to continuously change a field configuration in one component into a vacuum configuration of the vacuum sector. As a matter of fact, the finiteness of the energy also implies that the time derivative of the field must vanish at spatial infinity, as shown in (\ref{e:BC_time}). If that is not the case and the field transitioned from one sector to another, the energy of the field configuration would be infinite, thereby violating the energy conservation principle. In fact, this property allow us to introduce the so-called \textit{non-topological solitons}.
    \begin{definition}[Non-topological soliton]
        A non-topological soliton is a soliton homotopically equivalent to the vacuum configuration. In other words, they are solutions that are not topologically protected, and their possible stability is related to the conservation of other charges (e.g. Noether charges).
    \end{definition}

    This intuitive idea reveals the importance of a disconnected configuration space in order to have topological solitons. The possibility of a non-trivial homotopy group -- whether in a linear or non-linear theory -- and thus a disconnected configuration space, arises from spontaneous symmetry breaking. Physically, spontaneous symmetry breaking can be related to a phase transition from a symmetric phase to a broken phase. The arising of topological defects as a consequence of phase transitions is described by the \textit{Kibble-Zurek mechanism}. This mechanism was first introduced by Kibble in the context of high energies physics \cite{Kibble:1976,Kibble:1980}, and was further extended to condensed-matter systems by Zurek \cite{Zurek:1985,Zurek:1996}.

    Before finishing this section, we have to clarify that, although the topological protection enables us to determine the existence of non-dissipative solutions, it does not guarantee the stability of time-independent solutions for topological charges higher than one, even though the topological charge is conserved. Specifically, a non-trivial topology is a necessary but not sufficient condition for the existence of stable higher-charge states. For this purpose, we would need to verify if the time evolution of the system would lead to a stable field configuration. In the following sections, this issue will be examined via a linear stability analysis of the topological soliton and through the application of the so-called \textit{Bogomolny arrangement}.
    
\section{Linear stability}\label{s:Modes}

    We have seen that a further analysis is required to guarantee the stability of higher-charge topological solitons, as topological protection alone does not provide sufficient conditions for their stability. A common approach to assess stability is to examine the linear spectrum of perturbations around the topological soliton, searching for the presence of possible unstable modes.

    In order to do that, we have to perturb a field configuration $\tilde{\phi}^a = \phi^a + \delta\phi^a$. Substituting this new expression into the Euler-Lagrange equations (\ref{e:EL}), we obtain the following expansion up to first order in the perturbations
    \begin{equation}\label{e:Pert_EL}
            0 = \left[\partial_{\mu}\partial^{\mu}\phi^a + \dfrac{\partial U}{\partial\phi^a}\right] + \left[\delta_{ab}\partial_{\mu}\partial^{\mu} + \dfrac{\partial^2 U}{\partial\phi^a\partial\phi^b}\right]\delta\phi^b + o(\delta^2\phi)\,.
    \end{equation}
    If the perturbed field configuration $\phi^a$ is a static solution of the Euler-Lagrange equations (\ref{e:EL}), we already know that the zeroth-order term vanishes. Hence, let us analyse the first-order perturbation term. This contribution defines a linear operator
    \begin{equation}
        D := \delta_{ab}\partial_{\mu}\partial^{\mu} + \dfrac{\partial^2 U}{\partial\phi^a\partial\phi^b}\,,
    \end{equation}
    that is usually named as \textit{second-order fluctuation operator}. Assuming that the perturbation has the form of a normal mode $\delta\phi^a(t,\vec{x}) \propto e^{i\omega t}\delta\phi^a(\vec{x})$, we obtain the spectral problem for the second-order fluctuation operator
    \begin{equation}
        \mathcal{H}\,\delta\phi^a(\vec{x}) := \left[- \delta_{ab}\nabla^2 + \dfrac{\partial^2 U}{\partial\phi^a\partial\phi^b} \right]\delta\phi^b(\vec{x}) = \omega^2\,\delta\phi^a(\vec{x})\,.
    \end{equation}
    In this generally coupled Schrödinger equation, we could find modes with negative, zero and positive squared eigenfrequencies. The modes with negative eigenfrequency squared $\omega^2 < 0$ are related to instabilities: the system will never return to its original state, and an exponential growth of the perturbation will occur. On the other hand, fluctuations with positive eigenfrequency squared $\omega^2 > 0$ correspond to excitations bounded in time, and the system may return to its original state, exhibiting oscillatory behaviour. Finally, the so-called \textit{zero modes} are excitations related to the breaking of a continuous symmetry of the field theory that is no longer a symmetry of the field configuration, and identifies the existence of a family of field configurations $\vec{\phi}(\vec{x},\vec{c})$ with the same energy; for example, in Lorentz invariant theories, the zero-mode is usually related to the translational invariance. In general, each zero mode can be written as the derivative of the static profile with respect to the constants $c_i$ parametrising the family of solutions of equal energy. Indeed, if we identify the zero mode with $\delta \phi^b = \dfrac{\partial\phi^b}{\partial c_i}$, we obtain
    \begin{equation}\label{e:GeneralZeroMode}
        \left[- \delta_{ab}\nabla^2 + \dfrac{\partial^2 U}{\partial\phi^a\partial\phi^b} \right]\dfrac{\partial\phi^b}{\partial c_i} = \dfrac{\partial}{\partial c_i}\left[- \nabla^2\phi^a + \dfrac{\partial U}{\partial\phi^a} \right] = 0\,,   
    \end{equation}
    since $\phi^a$ is a static solution of the equations of motion.

    With this discussion, we conclude that the leading-order information about the stability of solitons is encoded in these linear modes. The main limitation of this approach is that the solution must be obtained either analytically or numerically. Consequently, the stability cannot be analysed without prior knowledge of the global structure of the topological soliton. 

\section{Bogomolny equations}\label{s:Bogomolny}

    In general, we have to solve the static Euler-Lagrange equations with the correct boundary conditions to search for topological solitons. This is not a simple task, specially in higher-dimensional field theories or theories with multiple fields. However, there exist special theories that possess an additional structure, where the Euler-Lagrange equations can be reduced to first-order differential equations, notably simplifying the process of finding topological solitons. 

    Originally, this possibility was found in the context of monopoles \cite{Prasad:1975}, instantons \cite{Belavin:1975}, and finally Abelian-Higgs vortices \cite{Vega:1976}. A significant breakthrough, however, was achieved through the seminal paper by Bogomolny \cite{Bogomolny:1976}. In that article, Bogomolny succeeded in rewriting the energy functional of the theories describing domain walls, vortex lines, and monopoles as the sum of a positive term and another which can be identified with a total derivative. This last term is said to be topological, because it depends solely on the asymptotic behaviour of the fields involved.
    
    As an illustrative example of this procedure, let us consider a $(1+1)$-dimensional scalar field theory given by (\ref{e:LagrangianDensity}) with a static field $\vec{\phi}: \mathbbm{R} \rightarrow \mathbbm{R}^m$. Let us assume that the potential function $U(\phi_1,\ldots,\phi_m)$ can be written in terms of a differentiable function $W: \mathbbm{R}^m \rightarrow \mathbbm{R}$ as follows
    \begin{equation}\label{e:Superpotential}
        U(\phi_1,\ldots,\phi_m) = \dfrac{1}{2}\sum_{a = 1}^m\left(\dfrac{\partial W}{\partial\phi^a} \right)^2\,.
    \end{equation}
    This differentiable function $W(\phi_1,\ldots,\phi_m)$ is usually referred to as \textit{superpotential}. The Bogomolny arrangement involves completing squares of the energy functional and bound it from below by a term that only depends on the asymptotic values taken by the field 
    \begin{align}\label{e:Bog_Arrangement}
        \displaystyle E[\,\vec{\phi}\,] &= \int_{-\infty}^{\infty} \left[\dfrac{1}{2}\sum_{a=1}^{m}\left(\dfrac{d\phi^a}{dx}\right)^2 + U(\phi_1,\ldots,\phi_m) \right]\, dx\nonumber\\
        &= \dfrac{1}{2}\int_{-\infty}^{\infty}\left[\sum_{a=1}^{m}\left(\dfrac{d\phi^a}{dx} \pm\dfrac{\partial W}{\partial\phi^a}\right)^2 \right]\, dx \mp \int_{-\infty}^{\infty}\,\dfrac{d\phi^a}{dx}\dfrac{\partial W}{\partial\phi^a}dx\nonumber\\
        &= \dfrac{1}{2}\int_{-\infty}^{\infty}\left[\sum_{a=1}^{m}\left(\dfrac{d\phi^a}{dx} \pm\dfrac{\partial W}{\partial\phi^a}\right)^2 \right]\, dx \mp \int_{-\infty}^{\infty}\,dW\,(\phi_1,\ldots,\phi_m)\nonumber\\
        &\geq \bigg |\lim_{x \rightarrow \infty} W(\phi_1,\ldots,\phi_m) - \lim_{x \rightarrow - \infty} W(\phi_1,\ldots,\phi_m) \bigg |\,.
    \end{align}
    This lower bound is usually referred to as BPS bound. Since it depends only on the asymptotic behaviour of the fields involved, it is reasonable to consider that this quantity is related to the topological charge of the map. Therefore, the minimal energy is proportional to the topological charge, with the proportionality constant being characteristic of the model under study, and identified with the mass of a one-charge topological soliton.

    The bound is saturated when the following first-order differential equations, called \textit{Bogomolny equations}, are satisfied
    \begin{equation}\label{e:Bogomolny_General}
        \dfrac{d\phi^a}{dx} \pm\dfrac{\partial W}{\partial\phi^a} = 0\,, \qquad a = 1,2, \ldots,m\,,
    \end{equation}
    which results from vanishing the squared terms in (\ref{e:Bog_Arrangement}). Due to the saturation of the bound, solutions of the Bogomolny equations (\ref{e:Bogomolny_General}) are global minima of the energy within a given topological sector and are referred to as \textit{self-dual} or \textit{BPS solitons}. Thus, they are critical points of the energy function and automatically static solutions of the second-order field equations. Indeed, it is a matter of algebra to verify that the static Euler-Lagrange equations can be derived from the Bogomolny equations. This reduction of order simplifies the search for topological solitons. The family of BPS solitons with topological charge $n$ forms a subspace of the $n$-th connected component of the configuration space that is called the $n$-soliton moduli space $\mathcal{M}_n$, that is, $\mathcal{M}_n \subset \mathcal{C}_n$. The dimensionality of this subspace is $\text{dim}(\mathcal{M}_n) = q\,n$, where $q$ counts the number of degrees of freedom of each topological soliton. Given an element of $\mathcal{M}_n$, we can generate the whole subspace by acting with the continuous symmetries of the theory, since these leave the potential energy invariant. This is not the case for discrete symmetries such as, for example, the parity transformation, which maps $\mathcal{M}_n$ to $\mathcal{M}_{-n}$. As we will show in more detail in \autoref{s:CCM}, the continuous transition through these equienergetic configurations will be useful in the description of soliton dynamics.   
         
    It is worth noting that the Bogomolny equations (\ref{e:Bogomolny_General}) correspond to the zero-pressure condition $T^{ii} = 0$. Indeed, by imposing that the $T^{11}$ component of the energy-momentum tensor (\ref{e:Energy_Momentum}) vanishes, we find in our illustrative example in one spatial dimension
    \begin{equation}
    T^{11} = \frac{\partial \mathcal{L}}{\partial (\partial_x \phi^a)}\, \partial^x \phi^a - \eta^{11} \mathcal{L} 
    = \tfrac{1}{2} (\partial_x \phi^a)^2 - U = 0\,,
    \end{equation}
    which is the sum of the product of the Bogomolny equations (\ref{e:Bogomolny_General}) upon applying the definition of the superpotential (\ref{e:Superpotential}).

    To conclude, remark that theories admitting a Bogomolny arrangement are intimately related to supersymmetric theories. E. Witten and D. Olive showed that in supersymmetric theories with topological solitons, the supersymmetric algebra presents a central extension, where the central charge is the topological charge \cite{Witten:1978}. In this context, the modification of the anticommutator between the supersymmetric charges by the topological charge and their Hermiticity is the reason behind the BPS bound of the energy. Moreover, the Bogomolny equations arise from the condition that some of the fermionic supersymmetry transformations vanish. The supersymmetric techniques are then useful tools to construct BPS theories. Despite its importance, we will not adopt this framework in our research.
     
\section{Derrick's theorem}\label{s:Derrick}

    At this point, it is natural to wonder whether there is an argument that can be used to prove that time-independent, finite-energy solitons exist in a given field theory. A crucial non-existence theorem in this context was established by Derrick \cite{Derrick:1964}, who realised that for a field configuration to be a stationary point of the energy, it should be stationary under arbitrary variations. In particular, the field configuration should be a stationary point of the energy under spatial rescaling, which fixes a characteristic size.
    
    The explanation of this theorem can be given in simple terms. To illustrate the result, let us assume a time-independent field configuration $\vec{\phi}(\vec{x})$ that solves the field equations (\ref{e:EL}) for a scalar field theory of the form (\ref{e:LagrangianDensity}). Applying a spatial rescaling $\vec{x} \rightarrow \vec{x}' = \lambda\,\vec{x}$, with $\lambda \in \mathbbm{R}^+$, the scalar field and its gradient transform as
    \begin{equation}
        \vec{\phi}(\vec{x}) \rightarrow \vec{\phi}(\lambda\,\vec{x})\,, \quad \quad
        \nabla\vec{\phi}(\vec{x}) \rightarrow \lambda \nabla\vec{\phi}(\lambda\,\vec{x})\,.
    \end{equation}
    Now, substituting the transformed scalar field into the energy functional and performing a change of variables, we obtain the following dependence of the energy on the scaling parameter $\lambda$,
    \begin{equation}\label{ScalingDerrick}
        E[\lambda] = \displaystyle \int_{\mathbbm{R}^{d}} \left[\dfrac{1}{2} \lambda^{2-d} \nabla' \vec{\phi}(\vec{x}') \cdot \nabla' \vec{\phi}(\vec{x}')  + \lambda^{-d}\,U(\vec{\phi}(\vec{x}')) \right]\,dx_1'\,\ldots\,dx_d' = \lambda^{2-d}E_2 + \lambda^{-d}E_0\,,
    \end{equation}
    where $E_2$ and $E_0$ are positive semi-definite. The condition of a stationary point at $\lambda = 1$ is
    \begin{equation}
    0 = \dfrac{dE}{d\lambda}\bigg|_{\lambda = 1} = (2 - d)E_2 - d\hspace{0.05cm}E_0\,.
    \end{equation}
    This expression suggests that the existence of stationary point depends on the dimensionality of the model.
    
    A quick inspection reveals that, when $d = 1$, the energy functional has a stationary point, and the analysis of the second derivative reveals that it corresponds to a minimum. Conversely, when $d = 2$, the contribution $E_0$ must vanish everywhere in order for the configuration to be a stationary point, which implies that the field must be a vacuum state\footnote{In this explanation, we have assumed that $E_0$ is initially non-zero. If instead $E_0 = 0$, the equation of motion is a homogeneous Poisson equation, which does not support static solitonic solutions.}. Finally, if $d \geq 3$, either $E_2$ or $E_0$ must be negative, but this condition cannot be fulfilled because both contributions are positive semi-definite by definition. As a result, the existence of non-trivial finite energy solutions can be summarised in the following theorem.
    
    \begin{theorem}[Derrick theorem \cite{Derrick:1964}]
        Let $\vec{\phi}: \mathbbm{R}^{1+d} \rightarrow \mathbbm{R}^{m}$ be a set of $m \in \mathbbm{N}$ scalar fields and $U(\phi^1,\phi^2,\ldots,\phi^m)$ be a potential which we will consider positive semi-definite. Let the dynamics of these fields be governed by
        \begin{equation*}
            \mathcal{L} = \dfrac{1}{2}\partial_{\mu}\vec{\phi}\cdot\partial^{\mu}\vec{\phi} - U(\phi^1,\phi^2,\ldots,\phi^m)\,.
        \end{equation*}
        Then, for $d \geq 2$, the only non-singular time-independent solutions of finite energy are the ground states.
    \end{theorem}
    
    Although it may seem that soliton theory is highly constrained by Derrick's theorem, there exist various mechanisms to evade it. A first reasonable possibility is to include additional terms that scale oppositely under rescalings. For instance, one can incorporate higher-order derivative terms of the scalar field, leading to the so-called \textit{baby Skyrmions} \cite{Bogolubskaya:1989,Bogolubskaya:1990} and \textit{Skyrmions} \cite{Skyrme:1961} in $d=2$ and $d=3$ spatial dimensions, respectively. This strategy of adding new terms can also be conducted by including gauge fields. The corresponding topological solitons are \textit{gauged vortices} \cite{Nielsen:1973} in $d=2$ dimensions and \textit{gauged monopoles} \cite{Polyakov:1974,Hooft:1974} in $d=3$ dimensions. A second possibility to evade Derrick’s theorem is to allow for time-dependent solutions, because Derrick's theorem only denies the existence of time-independent solutions. This is the case of the so-called \textit{Q-lumps} \cite{Leese:1991} or \textit{Q-balls} \cite{Coleman:1985}. Finally, one may also relax the requirement of finite energy, as is the case for the so-called \textit{global vortices} \cite{Vilenkin:2000}, or consider target spaces that are not metrically flat, as is the case of the so-called \textit{lumps} in the $O(3)$ sigma model \cite{Manton:2002} or the recently baptised \textit{vrochosons} in the $\mathbbm{S}^1 \times \mathbbm{S}^1$-Sigma model \cite{Balseyro:2022}.
    
\section{Topological solitons in one spatial dimension}\label{s:1D}

    In the previous sections, we established the essential ingredients required for the existence of stable topological solitons. We now turn our attention to the classically stable solutions of the equations of motion derived from the most prototypical field theories admitting topological solitons in one spatial dimension. The models under consideration are the $\phi^4$, $\phi^6$ and the sine-Gordon model. The topological solitons in this dimension are referred to as kinks\footnote{This terminology was first introduced by Finkelstein in \cite{Finkelstein:1966}.}. For simplicity, we will restrict ourselves to analysing these models assuming only one real scalar field.  

    \subsection{The \texorpdfstring{$\phi^4$}{phi4} model}\label{ss:phi4}
    
    The $\phi^4$ model was originally introduced as a phenomenological theory of second-order phase transitions \cite{Landau:1937}, and since then it has been often used as the paradigm of spontaneous symmetry breaking. This theory has found physical applications in ferroelectrics \cite{Khare:2006,Rowley:2014}, polymeric chains \cite{Rice:1980,Campbell:1982}, or even nuclear physics \cite{Campbell:1976}.

\begin{figure}[htb]
        \centering
        \noindent
        \begin{minipage}[t]{0.47\textwidth}
            \includegraphics[width=\linewidth]{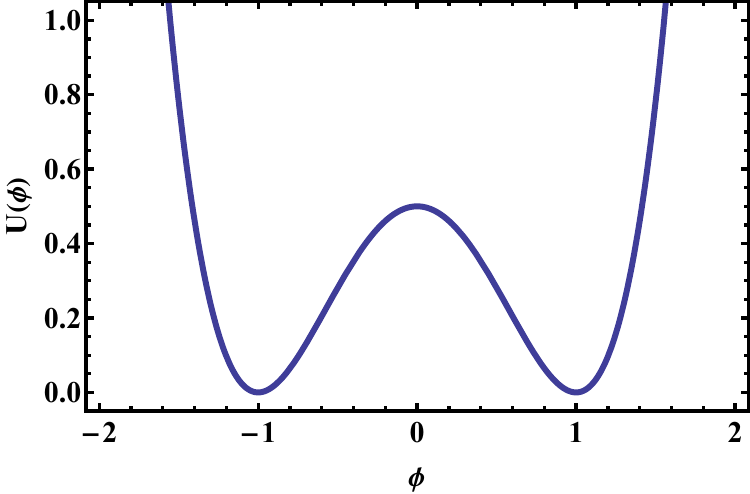}
        \end{minipage}
    
        \caption{Potential function of the $\phi^4$ model (\ref{e:potential_phi4}) for $m = 1$ and $\lambda = 1/2$.}
        \label{f:Phi4_Potential}
    \end{figure}

    The potential function defining the model is
    \begin{equation}\label{e:potential_phi4}
        U(\phi) =\lambda\left(m^2 - \phi^2\right)^2\,, \quad m,\lambda \in \mathbbm{R}^+.
    \end{equation}
    However, by a redefinition of the field and length units, the constants $m$ and $\lambda$ can be scaled to equal any given positive values. Therefore, we will henceforth assume the standard choice in the literature, namely $m = 1$ and $\lambda = 1/2$. We depict the potential profile (\ref{e:potential_phi4}) in \autoref{f:Phi4_Potential}. It is straightforward to verify that the vacuum manifold consists of two isolated points, given by $\mathcal{V} = \{\phi_v^- = -1,\,\phi_v^+ = 1\}\,.$
    
\begin{figure}[htb]
      \centering
      \noindent
      \begin{minipage}[t]{0.47\textwidth}
        \includegraphics[width=\linewidth]{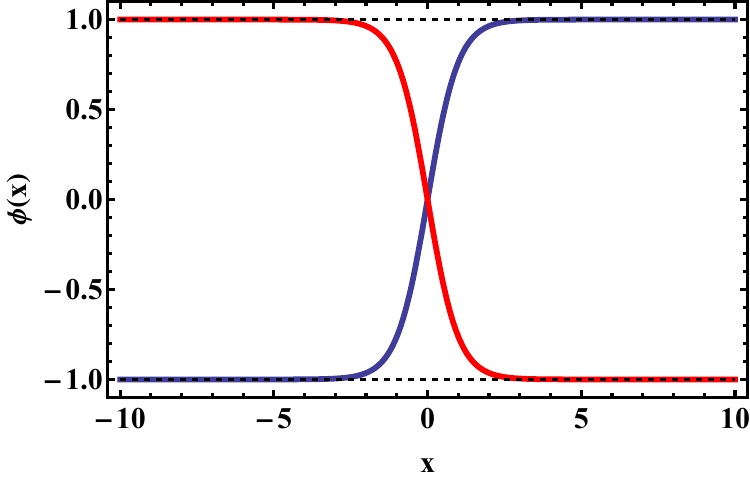}
      \end{minipage}%
      \hspace{0.8cm}
      \begin{minipage}[t]{0.46\textwidth}
        \includegraphics[width=\linewidth]{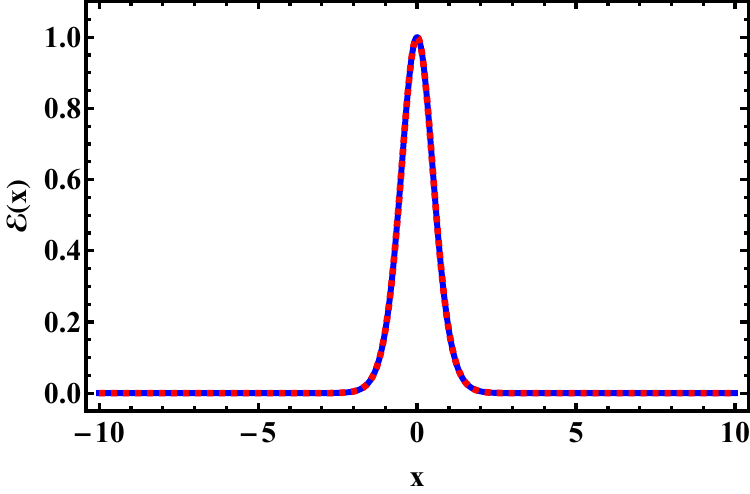}
      \end{minipage}
    
      \caption{$\phi^4$ kink and antikink solutions (\ref{e:Kink_phi4}) on the left and their energy density (\ref{e:Density_kink_phi4}) on the right. The vacuum configurations interpolated by the kink and the antikink are represented by black dashed lines. The red curve represents the antikink, while the blue curve corresponds to the kink. On the right panel, the solid blue line denotes the energy density of the kink, and the dashed red line represents that of the antikink. We have assumed $a = 0$.}
      \label{f:Phi4_Kink_ED}
    \end{figure}

    The Lagrangian density corresponding to the $\phi^4$ model is written as
    \begin{equation}\label{e:lagrangian_phi4}
        \mathcal{L} = \dfrac{1}{2}\partial_{\mu}\phi\partial^{\mu}\phi - \dfrac{1}{2}(1 -\phi^2)^2\,.
    \end{equation}
    Substituting the Lagrangian density (\ref{e:lagrangian_phi4}) into the general expression for the Euler-Lagrange equations (\ref{e:EL}), we obtain
    \begin{equation}\label{e:FE_phi4}
        \partial_{\mu}\partial^{\mu}\phi - 2(1 - \phi^2)\phi = 0\,.
    \end{equation}
    Assuming a static configuration, the field equations are reduced to
    \begin{equation}\label{e:SFE_phi4}
        \phi_{xx} + 2(1 - \phi^2)\phi = 0\,.
    \end{equation}
    The static solutions of these equations of motion have the following static energy \begin{equation}\label{e:Energy_phi4}
        E[\phi] = \int_{-\infty}^{\infty}\left[\dfrac{1}{2}\phi_x^2 + \dfrac{1}{2}\left(1 - \phi^2\right)^2\right]dx\,.
    \end{equation}
    Through a Bogomolny arrangement, we can rewrite the energy as 
    \begin{align}\label{e:BPS_bound_phi4}
        E[\phi] =&\hspace{0.1cm} \dfrac{1}{2}\int_{-\infty}^{\infty}\left[\phi_x \pm (1 - \phi^2) \right]^2dx \mp \int_{-\infty}^{\infty}\left[\phi_x (1 - \phi^2) \right]dx\nonumber\\
        \geq&  \,|Q|\displaystyle\int_{\phi_v^-}^{\phi_v^+}(1 - \phi^2)\,d\phi = \dfrac{4}{3}\,|Q|\,,
    \end{align}
    where we identify
    \begin{equation}
        Q = \dfrac{1}{2}\int_{-\infty}^{\infty}\phi_x\,dx = \dfrac{1}{2}\left(\phi(\infty) - \phi(-\infty)\right)\,,
    \end{equation}
    as the topological charge. The existence of only two vacua limits the possible values of the topological charge to $Q = \{-1,0,1\}$. The lower bound is saturated when the first term vanishes
    \begin{equation}\label{e:BPS_phi4}
        \phi_x = \pm(1 - \phi^2)\,.
    \end{equation}
    This first-order differential equation is the Bogomolny equation introduced in  \autoref{s:Bogomolny}. It can be easily verified that the non-trivial solutions of (\ref{e:BPS_phi4}) take the following form
    \begin{equation}\label{e:Kink_phi4}
        \Phi_K(x) = \tanh\left(x-a\right)\,, \quad \Phi_{AK}(x) = -\tanh\left(x-a\right)\,.
    \end{equation}
    Depending on the choice of sign, we refer to these solutions as kinks $\Phi_K$ (positive sign) or $\Phi_{AK}$ antikinks (negative sign). Interestingly, due to the reflection symmetry of the theory, these solutions are connected by a $\mathbbm{Z}_2$ transformation. Moreover, the parameter "$a$" is an integration constant that reflects the translational invariance of the theory. If one computes the energy density corresponding to the kink or antinkink 
    \begin{equation}\label{e:Density_kink_phi4}
        \mathcal{E}[\Phi_K] = \mathcal{E}[\Phi_{AK}] = \sech^4(x - a)\,,
    \end{equation}
    it is straightforward to verify that the maximum is attained at $x = a$. Usually, this position is considered as the centre of the kink. The kink and antikink profiles as well as their energy density are displayed in \autoref{f:Phi4_Kink_ED}. 
    Since the configurations (\ref{e:Kink_phi4}) have topological charge $Q = \pm1$, we can infer directly from the BPS bound (\ref{e:BPS_bound_phi4}) that their static energy is
    \begin{equation}\label{e:mass_phi4}
    E[\Phi_K] = E[\Phi_{AK}] = \dfrac{4}{3} \equiv M\,,
    \end{equation}
    which is identified with their mass. This static energy is independent of the position "$a$" of the kink or antikink. Then, it is straightforward to conclude that the moduli space of a single moving kink is simply $\mathcal{M}_1 = \mathbbm{R}$.

    Let us consider now the perturbation of a solution of the field equations by a normal mode of the form
    \begin{equation}
    \phi(x,t) = \phi(x) + \eta(x)e^{i \omega t}\,.
    \end{equation}
    Substituting this perturbation into the field equations (\ref{e:FE_phi4}) and expanding up to first order in the perturbation, we obtain the linear spectrum of fluctuations about the background solution
    \begin{equation}\label{e:linearised_phi4}
    -\eta_{xx}(x) + \left[6\,\phi(x)^2 - 2 \right]\eta(x) = \omega^2\eta(x)\,. 
    \end{equation}
    Let us assume that the background solution is a vacuum configuration, that is, $\phi \in \mathcal{V}$. In that case, the spectral problem is reduced to
    \begin{equation}\label{e:linearised_phi4_vacuum}
    -\eta_{xx}(x) + 4\,\eta(x) = \omega^2\eta(x)\,, 
    \end{equation}
    and the spectrum consists of a continuum of states above $\omega_{c}^2 = 4$. Although we will restrict ourselves to classical field theories, we remark that the excitations about the vacuum would correspond to the quanta of the associated quantum field theory, where these particles, called "mesons" for historical reasons, would have mass $m_v = 2$ \cite{Rajamaran:1987}.
    
\begin{figure}[htb]
      \centering
      \noindent
      \begin{minipage}[t]{0.47\textwidth}
        \includegraphics[width=\linewidth]{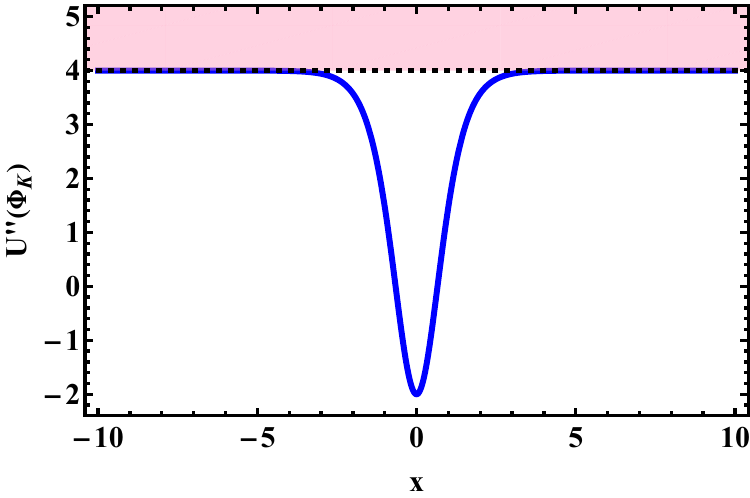}
      \end{minipage}
    
      \caption{Potential of the linearised equation of motion (\ref{e:linearised_phi4_kink}) for the $\phi^4$ kink. The shaded region indicates the continuum spectrum.}
      \label{f:Effec_Phi4_Potential}
    \end{figure}

    Conversely, if we now assume a kink (resp. antikink) as the background solution, the Schrödinger problem given by (\ref{e:linearised_phi4}) has a non-trivial well potential
    \begin{equation}\label{e:linearised_phi4_kink}
        -\eta_{xx}(x) + \left[6\,\tanh^2 x - 2 \right]\eta(x) = \omega^2\eta(x)\,. 
    \end{equation}
    The corresponding potential is depicted in \autoref{f:Effec_Phi4_Potential}. 
    Fortunately, this spectral problem is well known in the literature, as the potential well takes the form of a Pöschl–Teller potential \cite{Rosen:1932}. The general framework of Pöschl-Teller potentials allows us to predict the existence of two bound modes, the zero mode and a vibrational mode, apart from the continuum of states above the mass threshold
    \begin{eqnarray}\label{e:perturbations_phi4}
    \eta_{0}(x) \hspace{-0.2cm}&=&\hspace{-0.2cm} \sech^2 x\,, \qquad \omega_{0} = 0\,, \\
    \eta_{s}(x) \hspace{-0.2cm}&=&\hspace{-0.2cm} \sinh x\sech^2 x\,, \qquad \omega_{s} = \sqrt{3}\,,\label{e:shape_exp_phi4} \\ 
    \eta_{q}(x) \hspace{-0.2cm}&=&\hspace{-0.2cm}  (3 \tanh^2 x -q^2 - 1 - 3iq\tanh x) e^{iqx}\,, \qquad \omega_{q} = \sqrt{q^2 + 4}\,,
    \end{eqnarray}
    with $q \in \mathbbm{R}$. 
    
    Each of these modes has a simple physical interpretation. As anticipated in \autoref{s:Modes}, the linear spectrum of perturbations about the kink solutions contains a zero mode, related to the breaking of the translational invariance about kink solution. Therefore, this mode accounts for small rigid translations of the kink. Indeed, if we assume a small rigid translation of the kink (\ref{e:Kink_phi4}), we get the following Taylor expansion
    \begin{equation}\label{e:zero_phi4}
        \Phi_K(x - \epsilon) \approx \tanh x - \epsilon\,\sech^2 x + o(\epsilon^2)\,.
    \end{equation}
    As it is clearly visible, the first order correction to the translated kink profile is the zero mode. We conclude then that $\eta_0 \propto \Phi_K'(x)$. In fact, this result can be obtained by identifying the spectral problem (\ref{e:linearised_phi4_kink}) for $\omega_0^2 = 0$ with the equation resulting from differentiating the static Euler–Lagrange equation (\ref{e:SFE_phi4}) with respect to $x$, and then evaluating it at the kink solution $\Phi_K$
    \begin{equation}
       \Phi^{(3)}_K + \left(2 - 6\,\Phi_K^2\right)\Phi'_K = 0 \quad \Rightarrow \quad \eta''_0 + \left(2 - 6\,\Phi_K^2\right)\eta_0 = 0\,, \quad \eta_0 \propto \Phi'_K\,,
    \end{equation}
    as explained in a general framework in \autoref{s:Modes}.
    The $\phi^4$ kink also possesses a positive bound mode which is usually referred to as the \textit{shape-mode}. The reason behind this name is that this perturbation is bounded to the kink, and its effect is to modify the width of the kink while it oscillates in time with frequency $\omega_s^2 = 3$. Indeed, a small perturbation in the width of the kink (\ref{e:Kink_phi4}) leads to an expansion of the form
    \begin{equation}\label{e:shape_phi4}
        \Phi_K((1 + \epsilon)\,x) \approx \tanh x + \epsilon\,x\,\sech^2 x + o(\epsilon^2)\,.
    \end{equation}
    The first order correction to the kink profile resembles the expression (\ref{e:shape_exp_phi4}). Finally, the continuum of states above the mass threshold are the radiation or scattering modes. They represent the vacuum fluctuations in the non-trivial background created by the kink.

    \subsection{The \texorpdfstring{$\phi^6$}{phi6} model}\label{ss:phi6}

    The $\phi^6$ model has been used to describe first-order phase transitions in certain crystals \cite{Lindgard:1986, Dmitriev:1991}, in copolymers \cite{Furukawa:1989} and smectic $A$ to smectic $C$ phase transitions in liquid crystals \cite{Huang:1982}. It has also found applications in quantum phase transitions \cite{Bergner:2003} and first-order phase transition in the presence of gravity \cite{Kim:1996,Joy:2003}.

\begin{figure}[htb]
      \centering
      \noindent
      \begin{minipage}[t]{0.47\textwidth}
        \includegraphics[width=\linewidth]{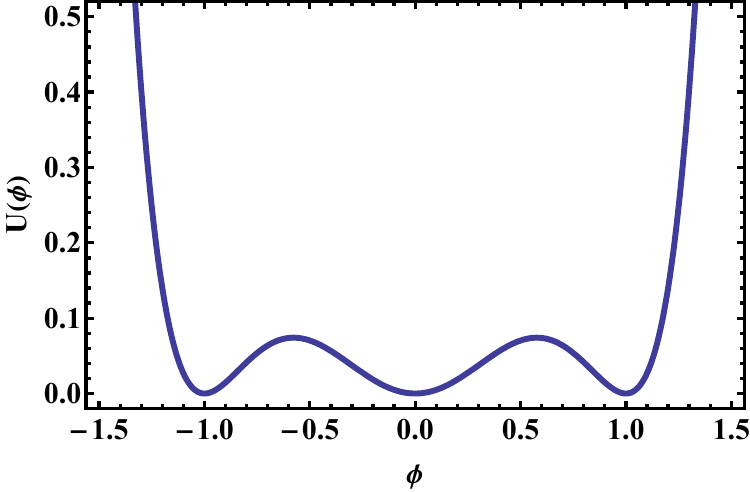}
      \end{minipage}
    
      \caption{Potential function of the $\phi^6$ model (\ref{e:lagrangian_phi6}).}
      \label{f:Phi6_Potential}
    \end{figure}

    The $\phi^6$ model is defined through the potential
    \begin{equation}
        U(\phi) = \dfrac{1}{2}\phi^2(1-\phi^2)^2\,,
    \end{equation}
    which is represented in \autoref{f:Phi6_Potential}. Unlike the $\phi^4$ potential, this sixth-order polynomial potential exhibits a more complex structure. In particular, it admits three degenerate minima. From the expression above, it follows that the vacuum manifold is given by $\mathcal{V} = \{\phi_v^- = -1,\,\phi_v^0 = 0,\, \phi_v^+ = 1\}\,$. The dynamics is governed by the Lagrangian density
    \begin{equation}\label{e:lagrangian_phi6}
        \mathcal{L} = \dfrac{1}{2}\partial_{\mu}\phi\partial^{\mu}\phi - \dfrac{1}{2}\phi^2(1-\phi^2)^2\,,
    \end{equation}
    and the equation of motion associated to this theory is 
    \begin{equation}\label{e:FE_phi6}
        \partial_{\mu}\partial^{\mu}\phi - (1 - \phi^2)(-1+3\phi^2)\phi = 0\,.
    \end{equation}
    The energy of the static solutions of the Euler-Lagrange equations (\ref{e:FE_phi6}) is given by
    \begin{equation}\label{e:Energy_phi6}
        E[\phi] = \int_{-\infty}^{\infty}\left[\dfrac{1}{2}\phi_x^2 + \dfrac{1}{2}\phi^2\left(1 -\phi^2\right)^2\right]dx\,.
    \end{equation}
    Once more, we can perform a Bogomolny arrangement to the energy functional
    \begin{align}
        E[\phi] =&\hspace{0.1cm} \dfrac{1}{2}\int_{-\infty}^{\infty}\left[\phi_x \pm \phi\,(1 - \phi^2) \right]^2dx \mp \int_{-\infty}^{\infty}\left[\phi\,\phi_x\,(1 - \phi^2) \right]dx\nonumber\\
        \geq&  \,|Q|\displaystyle\int_{\phi^{-}}^{\phi^{+}}\phi\,(1 - \phi^2)\,d\phi = \dfrac{1}{4}\,|Q|\,,
    \end{align}
    where we identify
    \begin{equation}
        Q =\int_{-\infty}^{\infty}\phi_x\,dx = \phi(\infty) - \phi(-\infty)\,,
    \end{equation}
    as the topological charge. Naturally, one may ask what the possible values of the topological charge $Q$ are, as we have now three vacua. The answer to this question is closely connected to the theory of differential equations. Due to the structure of the vacuum manifold, the maximum value of the topological charge could be $|Q| = 2$. However, Picard's theorem excludes such a possibility, as a solution with that topological charge would necessarily interpolate between the vacua $\phi_v^-$ and $\phi_v^+$ -- or vice versa -- thereby requiring it to pass through the intermediate vacuum $\phi_v^0$, which violates the uniqueness conditions of the theorem, since $\phi(x) = \phi_v^0$ is solution. 
    \begin{theorem}[Picard's theorem]\label{t:Picard}
        Let $f(x,y)$ be a $C^1(D)$ function, where $D \subset \mathbbm{R}^2$. Then, given a point $(x_0,y_0) \in D$, there exists only one solution $y = y(x)$ of the differential equation $y' = f(x,y)$, such that $y_0 = y(x_0)$. Such a solution is defined on a certain neighbourhood of $x_0$.
    \end{theorem}
     Therefore, the acceptable values of the topological charge are $Q = \{-1,0,1\}$.
    
\begin{figure}[htb]
    \centering
    \noindent
    \begin{minipage}[t]{0.47\textwidth}
    \includegraphics[width=\linewidth]{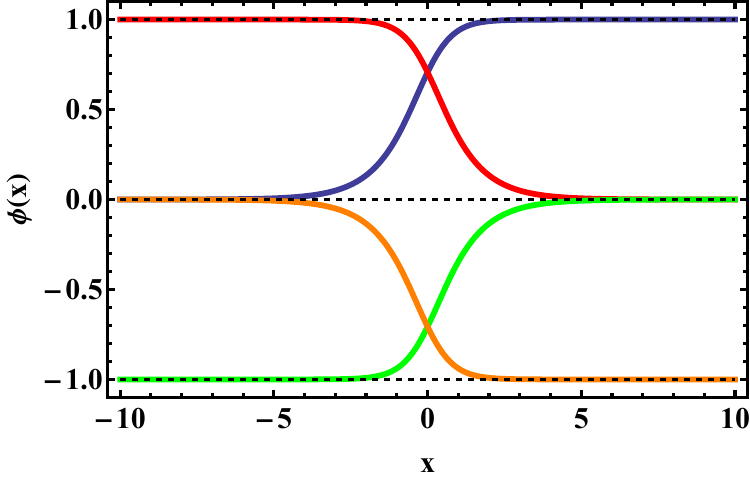}
    \end{minipage}%
    \hspace{0.8cm}
    \begin{minipage}[t]{0.47\textwidth}
    \includegraphics[width=\linewidth]{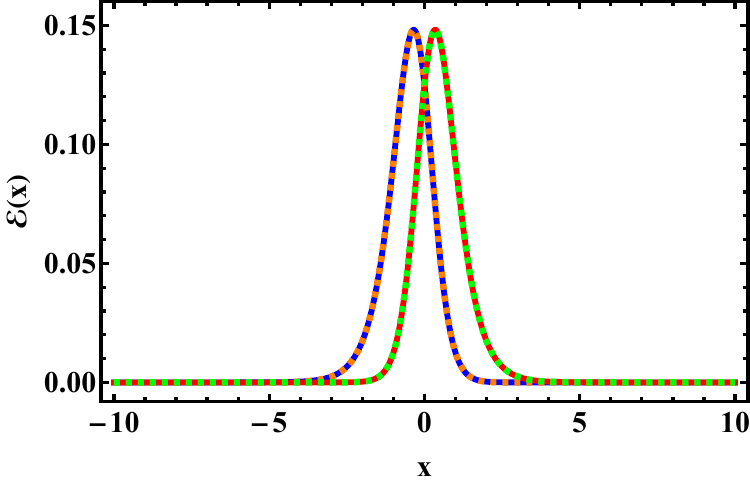}
    \end{minipage}
    
    \caption{Kink, antikink and their mirror solutions (\ref{e:Kink_phi6}) on the left and the corresponding energy density profiles (\ref{e:Density_kink_phi6}) on the right for the $\phi^6$ model. The vacuum configurations interpolated by the (rep. mirror) kinks and the (rep. mirror) antikinks are represented by black dashed lines. Due to the lack of reflection symmetry, (rep. mirror) kinks and (rep. mirror) antikinks do not have the maximum of their energy densities at the same position. We have assumed $a = 0$.}
    \label{f:Phi6_Kink_ED}
    \end{figure}

    Now, the BPS bound is saturated when the following Bogomolny equation is satisfied
    \begin{equation}\label{e:BPS_phi6}
        \phi_x = \pm\phi\,(1 - \phi^2)\,.
    \end{equation}
    The solutions belonging to the $|Q| = 1$ topological class are 
    \begin{equation}\label{e:Kink_phi6}
    \begin{split}
        \Phi_{K}(x) =& \sqrt{\dfrac{1 +\tanh(x - a)}{2}}\,, \quad  \Phi_{K}^*(x) = -\sqrt{\dfrac{1 - \tanh(x - a)}{2}}\,,\\
        \Phi_{AK}(x) =& \sqrt{\dfrac{1 - \tanh(x - a)}{2}}\,, \quad \Phi_{AK}^*(x) = - \sqrt{\dfrac{1 + \tanh(x - a)}{2}}\,.
    \end{split}
    \end{equation}
    We find a kink $\Phi_K$ and an antikink $\Phi_{AK}$ solution interpolating between the vacua $0$ and $1$ in opposite directions. Moreover, there is a mirror kink $\Phi_{K}^*$ and a mirror antikink $\Phi_{AK}^*$ solution interpolating between the vacua $-1$ and $0$, also in opposite directions. It is worthwhile to remark that the $\phi^6$ model does not exhibit reflection symmetry. For that reason, the kink and the antikink are not related by the transformation $\phi \rightarrow - \phi$, and the same argument applies to the mirror configurations. Specifically, this asymmetry is clearly visible if one computes the energy density corresponding to these configurations
    \begin{equation}
    \begin{split}\label{e:Density_kink_phi6}
        \mathcal{E}[\Phi_K] =& \hspace{0.1cm} \mathcal{E}[\Phi_{AK}^*] = \dfrac{\sech^4(x - a)}{8 (1 + \tanh (x - a))}\,,\\
        \mathcal{E}[\Phi_{AK}] =&  \hspace{0.1cm} \mathcal{E}[\Phi_{K}^*]  = \dfrac{\sech^4(x - a)}{8 (1 - \tanh (x - a))}\,.
    \end{split}
    \end{equation}
    The profile of the (mirror) kink and antikink and their corresponding energy densities are depicted in \autoref{f:Phi6_Kink_ED}. However, the saturation of the BPS bound guarantees that all of these configurations have the same static energy or mass
    \begin{equation}\label{e:mass_phi6}
        E[\Phi_K] = E[\Phi_{AK}] =E[\Phi_K^*] =E[\Phi_{AK}^*] =\dfrac{1}{4} \equiv M\,,
    \end{equation}
    and the moduli space of a single kink is again $\mathcal{M}_1 = \mathbbm{R}$.
    
    This lack of reflection symmetry has also a strong influence on the linear spectrum of perturbations. Once more, assuming a perturbation in terms of Fourier modes, we derive the following spectral problem 
    \begin{equation}\label{e:linearised_phi6}
        -\eta''(x) + \left[15\,\phi(x)^4 - 12\,\phi(x)^2+1 \right]\eta(x) = \omega^2\eta(x)\,. 
    \end{equation}
    Now, depending whether we consider the vacua $\phi_v^{\pm} = \pm1$ or the vacuum $\phi_v^0 = 0$, we have different mass thresholds. For the former case, $\omega_c^2 = 4$, and for the latter case, $\omega_c^2 = 1$. From the quantum point of view, the mass of the quanta is different depending on the vacuum. 

\begin{figure*}[htb]
    \centering
    \begin{subfigure}[t]{0.43\textwidth}
        \centering
        \includegraphics[width=\linewidth]{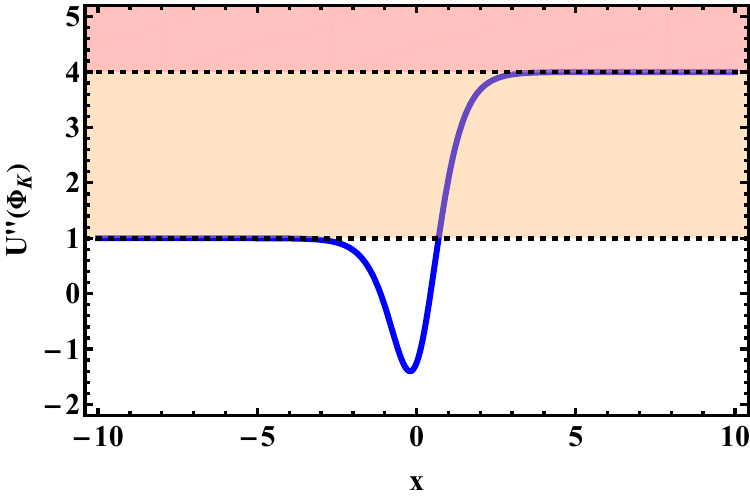}
        \caption{Linearised potential for the $\phi^6$ kink.}
        \label{f:Spectrum_K}
    \end{subfigure}
    
    \vspace{1.5em} 
    
    \begin{subfigure}[t]{0.43\textwidth}
        \centering
        \includegraphics[width=\linewidth]{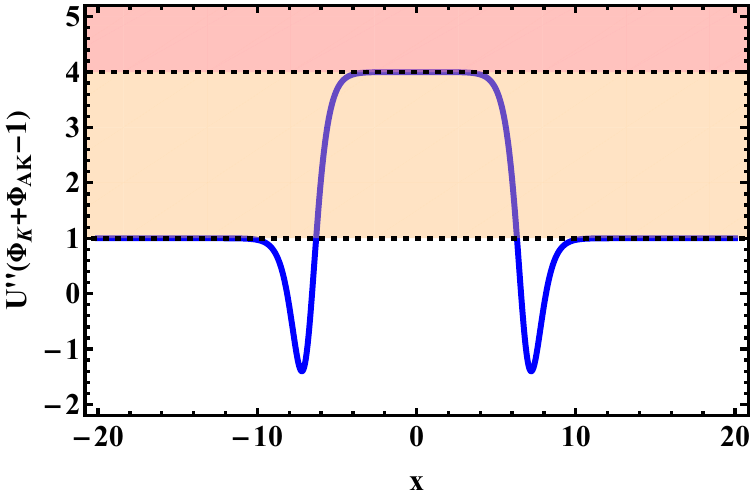}
        \caption{Linearised potential for the $\phi^6$ kink-antikink.}
        \label{f:Spectrum_KAK}
    \end{subfigure}
    \hfill
    \begin{subfigure}[t]{0.43\textwidth}
        \centering
        \includegraphics[width=\linewidth]{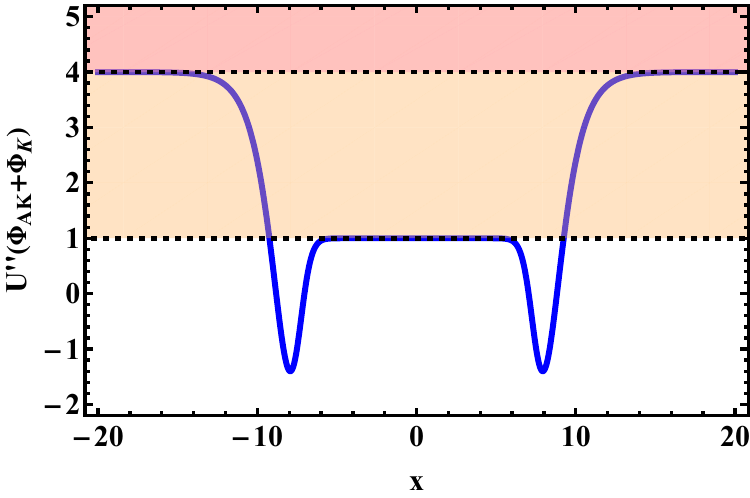}
        \caption{Linearised potential for the $\phi^6$ antikink-kink.}
        \label{f:Spectrum_AKK}
    \end{subfigure}

    \caption{\small Potential of the linearised equation of motion (\ref{e:linearised_phi6}) for the $\phi^6$ kink located at the origin ($a = 0$), as well as for well-separated kink–antikink and antikink–kink configurations (\ref{e:Kink_phi6}). We have assumed the constituent kinks to be localised at $a = \pm\, 7$. The shaded regions indicate the continuum spectrum associated with each vacuum.}
    \label{f:Phi6_KAK_AKK_Effec_Pot}
    \end{figure*}

    The situation becomes more subtle when a $\phi^6$ kink is taken as the background solution. In this case, the kink interpolates between vacua with different masses, and the associated spectral problem inherits this asymmetry. As a consequence, the effective potential that governs the fluctuations is only semi-bounded, as illustrated in \autoref{f:Spectrum_K}.
    Fortunately, this spectral problem can also be solved analytically and consists of the following modes \cite{Lohe:1979} 
    \begin{eqnarray}\label{e:perturbations_phi6}
        \eta_{0}(x) \hspace{-0.2cm}&=&\hspace{-0.2cm} \frac{\sech^2 x}{2 \sqrt{2}\sqrt{\tanh x + 1}}\,, \quad \omega_{0} = 0\,,\\
        \eta_{qn}(x) \hspace{-0.2cm}&=&\hspace{-0.2cm} \dfrac{e^{-(K_+ + \,i \,k_-)x/2}}{(e^x + x^{-x})^b}\,{}_2F_1\left(b - \dfrac{3}{2},b + \dfrac{5}{2}, K_+ +1; \dfrac{e^{-x}}{e^x + e^{-x}} \right)\,, \quad 1 \leq \omega_{qn}^2 \leq 4\,,\\
        \eta^{(1)}_q(x) \hspace{-0.2cm}&=&\hspace{-0.2cm} \dfrac{e^{i(k_+ - \,k_-)x/2}}{(e^x + x^{-x})^b}\,{}_2F_1\left(b - \dfrac{3}{2},b + \dfrac{5}{2}, 1 - i \,k_+; \dfrac{e^{-x}}{e^x + e^{-x}} \right)\,, \quad  \omega_q^2 \geq 4\,,
        \\
        \eta^{(2)}_q(x) \hspace{-0.2cm}&=&\hspace{-0.2cm} \dfrac{e^{i(k_+ - \,k_-)x/2}}{(e^x + x^{-x})^b}\,{}_2F_1\left(b - \dfrac{3}{2},b + \dfrac{5}{2}, 1 - i \,k_-; \dfrac{e^{x}}{e^x + e^{-x}} \right)\,, \quad  \omega_q^2 \geq 4\,,
    \end{eqnarray}
    where
    \begin{equation}
        k_{-} = \sqrt{\omega^2 - 1}\,, \quad k_{+} = \sqrt{\omega^2 - 4}\,, \quad K_+ = \sqrt{4 - \omega^2}\,, \quad b = -\dfrac{i}{2}(k_+ + k_-)\,.
    \end{equation}
    Here we distinguish a zero mode, $\eta_0(x)$, a new class of mode that we will refer to as \textit{quasi-bound mode}, $\eta_{qn}(x)$, and a continuum of scattering modes, $\eta^{(1)}_q(x)$ and $\eta^{(2)}_q(x)$. A quasi-bound mode refers to an oscillatory solution of the linearised field equations that is spatially localised around a potential well, but satisfies purely outgoing boundary conditions at spatial infinity. As a result, it is not strictly bounded. In this scenario, $\eta_{qn}(x)$ accounts physically for incoming vacuum fluctuations from the left that are completely reflected from the kink, while $\eta^{(1)}_q(x)$ and $\eta^{(2)}_q(x)$ correspond to vacuum fluctuations coming from the left and from the right, respectively, which are partially transmitted through the kink. 
      
    It is worth noting that, although a single kink or antikink does not support positive bound modes, there is another possibility related to the form of the effective potential. When considering a kink and an antikink interacting through the vacuum $\phi_v^+ = 1$, their opposing tails tend toward the vacuum $\phi_v^0 = 0$. In this case, the resulting effective potential resembles a barrier, and as such, the overall configuration does not support bound modes (see \autoref{f:Spectrum_KAK}). In contrast, when an antikink and a kink interact through the vacuum $\phi_v^0 = 0$ and their opposing tails tend toward the vacuum $\phi_v^+ = 1$, the effective potential takes the form of a well (see \autoref{f:Spectrum_AKK}). Importantly, the width of this well increases as the separation between the antikink and the kink becomes larger. This situation allows the system to support an increasing number of positive bound modes, which are not localised around each individual kink, but instead are extended across the region between them \cite{Dorey:2011}. In this context, these modes are referred to as \textit{delocalised modes} \cite{Huidobro:2022,Oles:2022}. 

\subsection{The sine-Gordon model}\label{ss:SG}

    The sine-Gordon model originally arose in the study of surfaces of a constant negative Gaussian curvature \cite{Bour:1862}. After that, it appeared as the continuous limit of the Frenkel–Kontorova model, describing dislocations in a crystal lattice \cite{Frenkel:1939}. It has also found applications in the description of long Josephson junctions \cite{Josephson:1865}. Moreover, a relationship between black holes in Jackiw–Teitelboim dilaton gravity and solitons in sine-Gordon field theory has been unveiled \cite{Gegenberg:1997}.

\begin{figure}[htb]
      \centering
      \noindent
      \begin{minipage}[t]{0.45\textwidth}
        \includegraphics[width=\linewidth]{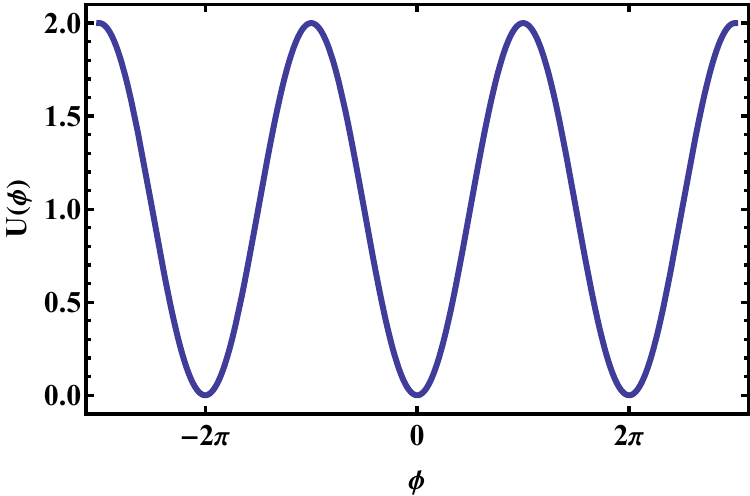}
      \end{minipage}
    
      \caption{Potential function of the sine-Gordon model (\ref{e:Potential_SG}).}
      \label{f:SG_Potential}
    \end{figure}

    The sine-Gordon model is defined through the following Lagrangian density
    \begin{equation}\label{e:potential_SG}
        \mathcal{L} = \dfrac{1}{2}\partial_{\mu}\phi\partial^{\mu}\phi - (1 - \cos \phi)\,,
    \end{equation}
    where the potential function
    \begin{equation}\label{e:Potential_SG}
        U(\phi) = 1 - \cos \phi\,,
    \end{equation}
    is periodic (see \autoref{f:SG_Potential}) and implies an infinite vacuum manifold $\mathcal{V} = \{\phi_v^k = 2\pi k\}$, where $k \in \mathbbm{Z}$. Note that the periodicity of the potential leads to the discrete symmetry $\phi \rightarrow \phi + 2\pi k$ with $k \in \mathbbm{Z}$.

    The associated Euler-Lagrange equation takes the form
    \begin{equation}\label{e:equation_SG}
        \partial_{\mu}\partial^{\mu}\phi + \sin \phi = 0\,.
    \end{equation}

    Analogously to the previous examples, we can rewrite the static energy functional
    \begin{align}
        E[\phi] = \int_{-\infty}^{\infty}\left[\tfrac{1}{2}\phi_x^2 + (1 - \cos \phi)\right]dx\,, 
    \end{align}
    using a Bogomolny arrangement and find the BPS bound 
    \begin{align} \label{e:Energy_SG}
        E[\phi] =& \dfrac{1}{2}\int_{-\infty}^{\infty}\left[\phi_x \pm 2\sin(\phi/2) \right]^2dx
       \mp 2\int_{-\infty}^{\infty}\left[\phi_x \sin(\phi/2) \right]dx \nonumber\\
    \geq& \,|Q|\int_{\phi^{-}}^{\phi^{+}}2\sin(\phi/2)\,d\phi = 8\,|Q|\,, 
    \end{align}
    where the topological charge is now given by
    \begin{equation}
        Q = \dfrac{1}{2\,\pi}\int_{-\infty}^{\infty}\phi_x\,dx = \dfrac{\phi(\infty) - \phi(-\infty)}{2\,\pi}\,.
    \end{equation}
    
\begin{figure}[htb]
    \vspace{-0.3cm}
    \centering
    \noindent
    \begin{minipage}[t]{0.457\textwidth}
        \includegraphics[width=\linewidth]{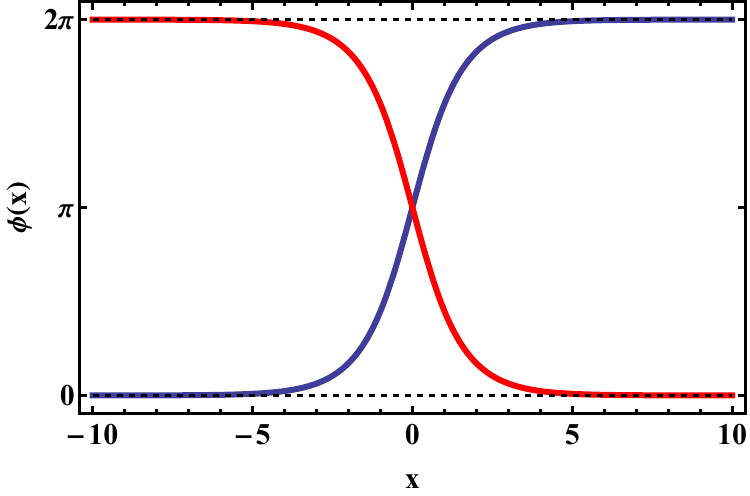}
    \end{minipage}%
    \hspace{0.8cm}
    \begin{minipage}[t]{0.448\textwidth}
    \includegraphics[width=\linewidth]{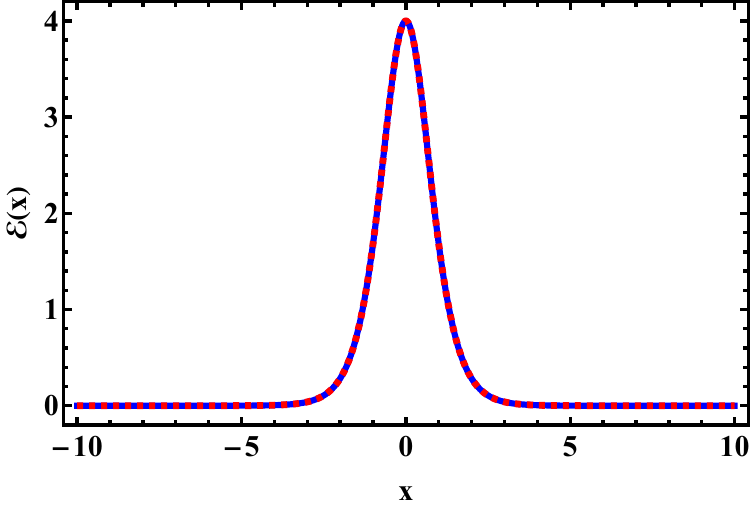}
    \end{minipage}
    
        \caption{Kink (blue) and antikink (red) solutions (\ref{e:kink_SG}) on the left and their energy density (\ref{e:density_SG}) on the right for the sine-Gordon model. The vacuum configurations interpolated by the kink and the antikink are represented by black dashed lines. On the right panel, the solid blue line denotes the energy density of the kink, and the dashed red line represents that of the antikink. We have assumed $a = 0$. Due to the discrete symmetry $\phi \rightarrow \phi \pm 2\pi$, the same results are obtained for $\phi \in \{0 + 2k\pi,2\pi + 2k\pi\}$ with $k \in \mathbbm{Z}$.}
      \label{f:SG_Kink_ED}
    \end{figure}

    The BPS bound (\ref{e:Energy_SG}) is attained when the following Bogomonly equation is satisfied
    \begin{equation}\label{e:BPS_SG}
        \phi_x = \pm 2\sin(\phi/2)\,.
    \end{equation}
    Once more, Picard's theorem introduced in \autoref{t:Picard} imposes that the only acceptable values of the topological charge are $Q = \{-1,0,1\}$. The kink and antikink configurations are given by
    \begin{equation}\label{e:kink_SG}
        \Phi_K(x) = 4\arctan(e^{x-a})\,, \quad \quad \Phi_{AK}(x) = - 4\arctan (e^{x-a})\,.
    \end{equation}
    The reflection symmetry of the theory allows us to connect the kink and the antikink through the replacement $\phi \rightarrow - \phi$. Then, they are centred at the same position, specifically $x = a$, as can be confirmed by examining the associated energy density 
    \begin{equation}\label{e:density_SG}
        \mathcal{E} = 4\sech^2(x- a)\,.
    \end{equation}
    The profiles of the kink and antikink and their energy density are depicted in \autoref{f:SG_Kink_ED}.
    Finally, the saturation of the bound allows us to determine trivially the mass of these non-trivial configurations
    \begin{equation}\label{e:energy_SG}
        E[\Phi_k] = E[\Phi_{AK}] = 8 \equiv M\,.
    \end{equation}
    \vspace{-0.3cm}
    
\begin{figure}[htb]
      \centering
      \noindent
      \begin{minipage}[t]{0.47\textwidth}
        \includegraphics[width=\linewidth]{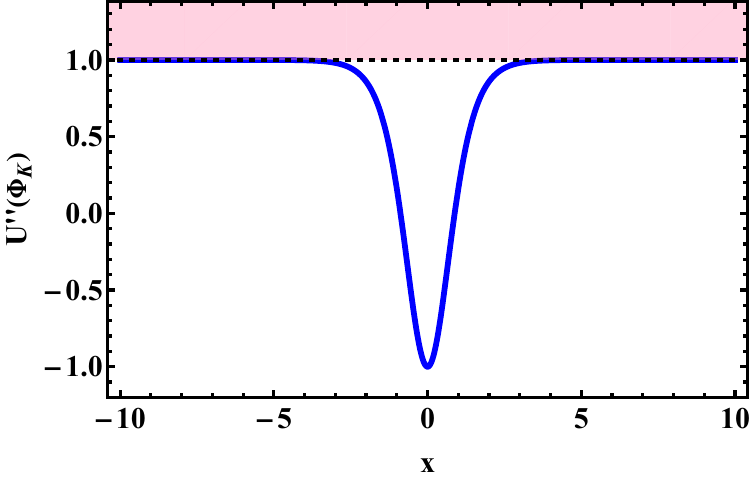}
      \end{minipage}
    
      \caption{Potential of the linearised equation of motion (\ref{e:linearised_SG_Kink}) for the sine-Gordon kink (\ref{e:kink_SG}). The shaded region indicates the continuum spectrum.}
      \label{f:Effec_SG_Potential}
    \end{figure}

    With respect to the spectrum of perturbations, the corresponding Schrödinger-like equation in the sine-Gordon model takes the following form    
    \begin{equation}\label{e:linearised_SG}
        -\eta''(x) + [\cos{\phi(x)}]\,\eta(x) = \omega^2\eta(x)\,. 
    \end{equation}
    Substituting the vacuum solution $\phi_v^k = \{2\pi k : k \in \mathbbm{Z}\}$, we deduce that  the continuum spectrum begins at $\omega_c^2 = 1$. In contrast, when perturbing around the sine-Gordon kink solution (\ref{e:kink_SG}), the spectral problem becomes
    \begin{equation}\label{e:linearised_SG_Kink}
        -\eta''(x) + [1- 2\sech^2{x}]\,\eta(x) = \omega^2\eta(x)\,, 
    \end{equation}
    where the effective potential has the form of a Pöschl–Teller potential. The potential is depicted in \autoref{f:Effec_SG_Potential}.    
    Applying the general framework of Pöschl-Teller potentials we deduce the explicit form of the zero mode and the continuum of states above the mass threshold
    \begin{eqnarray}\label{e:perturbations_SG}
    \eta_{0}(x) \hspace{-0.2cm}&=&\hspace{-0.2cm} \sech x\,, \qquad \omega_{0} = 0\,, \\
    \eta_{q}(x) \hspace{-0.2cm}&=&\hspace{-0.2cm}  (q - i \tanh x)\, e^{iqx}\,, \qquad \omega_{q} = \sqrt{q^2 + 1}\,,
    \end{eqnarray}
    where $q \in \mathbbm{R}$. Note that the sine-Gordon kink does not support any shape modes, which is a common feature of integrable models with solitons. However, as exemplified by the $\phi^6$ theory, the absence of positive bound modes does not necessarily imply that the model is integrable.

    As a final remark, it is noteworthy that, although Picard's theorem excludes static solutions of the Bogomonly equations (\ref{e:BPS_SG}) with $|Q| \geq 2$, it does not forbid time-dependent solutions of the Euler-Lagrange equations (\ref{e:potential_SG}) with topological charges higher than the unity. In fact, these solutions can be derived analytically. As briefly discussed in  \autoref{s:Solitons}, one can apply techniques as the Bäcklund transformations due to the integrability of the sine-Gordon model in one spatial dimension. 
    
    Originally, the first multi-soliton state described two kinks\footnote{A similar solution exists describing a kink-antikink collision \cite{Perring:1962}.} colliding elastically at $x = 0$ at time $t = 0$ (see reference \cite{Perring:1962}) \begin{equation}\label{e:SS_SG} 
        \Phi_{KK}(x,t;v) = 4 \tan^{-1}\left( \dfrac{v\sinh(\gamma x)}{\cosh(\gamma v t)} \right)\,, \quad \gamma = \dfrac{1}{\sqrt{1 - v^2}}\,. 
    \end{equation} 
    This result was later generalised in \cite{Gibbon:1972,Gibbon:1973} to describe the multiple collisions of $Q \geq 1$ kinks with different velocities. Furthermore, using the inverse-scattering method, a configuration consisting of a kink-antikink bound state oscillating periodically in time was constructed in \cite{Ablowitz:1973}. That solution is usually referred to as a \textit{breather} solution
    \begin{equation}\label{e:Breather_SG}
        \Phi_B(x,t;\omega) = 4 \tan^{-1}\left( \dfrac{\sqrt{1 - \omega^2}\cos(\omega t)}{\omega \cosh(\sqrt{1 - \omega^2} x)} \right)\,, \quad \omega^2 \leq 1\,. 
    \end{equation}
    
\section{Topological solitons in two spatial dimensions}\label{s:2D}

\begin{figure}[htb]
        \centering{\includegraphics[width=0.5\linewidth]{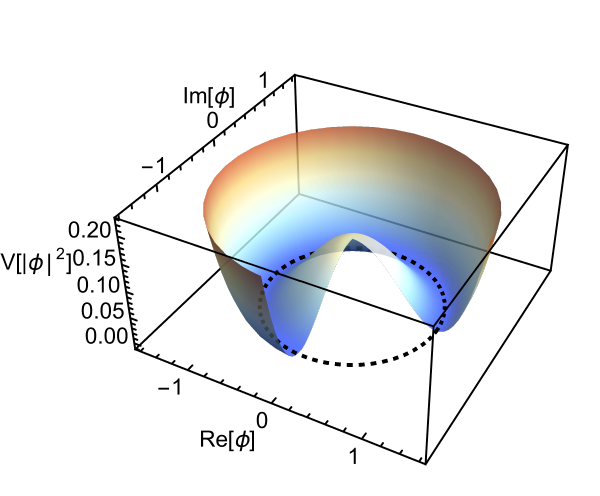}
        }
        \caption{Mexican hat potential for $\eta = 1$ and $\tilde{\lambda} = 1$. The black dashed curve represents the vacuum manifold $\mathcal{V} = \{\phi:\overline{\phi}\phi = 1\}$.}
        \label{f:Higgs_Potential}
    \end{figure}

    We exposed that, by virtue of Derrick's theorem, a standard scalar field theory defined on a flat space does not exhibit static, finite-energy solitons other than the vacuum configuration if the spatial dimension is higher than one. Indeed, consider the following model defined on $\mathbbm{R}^{1,2}$, involving only a single complex field 
    \begin{equation} \label{e:LagrangianDensity_GV}
        \tilde{\mathcal{L}} = \frac{1}{2}\partial_{\mu}\overline{\phi}\, \partial^{\mu}\phi-\frac{\tilde{\lambda}}{8}(\eta^2-\overline{\phi}\phi)^2, 
    \end{equation}
    where Minkowski metric is chosen with the sign convention $g_{\mu \nu} = {\rm diag}\{1,-1,-1\}$, $\overline{\phi}$ stands for the complex conjugate of $\phi$ and $\eta$ is a real parameter. Here, we have assumed the well-known \textit{"Mexican hat" potential}, which is depicted in \autoref{f:Higgs_Potential} along with its corresponding vacuum manifold. From the structure of the potential function, it is deduced that the vacuum manifold is given by the set $\mathcal{V} = \{\phi:\overline{\phi}\phi = \eta\} \cong \mathbbm{S}^1$. 
    
    Although the vacuum manifold has a non-trivial topological structure and the model admits topological solitons with non-zero topological charge, all such configurations possess infinite energy \cite{Vilenkin:2000,Manton:2002}, since the corresponding scaled energy functional does not exhibit a stationary point. These infinite energy solutions are the global vortices that we mentioned in \autoref{s:Derrick}. In that section, we noted that one way to evade Derrick's theorem is to introduce gauge fields, as they give rise to additional terms that can produce stationary points of the energy. In this section, we shall focus on that possibility. This procedure will lead us to the introduction of the so-called \textit{gauged vortices}.
    \vspace{-0.8cm}

    To proceed, note that the Lagrangian in (\ref{e:LagrangianDensity_GV}) is invariant under global $U(1)$ transformations of the form $\phi \rightarrow e^{iq\alpha}\,\phi$, with $q,\alpha \in \mathbbm{R}$. However, if we now promote this global symmetry to a local one, $\phi \rightarrow e^{iq\alpha(x^{\nu})}\,\phi$, the transformation is no longer a symmetry of the theory unless an additional vector field $A_{\mu}$ is introduced, transforming as $A_{\mu} \rightarrow A_{\mu} + (1/q)\partial_{\mu}\alpha(x^{\nu})$. Furthermore, the ordinary derivative must be replaced by a covariant derivative, $D_{\mu}\phi = \partial_{\mu}\phi - iqA_{\mu}\phi$. Then, $q$ represents the coupling constant between the scalar and the gauge field. Finally, for the vector field to be dynamical, we need to include gauge-invariant terms involving time derivatives. This can be achieved using the field strength tensor $F_{\mu \nu}=\partial_{\mu}A_\nu - \partial_{\nu}A_{\mu}$. This procedure is known as the \textit{gauge principle}. Applying this methodology to the Lagrangian density (\ref{e:LagrangianDensity_GV}), we obtain the so-called Abelian-Higgs model
    \begin{equation} \label{e:LagrangianDensity_AH_0}
        \tilde{\mathcal{L}}=-\frac{1}{4}F_{\mu\nu}F^{\mu\nu}+\frac{1}{2}\overline{D_{\mu}\phi} D^{\mu}\phi-\frac{\tilde{\lambda}}{8}(\eta^2 - \overline{\phi}\phi)^2\,,
    \end{equation}
    which describes the minimal coupling between a complex scalar (Higgs) field and a $U(1)$ gauge field. Here, the coupling constants $\tilde{\lambda}$ and $e$ account for the penetration lengths of the scalar and electromagnetic fields, respectively. In the literature, it is usual to rescale the fields and coordinates as follows
    \begin{equation}
        \phi\rightarrow \eta\, \phi\,,\quad A_\mu \rightarrow \eta \,A_\mu\,,\quad x^\mu \rightarrow x^\mu/\left(q\eta\right)\,.
    \end{equation}
    The Lagrangian density (\ref{e:LagrangianDensity_AH_0}) expressed in terms of the rescaled coordinates and fields is given by
    \begin{equation} \label{e:LagrangianDensity_AH}
        \mathcal{L}=q^2\eta^4\left(-\frac{1}{4}F_{\mu\nu}F^{\mu\nu}+\frac{1}{2}\overline{D_{\mu}\phi} D^{\mu}\phi-\frac{\lambda}{8}(1-\overline{\phi}\phi)^2\right)\,,
    \end{equation}
    where the covariant derivative is now defined as $D_\mu=\partial_\mu - i A_\mu$ and the self-coupling constant is $\lambda = \tilde{\lambda}/q^2$. 

    The Euler-Lagrange equations that govern the dynamics of this model are expressed as
    \begin{eqnarray}\label{e:EL_AH}
            D_0D_0 \phi \hspace{-0.2cm}&=&\hspace{-0.2cm} D_iD_i\phi+  \frac{\lambda}{2}(1-\overline{\phi}\phi)\phi\,,\\ 
            \partial_j F_{0j} \hspace{-0.2cm}&=&\hspace{-0.2cm} -\dfrac{i}{2}\left(\overline{\phi} D_0\phi-\phi \overline{D_0\phi}\right)\,,\label{e:Gauss_Law} \\ 
            \partial_{0}F_{0j} \hspace{-0.2cm}&=&\hspace{-0.2cm}   \partial_{k}F_{kj}-\frac{i}{2}\left(\overline{\phi} D_j\phi-\phi \overline{D_j\phi}\right)\,,
    \end{eqnarray}
    where $j,k = \{1,2\}$. It is worth noting that equation (\ref{e:Gauss_Law}) corresponds to Gauss's law. This equation is essential from a dynamical point of view, as it ensures that the evolution of the system is orthogonal to the gauge orbit through $(\phi,A)$. Otherwise, a pure gauge transformation would have associated kinetic energy, even though the physical state remains the same since they are connected by a gauge transformation. Moreover, Gauss's law can be rewritten as
    \begin{equation}\label{e:GaussLaw}
        (\nabla^2 - \overline{\phi}\phi)A_0 = \partial_i\partial_0A_i + \dfrac{i}{2}(\overline{\phi}\partial_0\phi - \phi \partial_0\overline{\phi})\,.
    \end{equation}
    This equation is not independent but is instead determined at each instant of time by $\phi$ and $A_i$. Interestingly, if we fix a gauge in which the right-hand side of (\ref{e:GaussLaw}) vanishes, we can impose the so-called temporal gauge, where $A_0 = 0$. Henceforth, we shall assume this gauge fixing.

    The Lorentz invariance of the theory implies the conservation of energy by virtue of Noether's theorem (see \autoref{t:Noether}). From the energy-momentum tensor
    \begin{equation}
        T_{\mu\nu} = - F_{\mu}^{\;\sigma}F_{\nu\sigma} + \dfrac{1}{2}\overline{D_\mu\phi}D_\nu\phi  + \dfrac{1}{2}\overline{D_\nu\phi}D_\mu\phi - \eta_{\mu\nu}\mathcal{L}\,,
    \end{equation}
    we can identify the energy functional $E$ corresponding to the Lagrangian density (\ref{e:EL_AH}) with the $T_{00}$ component. This energy functional can be decomposed into the sum of a kinetic $T$ and a potential $V$ term of the form
    \begin{eqnarray}
        T \hspace{-0.2cm} &=& \hspace{-0.2cm}\dfrac{1}{2}\int_{\mathbbm{R}^2}\left(\partial_0A_i\partial_0A_i + \partial_0\overline{\phi}\partial_0\phi \right)\,dx_1dx_2\,,\label{e:T_AH}\\ 
        V \hspace{-0.2cm} &=&\hspace{-0.2cm} \dfrac{1}{2}\int_{\mathbbm{R}^2}\left(B^2 + \overline{D_i\phi}D_i\phi + \frac{\lambda}{4}(1-\overline{\phi}\phi)^2 \right)\,dx_1dx_2\,, \label{e:StaticEnergy_AH}
    \end{eqnarray}
    where $B = F_{12}$ is the magnetic field. The potential term (\ref{e:StaticEnergy_AH}) corresponds to the free energy of the Ginzburg-Landau theory of superconductivity in the non-relativistic limit \cite{Ginzburg:1950}. In this theory, $\phi$ represents an effective wave function of the superconducting electrons and $|\phi(\vec{x})|^2$ accounts for the concentration of superconductor electrons in the bulk. In addition, the self-coupling constant $\lambda$ determines whether the material behaves as a Type I ($\lambda<1$) superconductor, where the range of the matter self-interaction exceeds that of the electromagnetic one, or Type II ($\lambda>1$) superconductor, where now the range of the electromagnetic self-interaction exceeds that of the matter one. The main difference between these two regimes is that, whilst the external magnetic field $H$ is completely expelled from the superconductor in a Type I superconductors below a critical field $H < H_c$ -- Meissner effect \cite{Meissner:1933} --, a Type II superconductor exhibits two critical fields delimitating an intermediate region -- the mixed state -- between the superconductor and the normal state $H_{c1} < H < H_{c2}$, where the magnetic flux penetrates the bulk gradually in tubes of magnetic field that carry the same magnetic flux, forming a regular lattice \cite{Abrikosov:1957}. These tubes of "quantised" magnetic field are referred to as \textit{vortices}. 

    These vortex lines appear in the $2+1$ dimensional Abelian-Higgs model as planar topological solitons. Note that the transformation properties of one-forms imply that the gauge field transforms as 
    \begin{equation}
        A_{\mu}(\vec{x}) \to \lambda A_{\mu}(\lambda \vec{x})\,,
    \end{equation}
    under a spatial scaling, consequently, the field strength tensor transforms as 
    \begin{equation}
        F_{\mu\nu}(\vec{x}) \to \lambda^2 F_{\mu\nu}(\lambda \vec{x})\,.
    \end{equation}
    Due to the presence of this last term, the energy functional develops a minimum, which ensures that finite-energy solutions are not ruled out by virtue of Derrick's Theorem. For a vortex to have finite energy, it must comply with the following asymptotic conditions at infinity \begin{equation}\label{e:BoundaryConditions_AH}
       \overline{\phi} \phi \, \big|_{\mathbbm{S}_\infty^1} = 1\,, \qquad D_i\phi \big|_{\mathbbm{S}_\infty^1} = 0\,, \qquad B \big|_{\mathbbm{S}_\infty^1} = 0~,
    \end{equation}
    deduced from the structure of the static energy functional $ V$. 
    
\begin{figure}[htb]
    \vspace{-2cm}\centering{\hspace{-0.2cm}\includegraphics[width=1.15\linewidth]{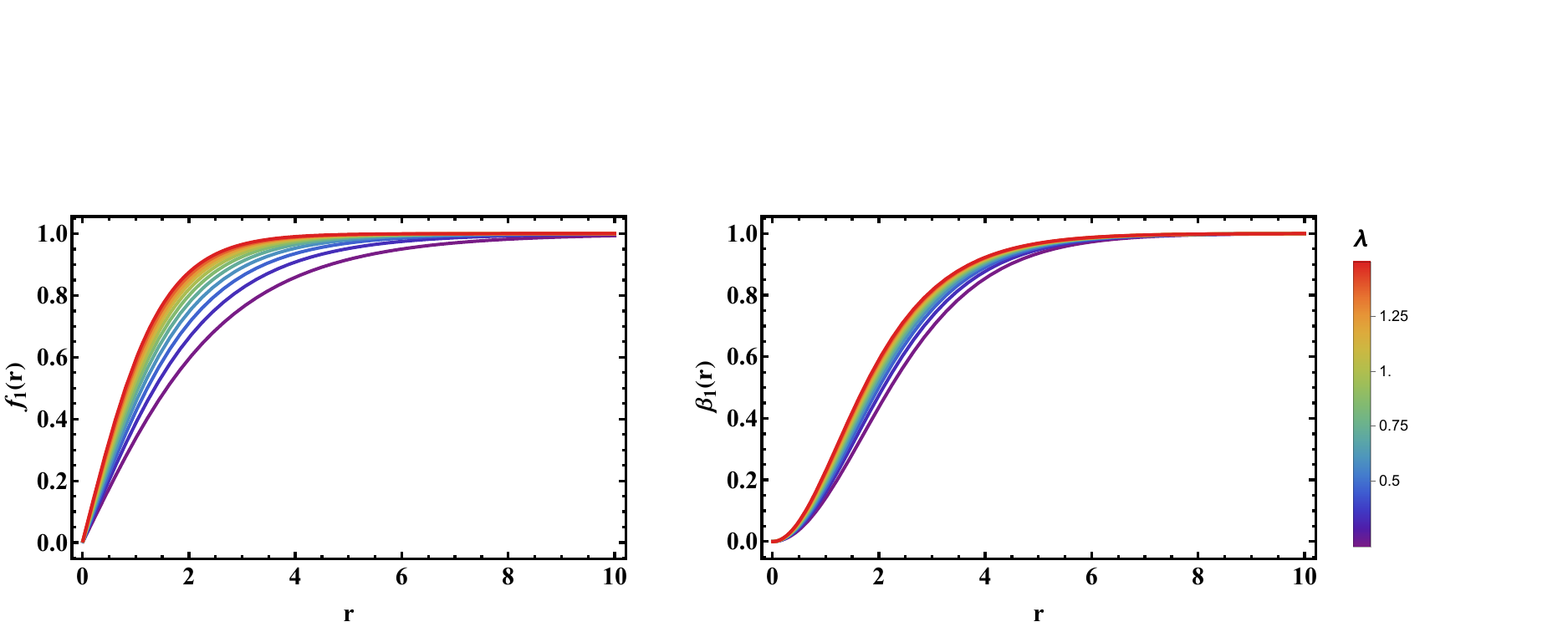}
        }
    \centering{\hspace{-0.2cm}\includegraphics[width=1.04\linewidth]{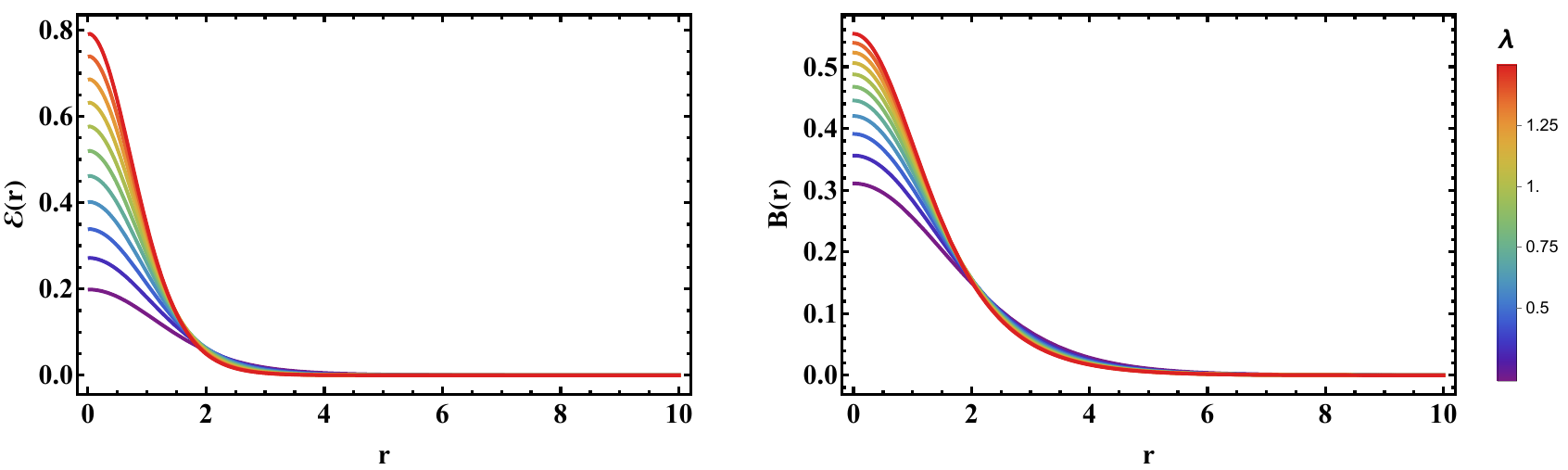}
        }
      
    \caption{Upper panel: Profiles of the scalar $f_n(r)$ and vector $\beta_n(r)$ components of the one-vortex. Lower panel: Radial profiles of the energy density $\varepsilon(r)$ and the magnetic field $B(r)$ associated with the one-vortex. The colour palette indicates different values of the self-coupling constant $\lambda$.}
        \label{f:profiles_AH}
    \end{figure}

    The first asymptotic condition suggests that the complex scalar field $\phi$ must satisfy 
    \begin{equation}
        \phi^\infty: =\lim_{||\vec{x}||\rightarrow\infty}\phi = e^{i\chi(\theta)},
    \end{equation}
    where $\theta$ is the angular coordinate. This means that vortices can be classified by the winding number $n$ of the map $\phi|_\infty : \mathbbm{S}_\infty^1 \rightarrow \mathbbm{S}^1$, which labels the homotopy classes of the first homotopy group $\pi_1(\mathbbm{S}^1) \cong \mathbbm{Z}$. Moreover, the second asymptotic condition suggests that the vector fields must satisfy \begin{equation}
    A_j^\infty: =\lim_{||\vec{x}||\rightarrow\infty}A_j = -ie^{-i\chi(\theta)}\partial_je^{i\chi(\theta)}.
    \end{equation}
    A direct consequence is that the magnetic flux of these configurations is classically quantised
    \begin{equation}
        \Theta = \int_{\mathbb{R}^2}B\, dx_1 dx_2= \oint_{\mathbbm{S}^1_{\infty}}\vv{A}\cdot \vv{dl} = \chi(2\pi) - \chi(0) =  2 \pi n\,, \quad n\in\mathbbm{Z}\,.
    \end{equation}
    In order to find static circularly-symmetric $n$-vortex solutions, we use the so-called \textit{Nielsen-Olesen ansatz} \cite{Nielsen:1973}
    \begin{equation}\label{e:AnsatzRadial_AH}
        \Phi^v(r,\theta)=f_n(r) e^{i n \theta}\,, \hspace{0.4cm}  A_r^v(r,\theta)=0 \,,  \hspace{0.4cm} A^v_\theta(r,\theta)= \frac{n\, \beta_{n}(r)}{r}\,,
    \end{equation} 
    where 
    \begin{equation}
        \lim_{r \to \infty} f_n(r) = 1\,, \quad \lim_{r \to \infty} \beta_n(r) = 1\,,
    \end{equation}
    to fulfil the asymptotic conditions (\ref{e:BoundaryConditions_AH}). Here, we have used the following convention 
    \begin{eqnarray}
        A_r = A_1\cos \theta + A_2\sin \theta\,, \qquad A_{\theta} = - A_1\sin \theta + A_2\cos \theta\,.
    \end{eqnarray}
    In addition, the so-called radial gauge $A_r = 0$ has been fixed. In terms of the Nielsen-Olesen ansatz (\ref{e:AnsatzRadial_AH}), the Euler-Lagrange equations (\ref{e:EL_AH}) are rewritten as 
    \begin{align}\label{e:EL_radial_AH}
        \dfrac{d^2 f_n}{d r^2}+\frac{1}{r}\dfrac{d f_n}{dr }-\frac{n^2}{r^2}(1-\beta_n)^2 f_n+\dfrac{\lambda}{2}(1-f_n^2)f_n&=0\,,\\
        \dfrac{d^2\beta_n}{d r^2}-\frac{1}{r}\frac{d \beta_n}{d r}+(1-\beta_n) f_n^2 &=0\,.
    \end{align}
    To guarantee the regularity of the solutions, the boundary conditions 
    \begin{equation}
        f_n(0) = 0\,, \quad \beta_n(0) = 0\,,
    \end{equation}
    must be imposed. By using the Nielsen-Olesen ansatz, the static energy and the magnetic field are now given by
    \begin{align}
            V =& \hspace{0.1cm} \pi \int_{0}^{\infty} \left(\left(\dfrac{df_n}{dr}\right)^2 + \dfrac{n^2}{r^2}\left(\dfrac{d\beta_n}{dr}\right)^2 + \dfrac{n^2}{r^2}(1 - \beta_n)^2f_n^2 + \dfrac{\lambda}{4}(1 - f_n^2)^2\right)\,r\,dr\,,\\
            B^v =& \hspace{0.1cm}  \dfrac{n \beta_n'(r)}{r}\,.
    \end{align}  
    Unfortunately, no analytical non-trivial solutions of (\ref{e:EL_radial_AH}) are known, and the system must be solved numerically (although analytical vortex solutions in closely related models have been found \cite{Bazeia:2018,Fuertes:2022}). For example, the vortex profiles can be obtained by adopting the gradient flow method. In this approach, we first assume an initial configuration that fulfils the boundary and asymptotic conditions. Then, the initial configuration is evolved using the time-dependent field equations with the second-order time derivatives replaced by first-order time derivatives, thereby introducing a damping mechanism that drives the system toward a static solution
    \begin{align}
        \dfrac{df_n}{dt}&= \dfrac{d^2 f_n}{d r^2}+\frac{1}{r}\dfrac{d f_n}{dr }-\frac{n^2}{r^2}(1-\beta_n)^2 f_n+\dfrac{\lambda}{2}(1-f_n^2)f_n\,,\\
        \dfrac{d\beta_n}{dt}&= \dfrac{d^2\beta_n}{d r^2}-\frac{1}{r}\frac{d \beta_n}{d r}+(1-\beta_n) f_n^2 
        \,.
    \end{align}
    The flow is in the direction where the potential energy $V$ decreases most steeply. Importantly, it can be proved that the gradient flow is orthogonal to gauge orbits, thereby satisfying Gauss's law. The profiles of the scalar and vector components of a vortex with $n=1$ are depicted in \autoref{f:profiles_AH} for different values of the coupling constant $\lambda$ as well as its energy density and magnetic field. 

\begin{figure}[htb]
    \centering{\hspace{-0.2cm}\includegraphics[width=1.0\linewidth]{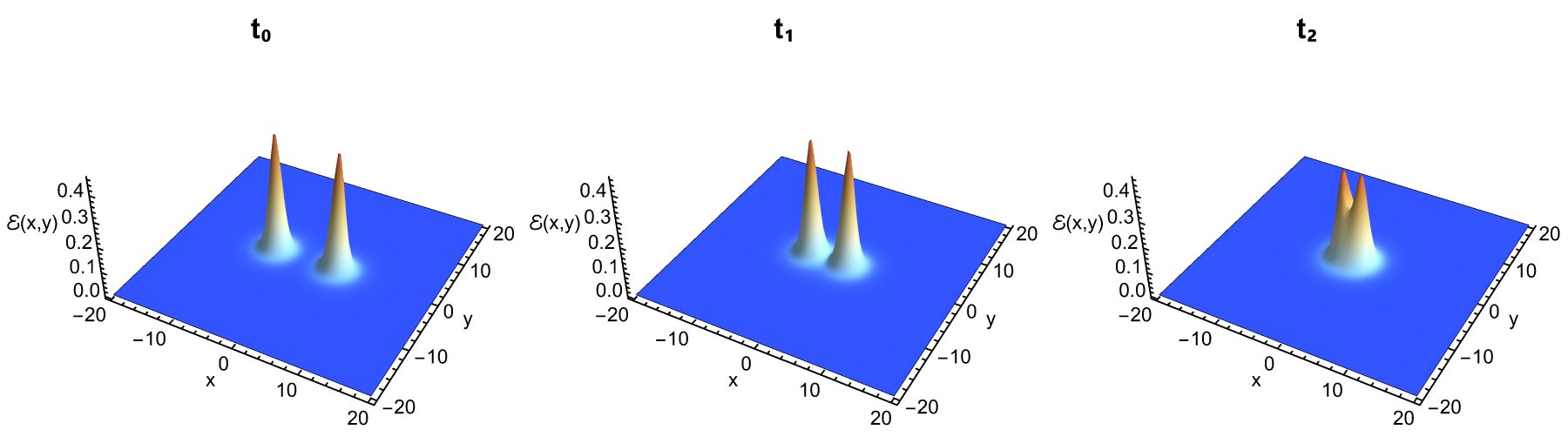}
        }
    \centering{\hspace{-0.2cm}\includegraphics[width=1.0\linewidth]{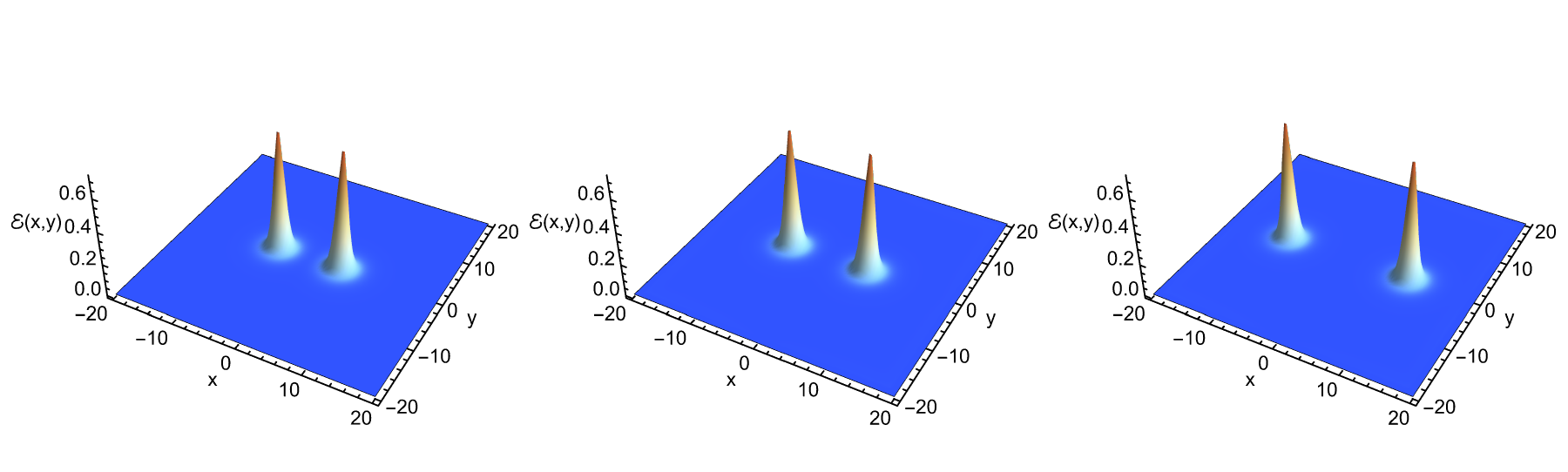}
        }
      
    \caption{Mutual interaction between two initially static one-vortices at three distinct times, $t_0 < t_1 < t_2$. Upper panel: Static attractive force between two one-vortices with $\lambda = 0.7$.
    Lower panel: Static repulsive force between two one-vortices with $\lambda = 1.3$. In both cases, the one-vortices are assumed to be initially located at $(x_0,y_0) = (\pm 5,0)$.}
    \label{f:StaticForce_Vortex}
    \end{figure}

    For a winding number $n \geq 2$, it has been shown that vortices attract each other for $\lambda < 1$ but they repel each other for $\lambda > 1$ \cite{Rebbi:1979}. The asymptotic interaction potential between two well-separated unit vortices has been computed by different routes in \cite{Bettencourt:1995} and \cite{Speight:1997} for any value of the self-coupling constant $\lambda$. Moreover, an explicit formula for the interaction energy of $n$ closely spaced vortices has been derived recently \cite{Speight:2025}. In \autoref{f:StaticForce_Vortex}, we present a representative example illustrating the static attraction and repulsion between two one-vortices. To generate the initial condition, we have assumed the so-called \textit{Abrikosov ansatz} or \textit{product ansatz}, which states that a multi-vortex solution for well-separated vortices can be approximated by
    \begin{equation}
        \hat{\Phi}^v(\vec{x}) = \prod_i\Phi^v(\vec{x} - \vec{x}_i)\,, \qquad \hat{A}^v_{\mu}(\vec{x}) = \sum_i A^v_{\mu}(\vec{x} - \vec{x}_i)\,.
    \end{equation}
    
    However, the limiting $\lambda = 1$ case is special. For that critical value, the magnetic repulsion and the scalar attraction are fully balanced, and vortices do not interact statically with each other. This suggests that the stress components of the energy-momentum tensor vanish \cite{Vega:1976}, and that a BPS bound has been saturated. Indeed, in that limit, the energy functional can be bounded from below by a topologically conserved quantity 
    \begin{align}
            V =& \hspace{0.1cm} \dfrac{1}{2}\int_{\mathbbm{R}^2}\left(\left[B^2 \pm \frac{1}{2}(1-\overline{\phi}\phi)^2\right]^2 + |D_1\phi \pm D_2\phi|^2\right)\,dx_1dx_2\nonumber\\
            &\pm \dfrac{1}{2}\int_{\mathbbm{R}^2}\left(B(1-\overline{\phi}\phi) + i\,\epsilon_{jk}\,\overline{D_j\phi} D_k\phi\right)\,dx_1dx_2\,.
    \end{align}
    Operating in the last term and applying the Stokes' theorem it can be verified that  
    \begin{align}\label{e:BPS_bound_AH}
            V &= \dfrac{1}{2}\int_{\mathbbm{R}^2}\left(\left[B^2 \pm \frac{1}{2}(1-\overline{\phi}\phi)^2\right]^2 + |D_1\phi \pm D_2\phi|^2\right)\,dx_1dx_2 \pm \dfrac{1}{2}\int_{\mathbbm{R}^2}B\,dx_1dx_2\nonumber\\
            &\geq \dfrac{1}{2}\bigg|\int_{\mathbbm{R}^2}B\,dx_1dx_2\bigg| = \pi\,|n|\,.
    \end{align}
    The saturation of the BPS bound leads us to a system of coupled Bogomolny equations
    \begin{equation}\label{e:BPS_AH}
        D_1\Phi^v \pm i D_2\Phi^v = 0\,, \quad B^v \pm \frac{1}{2}(1-\overline{\Phiv}\Phi^v) = 0\,.
    \end{equation}
    This system of first-order differential equations has been extensively analysed by Taubes in \cite{Taubes:1980,Taubes1:1980,Jaffe:1980}. Specifically, Taubes proved that, depending on the choice of sign, Bogomolny equations admit either $n$-vortex or $n$-antivortex configurations. However, such configurations cannot coexist statically, as vortices and antivortices attract each other even in the BPS limit. Furthermore, Taubes also made the important observation that the gauge vector field can be eliminated from the Bogomolny equations, combining both Bogomolny equations into a single expression. To see this, define the gauge-invariant field $h = \log |\Phi^v|^2$, so that $\Phi^v = |\Phi^v|e^{i \chi} = e^{\frac{1}{2}h + i \chi}$. Then, using the first Bogomolny equation to write the gauge field $A_{\mu}^v$ in terms of $\Phi^v$ and substituting these expressions into the second Bogomolny equation yields
    \begin{equation}
        \nabla^2h + 1 - e^h = 0\,.
    \end{equation}
    This equation holds everywhere except at the zeros of $\Phi^v$, where the new field $h$ develops logarithmic singularities. These singularities can be accounted for by including delta-function sources, leading to the modified equation  
    \begin{equation}\label{e:Taubes}
        \nabla^2h + 1 - e^h = 4\pi \sum_{i = 1}^n\delta^2(\vec{x} - \vec{x}_i)\,.
    \end{equation}
    
    In the BPS limit, Weinberg proved through an index theorem argument that, given a value of vorticity $n$, there is a $2n$-parameter family of $n$-vortex solutions, in other words, a $n$-vortex configuration exhibits $2|n|$ zero-modes \cite{Weinberg:1979}. As a result, the moduli space of solutions of the Bogomolny equations is a $2n$-dimensional manifold $\mathcal{M}_n$. The corresponding moduli space metric was investigated by Samols in \cite{Samols:1992} using Taubes's equation, and showed that this moduli space metric captures the translational and rotational invariance of the full field theory. The moduli space of BPS $n$-vortices may be identified with the space of $n$ unordered points $\{z_i\}_{i=1}^n$ on the complex plane, that is, $\mathcal{M}_n = \mathbbm{C}^n/S_n$, where $S_n$ denotes the permutation group of $n$ elements. However, this identification only provides good local coordinates, not a global parametrisation of $\mathcal{M}_n$, because singularities emerge once two or more vortices coincide. A good global system of coordinates on $\mathcal{M}_n$ is provided by the ordered set $\{p_i\}_{i=1}^n$ consisting of the coefficients of the following monic complex polynomial
    \begin{equation}
        p(z) = \prod_{i = 1}^n(z - z_i) = z^n + p_1z^{n - 1} + \ldots + p_n\,.
    \end{equation}
    With this choice, the moduli space is $\mathcal{M}_n = \mathbbm{C}^n$,  which is smooth and free of singularities.

    Now, let us investigate the spectrum of linear perturbations around the vortex solution. To proceed, let us consider a perturbation of the form
    \begin{align}\label{e:Perturbation_AH}
        \phi_j(\vec{x}) &= \Phi^v_j(\vec{x}) + \delta\,\varphi_j(\vec{x})e^{i\omega\theta}e^{i\omega t}\,,\\
        A_j(\vec{x}) &= A_j^v(\vec{x}) + \delta \,a_j(\vec{x})e^{i\omega t}\,,
    \end{align}
    where $\delta$ denotes the amplitude of the fluctuations. It is usual to impose the background gauge condition 
    \begin{equation}\label{e:Background_Gauge}
        \partial_{j}a_j(\vec{x}) - (\Phi^v_1(\vec{x})\varphi_2(\vec{x}) - \Phi^v_2(\vec{x})\varphi_1(\vec{x})) = 0\,,
    \end{equation}
    to discard pure gauge fluctuations and fix the gauge condition on the fluctuation modes. Upon substituting the perturbed configuration (\ref{e:Perturbation_AH}) into the equations of motion (\ref{e:EL_AH}), we expand up to first order in the amplitude $\delta$, and taking into account the gauge condition (\ref{e:Background_Gauge}) we arrive at the following spectral problem
    \begin{equation}\label{e:SpectralProblem_AH}
    \mathcal{H}
    \begin{pmatrix}
        \varphi_1(r) \\
        \varphi_2(r) \\
        a_1(r) \\
        a_2(r) \\
    \end{pmatrix}
    = \omega_{n,m}^2
    \begin{pmatrix}
        \varphi_1(r) \\
        \varphi_2(r) \\
        a_1(r) \\
        a_2(r) \\
    \end{pmatrix},
    \end{equation}
    where the second-order fluctuation operator $\mathcal{H}$ reads
    \begin{equation}
        \resizebox{\textwidth}{!}{$
        \mathcal{H} = \left( \begin{array}{cccc}
        -\nabla^2 + |\Phi^v|^2 & 0 & -2 \widetilde{D}_1 \Phi^v_2 & 2 \widetilde{D}_1 \Phi^v_1 \\
        0 & -\nabla^2 + |\Phi^v|^2 & -2 \widetilde{D}_2 \Phi^v_2 & 2 \widetilde{D}_2 \Phi^v_1 \\
        -2 \widetilde{D}_1 \Phi^v_2 & -2 \widetilde{D}_2 \Phi^v_2 & -\nabla^2 + \frac{3\lambda}{2} (\Phi^v_1)^2 + \left(1 + \frac{\lambda}{2}\right) (\Phi^v_2)^2 - \frac{\lambda}{2} + A^v_j A^v_j & -2 A^v_j \partial_j + (\lambda - 1) \Phi^v_1 \Phi^v_2 \\
        2 \widetilde{D}_1 \Phi^v_1 & 2 \widetilde{D}_2 \Phi^v_1 & 2 A^v_j \partial_j + (\lambda - 1) \Phi^v_1 \Phi^v_2 & - \nabla^2 + \left(1 + \frac{\lambda}{2}\right) (\Phi^v_1)^2 + \frac{3\lambda}{2} (\Phi^v_2)^2 - \frac{\lambda}{2} + A^v_j A^v_j
        \end{array} \right),$} \label{e:H_operator_AH}
    \end{equation}
    with $\widetilde{D}_i\,\Phi^v_j = \partial_i\,\Phi^v_j + \epsilon^{jk}A^v_i\Phi^v_k$\,. Here, the subscript $n$ denotes the winding number of the configuration and $m$ labels the mode. 

\begin{figure}[htb]
        \centering{\includegraphics[width=0.5\linewidth]{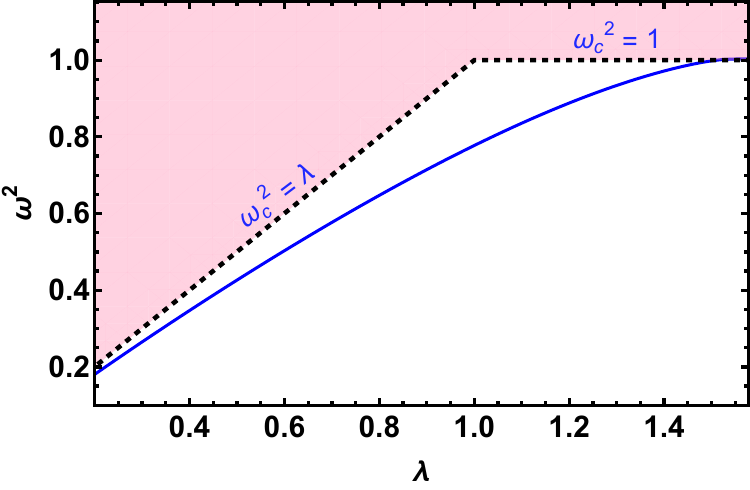}
        }
        \caption{Spectral flow of the radial mode supported by a one-vortex. The blue curve denotes the frequency $\omega^2$ of the radial mode and the shaded red area represents the continuous spectrum.}
        \label{f:Spectrum_AH_n1}
    \end{figure}

\begin{figure}[htb]
        \vspace{-2.5cm}\hspace{-0.1cm}\centering{\hspace{-0.1cm}\includegraphics[width=1.15\linewidth]{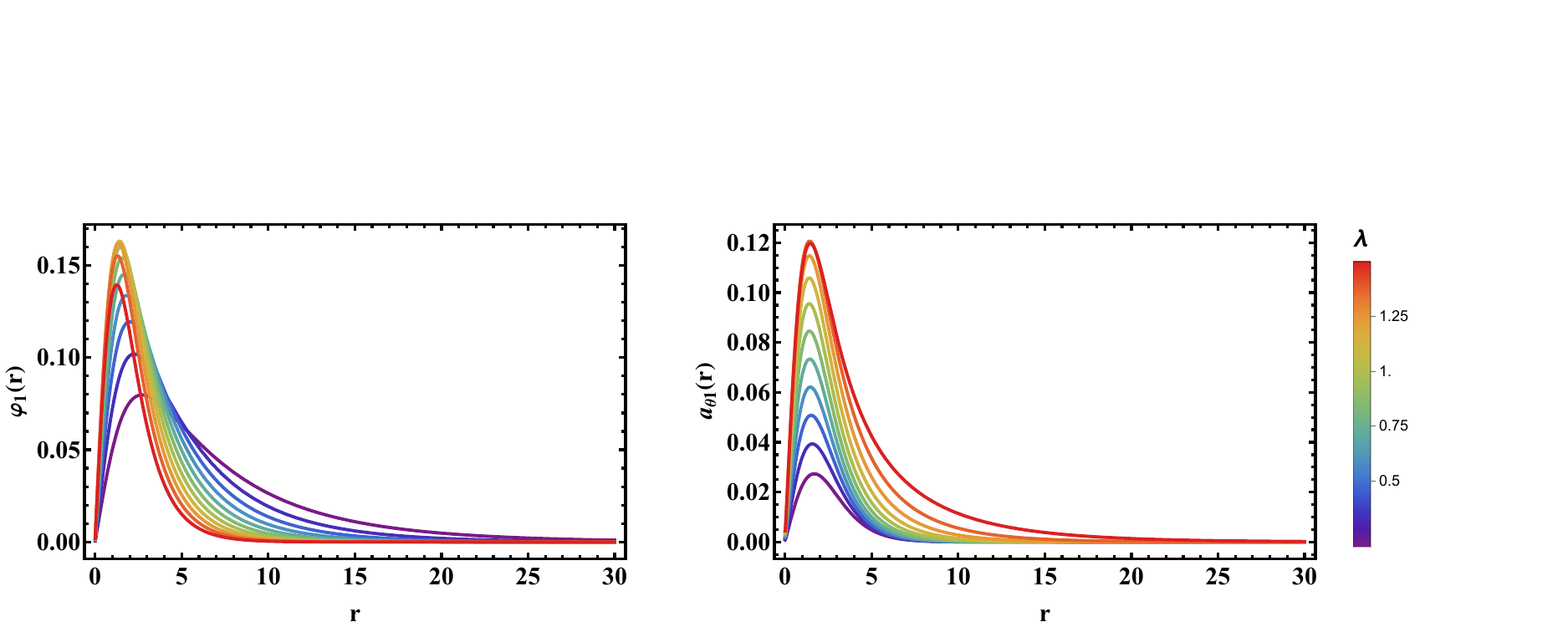}
        }
      
        \caption{Scalar (left) and vector (right) component profiles of the one-vortex radial bound mode for different values of the self-coupling constant $\lambda$. The modes are normalised using the $L^2(\mathbbm{R}^2)\bigoplus\mathbbm{R}^4$ norm.}
        \label{f:BoundMode_AH}
    \end{figure}

    The complexity of the spectral problem (\ref{e:SpectralProblem_AH})-(\ref{e:H_operator_AH}) entails that the discrete spectrum of (\ref{e:H_operator_AH}) must be obtained numerically. This spectral problem was originally studied in \cite{Goodband:1995} for any winding number $n$ and self-coupling value $\lambda$. Expanding the scalar and vector perturbations in terms of angular and total angular momentum states, respectively, the authors analysed vortex excitations with different angular dependencies. Remarkably, this spectral problem has recently been further investigated in \cite{Izquierdo:2024}, and the authors have achieved to identify the precise angular dependence of the eigenfunctions. There, it was shown that the lowest positive bound mode for a $n$-vortex exhibits radial symmetry. Specifically, a vortex with winding number $n = 1$ only admits one positive bound mode, in agreement with previous results \cite{Goodband:1995,Arodz:1991,Kojo:2007}. For modes with radial symmetry, the spectral problem  (\ref{e:SpectralProblem_AH}) reduces to 
    \begin{equation}\label{e:SpectralProblem_AH_radial}
    \mathcal{H}_r
    \begin{pmatrix}
        \varphi(r) \\
        a_{\theta}(r) \\
    \end{pmatrix}
    = \omega_{n,m}^2
    \begin{pmatrix}
        \varphi(r) \\
        a_{\theta}(r) \\
    \end{pmatrix},
    \end{equation}
    where
    \begin{equation} \label{e:OperatorH_AH}
        \resizebox{\textwidth}{!}{$
        \mathcal{H}_r =
        \left( \begin{array}{cc}
             -\dfrac{d^2}{dr^2}  - \dfrac{1}{r}\dfrac{d}{dr}  + \left(\dfrac{3}{2}\lambda f_{n}(r)^{2} -  \dfrac{\lambda}{2} + \dfrac{n^{2}}{r^{2}} - \dfrac{n^{2}\beta_{n}(r)}{r^{2}}\left( 2 - \beta_{n}(r) \right)\right) 
             & - \dfrac{2 n f_{n}(r)}{r}\left( 1 - \beta_{n}(r)\right)\\
             - \dfrac{2 n f_{n}(r)}{r}\left( 1 - \beta_{n}(r)\right) 
             & - \dfrac{d^2}{dr^2}  - \dfrac{1}{r}\dfrac{d}{dr}  + \left(f_{n}(r)^2 + \dfrac{1}{r^2}\right)
        \end{array} \right),
    $}
    \end{equation}
    $\varphi$ and $a_{\theta}$ denoting respectively the fluctuations of the complex scalar field $\phi$ and of the angular component $A_\theta$ of the vector field. As an illustrative example, we show the spectrum in  \autoref{f:Spectrum_AH_n1} the spectral flow with the parameter $\lambda$ of the radial mode held by a one-vortex and we depict the profile of the corresponding mode in \autoref{f:BoundMode_AH}.

    The spectral problem (\ref{e:SpectralProblem_AH}) has been examined in detail in the critical case $\lambda = 1$. In \cite{Vega:1976} the authors pointed out that the critical value of the self-coupling constant should be connected with the supersymmetric extension of the model. Indeed, in \cite{Edelstein:1994} the authors proved that the critical value of the self-coupling constant $\lambda$ is the necessary condition to extend the $\mathcal{N} = 1$ supersymmetric Abelian-Higgs model in $2+1$ dimensions to a $\mathcal{N} = 2$ supersymmetric theory, thereby recovering the BPS bound and the Bogomolny equations within that formalism. Furthermore, the study of the spectral problem is significantly simplified in that limit thanks to the supersymmetric structure. In \cite{Izquierdo:2016,Izquierdo1:2016} the authors derived a supersymmetric partner of the second-order fluctuation operator, whose eigenfunctions and eigenfrequencies related to those of the original operator. The advantage of this construction is that the associated spectral problem becomes significantly simpler.
    
    To conclude, we note that the Abelian-Higgs model is not the only planar model featuring gauge vortices. Such vortices also appear in the Chern-Simons-Higgs \cite{Jackiw:1990,Lee:1990} and Maxwell-Chern-Simons-Higgs \cite{Min:1990} models. Moreover, unlike Abelian-Higgs vortices, the vortices in these latter theories carry electric charge. Notably, these models also exhibit a BPS limit. Although we have not considered other gauge symmetries here, it is worth emphasising that gauge vortices have also been found in models with different symmetry groups, for example, those with $CP(2)$ symmetry \cite{Casana:2017,Andrade:2019}.

\section{Collective coordinate method}\label{s:CCM}

    So far, we have focused our efforts on describing the topological soliton solutions arising in prototypical models. In particular, we have examined some of their fundamental properties and analysed their spectral structure. However, we have not yet addressed their dynamical behaviour. A significant challenge lies in the nonlinearity of these theories, as well as in the fact that most of them are non-integrable. These characteristics, combined with the infinitely many degrees of freedom inherent to field theories, make the analysis of soliton dynamics particularly difficult. Most of the times, one has to resort to numerical simulations.  

    Despite this difficulty, a standard approach to address partially this issue is to use the so-called \textit{collective coordinate method}, also known as the \textit{moduli space dynamics}, which allows for a semi-analytical treatment. The main idea behind this method is to replace the infinite-dimensional configuration space by a restricted set of field configurations, which are spanned by a finite number of parameters $X = \{X_i\}_{i=1}^N$, usually referred to as \textit{moduli}. These moduli parametrised a \mbox{submanifold} known as the \textit{moduli space} $\mathcal{M}_n = \{\Phi(\vec{x};X)\}$, with dimension $\text{dim}(\mathcal{M}_n) = N$. When these moduli are identified, they are promoted to time-dependent variables $X_i \rightarrow X_i(t)$. After that, the field configurations $\Phi(\vec{x};X)$ are inserted into the Lagrangian density of the field theory. Upon integrating over the space coordinates, the field theory Lagrangian is truncated to a mechanical Lagrangian. As an illustrative example, let us consider a scalar field theory in $1+1$ dimensions. Then, the mechanical Lagrangian reads
    \begin{equation}\label{e:Mech_Lag}
        L[X] = \int_{\mathbbm{R}} \mathcal{L}[\Phi(x,X)]\,dx = \dfrac{1}{2}g_{ij}(X)\dot{X}^i\dot{X}^j - V(X)\,,
    \end{equation}
    where the $g_{ij}(X)$ and $V(X)$ terms are given by
    \begin{eqnarray}
        g_{ij}(X) = \int_{\mathbbm{R}} \dfrac{\partial\Phi}{\partial X^i}\dfrac{\partial\Phi}{\partial X^j}\,dx\,, \quad \quad
        V(X) = \int_{\mathbbm{R}} \left[\dfrac{1}{2}\left(\dfrac{\partial \Phi}{\partial x}\right) + U(\Phi)\right]\,dx\,,
    \end{eqnarray}
    which come from the kinetic and potential terms of the original field theory, respectively. Here, the $g_{ij}(X)$ term accounts for the metric of the moduli space, whilst $V(X)$ represents an effective potential function defined on it. Generally, the metric is not constant, which implies that the moduli space is curved. The mechanical Euler-Lagrange equations related to (\ref{e:Mech_Lag}) read
    \begin{equation}\label{e:EL_Mech}
        g_{ij}\left[\ddot{X}^j + \Gamma^{j}_{kl}\dot{X}^k\dot{X}^l\right] + \dfrac{\partial V}{\partial X^i} = 0\,,
    \end{equation}
    where $\Gamma^{j}_{kl}$ is the Levi-Civita connection of the metric
    \begin{equation}
        \Gamma^{j}_{kl} = \dfrac{1}{2}g^{ij}\left(\dfrac{\partial g_{li}}{\partial X^k} + \dfrac{\partial g_{ki}}{\partial X^l} - \dfrac{\partial g_{kl}}{\partial X^i} \right)\,.
    \end{equation}
    By multiplying (\ref{e:EL_Mech}) by $\dot{X}_i$ and summing over the index $i$, the energy conservation law is obtained
    \begin{equation}
         \dfrac{d}{dt}\left(\dfrac{1}{2}g_{ij}\dot{X}^i\dot{X}^j + V \right) = \dfrac{dE}{dt} = 0\,.
    \end{equation}

    A special case occurs when the moduli considered define the set of energetically equivalent BPS configurations \cite{Manton:1982}. In that scenario, there are no static forces, and the effective potential $V$ is a constant that acts as a holonomic constraint. Indeed, the effective potential attains its absolute minimum in the relevant topological sector. As a result, the equations of motion (\ref{e:EL_Mech}) are reduced to
    \begin{equation}
        \ddot{X}^j + \Gamma^{j}_{kl}\dot{X}^k\dot{X}^l = 0\,.
    \end{equation}
    This equation corresponds to a geodesic on the moduli space, which, in these cases, it is usually referred to as \textit{canonical moduli space} in the literature. At this point, it is crucial to note that, when the initial conditions in the original field theory are tangent to $\mathcal{M}_n$ and the solitons move at slow speeds, the motion must remain close to $\mathcal{M}_n$ due to the conservation of the energy. This picture explains the triumph of this approach in capturing the low-speed motion in BPS theories. In particular, it has successfully described the $\pi/2$ scattering of two $SU(2)$ monopoles \cite{Atiyah:1985,Atiyah:1988} and also of two Abelian-Higgs vortices \cite{Ruback:1988,Ruback1:1988,Rebbi:1992, Thatcher:1997}, or the scattering of CP$^1$ lumps \cite{Ward:1985} and Skyrmions \cite{Atiyah:1993,Gisiger:1994}. Moreover, the head-on collisions of BPS $3$-vortex solutions 
    with equidistant collinear and equilateral arrangements have been described very recently \cite{Rees:2025}. 

    However, in order to describe most phenomena involving solitons, it is not sufficient to restrict the dynamics to energetically equivalent field configurations. For instance, kink–antikink annihilation cannot be captured under this assumption. To accurately describe the dynamics in such cases, one must go beyond geodesic motion and introduce additional moduli, which in turn lead to the emergence of a non-constant effective potential $V$ \cite{Manton:1988}. The main challenge lies in the absence of a canonical choice of moduli or a unique field ansatz. The only guiding principle is that, to construct a suitable family of field configurations, the ansatz should approximate as closely as possible the actual field profiles occurring during the process under study\footnote{Here, we paraphrase Richard Feynman's quote on the scientific method for discovering a new physical law: \textit{"First, we guess it. Then, we compute the consequences of the guess to see what it would imply, and then we compare those computation results to nature. If it disagrees with the experiment, it is wrong. And that simple statement is the key to science."}}. Typically, the general strategy involves starting with the massless modes and subsequently including the massive excitations. Naturally, the relevant collective excitations will depend on the specific configuration; for example, whether one is dealing with a single soliton or a multi-soliton state. This generalised moduli space approximation has succeeded in capturing the Lorentz contraction of a single kink \cite{Rice:1983} or in recovering the existence of a fractal structure in the final state formation in short-range kink collisions \cite{Sugiyama:1979,Campbell:1983,Weigel:2019,Wereszczynski:2021,Manton:2021,Manton:2022,Huidobro:2022,Ciurla:2023} or in long-range kinks collisions \cite{Campos:2025}, as well as in the scattering of excited vortices \cite{Manton:2024}. This study has also been extended to describe kink collisions with a moving center of mass \cite{Halcrow:2023}. In addition, this approach has allowed to explain the spectral wall phenomenon triggered by normal or quasi-normal modes \cite{Adam:2019,Oles:2020,Oles:2021,Queiruga:2025}. Interestingly, this technique has recently been applied in the context of compact kink scattering \cite{PawelKlimas:2024}.

    Although not treated in this thesis, the collective coordinate method is also important to quantise solitons \cite{Christ:1975,Gervals:1976,Gervais:1976,Jevicki:1980,Rajamaran:1987}. Without going into detail, the procedure involves expanding the solitons in terms of their normal modes and promoting the collective coordinates to time-dependent variables, paying special attention to the zero mode, as its associated zero frequency leads to divergences. The corresponding effective Lagrangian is then computed, from which the associated Hamiltonian is derived. Finally, canonical commutation relations are applied to the amplitudes and their conjugate momenta, after promoting them to operators. 

    Throughout this thesis, we adopt the collective coordinate method as one of the main tools for gaining insight into the models under consideration. It is important to emphasise that obtaining analytical solutions is not always feasible. As previously mentioned, this is a semi-analytical approach: although the coefficients of the effective Lagrangian can be determined analytically, the resulting equations remain highly non-linear and thus require a numerical solver to determine their time evolution. In spite of that, we have been able to extract valuable information through this method, as well as predicting some features well-established in the literature. In the following two chapters, we will apply the collective coordinate method in the context of kinks and vortices respectively. In \autoref{c:Radiation}, we will include genuine radiation modes within this context for the first time in the $1+1$ dimensional $\phi^4$ model, and in \autoref{c:Vortex} we will generalise the moduli space metric of a single vortex by including its unique positive bound mode. Moreover, after introducing the concept of the sphaleron in \autoref{c:Sphalerons}, we shall investigate in \autoref{c:BPS_Sph} whether the geodesic motion remains a valid approximation for describing the simplest dynamics of this unstable solution within a model admitting a BPS state. Finally, this technique will also be employed in \autoref{c:Sph_False} to analyse the evolution of the oscillon that emerges following the collapse of the sphaleron.

        \chapter{Effective description of radiation in soliton dynamics}\label{c:Radiation}

This chapter is adapted from \cite{Navarro:2023}:
    \hspace{-2.5cm}
    \vspace{-0.3cm}
    \begin{figure}[H]
        \hspace{-2.5cm}
        \vspace{-0.3cm}
        \centering{\includegraphics[width=1.1\linewidth]{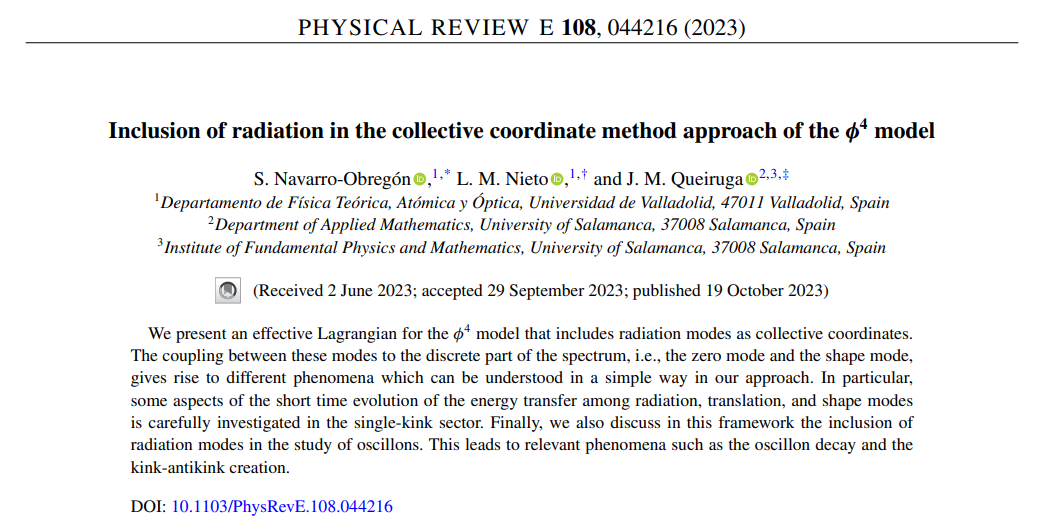}
        }
        \label{f:Thumbail_Radiation}
        \hspace{-2.5cm}
    \end{figure}

\section{Introduction}

    In \autoref{s:Topology} we explained the notion of a topological soliton. Among all of them, kinks are the prototypical examples. They appear in scalar field theories formulated in $1+1$ dimensions, which makes simpler their analytical treatment as well as the numerical computations. A particularly interesting model is the $\phi^4$ theory introduced in \autoref{ss:phi4}. This model possesses a kink solution that can be determined along with its linear spectrum of perturbations analytically. However, the analysis of the dynamics in the full non-linear theory is extremely involved, which often means that the equations have to be solved numerically. The origin of this complexity lies partly in the presence of multiple interaction channels and partly in the non-integrability of the model.

    In \autoref{s:CCM} we exposed that one method to reduce the complexity of the topological soliton dynamics is the collective coordinate method. Within this approach, the field theory Lagrangian is reduced to a mechanical one with a finite number of degrees of freedom. The standard strategy consists in introducing the lightest modes. Nevertheless, this effective point of view can be improved by introducing new coordinates which take into account internal degrees of freedom. But, even in the apparently simple case of the $\phi^4$ model, there is a complicated pattern of final states in scattering processes related to the non-integrability of the model, which cannot be explained satisfactorily by a simple choice of translational and internal oscillatory degrees of freedom \cite{Sugiyama:1979, Campbell:1983}.
    
    Recently, an important improvement has been made by means of the introduction of the \textit{Perturbative Relativistic Moduli Space} \cite{Manton:2022,Ciurla:2023}. This approach, unlike the standard collective coordinate method, can also accommodate some relativistic degrees of freedom. As shown in \cite{Manton:2022}, this quantitatively improves the agreement between the effective model and the field theory. However, neither of these approaches considers radiation degrees of freedom as generalised coordinates, i.e., they cannot describe radiation. The effect of radiation in soliton dynamics can be very relevant in certain violent processes, such as the kink-antikink annihilation, but it is also determinant in long-time dynamics \cite{Romanczukiewicz:2017,Dorey:2023} and may contribute to the fractal structure of the kink scattering. One of the purposes of this work is to introduce the radiation modes as generalised coordinates and study their role in certain dynamical processes. As we shall see, our approach is able to describe some relevant aspects of the short-time evolution of soliton dynamics and extract analytical insight. 

    Apart from the topologically non-trivial solutions of the $\phi^4$ model, there are other time dependent soliton-like structures that deserve special attention, the oscillons\footnote{Originally, they were referred to as "pulsons" in the literature.} \cite{Dashen:1974, Kudryavtsev:1975, Bogolyubskii:1976}. They are long-lived, oscillatory solutions that are topologically trivial, and unlike other time-dependent solitons such as Q-balls, they are not associated with any conserved charge. They are ubiquitous in a wide range of models from one to three dimensions \cite{Segur:1987, Salmi:2006, Gleiser:2007, Fodor1:2009, Salmi:2012, Manton:2023}, and they have found applications in many scenarios in theoretical physics, from dark matter \cite{Olle:2020,Kawasaki:2020,Arvanitaki:2020} to cosmology \cite{Hindmarsh:2008,Gorghetto:2021,Blanco:2021,Aurrekoetxea:2023,Sibiryakov:2024,Lozanov:2025}. Some of their characteristics such as profiles and life-time have been studied, mostly numerically, in the literature \cite{Honda:2002,Gleiser:2004,Gleiser:2008,Gleiser:2009,Gleiser:2019,Pujolas:2021,Barashenkov:2023}. Much less is known about their internal structure; see \cite{Dissel:2023,Blaschke:2024,Blaschke:2025,Blaschke1:2025,Evslin:2025} for recent publications. We will introduce radiation degrees of freedom in an effective model for the $\phi^4$ oscillon. To some extent, they are able to provide a decay channel for oscillons below the critical amplitude. In addition, the scattering modes, or more precisely, an effective version of them, are able to describe qualitatively some features of the internal modes hosted by the oscillon, including the decay into kink-antikink pairs. 

    This chapter is organised as follows. In \autoref{s:AnsatzRad}, we briefly review the state of the art in the effective description of radiation and introduce a new possible alternative. In \autoref{s:wobbling}, we introduce the radiation modes as collective coordinates and analyse the radiation emitted by a wobbling kink. After that, in \autoref{s:Oscil_Radi} we explore some analytic solutions to the lowest order in perturbation theory involving radiation. In \autoref{s:Trans_Oscil_Radi}, we introduce the zero mode as a collective coordinate and study its interaction with the rest of the modes. In \autoref{s:oscillon}, we extend our approach to describe oscillon dynamics. Finally, \autoref{s:conclusions_Radiation} contains our conclusions and further comments.  

\section{Inclusion of radiation in the collective coordinate method}\label{s:AnsatzRad}

    Until now, the effective description of radiation in the literature has been performed through the perturbative relativistic moduli space. This approach consists in treating the Derrick mode perturbatively. The Derrick mode arises due to the scale transformation $x \to \lambda x$ and reflects the ability of the soliton to be compressed or expanded. Interestingly, in the case of kinks, the Derrick mode often reproduces very well the first shape mode \cite{Manton:2021}. This was observed, e.g., in the $\phi^4$, the Christ-Lee and the double sine-Gordon model \cite{Ciurla:2023}. Moreover, Derrick modes serve as degrees of freedom that can approximate the Lorentz contraction of a boosted soliton \cite{Rebbi:1979,Manton:2022}. Therefore, collective coordinate models based on Derrick modes are also useful for describing multi-kink scatterings.

    In the perturbative relativistic moduli space approach, the scaling factor $\lambda$ is written as $\lambda = 1 + \epsilon$. Thus, by expanding the kink profile $\Phi_K(\lambda(x - a))$ around $\epsilon = 0$, the following field configuration is obtained
    \begin{equation}
        \Phi(x;a,\epsilon) = \Phi_K(x - a) + \sum_{k=1}^n\dfrac{\epsilon^k}{k!}(x - a)^k\dfrac{d^k\Phi_K(x -a )}{dx^k}\,, 
    \end{equation}
    where $a$ accounts for the position of the kink. Finally, the amplitudes $\epsilon^k$ of the $k$-th Derrick modes are promoted to new independent moduli $C_k(t)$ along with the position of the kink $a(t)$. This idea can be readily applied to kink-antikink configurations. Remarkably, the  $k$-th Derrick modes become more extended for higher values of $k$, and their frequencies can lie above the continuum threshold $\omega_c$ of the theory. As a consequence, they effectively behave as radiation modes. However, this description is only valid over short time scales, since the modes remain confined to the kink and the energy they carry can never be radiated away to infinity.  

    In this section, we shall introduce an alternative to describe radiation based on genuine radiation modes. In order to do so, recall that the linear spectrum of perturbations about the kink solution in the $\phi^4$ model can be obtained analytically, as seen in \autoref{ss:phi4}. By virtue of the following theorem, we conclude that the eigenstates of the spectral problem (\ref{e:linearised_phi4_kink}) form an orthogonal basis, where $p(x) = 1$, $q(x) = U''(\Phi_K(x))$ and $\lambda = 1$ \footnote{The theorem as stated applies to a finite interval $x \in [a,b]$. However, it can be generalised to the whole real line with essentially the same implications. The only difference is that the modes above the continuum are not $L^2(\mathbbm{R})$-normalisable; instead, they are delta-normalisable, as discussed later.}
    \begin{theorem}[Sturm-Liouville theorem]\label{t:Sturm}
        For a second-order differential equation of the form
        \begin{equation}
            - \dfrac{d}{dx}\left[p(x)\dfrac{d\eta(x)}{dx}\right] - q(x)\eta(x) = \omega^2\,\lambda(x)\eta(x)\,,
        \end{equation}
        on an interval $x \in [a,b]$, the following properties are satisfied
        \begin{enumerate}
            \item The eigenvalues $\omega_i$ are real and can be ordered as $\omega^2_1 < \omega^2_2 < \ldots < \omega^2_n$.
            \item For each eigenvalue $\omega^2_i$ there is one eigenfunction $\eta_i(x)$ with exactly $i - 1$ nodes.
            \item The set of normalised eigenfunctions $\{\eta_i\}_{i= 1}^n$ forms an orthonormal basis, that is, $ \langle \eta_i | \eta_j \rangle= \delta_{ij}$.
        \end{enumerate}
    \end{theorem}
    Hence, the set $\mathcal{B} = \left\lbrace \eta_{0}(x), \eta_{s}(x), \eta_{q}(x) \right\rbrace $ may be used to build a general configuration belonging to the linearised field configuration space. As a consequence, a general field configuration close to the kink solution $\Phi_{K}(x)$ can be expanded as follows
    \begin{equation}\label{e:general_ansatz}
        \Phi (x) = \Phi_{K}(x) + c_{0}\,\eta_{0}(x) + c_{s}\,\eta_{s}(x) + \int_{\mathbb{R}} dq\, c_{q}\,\eta_{q}(x)\,. 
    \end{equation}
    \enlargethispage{2\baselineskip}
    This natural assumption contains all possible degrees of freedom of the kink: the zero mode $\eta_{0}(x)$ is responsible for the infinitesimal rigid translation of the kink, the shape mode $\eta_{s}(x)$ is responsible for the modification of the width of the kink, and the radiation or scattering modes $\eta_{q}(x)$ are related to the continuum of perturbative fluctuations around the vacuum, that propagate freely to infinity. Henceforth, we will assume that the linear modes are given by
    \begin{eqnarray}\label{e:linear_perturbations_phi4_normalised}
        \eta_{0}(x) \hspace{-0.2cm}&=&\hspace{-0.2cm} \frac{\sqrt{3}}{2}\sech^2 x\,,\\
        \eta_{s}(x) \hspace{-0.2cm}&=&\hspace{-0.2cm} \sqrt{\dfrac{3}{2}}\sinh x\sech^2 x\,, \\
        \eta_{q}(x) \hspace{-0.2cm}&=&\hspace{-0.2cm}  \dfrac{ 3 \tanh^2 x -q^2 - 1 - 3iq\tanh x}{\sqrt{(q^2+1)(q^2+4)}} e^{iqx}\,,
    \end{eqnarray}
    with $q \in \mathbb{R}$. 
    
    With this choice of constants, the bound modes are normalised with respect to the $L^2[\mathbbm{R}]$ norm, and the scattering modes are "delta-normalised" in such a way that, asymptotically, they are plane waves of amplitude one. Altogether, the following orthogonality relations are satisfied
    \begin{equation}
    \begin{split}
        \langle \eta_{0}(x), \eta_{s}(x) \rangle &= \langle \eta_{0}(x), \eta_{q}(x) \rangle = \langle \eta_{s}(x), \eta_{q}(x) \rangle = 0\,, \\
        \langle \eta_{q}(x), \eta_{q'}(x) \rangle &= 2\pi\, \delta(q - q')\,.
    \end{split}
    \label{e:ortho_etaq}
    \end{equation}
    The general ansatz (\ref{e:general_ansatz}) seems to codify the kink dynamics when radiation is involved, and we will use it throughout the subsequent study.

\section{Leading radiation from the wobbling kink}\label{s:wobbling}

    Although the shape mode is an exact solution of the linearised equation of motion, it is not a solution of the full non-linear field theory. Indeed, the non-linear terms couple the shape mode to the scattering modes. As a consequence, the energy of an initially excited shape mode is gradually radiated away to infinity, leading to a decay in its amplitude. An approximate expression describing this decay has been derived through two different approaches in \cite{Manton:1997,Oxtoby:2009}.

    In this section, we validate our effective description of radiation by comparing the results obtained through our description with those presented in \cite{Manton:1997,Oxtoby:2009}. To proceed, we assume the kink is at rest at the origin, thereby neglecting the translational degree of freedom, $\eta_0(x)$. Promoting the coefficients from (\ref{e:general_ansatz}) to time-dependent variables we are left with the following simplified ansatz   
    \begin{equation}\label{e:ansatz_oscill}
        \Phi (x,t) = \Phi_{K}(x) + c_{s}(t)\eta_{s}(x) + \int_{\mathbb{R}} c_{q}(t)\eta_{q}(x)\,dq\,. 
    \end{equation}
    The exactness of the shape mode at linear order in perturbation theory ensures that, when this mode is excited, it does not decay at that order. Assuming the shape mode is the sole source of radiation, it is natural to expect that if $c_s \sim \mathcal{O}(A_0)$, then $c_{q}(t) \sim \mathcal{O}(A_0^2)$. This is equivalent to the assumption that there is no radiation initially. Substituting (\ref{e:ansatz_oscill}) into (\ref{e:FE_phi4}) we obtain, at linear order in $c_{s}(t)$, the following expression for the shape mode amplitudes
    \begin{equation}
        \eta_{s}(x)\left[\ddot{c}_{s}(t) + \omega_{s}^{2}c_{s}(t)\right] = 0\,.
    \end{equation}
    Consequently, the shape mode oscillates with frequency $\omega_{s}$, i.e.,
    \begin{equation}\label{e:harmon_oscill}
        \ddot{c}_{s}(t) + \omega_{s}^{2}c_{s}(t) = 0 \Rightarrow c_{s}(t) = A_{0}\cos (\omega_{s}t)\,.
    \end{equation}
    This is the expected behaviour at linear order, as the shape mode is an exact solution of the equation of motion at this order. If we expand now up to second order in $c_{s}(t)$, we are left with
    \begin{equation}\label{e:second_order}
        \eta_{s}(x)\left[\ddot{c}_{s}(t) + \omega_{s}^{2}c_{s}(t) \right] + \int_{\mathbb{R}} \eta_{q}(x)\left[ \ddot{c}_{q}(t) + \omega_{q}^{2}c_{q}(t)\right]\,dq + 6\,c_{s}^{2}(t)\,\Phi_{K}(x)\,\eta_{s}^{2}(x) = 0\,.
    \end{equation}
    Projecting onto $\eta_{s}(x)$ and imposing that the shape mode and the scattering modes are orthogonal by virtue of \autoref{t:Sturm}, the previous expression reduces to the equation of an anharmonic oscillator corrected by a quadratic term
    \begin{equation}\label{e:anharmon_oscill}
        \ddot{c}_{s}(t) + \omega_{s}^{2}c_{s}(t) + \frac{9\pi \sqrt{6}}{32}\, c_{s}^{2}(t) = 0\,.
    \end{equation}
    If we now project onto a scattering mode $\eta^{\ast}_{q'}(x)$ and impose again the orthogonality relations (\ref{e:ortho_etaq}), the equation (\ref{e:second_order}) reads as
    \begin{equation}\label{e:etas_1}
        \ddot{c}_{q}(t) + \omega_{q}^{2}\,c_{q}(t) + \dfrac{3}{\pi}c_{s}^{2}(t)\int_{\mathbb{R}} \eta^{\ast}_{q}(x)\, \Phi_{K}(x)\,\eta_{s}^{2}(x)\,dx = 0\,,
    \end{equation}
    where the last term can be interpreted as the overlap between the scattering state of frequency $\omega_{q}$ with the combination $\Phi_{k}(x)\,\eta_{s}^2(x)$. Such a term can be computed exactly
    \begin{equation}
        \mathcal{F}(q) := \int_{\mathbb{R}} \eta^{\ast}_{q}(x)\, \Phi_{K}(x)\,\eta_{s}^{2}(x)\,dx = - \dfrac{i \pi}{32} \sqrt{\dfrac{q^2 + 4}{q^2 + 1}}  \dfrac{q^2 (q^2 - 2)}{\sinh \left( \pi q/2 \right)}\,.  
    \end{equation}
    Therefore, the expression (\ref{e:etas_1}) can be written as
    \begin{equation}\label{e:etas_2}
        \ddot{c}_{q}(t) + \omega_{q}^{2}\,c_{q}(t) + \dfrac{3}{\pi}\,c_s^2(t)\,\mathcal{F}(q) = 0\,.
    \end{equation}
    As we are considering that the shape mode is the only source of radiation, there is no radiation in absence of shape mode. We will take this into account imposing the initial conditions $c_{q}(0) = 0$ and $\dot{c}_{q}(0) = 0$. Following \cite{Manton:1997}, we assume that the amplitude of the shape mode $c_{s}(t)$ is given by its linear approximation, i.e., $c_{s}(t) = A_{0}\cos(\omega_{s} t)$. In other words, we are regarding that there is no back-reaction of the scattering modes on the shape mode. With this choice, the general solution of (\ref{e:etas_2}) takes the form
    \begin{equation}\label{e:cq}
        c_{q}(t) = - \dfrac{3}{2\pi}\dfrac{(4\omega_{s}^{2} - \omega_{q}^{2}) - \omega_{q}^{2}\cos(2 \omega_{s} t) - (4\omega_{s}^2 - 2\omega_{q}^{2})\cos(\omega_{q} t)}{\omega_q^2 (4\omega_s^2 - \omega_q^2)}\mathcal{F}(q)\,.
    \end{equation}
    This expression provides the time-dependent amplitudes of the radiation modes. As a consequence, the radiation emitted by an oscillating kink is 
    \begin{equation}\label{e:radiation}
        R(x,t) = \int_{\mathbb{R}}\, c_{q}(t)\eta_{q}(x)\,dq\,,
    \end{equation}
    with $c_q(t)$ given by (\ref{e:cq}). Remarkably, this is the exact form of the radiation at leading order for a static wobbling kink, and it is valid at all distances and times. Notice that the structure of $\mathcal{F}(q)$ indicates that some frequencies are suppressed in the radiation. It has a maximum at $q\approx 2\sqrt{2}$, i.e., $\omega_q^2 \approx 2\sqrt{3} = 2\,\omega_s^2$, which is consistent with the fact that the shape mode is the quadratic source for radiation.

    The radiation term (\ref{e:radiation}) allows for an approximate analytical treatment. To compute it, we begin by splitting (\ref{e:radiation}) as follows
    \begin{equation}\label{e:rad}
        R(x,t) = R_{1}(x) + R_{2}(x,t)\,,
    \end{equation}
    where 
    \begin{eqnarray}
        R_{1}(x) \hspace{-0.2cm}&=&\hspace{-0.2cm}  i \int_{\mathbb{R}}\, R_{0}(q)\,\left(4\omega_{s}^{2} - \omega_{q}^{2}\right)\, \eta_{q}(x)\,dq\,, \label{e:R1}\\
        R_{2}(x,t) \hspace{-0.2cm}&=&\hspace{-0.2cm} - i \int_{\mathbb{R}}\, R_{0}(q)\, \left(\omega_{q}^{2}\cos(2 \omega_{s} t) + (4\omega_{s}^2 - 2\omega_{q}^{2})\cos(\omega_{q} t)\right) \eta_{q}(x)\,dq\,, \label{e:R2}
    \end{eqnarray}
    and
    \begin{equation}
        R_{0}(q) = \dfrac{3 A_{0}^2}{64}\dfrac{ q^2 \left( q^2 - 2 \right)}{ \sqrt{q^2 + 1}\, \omega_{q} (4\omega_{s}^{2} - \omega_{q}^{2})\sinh(\pi q / 2)}\,.
    \end{equation}
    Therefore, the radiation profile decomposes into two distinct parts: a time-independent component and a time-dependent one.

\subsection{Evaluation of the function \texorpdfstring{$R_{1}(x)$}{R1(x)}}

\begin{figure}[htb]
        \centering{
        \includegraphics[width=0.495\linewidth]{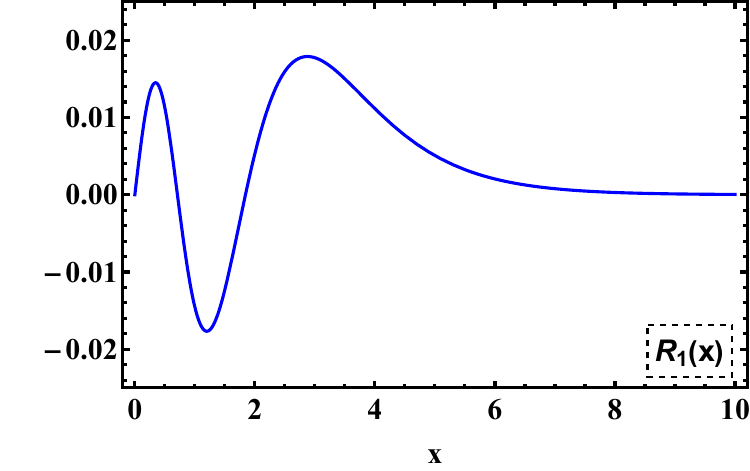}}
        \caption{Representation of the $R_1(x)$ term given by (\ref{e:finalR1}) for $A_0 = 1$.}
        \label{f:R1_term}
    \end{figure}

    We begin by evaluating the integral $R_1(x)$. Note that the integrand of (\ref{e:R1}) can be divided into an odd and even contribution. Due to the symmetric interval of integration, only the even contribution is not null, which results in 
    \begin{equation}
        R_{1}(x) =  - 2 \int_{0}^{\infty}\, R_{0}(q)\,(4\omega_{s}^{2} - \omega_{q}^{2})\operatorname{Im}[\eta_{q}(x)]\,dq\,,
    \end{equation}
    where 
    \begin{equation}
        \operatorname{Im}(\eta_{q}(x)) = - (q^2 +1) \sin qx - 3\, q \tanh x \cos qx + 3 \tanh^2 x \sin q x\,.
    \end{equation}
    This integral can be decomposed as follows
    \begin{align}\label{e:R1_integrals}
        R_{1}(x) = \frac{3 A_{0}^{2}}{32}  \bigg(& \int_{0}^{\infty} \dfrac{q^2\,(q^2 - 2)\sin qx}{(q^2 + 4)\sinh(\pi q /2)}\,dq + 3\tanh x \int_{0}^{\infty}\dfrac{q^3\,(q^2 - 2)\cos qx }{(q^2 + 4)(q^2 + 1)\sinh(\pi q / 2)}\,dq \nonumber\\ 
        & - 3 \tanh^2 x \int_{0}^{\infty}\dfrac{q^2\,(q^2 - 2)\sin qx}{(q^2 + 4)(q^2 + 1)\sinh(\pi q / 2 )}\,dq  \bigg)\,.
    \end{align} 
    Each of the terms can be solved separately by transforming the computation of the integrals into the calculation of the solution of an associated inhomogeneous differential equation. Specifically, let us write (\ref{e:R1_integrals}) as
    \begin{eqnarray}\label{e:R1_parts}
        R_{1}(x) =  \frac{3 A_{0}^{2}}{32}  \bigg(\alpha(x) + 3\tanh x \, \beta(x) - 3 \tanh ^2 x \, \gamma(x) \bigg)\,,
    \end{eqnarray} 
    where $\alpha(x), \beta(x)$ and $\gamma(x)$ denote the respective integrals. It can be verified that the function $\alpha(x)$ satisfies the following differential equation
    \begin{equation}
        \alpha''(x) - 4 \alpha(x) = - \int_{0}^{\infty} \dfrac{ q^2 (q^2 - 2) \sin qx}{\sinh(\pi q / 2)}\,dq = - 6 \, (3 - \cosh 2x) \tanh x \sech^4 x\,.
    \end{equation}
    The requirement that the function $\alpha(x)$ does not diverge as $|x| \rightarrow \infty$ and the obvious condition $\alpha(0) = 0$ allow us to determine that the solution of the inhomogeneous differential equation is
    \begin{equation}\label{e:alpha:R1}
        \alpha(x) = 6 e^{2 x} \log \left( 1+e^{-2 x} \right) - 6 e^{-2 x} \log \left(  1 + e^{2 x}\right) - 2 \tanh  x \left(3 - \sech^2 x  \right)\,.
    \end{equation}
    Regarding the function $\beta(x)$, it fulfils a fourth-order differential equation of the form
    \begin{equation}
        \beta^{(4)}(x) - 5\beta''(x) + 4\beta(x) = \int_{0}^{\infty} \dfrac{ q^3 (q^2 - 2) \cos qx}{\sinh(\pi q / 2)}\,dq  = 3 \, (21 - 18 \cosh 2x + \cosh 4x ) \sech^6 x\,. 
    \end{equation}
    Now, the trivial conditions are $\beta'(0)= 0$ and $\beta^{(3)}(0) = 0$. Employing those conditions along with the requirement of the finiteness as $|x| \rightarrow \infty$ we deduce 
    \begin{eqnarray}\label{e:beta:R1}
        \beta(x) = \hspace{-0.2cm} &-& \hspace{-0.2cm} 4  e^{2 x} \log  ( 1 + e^{-2 x}  ) - 4 e^{- 2 x} \log  ( 1 + e^{2 x}   ) + 5 - \dfrac{4}{ (1 + e^{2 x}  )^2} + \frac{\pi  e^x}{2} - 2 \tanh x \nonumber\\
    &-&\hspace{-0.2cm} 2 \sinh x \arctan e^x\,.
    \end{eqnarray}   
    Finally, the $\gamma(x)$ term of $R_{1}(x)$ will be the solution of the differential equation
    \begin{equation}
        \gamma^{(4)}(x) - 5\gamma''(x) + 4\gamma(x) = \int_{0}^{\infty} \frac{q^2 (q^2 - 2)}{\sinh( \pi q / 2)}\sin qx\,dq = 6 \, (3 - \cosh 2x) \tanh x \sech^4 x\,, 
    \end{equation}
    with conditions $\gamma(0) = 0$ and $\gamma''(0) = 0$. The well-behaved solution at $|x| \rightarrow \infty $ is
    \begin{align}\label{e:gamma:R1}
        \gamma(x) =\hspace{0.1cm} & 2 e^{-2 x} \log(1+ e^{2 x}) - 2 e^{2 x} \log(1+  e^{-2 x}) - e^{-2 x} ( 1 + \tanh x ) + 1 + \frac{\pi  e^x}{2} \notag\\
        &- 2 \cosh x  \arctan e^x\,.
    \end{align}
    Substituting (\ref{e:alpha:R1}), (\ref{e:beta:R1}) and (\ref{e:gamma:R1}) into (\ref{e:R1_parts}), we obtain the complete expression of $R_{1}(x)$ 
    \begin{eqnarray}\label{e:finalR1}
        R_{1}(x) = \dfrac{3 A_{0}^{2}}{64 \cosh^2 x} \left( 3\pi \sinh x + 16 \tanh x - 24 x \right)\,.
    \end{eqnarray}
    A plot of this first time-independent contribution to the radiation profile is shown in \autoref{f:R1_term}.

    \vspace{-0.5cm}
    
\subsection{Evaluation of the function \texorpdfstring{$R_{2}(x)$}{R2(x)}}

    The evaluation of the contribution $R_2(x,t)$ seems to be more challenging although, as we shall verify, it is the most crucial term. A complete analytical computation does not seem possible. However, some properties of the radiation can be obtained by certain approximations. Specifically, we will focus now our efforts on describing the asymptotic radiation emitted by the excited kink. As we have seen, the $R_1(x)$ contribution is exponentially suppressed for large $x$, therefore the asymptotic radiation is entirely given by the $R_{2}(x,t)$ contribution.
    
    Through some elemental algebra, it is straightforward to verify that the integrand of the $R_{2}(x,t)$ term (\ref{e:R2}) can be written as 
    \begin{equation}\label{e:term_parenthesis:2}
        (4\omega_{s}^2 - \omega_{q}^2)\bigg(- 2\sin(\Omega_q^{+}t)\sin(\Omega_q^{-}t)+ \cos(\omega_q t) - 4\,\omega_{s}^2\dfrac{\cos(\omega_q t) - \cos(2\omega_s t)}{4\omega_{s}^2 - \omega_{q}^2} \bigg)\,,
    \end{equation}
    where we have removed temporarily a $-i R_0(q)\,\eta_q(x)$ factor and defined the following variables
    \begin{equation}
        \Omega_q^{+} = \dfrac{\omega_q + 2\omega_s}{2}\,, \qquad \Omega_q^{-} = \dfrac{\omega_q - 2\omega_s}{2}\,.
    \end{equation}
    As it is well known, there are some sequence of functions that converge weakly to the Dirac delta. One of them is  
    \begin{equation}
        f_n(x) = \dfrac{\sin(n x)}{\pi x} \overset{n \rightarrow \infty}{\longrightarrow} \delta(x)\, .
    \end{equation}
    Then, multiplying and dividing the first term in (\ref{e:term_parenthesis:2}) by $\pi\,\Omega_q^{-}$ we can identify 
    \begin{equation}\label{e:dirac_Approx}
        \dfrac{\sin(\Omega_q^{-} t)}{\pi \Omega_q^{-}} \overset{t \rightarrow \infty}{\longrightarrow} \delta(\Omega_q^{-}) = \sqrt{6}\left( \delta(q - 2\sqrt{2}) + \delta(q + 2\sqrt{2}) \right)\,,
    \end{equation}
    where it has been used that $\Omega_q^{-}$ only has simple zeros so that we can apply
    \begin{equation}
        \delta(g(x)) = \sum_k \dfrac{\delta(x - x_k)}{|g'(x_k)|}\,,
    \end{equation}
    with $k$ labelling the zeros of $g(x)$. Substituting (\ref{e:dirac_Approx}) into (\ref{e:term_parenthesis:2}) and integrating, we realise that the contribution from the first term in (\ref{e:term_parenthesis:2}) vanishes. Indeed, the numerical computation of this terms shows that such a contribution is suppressed in time. The second term may be computed as follows: first let us call such contribution as  
    \begin{equation}\label{e:I_Integral}
        \Psi(x,t) = i \int_{\mathbb{R}}\, dq\, R_{0}(q)\, (4\omega_{s}^2 - \omega_{q}^2)\cos(\omega_{q} t)\eta_{q}(x)\,.
    \end{equation}
    To facilitate the calculation, let us divide $\Psi(x,t)$ in the same way as we did for $R_{1}(x)$
    \vspace{0.1cm}
    \begin{align}
        \Psi(x,t) = &\hspace{0.1cm} \frac{3 i A_{0}^{2}}{64}  \bigg( \int_{\mathbb{R}} \dfrac{q^2\,(q^2 - 2)\cos(\omega_{q} t)}{(q^2 + 4)\sinh(\pi q /2)}e^{i q x}\,dq + 3i\tanh x \int_{\mathbb{R}} \dfrac{q^3\,(q^2 - 2)\cos(\omega_{q} t)}{(q^2 + 1)(q^2 + 4)\sinh(\pi q / 2)}e^{i q x}\,dq\nonumber\\ 
        &- 3 \tanh^2 x \int_{\mathbb{R}} \dfrac{q^2\,(q^2 - 2)\cos(\omega_{q} t)}{(q^2 + 1)(q^2 + 4)\sinh(\pi q / 2 )}e^{i q x}\,dq \bigg)\,.
    \end{align} 
    \vspace{0.1cm}
    Each of the terms from $\Psi(x,t)$ can be conceived as wave packets where the dispersion relation is non-linear and where the constituent amplitudes are exponentially suppressed in $q$
    \vspace{0.1cm}
    \begin{align}\label{e:def_I}
        \Psi(x,t) =& \hspace{0.1cm} \frac{3 i A_{0}^{2}}{138} \bigg( \int_{\mathbb{R}} \dfrac{q^2\,(q^2 - 2)}{(q^2 + 4)\sinh(\pi q /2)}\left[e^{i (q x + \omega_{q}t)} + e^{i (q x - \omega_{q}t)} \right]\,dq\nonumber\\
        &+  3i\tanh x \int_{\mathbb{R}} \dfrac{q^3\,(q^2 - 2)}{(q^2 + 1)(q^2 + 4)\sinh(\pi q / 2)}\left[e^{i (q x + \omega_{q}t)} + e^{i (q x - \omega_{q}t)} \right]\,dq\nonumber\\ 
        &- 3 \tanh^2 x \int_{\mathbb{R}} \dfrac{q^2\,(q^2 - 2)}{(q^2 + 1)(q^2 + 4)\sinh(\pi q / 2 )}\left[e^{i (q x + \omega_{q}t)} + e^{i (q x - \omega_{q}t)} \right]\,dq \bigg)\,.
    \end{align} 
    \vspace{0.1cm}
    In the literature, the problem of the calculation of wave packets is well known when the amplitude is a Gaussian function, that is, $\sigma(q) = e^{-\alpha^2(q - q_{c})^2}$. For that case, it is a good approximation to Taylor expand the dispersion relation around $q_{c}$ as the amplitude is exponentially suppressed away from the maximum. Therefore, we can write  $\omega(q) \approx q_{c}v_{p} + (q - q_{c})v_{g} + \dfrac{1}{2}\Gamma(q - q_{c})^2 + \dots$ where $v_{p}$ is the phase velocity, $v_{g}$ is the group velocity, and $\Gamma$ is the dispersion parameter. The final result of this approximation is collected by the following expression
    \vspace{0.1cm}
    \begin{equation}\label{e:wave_packet}
        \int_{\mathbb{R}} \sigma(q)e^{i(qx \mp \omega(q)t)}\,dq \approx \dfrac{\sqrt{2 \pi }}{\sqrt{2\alpha^2 \pm i\Gamma  t}} \exp \left(-\dfrac{1}{2} \left( \dfrac{x \mp v_{g}t}{\sqrt{2\alpha^2 \pm i \Gamma t}} \right)^2 \right)\exp\left(i q_{c}(x \mp v_{p}t)\right)\,.
    \end{equation} 
    \vspace{0.1cm}
    It is clearly visible that this integral is also suppressed over time, a fact confirmed by the numerical simulations. Applying the approximation (\ref{e:wave_packet}) to each of the components in (\ref{e:def_I}), and in particular around the extrema of the different components, we deduce that the dominant contribution arises from the last term in (\ref{e:term_parenthesis:2}).
    
\begin{figure}[H]
    \centering{
    \includegraphics[width=0.55\linewidth]{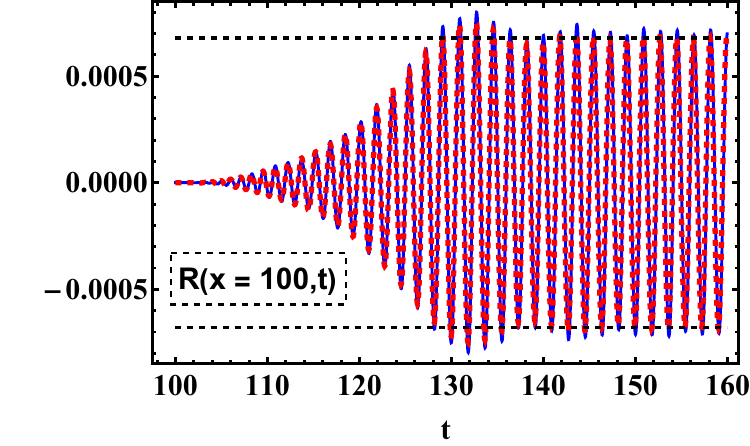}}
    \caption{Radiation field at $x = 100$ obtained through (\ref{e:radiation}) (dashed red line) and through field theory (solid blue line). The initial conditions are set as $\phi(x, 0) = \Phi_K(x) + A_0\,\eta_s(x)$ and $\dot{\phi}(x, 0) = 0$ with $A_0 = 0.1$. The horizontal dashed line represents the amplitude of (\ref{e:rad_inf}).}
    \label{f:Radiation}
    \end{figure}

    Indeed, we can recast the last term of (\ref{e:term_parenthesis:2}) to apply the Dirac delta approximation (\ref{e:dirac_Approx}), yielding the only non-vanishing contribution at large times
    \begin{align}
        R_{2}(x,t) =& \hspace{0.1cm} \dfrac{9 \pi A_0^2}{2 \sqrt{8} \sinh(\sqrt{2} \pi)}\sin(2 \sqrt{3}\,t) \sin (2 \sqrt{2}\,x) + \dfrac{3 \pi A_0^2}{2 \sinh(\sqrt{2}\pi)} \sin(2\sqrt{3}\,t) \cos(2 \sqrt{2}\,x) \tanh x\nonumber\\
        &- \dfrac{3\, \pi A_0^2}{2 \sqrt{8}\sinh(\sqrt{2} \pi)} \sin (2 \sqrt{3}\,t) \sin (2 \sqrt{2}\,x) \tanh^2 x\, . 
    \end{align} 
    In the asymptotic spatial regime, the previous expression reduces to
    \begin{eqnarray}
        R_{2}(x,t) =  \dfrac{3\,\pi A_0^2}{2 \sinh(\sqrt{2}\pi)}\sqrt{\dfrac{3}{8}}\bigg(\cos\big(2\sqrt{3}\,t \mp 2\sqrt{2}\,|x| \mp \delta\big) - \cos\big(2\sqrt{3}\,t \pm 2\sqrt{2}\,|x| \pm \delta\big) \bigg)\,,
    \end{eqnarray} 
    with
    \begin{equation}
        \delta = \arctan\sqrt{2}\,.
    \end{equation}
    It is not surprising that we obtain a superposition of two travelling waves with the same frequency but opposite directions, that is, a standing wave, since the Dirac delta approximation contributes only with a single frequency in the $q$-integral.
    \enlargethispage{2\baselineskip}
    An alternative derivation consists in splitting the last term in (\ref{e:term_parenthesis:2}) as follows
    \begin{eqnarray}\label{e:I_main}
        I(x,t) \hspace{-0.2cm} &=& \hspace{-0.2cm} -4 i \omega_s^2\int_{-\infty}^0  \tilde{R}_0(q)\tilde{\eta}_q(x)\frac{\cos\left(\omega_q t\right)-\cos\left(2\omega_s t\right)}{4\omega_s^2-\omega_q^2}e^{i q x}\,dq \nonumber\\
        &&\hspace{-0.2cm}-4 i \omega_s^2\int_{0}^\infty \tilde{R}_0(q)\tilde{\eta}_q(x)\frac{\cos\left(\omega_q t\right)-\cos\left(2\omega_s t\right)}{4\omega_s^2-\omega_q^2}e^{i q x} \,dq \equiv I_1+I_2\,,
    \end{eqnarray}
    where
    \begin{eqnarray}
        \tilde{R}_0(q) \hspace{-0.2cm}&=& \hspace{-0.2cm} \dfrac{3 A_{0}^2}{64}\sqrt{\dfrac{q^2 + 4}{q^2 + 1}}\dfrac{ q^2 \left( q^2 - 2 \right)}{\omega_{q}^2 \sinh(\pi q / 2)}\,, \\ 
        \tilde{\eta}_q(x) \hspace{-0.2cm}&=& \hspace{-0.2cm}\dfrac{ 3 \tanh^2 x -q^2 - 1 - 3iq\tanh x}{\sqrt{(q^2+1)(q^2+4)}}\,. 
    \end{eqnarray}
    The main contribution to the $I_1$ and $I_2$ integrals at large $t$ comes from a neighbourhood of $q_\pm=\pm 2\sqrt{2}$ because the modes corresponding to $q=q_\pm$ grow linearly with time. Therefore, the frequency $\omega_q$ can be linearly approximated by
    \begin{equation}
        \omega_{q_{\pm}} \approx 2\sqrt{3}\pm \sqrt{\frac{2}{3}}\left(q\mp 2\sqrt{2}\right) +   \mathcal{O}\left(q\mp 2\sqrt{2}\right)^2\,.
    \end{equation}
    As a result, we may approximate $I_1$ by the following expression
    \begin{equation}
        I_1(x,t)\approx 4\, i\,  \tilde{R}_0(q_+)\,\tilde{\eta}^\ast_{q_+}(x)\, \omega_s^2\int_{0}^\infty \frac{\cos\left(\widetilde{\omega}_q t\right)-\cos\left(2\omega_s t\right)}{4\omega_s^2-\omega_q^2}e^{-i q x}\,dq\,,
    \end{equation} 
    where $\widetilde{\omega_q} = 2\sqrt{3} +  \sqrt{\dfrac{2}{3}}\left(q - 2\sqrt{2}\right)$. Here, we are assuming implicitly that in a neighbourhood of $q = 2\sqrt{2}$, both $\tilde{R}_0(q)$ and $\tilde{\eta}_{q}(x)$ are approximately constant in $q$. Similarly, a straightforward manipulation leads to the following expression for $I_2$
    \begin{equation}
        I_2(x,t)\approx -4 \,i\,  \tilde{R}_0(q_+)\,\tilde{\eta}_{q_+}(x)\, \omega_s^2\int_{0}^\infty \frac{\cos\left(\widetilde{\omega}_q t\right)-\cos\left(2\omega_s t\right)}{4\omega_s^2-\omega_q^2}e^{i q x}\,dq\,,
    \end{equation}
    Finally, we are left with the following approximation for the $I(x,t)$ integral
    \begin{equation}\label{e:rad_I_1}
        I(x,t) = 8\, \omega_s^2\,\tilde{R}_0(q_+)\left(\text{Re}\left[\tilde{\eta}_{q_+}(x)\right]\mathcal{F}_s(t)+\text{Im}\left[\tilde{\eta}_{q_+}(x)\right]\mathcal{F}_c(t)\right)\,,
    \end{equation}
    where
    \vspace{-0.4cm}
    \begin{eqnarray}\label{e:four_1}
        \mathcal{F}_s(t) \hspace{-0.2cm}&=&\hspace{-0.2cm} \int_0^\infty \frac{\cos\left(\widetilde{\omega}_q t\right)-\cos\left(2\omega_s t\right)}{4\omega_s^2-\omega_q^2}\sin\left(q x\right)\,dq\,,\\\label{e:four_2}
        \mathcal{F}_c(t) \hspace{-0.2cm}&=&\hspace{-0.2cm} \int_0^\infty \frac{\cos\left(\widetilde{\omega}_q t\right)-\cos\left(2\omega_s t\right)}{4\omega_s^2-\omega_q^2}\cos\left(q x\right)\,dq\,.
    \end{eqnarray}
    Remarkably, the integrals (\ref{e:four_1}) and (\ref{e:four_2}) can be computed analytically for all $x$, but the expressions are not particularly illuminating. For $x \gg 0$ and large $t$, we have determined that
    \vspace{-0.15cm}
    \begin{eqnarray}\label{e:fsin}
        \mathcal{F}_s(t) \hspace{-0.2cm}&=&\hspace{-0.2cm} - \dfrac{\pi  \cos \left(2 \sqrt{3}\,t-2 \sqrt{2}\, x\right)}{4 \sqrt{2}}\,,\\\label{e:fcos}
        \mathcal{F}_c(t) \hspace{-0.2cm}&=&\hspace{-0.2cm} - \dfrac{\pi  \sin \left(2 \sqrt{3}\,t-2 \sqrt{2}\, x\right)}{4 \sqrt{2}}\,.
    \end{eqnarray}
    Substituting (\ref{e:fsin}) and (\ref{e:fcos}) into (\ref{e:rad_I_1}) we obtain 
    \begin{equation}\label{e:rad_inf}
        I(x,t)= \dfrac{3\,\pi A_0^2}{2 \sinh(\sqrt{2}\pi)}\sqrt{\dfrac{3}{8}}\cos\big(2\sqrt{3}\,t - 2\sqrt{2}\,x - \arctanh \sqrt{2} \big)\,.
    \end{equation}
    This expression agrees with the one obtained in \cite{Manton:1997,Oxtoby:2009} except for a $3/2$ factor in the amplitude $A_0$, that arises from our normalisation of the shape mode. We have solved numerically (\ref{e:radiation}) for large values of $x$, illustrating the temporal evolution of radiation in \autoref{f:Radiation}. We observe the radiation gradually converging towards the profile described by (\ref{e:rad_inf}), as a sufficient amount of time has passed. Similar results were also obtained in \cite{Romanczukiewicz:2004}.

    \enlargethispage{1\baselineskip}
    Following \cite{Manton:1997}, the decay of the amplitude of the shape mode into radiation can be determined directly from the expression (\ref{e:rad_inf}). In a first step, we compute the average energy flux carried by the wave (\ref{e:rad_inf}). For a wave of the form $\eta = C\cos(\omega t - kx)$, the average power radiated in a period is $\langle P\rangle = - \frac{1}{2}C^2\omega\,k$. As radiation is emitted both to the left and to the right, the total average power is $\langle P_{tot}\rangle = - C^2\omega\,k$. In our case,
    \begin{equation}
        \langle P_{tot}\rangle = - 2\int_{\mathbbm{R}}\langle T_{0x} \rangle\,dx = - \dfrac{27\,\pi^2 A_0^4}{4 \sinh^2\sqrt{2}\pi}\sqrt{\dfrac{3}{2}}\,.   
    \end{equation}
    Note that until now we have assumed that the amplitude of the shape mode is time-independent. However, this is not true in an isolated system, as the kink is radiating away the energy of the shape mode. Therefore, let us introduce the time dependence of the shape mode amplitude as $c_s(t) = \hat{A}(t)\cos(\omega_s t)$ with $\hat{A}(0) = A_0$.
    Following this, we compute the energy of the excited kink with amplitude $\hat{A}$ at an instant of time $t$
    \begin{equation}
        E = M_K + \dfrac{1}{2}\omega_s^2\hat{A}(t)^2\,.
    \end{equation}
    Finally, by equating the rate of change of the energy of the excited kink, $dE/dt$, to the total average power, we derive the following decay law for the shape mode amplitude
    \begin{eqnarray}\label{eq:manton_law}
        \hat{A}(t) = \dfrac{1}{\sqrt{ A_{0}^{-2} + 0.03\, t}}\,,
    \end{eqnarray}
    where $A_0$ is the initial amplitude of the shape mode. After introducing the normalisation constant, this decay law becomes identical to the formulas derived in \cite{Manton:1997,Oxtoby:2009}.

    One of the advantages of our general collective coordinate model is that it offers a broader scenario to the study of different phenomena regarding the kink dynamics. So far, we have described the radiation emitted by an excited kink, and our results are in complete agreement with those obtained previously in the literature \cite{Manton:1997,Romanczukiewicz:2004,Oxtoby:2009}. Moreover, under certain approximations, we have succeeded in obtaining an analytical expression for the radiation profile valid for all $x$. Nevertheless, as we will see in the following sections, our framework can be easily extended to describe more phenomena involving potentially any degree of freedom in the single kink sector. For example, expressions (\ref{e:etas_2}) and (\ref{e:rad_inf}) suggest that it is possible to excite resonantly the shape mode with radiation of the appropriate frequency. In order to obtain these results, we have to consider the shape mode amplitude as a free collective coordinate interacting with the radiation coordinates. We will study these issues in detail in the next section.

\section{Interaction of radiation and the shape mode}\label{s:Oscil_Radi}

    In this section, we derive the effective Lagrangian within the collective coordinate approach to capture the mutual interaction between the shape mode and radiation. Both $c_s(t)$ and $c_q(t)$ are treated as genuine collective coordinates representing the dynamics of the shape mode and radiation, respectively.
    
    Therefore, let us consider once more the ansatz (\ref{e:ansatz_oscill}). Substituting it into the Lagrangian density (\ref{e:lagrangian_phi4}) and integrating over space, we obtain the following effective Lagrangian, retaining terms up to third order
    \begin{align}\label{e:effective_lagran:oscill}
        L_{s,q}[c_s,c_q] = &- \dfrac{4}{3} + \dfrac{1}{2}\left(\dot{c}_{s}^{2}(t) - \omega_{s}^{2}c_{s}^{2}(t)\right) + \pi \int_{\mathbb{R}} \left( \dot{c}_{q}(t)\dot{c}_{-q}(t) - \omega_{q}^{2}c_{q}(t)c_{-q}(t) \right)\,dq\nonumber\\ 
        &- \dfrac{3 \pi}{16}\sqrt{\dfrac{3}{2}}c_{s}^{3}(t) - \dfrac{3 \pi i}{16} \int_{\mathbb{R}}\, \sqrt{\dfrac{q^2 + 4}{q^2 + 1}}\dfrac{q^2 (q^2 - 2)}{\sinh (\pi q/2)} c_{s}^{2}(t)\,c_{q}(t)\,dq\nonumber\\
        &+ c_s(t)\int_{\mathbb{R}^2}f_{sq}(q,q')\, c_q(t)\,c_{q'}(t)\,dq\,dq'\,,
    \end{align}
    where
    \begin{eqnarray}\label{e:cs_cq_cq}
        f_{sq}(q,q') = \frac{\pi}{16}\sqrt{\frac{3}{2}}\dfrac{17+17q^2+17 q'^2+10 q^2q'^2-q^4-q'^4+q^2q'^4+q^4q'^2-q^6-q'^6}{\sqrt{(q^2+1)(q^2+4)}\sqrt{(q'^2+1)(q'^2+4)}\cosh(\frac{\pi}{2}(q+q'))}\,. 
    \end{eqnarray}
    Note that the mechanical Lagrangian (\ref{e:effective_lagran:oscill}) describes a system of harmonic oscillators coupled by the last two terms. The equations of motion governing the evolution of $c_{q}(t)$ and $c_{s}(t)$ are yielded by
    \vspace{-0.1cm}
    \begin{eqnarray}
        \ddot{c}_{-q}(t) \hspace{-0.25cm} &+& \hspace{-0.25cm}\omega_{q}^{2}\,c_{-q}(t) +  \dfrac{3 i}{32}\sqrt{\dfrac{q^2 + 4}{q^2 + 1}}\dfrac{q^{2}(q^2-2)}{\sinh\left( \pi q / 2 \right)}c_{s}^{2}(t) - \dfrac{1}{\pi} c_s(t)\int f_{sq}(q,q')\,c_{q'}(t)\,dq' = 0\,, \label{e:effec_eq:cq} \\  
        \ddot{c}_{s}(t) \hspace{-0.25cm} &+& \hspace{-0.25cm}\omega_{s}^{2}\,c_{s}(t) + \dfrac{9\pi}{16}\sqrt{\dfrac{3}{2}}c_{s}^{2}(t) + \dfrac{3\pi}{8}i \int_{\mathbb{R}} \sqrt{\dfrac{q^2 + 4}{q^2 + 1}}\dfrac{q^2 (q^2 - 2)}{\sinh( \pi q/2)} \,c_{s}(t)\,c_{q}(t)\,dq \nonumber\\ 
        &-&\hspace{-0.2cm} \int_{\mathbb{R}^2}f_{sq}(q,q') \,c_q(t)\,c_{q'}(t)\,dq\,dq' = 0\,.\label{e:effec_eq:cs}
    \end{eqnarray}
    The physical interpretation of these coupled systems is straightforward: the equation for the amplitudes of the scattering modes (\ref{e:effec_eq:cq}) is a forced harmonic oscillator of fundamental frequency $\omega_q$ with external force proportional to $c_s^2(t)$. Notice that this equation was derived in the previous section, corresponding to (\ref{e:etas_2}), but now we will allow for back-reaction of radiation in the shape mode. Regarding the equation for the shape mode amplitude (\ref{e:effec_eq:cs}), its structure is more involved. Specifically, it describes an anharmonic oscillator of fundamental frequency $\omega_s$ coupled linearly to $c_q(t)$. Note that, for sufficiently small amplitudes of $c_s(t)$, this expression reduces to the well-known Hill's equation provided that $c_q(t)$ is periodic. 

    In order to solve the system (\ref{e:effec_eq:cq})-(\ref{e:effec_eq:cs}) numerically, we have to choose a discretisation in $q$, i.e., we have to select $N$ scattering modes labelled by $q_i$ and solve the coupled system of $N+1$ ordinary differential equations. Note that the discretisation of the integrals in $q$ gives rise to a Hamiltonian system that conserves energy. Then, the discretisation fixes a time cut-off of order $t_c = 1/\Delta q$ (where $\Delta q=q_i-q_{i-1}$) beyond which our computations are no longer reliable. 

\begin{figure}[htb]
    \begin{center}
        \includegraphics[scale=0.64]{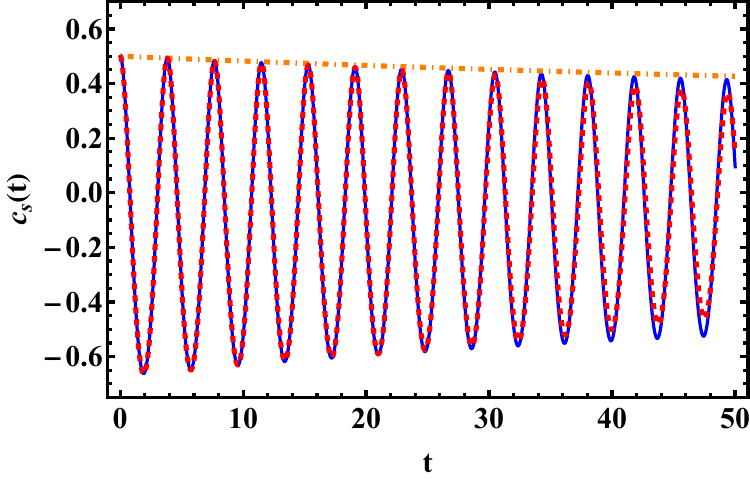}    
        \caption{Decay of the shape mode in field theory (solid blue line) and in the effective model (\ref{e:effective_lagran:oscill}) (dashed red line) compared to the analytical decay law given by (\ref{eq:manton_law}) (dotted-dashed line). For the computation, we have chosen $n = 20$ equidistant scattering modes within the interval $q \in [-3,3]$.}
    \label{f:manton_Decay}
    \end{center}
    \end{figure}

    In our first numerical experiment, we shall show that the Lagrangian (\ref{e:effective_lagran:oscill}) accurately describes the decay of the shape mode for $t \lesssim t_c$. We will choose the following initial conditions
    \begin{equation}
        c_s(0)=A_0\,,\quad c_s'(0)=0\,,\quad c_q(0)=0\,,\quad \text{and}\quad c_q'(0)=0\,.
    \end{equation} 
    These initial conditions describe a kink with its shape mode initially excited with an amplitude $A_0$ and no radiation at all. Due to the addition of the radiation degrees of freedom, the effective system (\ref{e:effec_eq:cq})-(\ref{e:effec_eq:cs}) naturally allows the decay of the shape mode. In \autoref{f:manton_Decay} we compare the decay of the shape mode obtained from field theory with the solution of the effective system (\ref{e:effec_eq:cq})-(\ref{e:effec_eq:cs}). 
    \vspace{0.2cm}
    
    Naturally, the bigger the number of scattering modes added to the effective model, the better the fit to field theory. We have observed that, for times $t_c < 1/\Delta q$ and a number of scattering modes $N > 5$, both solutions agree with great accuracy. 

\begin{figure}[htb]
    \begin{center}
        \includegraphics[scale=0.64]{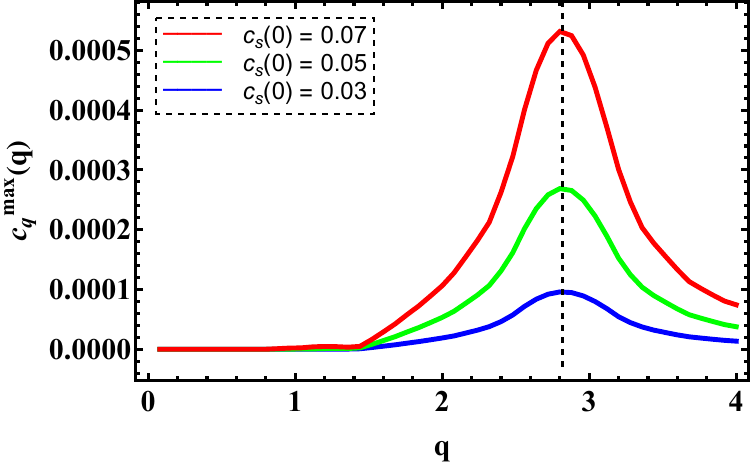}    
        \caption{Frequency spectrum of the radiation emitted by a wobbling kink obtained from (\ref{e:effec_eq:cq})-(\ref{e:effec_eq:cs}). We have taken $n = 100$ equidistant scattering modes in the interval $q \in [-4,4]$, the time of simulation as $t = t_c$, and the initial values of the shape mode are $c_s(0) = 0.03,\, 0.05$ and $0.07$. It can be appreciated that the maxima take place approximately at $\omega_q = 2\,\omega_s$ for small shape mode amplitudes.}
        \label{f:excitation_rad}
    \end{center}
    \end{figure}

    From \autoref{f:excitation_rad} we can observe that the predominant frequency of the radiation emitted by the excited kink corresponds to $\omega \approx 2\,\omega_s$. This observation is consistent with the value predicted in (\ref{e:rad_inf}). Remarkably, this specific frequency can be explained by referring to the expression (\ref{e:effec_eq:cq}). As previously mentioned, equation (\ref{e:effec_eq:cq}) represents the equation of a forced harmonic oscillator. Examining the maximum value of the forced term, it is easy to verify that it occurs at $\omega \approx 2\omega_s$. Consequently, the frequencies in the vicinity of $2\omega_s$ are strongly excited due to the resonance phenomenon, resulting in the transfer of the energy of the shape mode to infinity through these modes.
    \vspace{-0.35cm}
    
    In our second numerical experiment, we will explore the excitation of the shape mode when the kink is illuminated with radiation of frequency $\omega_{q_0}$. In order to extract some analytical insight, let us assume that there is no back-reaction of the shape mode in the radiation modes. In particular, let us assume the following choice for the radiation amplitudes
    \begin{equation}\label{e:cq_lin_rad}
        c_q(t)=A_q \,e^{i\omega_q t}\,\delta(q-q_0) + A_q \,e^{-i\omega_q t}\,\delta(q+q_0)\,.
    \end{equation}
    These radiation amplitudes describe the superposition of a kink with a combination of scattering modes of frequency $\omega_{q_0}$. Therefore, the corresponding radiation profile is given by
    \begin{align}\label{e:lin_rad}
        R(x,t) =& \hspace{0.1cm} \int_{\mathbb{R}} c_q(t)\, \eta_q(x)\,dq = \frac{A_{q_0}}{\sqrt{(q_0^2+1)(q_0^2+4)}} \left[6\tanh(x)( \cos(\omega_{q_0} t + q_0 x)\tanh(x) \right.\nonumber\\ 
        &+ q_0 \sin (\omega_{q_0} t + q_0 x)) 
        -\left.  2 (1 + q_0^2)\cos(\omega_{q_0} t + q_0 x)\right]\,.
    \end{align}

    Note that, although the asymptotic form of (\ref{e:lin_rad}) is a plane wave of frequency $\omega_{q_0}$, near the origin it becomes distorted due to the presence of the kink. Assuming small shape mode amplitudes and radiation frequencies close to $2\omega_s$, and adopting a monochromatic wave ansatz as in (\ref{e:cq_lin_rad}), equation (\ref{e:effec_eq:cs}) reduces to
    \begin{eqnarray}\label{e:mathieu}
        \ddot{c}_s(t) + \left(\omega_s^2 + f(q_0)\sin(\omega_{q_0} t)\right) \,c_s(t)=0\,,
    \end{eqnarray} 
    with
    \begin{eqnarray}
    f(q_0) = - \frac{3\pi A_{q_0}}{4} \dfrac{q_0^2 (q_0^2 - 2)}{\sinh \left( \pi  q_0 / 2 \right)}\sqrt{\dfrac{q_0^2 + 4}{q_0^2 + 1}}\,. 
    \end{eqnarray} 
    Notably, the equation (\ref{e:mathieu}) constitutes a Mathieu equation. This differential equation is known to present instability regions in the parameter space. By means of Floquet theory, one can deduce the stability chart of the Mathieu equation through a condition on the trace of the monodromy matrix \cite{Kovacic:2018}. In our particular case, it can be verified that the instability regions, for small $A_{q_0}$, occur in a neighbourhood of $\omega_s/\omega_{q_0} = k/2$, where $k \in \mathbb{Z}$. Moreover, as $A_{q_0}$ increases, the instability bands broaden. However, we are mainly interested in small excitations. From the above condition, the instability appears when the frequency of the radiation is twice the frequency of the shape mode if we assume $k=1$. Hence, the radiation triggers resonantly the shape mode and one should expect an exponential amplification of its amplitude. Naturally, once back-reaction is taken into account, energy conservation implies that as $c_s(t)$ grows, the third term in (\ref{e:effec_eq:cq}) transfers energy to the radiation modes, causing the exponential growth to cease. Indeed, it has been observed numerically in \cite{Romanczukiewicz:2008} that when the kink is illuminated with radiation of frequency $\omega \approx 2\omega_s$, it undergoes a significantly smaller acceleration compared to nearby frequencies, suggesting that this resonance peak is associated with energy transfer to the shape mode. This effect is naturally explained by the instability of the Mathieu equation (\ref{e:mathieu}). Furthermore, this resonant transfer mechanism between internal modes has also been observed in two and three-dimensional solitons \cite{Blanco:2023,Blanco1:2023}. 

\begin{figure}[H]
    \centering
    \subfloat[Linear radiation with $q = 2.0$.]{
    \includegraphics[scale=0.6]{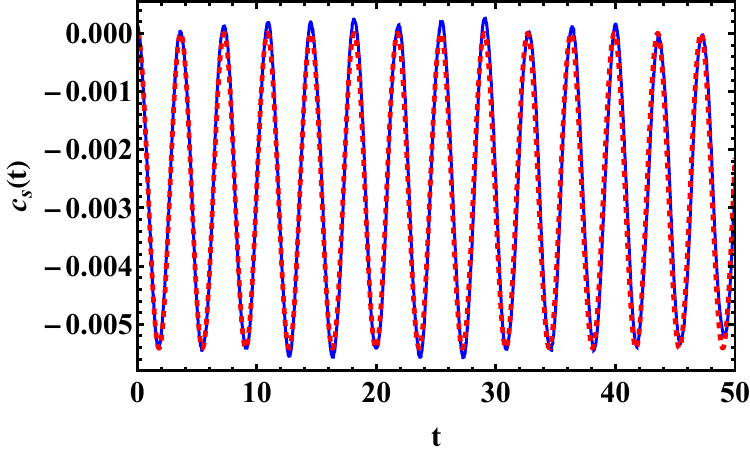}   
    }
    \subfloat[Linear radiation with $q = 3.0$.]{
    \includegraphics[scale=0.6]{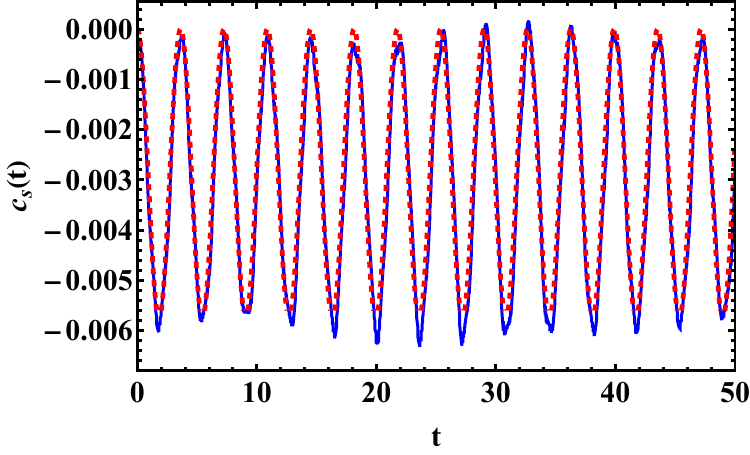} 
    }
    \caption{Excitation of the shape mode by external radiation of different frequencies and the same amplitude $A_q = 0.02$. The figure compares the results obtained from the full field theory (solid blue line) with those from the approximate solution given by equation (\ref{e:eq_rad_appr}) (dashed red line).}
    \label{f:excitation_shape2}
    \end{figure}

    Remarkably, we can also obtain an approximate analytical solution for the shape mode in the background of radiation for frequencies different from the resonance frequency. For small radiation amplitudes, the relevant terms in (\ref{e:effec_eq:cs}) are the harmonic oscillator part and the last term, representing the radiation source. This equation has the form of a forced harmonic oscillator. Regarding the initial conditions derived from (\ref{e:cq_lin_rad}) along with an initially unexcited shape mode $c_s(0)=0,\, c_s'(0)=0 $, we get
    \vspace{0.1cm}
    \begin{align}\label{e:eq_rad_appr}
        c_s(t) = \Omega(q_0)\left(\dfrac{1}{\omega_s^2} + \dfrac{\bigl(4\,\omega_{q_0}^2-(\sech(\pi q_0)+1)\,\omega_s^2\bigr) \cos(\omega_s t)}{\omega_s^2\left(\omega_s^2-4\,\omega_{q_0}^2\right)}+ \dfrac{\sech(\pi q_0)\,\cos(2\,\omega_{q_0}t)}{\omega_s^2-4\,\omega_{q_0}^2}\right),
    \end{align}
    where
    \vspace{0.1cm}
    \begin{equation}
    \Omega (q_0)= - A_{q_0}^2\dfrac{3 \sqrt{\dfrac{3}{2}} \pi  \left(8 \,q_0^4 + 34\,q_0^2+17\right)}{4 \left(q_0^4+5 q_0^2+4\right)}\,.
    \end{equation}    
    \vspace{0.04cm}
    For larger radiation amplitudes, this expression is no longer valid, and new phenomena may emerge. Similar results concerning kink-antikink pair creation induced by radiation were reported in \cite{Romanczukiewicz:2006}. A comparison between the approximate analytical solution and the full field theory results is presented in \autoref{f:excitation_shape2}, where the close agreement between both can be appreciated. Notably, this analytical expression also explains the excitation of the shape mode with negative amplitude.

    So far, the translational degree of freedom has not been included in our model. However, the translational mode can interact with radiation and lead to unexpected phenomena, such as negative radiation pressure \cite{Romanczukiewicz:2008}. In the following section, we explore key aspects of the interaction between the translational mode and radiation within our framework.

\section{Adding the translational mode}\label{s:Trans_Oscil_Radi}

    In order to generalise the previous approach allowing for the rigid motion of the kink, we require to introduce a collective coordinate accounting for the translation of the kink. A first naive possibility consists in considering the ansatz introduced in (\ref{e:general_ansatz}). However, that field configuration includes the zero mode, that only can describe infinitesimal translations. To address that issue and describe a continuous and finite motion, we consider the following field configuration
    \begin{equation}\label{e:ansatz:trans}
        \Phi (x,t) = \Phi_{K}\big( x - a(t) \big) + c_{s}(t)\eta_{s}\big( x - a(t) \big) + \int_{\mathbb{R}} dq\, c_{q}(t)\eta_{q} \big( x - a(t) \big)\,,
    \end{equation}
    where $a(t)$ is the collective coordinate that describes the position of the kink. We will show that the corresponding effective model exhibits a richer structure and enables new couplings among the different collective coordinates.
    
    \enlargethispage{3\baselineskip}
    Once more, we have to substitute the field configuration ansatz into the Lagrangian density of the full theory (\ref{e:lagrangian_phi4}) in order to obtain the corresponding collective coordinate model. Notice that (\ref{e:ansatz:trans}) only adds new contributions to the kinetic part with respect to (\ref{e:ansatz_oscill}), whereas the potential terms remain the same. The invariance of the potential term under the translation of the kink is related to the Lorentz invariance of the theory. The additional contributions to the effective Lagrangian due to the presence of the zero mode are given by 
    \begin{align}\label{e:expansion_action:kinetic_trans}
        L_{t}[a] = & \hspace{0.1cm} \dfrac{1}{2}\int_{\mathbb{R}} dx\, \bigg( \dot{a}^{2}(t) \bigg[ \int_{\mathbb{R}} \, c_{q}(t)\,c_{q'}(t)\,\eta'_{q}(x)\,\eta'_{q'}(x)\,dq\, dq' + \Phi'\hspace{0.01cm}_{K}^{2}(x) + c_{s}^{2}\,(t)\eta'\hspace{0.01cm}_{s}^{2}(x)\nonumber \\
        &+ 2\,\Phi'_{K}(x)\,c_{s}(t)\,\eta'_{s}(x) + 2\,\Phi'_{K}(x)\int_{\mathbb{R}} c_{q}(t)\eta'_{q}(x)\,dq + 2 c_{s}(t)\,\eta'_{s}(x)\int_{\mathbb{R}} c_{q}(t)\,\eta'_{q}(x)\,dq\bigg]\nonumber\\
        &- \dot{a}(t) \bigg[ 2\dot{c}_{s}(t)\eta_{s}(x)\int_{\mathbb{R}} c_{q}(t)\eta'_{q}(x)\,dq + 2\,c_{s}(t)\,\eta'_{s}(x)\int_{\mathbb{R}} \dot{c}_{q}(t)\,\eta_{q}(x)\,dq\nonumber\\
        &+ 2\int_{\mathbb{R}} c_{q'}(t)\,\dot{c}_{q}(t)\,\eta_q(x)\,\eta'_{q'}(x)\,dq \,dq'\, \bigg] \bigg)\,.
    \end{align}
    From now on, we will assume that $|\dot{a}(t)| \ll 1$. Consequently, we will retain only the terms involving $\dot{a}(t)$ linearly, as well as the contributions that couple $\dot{a}^2(t)$ with the other collective coordinates at lowest order.
    \newpage
    Integrating over the $x$-coordinate and adding the Lagrangian term $\mathcal{L}_{s,q}$, we finally obtain the following effective Lagrangian
    \vspace{0.1cm}
    \begin{align}\label{e:Lag_sqt}
        L_{s,q,t}[a,c_s,c_q]= &- \dfrac{4}{3} + \dfrac{1}{2}\left(\dot{c}^2_s(t)-\omega_s^2 \,c_s^2(t)\right) + \pi \int  \left(\dot{c}_{q}(t)\,\dot{c}_{-q}(t) - \omega_q^2\, c_{q}(t)\,c_{-q}(t)\right)\,dq\nonumber\\
        &- \dfrac{3 \pi}{16}\sqrt{\dfrac{3}{2}}c_{s}^{3}(t) + c_s^2(t)\int f_s(q)\,c_q(t)\,dq + c_s(t)\int f_{sq}(q,q')\,c_q(t)\,c_{q'}(t)\,dq\,dq'\nonumber\\
        &+ \dfrac{2}{3}\dot{a}^2(t) + \dfrac{\pi}{4}\sqrt{\dfrac{3}{2}}\dot{a}^2(t)c_s(t) +\dot{a}^2(t)\int f_{aa}(q)\,c_q(t)\,dq\nonumber\\
        &+ \dot{a}(t)\int f_{as}(q)\left(\dot{c}_s(t)\,c_q(t)-c_s(t)\,\dot{c}_q(t)\right)\,dq\nonumber\\
        &+ \dot{a}(t)\int f_a(q,q')\, \dot{c}_{q}(t)\,c_{q'}(t)\,dq\,dq'\,.
    \end{align}
    \vspace{0.1cm}
    The couplings between modes are collected in the following functions
    \vspace{0.1cm}
    \begin{align}
        f_s(q) &= - \frac{3 i \pi}{16}\sqrt{\frac{q^2+4}{q^2+1}}
                  \frac{q^2(q^2-2)}{\sinh\!\left(\frac{\pi q}{2}\right)}\,, \\[2mm]
        f_{as}(q) &= - \frac{\pi}{4}\sqrt{\frac{3}{2}}
                    \sqrt{\frac{q^2+1}{q^2+4}}
                    \frac{q^2 + 3}{\cosh\!\left(\frac{\pi q}{2}\right)}\,, \\[2mm]
        f_{aa}(q) &= - i \frac{\pi}{4}\sqrt{\frac{q^2+4}{q^2+1}}
                    \frac{q^2}{\sinh\!\left(\frac{\pi q}{2}\right)}\,, \\[2mm]
        f_a(q,q') &=
        \begin{aligned}[t]
          &\frac{3 i \pi}{4}
            \frac{4+q^2+q'^2}{\sqrt{(q^2+1)(q^2+4)}\sqrt{(q'^2+1)(q'^2+4)}}
            \frac{q^2 - q'^2}{\sinh\!\left(\tfrac{\pi}{2}(q + q')\right)}
        \end{aligned}\nonumber\\[2mm]
        &\quad - \frac{2 i\pi q' \left(4-9 q q' - 2 q^2 - 2 q'^2 + q^2 q'^2\right)}
            {\sqrt{(q^2+1)(q^2+4)}\sqrt{(q'^2+1)(q'^2+4)}} \delta(q+q')\,, \label{eq:fa_second}\\[2mm]
        f_{sq}(q,q') &= -\frac{3\pi}{8}\sqrt{\frac{3}{2}}
            \frac{17+17q^2+17 q'^2+10 q^2q'^2-q^4-q'^4+q^2q'^4+q^4q'^2-q^6-q'^6}
            {\sqrt{(q^2+1)(q^2+4)}\sqrt{(q'^2+1)(q'^2+4)}\cosh\!\left(\frac{\pi}{2}(q+q')\right)}\,.
    \end{align}
    \vspace{0.1cm}
    The quadratic terms in (\ref{e:Lag_sqt}) correspond to a collection of harmonic oscillators plus the kinetic term of the kink. In contrast, the cubic terms describe the interactions between the modes. The field equations for the radiation modes, derived from (\ref{e:Lag_sqt}), are given by
    \vspace{0.1cm}
    \begin{align}\label{e:eq_rad_kink}
        \ddot{c}_{-q}(t) &+ \omega_{q}^2\, c_{-q}(t) - \dfrac{1}{2\pi}c^2_s(t) f_s(q) - \dfrac{1}{\pi} c_s(t)\int  f_{sq}(q,q')\,c_{q'}(t)\,dq' - \dfrac{1}{2\pi}\dot{a}^2(t)f_{aa}(q)\nonumber\\
        &- \dfrac{1}{2\pi}\ddot{a}(t)\,f_{as}(q)\,c_s (t) - \dfrac{1}{\pi}\dot{a}(t)\,f_{as}(q)\,\dot{c}_s(t) +\dfrac{1}{2\pi}\dot{a}(t)\int  \dot{c}_{q'}(t)\left(f_a(q,q')-f_a(q',q)\right)\,dq'\nonumber\\
        &+ \dfrac{1}{2\pi}\ddot{a}(t)\int f_a(q,q')\,c_{q'}(t)\,dq' = 0\,,
    \end{align}
    \vspace{0.1cm}
    whilst the equations of motion determining the evolution of $c_s(t)$ and $a(t)$ take the form
    \vspace{0.1cm}
    \begin{align}\label{e:eq_shape_kink}
        \ddot{c}_s(t) &+ \omega_s^2\, c_s(t) + \dfrac{9\pi}{16}\sqrt{\dfrac{3}{2}}c_{s}^{2}(t) - 2 \,c_s(t)\int f_s(q)\,c_q(t)\,dq - \int f_{sq}(q,q')\,c_q(t)\,c_{q'}(t)\, dq\,dq'\nonumber \\
        &- \frac{\pi}{4}\sqrt{\frac{3}{2}}\,\dot{a}^2(t) + 2\,\dot{a}(t)\int f_{as}(q)\,\dot{c}_{q}(t)\, dq + \ddot{a}(t)\int f_{as}(q)\,c_q(t)\, dq = 0\,,
    \end{align}
    and
    \vspace{0.1cm}
    \begin{align}\label{e:eq_trans_kink}
        \frac{4}{3}\ddot{a}(t)&+\dfrac{\pi}{2}\sqrt{\dfrac{3}{2}}\left(\ddot{a}(t)c_s(t) + \dot{a}(t)\dot{c}_s(t)\right) + 2\int f_{aa}(q)\left(\ddot{a}(t)c_q(t) + \dot{a}(t)\dot{c}_q(t)\right)\,dq\nonumber \\
         &+\int f_{as}(q)\left(\ddot{c}_s(t)c_q(t) - c_s(t)\ddot{c}_q(t)\right)\,dq + \frac{d}{dt}\int f_a(q,q')\, \dot{c}_{q}(t)\,c_{q'}(t)\,dq\,dq' = 0\,.
    \end{align}
    
    The first configuration we analyse is the translating kink without any excitations. In the non-relativistic regime, this corresponds to $a(t) = x_0 + v t$, with all other modes vanishing. However, a straightforward inspection of the system (\ref{e:eq_rad_kink})–(\ref{e:eq_trans_kink}) reveals a surprising feature: if $\dot{a}(t)\neq0$, the terms proportional to $\dot{a}(t)^2$ in equations (\ref{e:eq_rad_kink}) and (\ref{e:eq_shape_kink}) act as sources for the shape mode and radiation. At first glance, this appears contradictory, since the Lorentz invariance of the model guarantees that the boosted kink is an exact solution. Indeed, the Lorentz-boosted kink solution takes the form
    \begin{equation}\label{e:boosted}
        \phi(x,t) = \tanh\left(\frac{x-v t}{\sqrt{1-v^2}}\right).
    \end{equation}
    Clearly, this solution does not contain any radiation. This apparent inconsistency in the effective model can be readily resolved. The standard collective coordinate method is not Lorentz invariant at leading order, which is evident from the asymmetry between time and spatial coordinates in the ansatz (\ref{e:ansatz:trans}). Nevertheless, when higher-order terms are taken into account, approximate Lorentz-invariant solutions emerge. In fact, the aforementioned source terms for $c_s(t)$ and $c_q(t)$ are responsible for capturing the Lorentz contraction of a moving kink. Let us consider a moving kink with velocity $v \ll 1$ initially located at the origin so that $\dot{a}(t)=v$. Assuming that $c_s(t)$ and $c_q(t)$ do not depend on time, an approximate solution of (\ref{e:eq_shape_kink}) and (\ref{e:eq_rad_kink}) is given by
    \begin{equation}\label{e:sol_boost}
        c_s(t)=\frac{\pi }{4\sqrt{6}}v^2\,,\quad c_q(t)=-\frac{i q^2\csch(\pi q/2)}{8\sqrt{(q^2+1)(q^2+4)}}v^2\,,
    \end{equation}
    where we disregard corrections of order $\mathcal{O}(v^4)$. On the other hand, if we expand (\ref{e:boosted}) with respect to $v$, at $t=0$ we obtain
    \begin{equation}
        \phi(0,x) = \tanh(x)+\frac{1}{2}\left(x-x\tanh^2(x)\right)v^2+\mathcal{O}(v^4)\,.
    \end{equation}
    Denoting the first-order correction to the Lorentz contraction by $\phi^{(1)}(x)=\frac{1}{2}\left(x-x\tanh^2(x)\right)v^2$ we observe that its projection onto the spectral modes yields
    \begin{align}
        \langle \phi^{(1)}(x),\eta_s(x) \rangle&=\frac{\pi }{4\sqrt{6}}v^2\,, \label{e:projection_cs} \\
        \langle \phi^{(1)}(x),\eta_q(x) \rangle&=-\frac{i \pi q^2\csch(\pi q/2)}{4\sqrt{(q^2+1)(q^2+4)}}v^2\,. \label{e:projection_cq}
    \end{align}
    Comparing (\ref{e:sol_boost}) and (\ref{e:projection_cs})-(\ref{e:projection_cq}) it is clear that the only difference is an additional $2\pi$ factor for the constant $c_q$ amplitudes that arises due to the normalisation of the scattering modes. We conclude that, already at quadratic order, our "non-relativistic" collective coordinate approach describes relativistic effects, although these effects require in general, the excitation of scattering modes. Specifically, the solution (\ref{e:sol_boost}) can be interpreted as a first-order Lorentz boost, which in terms of the coordinates of our model takes the form
    \begin{align}
        \Phi_{\gamma}(x,t)&=\tanh(x- v t)+\phi^{(1)}(x-v t)+\mathcal{O}(v^4)\nonumber \\
        &=\tanh(x-v t)+v^2\frac{\pi }{4\sqrt{6}} \eta_s(x-v t)-v^2\int \frac{i q^2\csch(\pi q/2)}{8\sqrt{(q^2+1)(q^2+4)}}\eta_q(x-v t)\,dq\,,
    \end{align}
    for $x_0 = 0$. For a perturbative expansion of the Lorentz-boosted version of the excited kink plus radiation see \cite{Oxtoby:2009}. However, if one aims to describe relativistic processes, this approach becomes inadequate, as even a simple boosted solution requires the excitation of a large number of modes. In fact, there is a simpler ansatz that describes the exact Lorentz boost with only two degrees of freedom given by the following expression
    \begin{equation}\label{e:rel_ansatz}
        \phi(x,t)=\Phi_K\left(\frac{x-a(t)}{\delta(t)}\right)\,.
    \end{equation}
    This type of ansatz was first considered in \cite{Rice:1983} and has been used recently in \cite{Manton:2022}. However, after the introduction of the scattering modes, it seems not possible to obtain analytical results; thus we will not pursue this line in this section. Nevertheless, as we shall see in \autoref{s:oscillon}, some small generalisations of (\ref{e:rel_ansatz}) can be used to effectively describe dissipative degrees of freedom. 

\begin{figure}[htb]
        \begin{center}
        \includegraphics[scale=0.6]{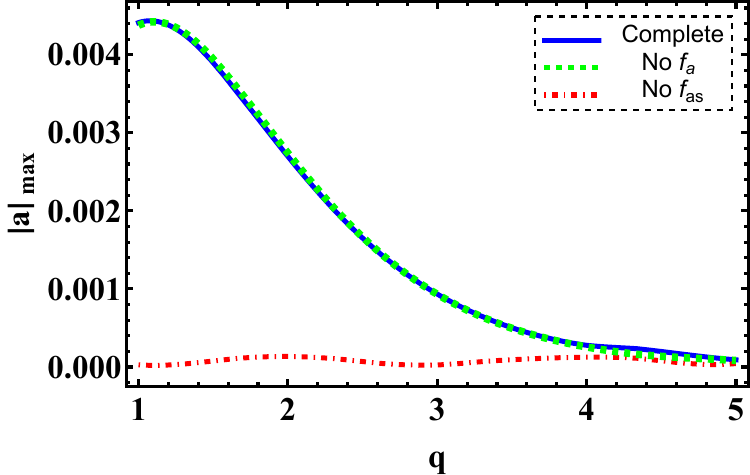}    
        \caption{Maximum displacement of the kink due to radiation of different frequencies and the same amplitude $A_q = 0.05$. The solid line represents the complete effective model, the dashed line represents the simulation when only the $f_{as}$ source is considered, and the dotted dashed line represents the simulation when only the source $f_{a}$ is taken into account. We have employed $n = 30$ equidistant scattering modes in the interval $q \in [-5,5]$. The time of simulation is $t = t_c$.}
        \label{f:displacement}
        \end{center}
    \end{figure}

    In our second experiment, we study the excitation of the translational mode by radiation. Note that the first and last terms in (\ref{e:eq_trans_kink}) resemble Newton’s second law, where the factor $4/3$ can be identified with the mass of the kink, and the integral represents the momentum carried by the radiation wave, given by
    \begin{equation}
        P(t)=\int_{\mathbbm{R}}\dot{R}(x,t)R'(x,t)\,dx\,.
    \end{equation}
    For initial conditions corresponding to a kink at rest illuminated by linear radiation, as described in (\ref{e:lin_rad}), the rate of change of the total radiation momentum governed by the last term in (\ref{e:eq_trans_kink}) vanishes, and the kink remains at rest. In fact, it is well known that the $\phi^4$ kink is transparent to linear radiation, as emphasised in \cite{Romanczukiewicz:2008}. The $\phi^4$ kink is only accelerated at fourth order in perturbation theory. Specifically, the kink is accelerated in the direction of the incoming wave, leading to the so-called negative radiation pressure phenomenon \cite{Romanczukiewicz:2004,Romanczukiewicz:2008}. As the phenomenon of negative radiation pressure only arises at fourth order in perturbation theory, the system of equations (\ref{e:eq_rad_kink})–(\ref{e:eq_trans_kink}), truncated at cubic order, cannot capture this effect, and analysing such higher-order phenomena within our framework is challenging. A more effective strategy, as proposed in \cite{Romanczukiewicz:2008}, involves extracting quartic-order corrections to the radiation from the asymptotic behaviour of the quadratic solution. Nonetheless, extending our approach to simultaneously describe both the negative radiation pressure and the energy transfer to the shape mode within a unified effective model would be of great interest. On the other hand, we want to emphasise that the shape mode may act as an intermediary between the translational mode and radiation. Specifically, a non-trivial interplay between the modes could occur as follows: radiation excites the shape mode via the term proportional to $f_{sq}(q,q')$ in (\ref{e:eq_shape_kink}). Once $c_s(t)$ is excited, there is a source for $a(t)$ in (\ref{e:eq_trans_kink}) given by the term proportional to $f_{as}(q)$. The resulting excitation of $a(t)$ is approximately of order $A_0A_q$, where $A_0$ is the amplitude of the shape mode and $A_q$ is the amplitude of the $q$-scattering mode. Since from (\ref{e:eq_rad_appr}) $A_0$ is of order $A_q^2$, the excitation of the zero mode due to this mechanism appears at order $\mathcal{O}(A_q^3)$. However, unlike the negative radiation pressure effect, this mechanism does not lead to a net momentum transfer at this order; instead, the kink undergoes small oscillations about its rest position. This can be seen directly from (\ref{e:eq_trans_kink}) by neglecting the last term and retaining only terms of order $\mathcal{O}(A_q^3)$ order. In \autoref{f:displacement}, we present the maximum amplitude of the translational mode as a function of the incident radiation frequency. The results confirm that, for small values of $q$, the term proportional to $f_a$ is indeed subleading. 

    For larger times, higher-order corrections to radiation start to play a role and the last term in (\ref{e:eq_trans_kink}) cannot be neglected anymore. In fact, this term transfers a net momentum to the kink \cite{Romanczukiewicz:2008}, causing it to be pulled in the direction opposite to that of the incoming radiation.

    Thus far, we have discussed only the single-kink sector. Of course, one expects that the combination of translational and scattering modes should play an important role in the kink-antikink sector, allowing for energy dissipation in scattering processes or even describing kink-antikink annihilation. This is a more ambitious goal that we leave for future research. However, during a kink-antikink scattering process, when the solitons completely overlap, an intermediate state that resembles the profile of an oscillon is formed. This observation suggests that the study of oscillons, under suitably chosen initial conditions, may offer valuable insights into such violent processes. This perspective will be adopted in the following section.

\section{Effective model for the radiating oscillon} \label{s:oscillon}

    The previous sections have confirmed the importance of the scattering modes in the dynamics of a single kink. In this section, we aim to illustrate the importance of such modes for the description of the dynamics of non-topological solitons. In particular, we will focus our attention to oscillon configuration in the one-dimensional $\phi^4$ theory. We will describe the decay of oscillons below the critical amplitude, the possibility of internal modes, and the kink-antikink formation. 
    
    Oscillons have been observed starting from rather generic initial data. Hence, we have decided to follow \cite{Fodor:2008,Fodor:2009} and take the ansatz
    \begin{equation}\label{e:KAK_ansazt_1}
        \Phi_\text{o}(x;a,R) = -1 + a\sech(x/R)\,, 
    \end{equation}
    where $a$ is identified with the amplitude of the oscillon and $R$ accounts for its size. Following the collective coordinate method, we promote the amplitude $a$ to a time-dependent variable and introduce (\ref{e:KAK_ansazt_1}) into the Lagrangian density (\ref{e:lagrangian_phi4}). Integrating over the spatial variable we are left with the associated effective Lagrangian
    \begin{equation}\label{e:Lag_eff_osci}
        L^{o}[a] = R\left(\dot{a}^2(t)-\dfrac{1}{3}\left[12+\dfrac{1}{R^2}\right]\,a^2(t)+\pi\, a^3(t)-\dfrac{2}{3}a^4(t)\right)\,.
    \end{equation}
    Note that the size $R$ is treated as a fixed constant throughout the analysis. The mechanical Lagrangian (\ref{e:Lag_eff_osci}) corresponds to an anharmonic oscillator of frequency $\omega_o = \sqrt{\frac{1}{3}(12+\frac{1}{R^2})}$. When small amplitudes are considered, the frequency of $a(t)$, $\omega_0$, is above the mass threshold of the theory $\omega_c = 2$. As a result, $a(t)$ couples directly to the continuum and collapses into radiation. Thus, the initial data (\ref{e:KAK_ansazt_1}) does not evolve into an oscillon when $a$ is sufficiently small. However, for sufficiently large values of $a(t)$, the non-linear terms in (\ref{e:Lag_eff_osci}) act to lower the effective frequency of oscillation, thereby avoiding the direct coupling to radiation. Now, the oscillon couples weakly to radiation and its amplitude decrease very slowly. We want to emphasise that the boundary between both regimens depends on the specific size $R$ of the oscillon. Naturally, the details of this coupling are not described through (\ref{e:Lag_eff_osci}), which requires the inclusion of radiation degrees of freedom. This can be achieved as in the previous sections considering an ansatz of the form
    \begin{equation}\label{e:KAK_ansazt_rad}
        \Phi_\text{o,rad}(x;a,c_q)= - 1 + a(t) \sech(x/R)+\int_\mathbb{R} c_q(t)\,\eta_q(x/R)\,dq\,.
    \end{equation}
    Although the scattering modes $\eta_q(x)$ are not part of the spectrum of perturbations around the oscillon background, they can still be used to represent the radiative degrees of freedom. Moreover, we have scaled these modes by the size of the oscillon to take into account that the source of radiation is not the kink. As we are interested in the description of the oscillon coupled to the slow-amplitude emitted radiation, it is enough to truncate the effective Lagrangian to second order in $c_q(t)$. Conversely, it is crucial to retain all orders in $a(t)$, since the existence of the oscillon is linked to the non-linear structure of the Lagrangian. 
    
\begin{figure}[H]
        \centering
        \subfloat[$n = 30$, $a_0 = 0.2$, $R = 1.5.$]{
        \includegraphics[scale=0.55]{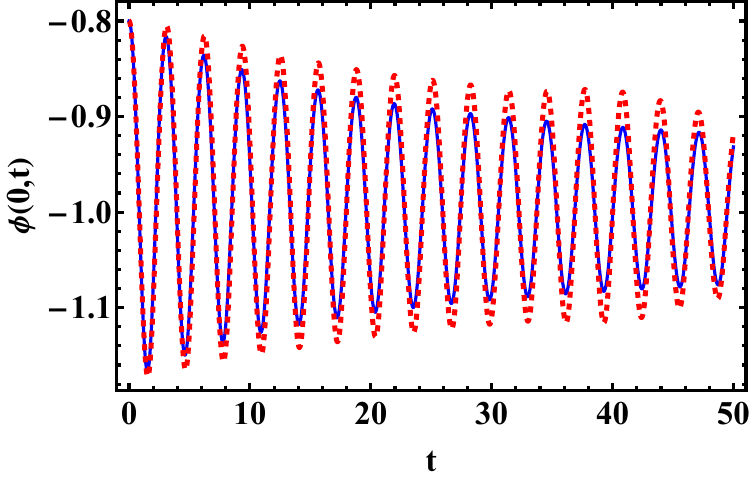}   
        }
        \subfloat[$n = 20$, $a_0 = 0.3$, $R = 1.5.$]{
        \includegraphics[scale=0.55]{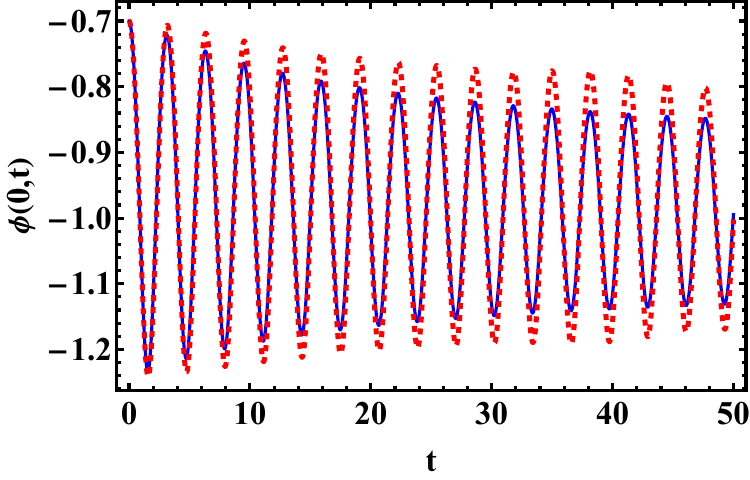}   
        }\\
        \subfloat[$n = 10$, $a_0 = 0.4$, $R = 2.$]{
        \includegraphics[scale=0.55]{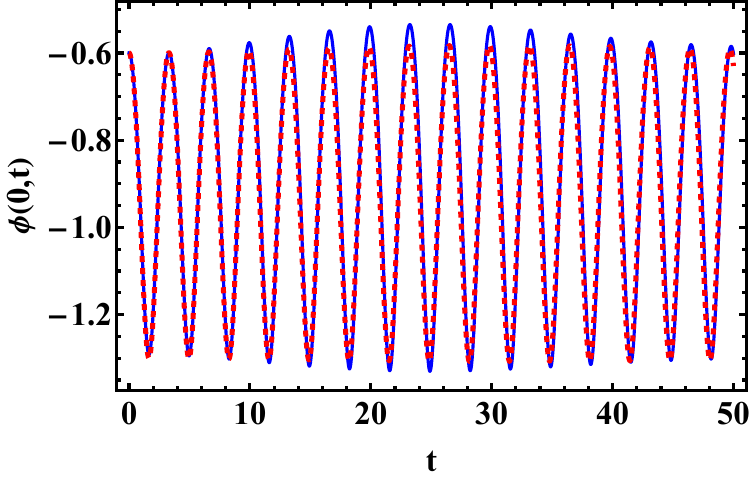}   
        }
        \subfloat[$n = 10$, $a_0 = 0.5$, $R = 2.$ \label{f:oscillon_radiation_d}]{
        \includegraphics[scale=0.55]{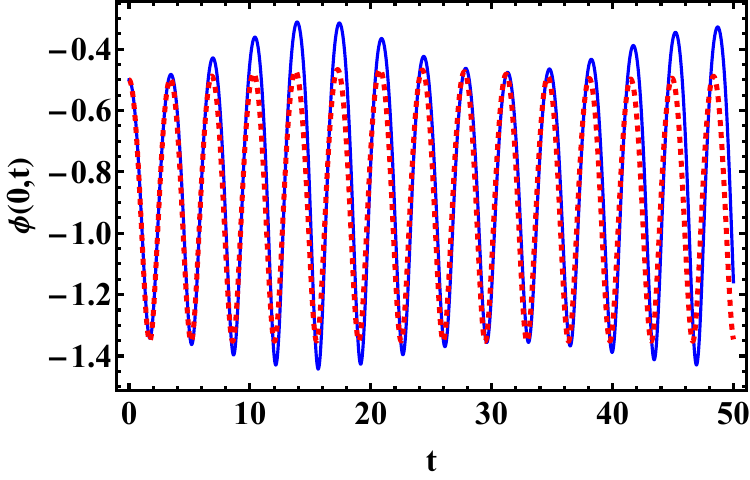}    
        }
        \caption{Comparison of oscillon decay and the formation of long-lived oscillons between the full field theory simulation (solid blue line) and the effective model (dashed red line) based on (\ref{e:lag_oscill_rad}). The scattering modes have been taken in the interval $q \in [-5,5]$.}
        \label{f:oscillon_radiation}
    \end{figure}

\begin{figure}[htb]
        \centering
        \includegraphics[scale=0.62]{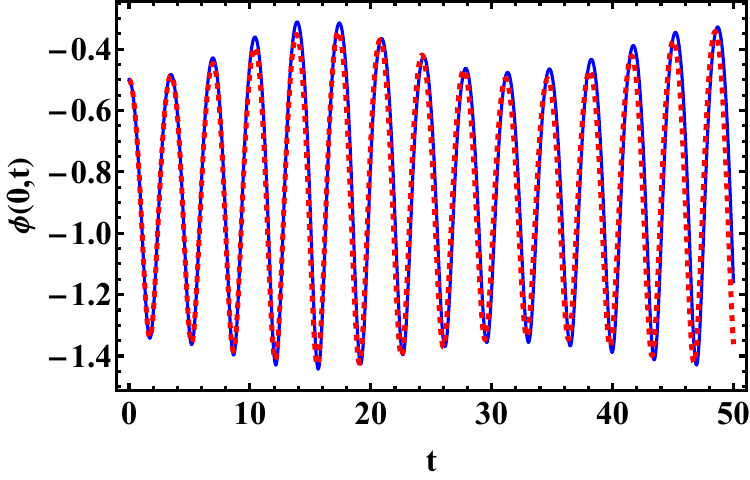} 
        \caption{Comparison of modulated oscillon between the full field theory simulation (solid blue line) and the effective model (dashed red line) based on (\ref{e:KAK_ansazt_size_1}). We have assumed the initial conditions $a_0 = 0.5,\, \delta_0 = 0$, and $R = 2$.}
        \label{f:oscillon_shape}
    \end{figure}

    Proceeding as in the previous examples, it is straightforward to verify that the effective Lagrangian corresponding to (\ref{e:KAK_ansazt_rad}) reads as
    \begin{align}\label{e:lag_oscill_rad}
        L^{o}_{r}[a,c_q] =&\hspace{0.1cm} \pi \int_{\mathbb{R}}\left( \dot{c}_q(t)\,\dot{c}_{-q}(t) -  w_{q,R}^2\, c_q(t)\,c_{-q}(t)\right)\,dq  + \dot{a}^{2}(t) - \omega_o^2\, a^{2}(t) + \pi\, a^{3}(t) - \dfrac{2}{3}a^{4}(t)\nonumber\\
        &+ \int_{\mathbb{R}^2} f_1(q,q',R)\, c_{q}(t)\,c_{q'}(t)\,dq\,dq' + \dot{a}(t)\int_{\mathbb{R}} f_2(q)\,\dot{c}_q(t)\,dq + a(t)\int_{\mathbb{R}}  f_3(q,R)\,c_q(t)\,dq\nonumber\\
        &+ a(t)\int_{\mathbb{R}^2} f_4(q,q')\,c_q(t)\,c_{q'}(t)\,dq\,dq'\,  + a^2(t)\int_{\mathbb{R}} f_{5}(q,q')\,c_q(t)\,c_{q'}(t)\,dq\,dq'\,\nonumber \\
        &+ a^{3}(t)\int_{\mathbb{R}}  f_6(q)\,c_q(t)\,dq\,,
    \end{align}
    with $w_{q, R} = \sqrt{ \left(q/R\right)^2 + 4}$ and where
    \begin{align}
        f_1(q, q', R) &= - \dfrac{3 \pi}{5 R^2} \dfrac{ 4 q + 4 q' + 5 q^3  + 5 q'^3  + q^5 + q'^5 }{\sqrt{\left(q^2+1\right) \left(q^2+4\right)}\sqrt{\left(q'^2+1\right) \left(q'^2+4\right)}} \csch\left(\dfrac{\pi}{2}(q + q')\right)\,,\\
        f_2(q) &= \dfrac{\pi}{2}\dfrac{\sqrt{q^2 + 1}}{\sqrt{q^2 + 4}} \sech\left(\dfrac{\pi 
        q}{2}\right)\,,\\
        f_3(q, R) &= \pi\dfrac{\sqrt{q^2 + 1}}{\sqrt{q^2 + 4}}\left( \dfrac{q^2 + 3}{4 R^2} - 2\right)\sech\left(\dfrac{\pi q}{2}\right)\,,\\
        f_4(q, q') &= \dfrac{3\pi}{4}\dfrac{11 + 14q^2 + 14q'^2 + 2q^2q'^2 + 3q^4 + 3q'^4}{\sqrt{\left(q^2+1\right) \left(q^2+4\right)}\sqrt{\left(q'^2+1\right) \left(q'^2+4\right)}}\sech\left(\dfrac{\pi}{2} (q + q')\right)\,,\\
        f_5(q, q') &= - \dfrac{3 \pi}{5}  \dfrac{ 4 q + 4 q' + 5 q^3 + 5 q'^3 + q^5 + q'^5 }{\sqrt{\left(q^2+1\right) \left(q^2+4\right)}\sqrt{\left(q'^2+1\right) \left(q'^2+4\right)}} \csch\left(\dfrac{\pi}{2}(q + q')\right)\,,\\
        f_6(q) &= \dfrac{\pi}{4}\dfrac{(q^2 + 1)^{3/2}}{\sqrt{q^2 + 4}} \sech\left(\dfrac{\pi 
        q}{2}\right)\,.
    \end{align}
    We have omitted a global factor $R$ in (\ref{e:lag_oscill_rad}) since it does not have any effect on the equations of motion. In order to obtain the two main evolutions of the oscillon, we give different suitable initial conditions for the amplitude and the size. In \autoref{f:oscillon_radiation} we illustrate the evolution of the oscillon profile at $x = 0$ in the two possible regimens.
    In the upper panel we show two configurations with small initial amplitudes. This results in a fast decay of the initial configuration into radiation. It is worth mentioning that even with a large number of radiation modes ($n>20$), the amplitude in the effective model decays more slowly than in the field-theoretical simulation. This suggests that the decay mechanism provided by the set of scattering modes is actually not very efficient. Although our approach is effective in describing the decay of initial configurations such as (\ref{e:KAK_ansazt_1}), or other similar profiles, into either a true oscillon or the vacuum when the initial amplitude is sufficiently small, it seems inadequate for capturing the radiation emission by a true oscillon. In fact, this is a very small effect that has to be measured at large times. The computation beyond-all-orders has been carried out in \cite{Fodor:2009}. The beyond-all-orders behaviour suggests that it is not expected to be predicted by the perturbative approach used here. In the lower panel, we present a genuine oscillon configuration. The numerical simulation reveals that the oscillon supports an internal, non-dissipative mode, as evidenced by the amplitude modulation observed during its time evolution. This internal mode can be associated with variations in the oscillon size. Indeed, this phenomenon finds a simple explanation in our approach. To verify that, let us consider a small variation of the oscillon width as follows
    \begin{equation}\label{e:KAK_ansazt_size}
        \Phi_\text{o}(x;a,\delta)= -1+ a\sech\left(\frac{x}{R+\delta}\right).
    \end{equation}
    If $\delta \ll 1$ we may expand about $\delta=0$ up to first order 
    \begin{equation}
        \Phi_\text{o}(x;a,\delta)=-1+a \sech(x/R)+\frac{a \delta}{R^2} \,x\, \sech(x/R)\tanh(x/R)\,.
    \end{equation}
    The additional term to the unperturbed oscillon corresponds to the so-called Derrick mode. This correction should codify the possible changes of the oscillon size, and luckily represent the behaviour expected from the full numerical result. However, this simple choice has a problem. At $a = 0$, the ansatz (\ref{e:KAK_ansazt_size}) does not depend on $\delta$. This implies that the moduli metric associated to $(a, \delta)$ is not well-defined at this point, leading to a null vector problem \cite{Wereszczynski:2021}. In order to cure this issue we may perform a simple change of coordinates $\delta\rightarrow \delta/a $. Finally, we get
    \begin{equation}\label{e:KAK_ansazt_size_1}
        \Phi_\text{o,s}(x;a,\delta)=-1+ a\sech(x/R)+\frac{\delta}{R^2}\, x\, \sech(x/R)\tanh(x/R)\,.
    \end{equation}
    Treating the coefficients $a$ and $\delta$ as collective coordinates whilst $R$ remains fixed,  we can compute the following effective Lagrangian
    \begin{align}\label{e:lag_osc_internal}
        L^{o}_{s}[a,\delta] =& \hspace{0.1cm} R \,\dot{a}(t)^2 - \omega_o^2\, a(t)^2 + \dfrac{1}{R}\left(\dfrac{\pi ^2}{36} + \dfrac{1}{3} \right) \dot{\delta}(t)^2 - \dfrac{1}{R}\left( \pi ^2\left(\dfrac{1}{9} + \dfrac{7}{180 R^2} \right) + \dfrac{4}{3}  \right) \delta (t)^2\nonumber\\
        &+ \pi\,  R\, a(t)^3 - \dfrac{2}{3} R\, a(t)^4 + \dfrac{\pi}{R^2}\left( \dfrac{11 \pi ^2}{80} - 1 \right) \delta (t)^3 - \dfrac{1}{R^3}\left(\dfrac{\pi ^4}{600} + \dfrac{\pi ^2}{90} - \dfrac{2}{15}\right) \delta (t)^4\nonumber\\
        &+ \dot{a}(t)\, \dot{\delta }(t) + \left(\dfrac{1}{3 R^2}-4\right) a(t)\,\delta (t) + \dfrac{\pi}{R}\left(\dfrac{3 \pi ^2}{16} - 1 \right) a(t) \,\delta (t)^2\nonumber \\
        &- \dfrac{1}{R^2}\left( \dfrac{7 \pi ^2}{90} - \dfrac{1}{3}\right) a(t) \,\delta (t)^3 + \pi\,  a(t)^2\, \delta(t) - \dfrac{2}{3} a(t)^3 \,\delta (t) - \dfrac{\pi ^2}{15 R} a(t)^2\, \delta (t)^2. 
    \end{align}
    The associated equations of motion are as follows
    \begin{align}
        \ddot{ a}(t) &+ \dfrac{1}{3}\left(\dfrac{1}{R^2} + 12\right) a(t) - \dfrac{3}{2} \pi  a(t)^2 + \dfrac{4}{3}a(t)^3 + \dfrac{1}{2 R}\ddot{ \delta }(t) + \dfrac{\left(12 R^2 - 1\right)}{6 R^3}\delta (t)\nonumber\\
        &- \dfrac{\pi  \left(3 \pi ^2-16\right)}{32 R^2}\delta (t)^2 + \dfrac{\left(7 \pi ^2 - 30\right)}{180 R^3} \delta (t)^3 - \dfrac{\pi}{R}a(t) \delta (t) + \dfrac{\pi ^2}{15 R^2}a(t) \delta (t)^2\nonumber\\
        &+ \dfrac{1}{R}a(t)^2 \delta (t) = 0\,,
    \end{align}
    and
    \begin{align}
        \ddot{ \delta }(t) &+ \dfrac{\left(\pi ^2 \left(20 R^2 + 7\right) + 240 R^2 \right)}{5 \left(\pi ^2 + 12\right) R^2} \delta (t) -\dfrac{27 \pi  \left(11 \pi ^2 - 80\right)}{40 \left(\pi ^2 + 12\right) R}\delta (t)^2 +\dfrac{\left(3 \pi ^4 + 20 \pi ^2 - 240\right)}{25 \left( \pi ^2 + 12 \right) R^2}\delta (t)^3\nonumber\\
        &+ \dfrac{18 R}{\pi ^2 + 12}\ddot{ a}(t) + \dfrac{\left(72 R^2-6\right)}{\left(\pi ^2 + 12\right) R}a(t) -\dfrac{18 \pi  R}{\pi ^2 + 12}a(t)^2 +\dfrac{12 R}{\pi ^2 + 12}a(t)^3\nonumber \\
        &- \dfrac{9 \pi  \left(3 \pi ^2-16\right)}{4 \left(\pi ^2 + 12\right)}a(t) \delta (t) + \dfrac{3 \left(7 \pi ^2-30\right)}{5 \left(\pi ^2 + 12\right) R}a(t) \delta (t)^2 +\dfrac{12 \pi ^2}{5 \pi ^2 + 60}a(t)^2 \delta (t) = 0\,.
    \end{align}
    The fundamental frequency of $\delta(t)$, given by its linear term, lies above the continuum threshold, which implies that excitations of this mode should, in principle, dissipate quickly due to radiation coupling. Nevertheless, similar to the behaviour observed in the amplitude of the oscillon itself, sufficiently large-amplitude excitations of $\delta(t)$ can lead to non-linear effects that reduce its frequency below $\omega_c$, effectively trapping the mode and preventing energy loss through radiation. To investigate the accuracy of our new effective model with field theory, we assume the same initial conditions as for \autoref{f:oscillon_radiation_d}. The corresponding comparison is depicted in \autoref{f:oscillon_shape}.
    
    Note that the prediction of the effective model (\ref{e:lag_osc_internal}) and field theory agree with great precision. Therefore, we can confirm that the previous quasi-periodic behaviour is related to the existence of an internal state bounded to the oscillon. Despite the accuracy of the model, we can go beyond and include the radiation modes as well, since these degrees of freedom may be significant for certain field configurations. However, as shown in  \autoref{f:oscillon_radiation}, the addition of genuine radiation modes to the effective model does not seem to dissipate efficiently the energy. 
    
    If one is interested in the dynamics close to the oscillon core and for moderately short time scales, one may instead consider adding modes that resemble scattering modes, i.e., modes with spatial frequencies above the mass threshold, but spatially confined within the oscillon. Following this strategy we consider the following ansatz
    \begin{equation}\label{e:KAK_ansazt_size_2}
        \Phi_\text{o, rad}(x;a,\delta) = - 1+ a(t) e^{-\left(\frac{x}{R}\right)^2}+\sum_{k=1}^n \delta_k(t)\, h_k(x/r)\,.
    \end{equation}
    For simplicity, we have changed the initial oscillon profile to a Gaussian profile \cite{Salmi:2012}. Although (\ref{e:KAK_ansazt_1}) seems to model better the oscillon  tails, the profile (\ref{e:KAK_ansazt_size_2}) greatly simplifies the calculations. We choose the modes as follows
    \vspace{0.2cm}
    \begin{equation}
        h_k(x/r) = \dfrac{1}{k!}\frac{d^k}{dr^k}e^{-\left(\frac{x}{r}\right)^2}.
    \end{equation}
    The first term is, in fact, the Derrick mode associated to the Gaussian profile. The precise choice of the rest of the modes is not very relevant as long as they have increasing spatial frequencies around the oscillon core. The reason for choosing a different scale $r$ for the modes mimicking radiation, distinct from the oscillon scale $R$, will become clear later on.
    \enlargethispage{2\baselineskip}
    The effective Lagrangian corresponding to (\ref{e:KAK_ansazt_size_2}) can be expressed symbolically in a particularly simple form
    \begin{equation}\label{e:Lag_os_new}
        L^{o}_{r}[\xi_k]=\sum_{k,l = 0}^{n}m_{k,l} \,\dot{\xi}_k(t)\dot{\xi}_l(t)-\sum_{k,l = 0}^{n}\omega^2_{k,l}\, \xi_k(t)\xi_l(t)-V(\xi_k(t))\,,
    \end{equation}
    where $\xi_0(t)=a(t)$ and $\xi_k(t)=\delta_k(t)$ for $k=1,\ldots,n$, the matrices $m_{k,l}$ and $\omega^2_{k,l}$ are constant, and $V(\xi_k(t))$ is a potential that couples non-linearly all the modes. 
    
    Using standard techniques we can diagonalise simultaneously $m_{k,l}$ and $\omega_{k,l}$ and rewrite (\ref{e:Lag_os_new}) in normal coordinates $\eta_k(t)$ as follows
    \begin{equation}\label{e:Lag_os_final}
        L^\text{o}_\text{r}[\eta_k]=\sum_{k=0}^{n} m_k\, \dot{\eta}_k^2 (t)-\sum_{k=0}^{n} \omega_k^2\, \eta_k^2 (t)-V(\eta_k(t))\,.
    \end{equation}
    Therefore, in our approach, the oscillon is described by $n+1$ anharmonic oscillators of proper frequencies $\omega_k$ coupled non-linearly by the potential $V(\eta_k(t))$. Although an explicit expression of the coefficients of the model seems not to be possible for a generic number of modes $n$, we can guarantee that all the non-zero spatial integrations involved during the computation of the effective model can be obtained analytically through
    \begin{equation}
        \bigintsss_{\mathbb{R}}\, x^{2a}e^{-b(x/R)^2}e^{-c(x/r)^2}\,dx = \Gamma \left(\dfrac{1 + 2a}{2}\right)\left(\dfrac{b}{R^2} + \dfrac{c}{r^2}\right)^{-\dfrac{1}{2}\big(1 + 2a\big)}\,,
    \end{equation}
    where $a,b,c \in \mathbbm{Z}^+$ and $\Gamma$ denotes the Gamma function. The system (\ref{e:Lag_os_final}) is conservative, therefore it cannot dissipate energy. However, the modes $\delta_k(t)$ act effectively as radiation degrees of freedom, storing energy from the amplitude $a(t)$. The energy transfer mechanism works actually very efficiently as we will show. As a consequence, for not very large times, the model is able to describe how the initial data decays into radiation.
 
\begin{figure}[H]
    \centering
    \subfloat[Initial condition: $a_0 = 0.1$, $R=4$, $r = 4$, $n=7$.\label{f:oscillon_internal_full:a}]{
        \includegraphics[width=0.45\textwidth]{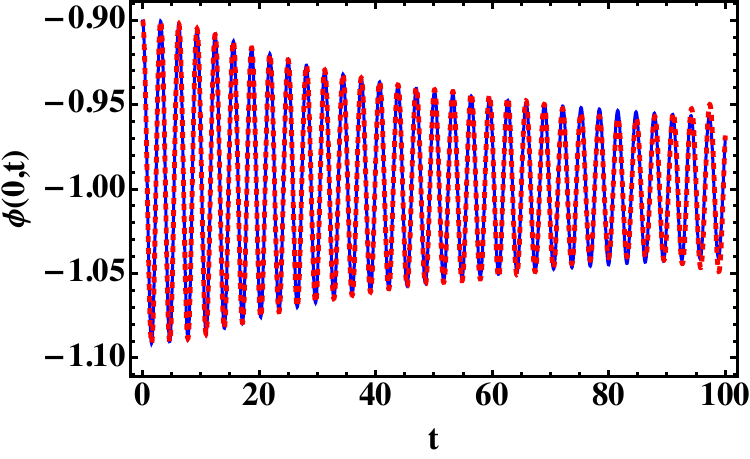}
    }
    \hfill
    \subfloat[Initial condition: $a_0 = 0.5$, $R=4$, $r = 4$, $n=7$.\label{f:oscillon_internal_full:b}]{
        \includegraphics[width=0.45\textwidth]{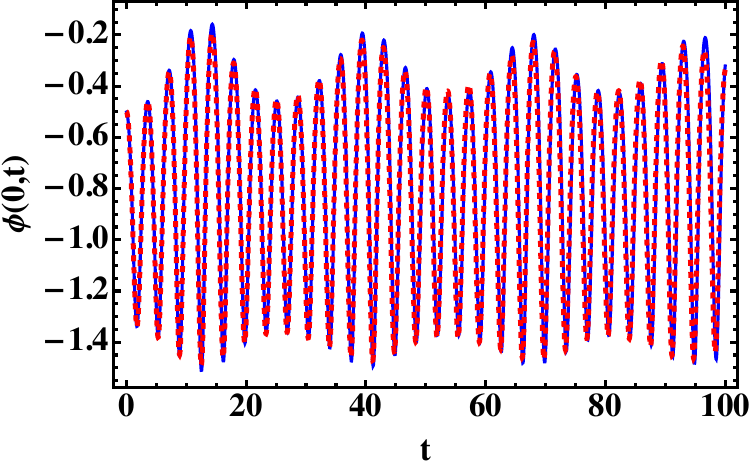}
    }
    \caption{Time evolution of $\phi(0,t)$ for initial data given by (\ref{e:KAK_ansazt_size_3}), shown for various initial amplitudes. Solid blue lines correspond to full numerical simulations, while dashed red lines represent the results from the effective model derived in (\ref{e:Lag_os_final}).}
    \label{f:oscillon_internal_full}
    \end{figure}

    In our following experiment we study the decay of an initial configuration of the form 
    \begin{equation}\label{e:KAK_ansazt_size_3}
        \Phi_\text{o}(x;a_0)=-1+ a_0\, e^{-\left(\frac{x}{R}\right)^2}\,.
    \end{equation}
    Below a critical value of $a_0$, the initial configuration given by (\ref{e:KAK_ansazt_size_3}) decays into radiation as we discussed. This value can be taken as the minimal $a_0$ such that the proper frequency of $a(t)$ coincides with the mass threshold frequency. To compute the critical amplitude, we consider a truncated version of the effective model (\ref{e:Lag_os_new}), retaining only the Gaussian profile. The resulting simplified effective model takes the form
    \begin{equation}\label{e:Lag_effec_a}
        L^o[a]=\sqrt{\frac{\pi }{2}} R \left(\frac{1}{2} \dot{a}(t)^2-\frac{\left(4 R^2+1\right) a(t)^2}{2R^2}+2 \sqrt{\frac{2}{3}} a(t)^3-\frac{a(t)^4}{2 \sqrt{2}}\right)\,,
    \end{equation}
    where the corresponding equation of motion is given by
    \begin{equation}
        \ddot{a}(t) + \left(4 + \dfrac{1}{R^2}\right)\, a(t) - 2\sqrt{6} \, a(t)^2 + \sqrt{2}\, a(t)^3 = 0\,.
    \end{equation}
    We now apply the Poincaré–Lindstedt method to solve the differential equation. To this end, we introduce a rescaled time variable $\tau = \omega t$, and expand both the amplitude $a(t)$ and the modified frequency $\omega$ as follows
    \begin{eqnarray}
    a \hspace{-0.2cm}&=&\hspace{-0.2cm} a^{(0)} + a^{(1)} + a^{(2)} + \dots\,,\\
    \omega^2 \hspace{-0.2cm}&=&\hspace{-0.2cm} \omega_0^2 +  \omega^{(1)} + \omega^{(2)} + \dots\,,\label{e:eq_omega}
    \end{eqnarray}
    with $\omega_0 = \sqrt{ 4 + 1/R^2}$. 
    By solving the equation order by order and imposing the vanishing of secular terms at each step, we obtain corrections to the frequency arising from higher-order effects. The first non-vanishing correction is given by
    \begin{equation}
        \omega^{(2)} = a_0^2\,\dfrac{80 - 3 \sqrt{2} \,\omega^2}{4\, \omega^2}\,.
    \end{equation}
    Finally, solving for $\omega$ in (\ref{e:eq_omega}) and equating $\omega = \omega_c = 2$, we are left with the critical amplitude at second order
    \begin{equation}
        a_0^\text{crit}(R) = \dfrac{\sqrt{\dfrac{2}{191} \left(20+3 \sqrt{2}\right)}}{R}\,.
    \end{equation}
    For a given initial amplitude $a_0$, there exists an interval $I$ such that, for $R\in I$, the initial configuration (\ref{e:KAK_ansazt_size_3}) evolves into an oscillon state \cite{Copeland:1995, Salmi:2012, Salmi:2006}. In the following figures, we select different values of $R$ within this interval to illustrate the different dynamical behaviours that can arise. One situation with $a_0<a_0^\text{crit}(R)$ is illustrated in  \autoref{f:oscillon_internal_full:a}. The effective model mimics the decay accurately for $t \lesssim 60$. However, for $t > 60$ the energy stored in the internal modes is transferred back to the amplitude. Moreover, in \autoref{f:oscillon_internal_full:b} we show a genuine oscillon with an internal mode excited. 
    
\begin{figure}[htb]
    \begin{center}
        \includegraphics[scale=0.63]{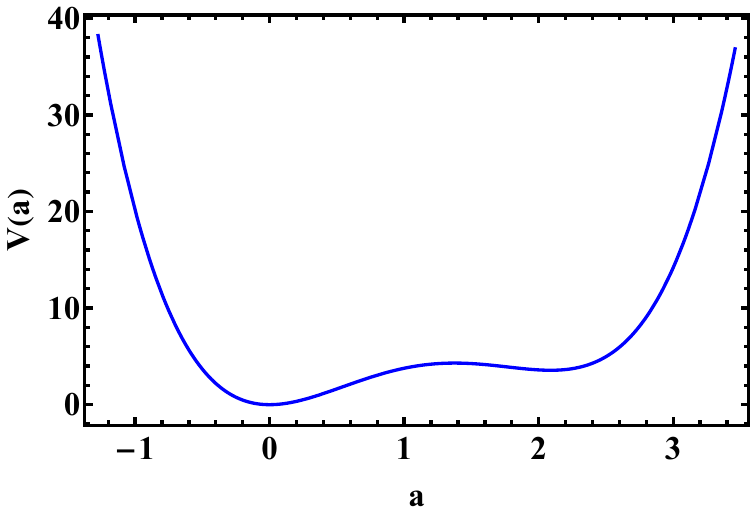}    
        \caption{Effective potential derived from the Lagrangian (\ref{e:Lag_effec_a}) for $R = 4$.}
        \label{f:effective_pot}
    \end{center}
    \end{figure}

\begin{figure}[htb]
    \centering
        \subfloat[Full numerics: $R=4$.]{
        \includegraphics[scale=0.7]{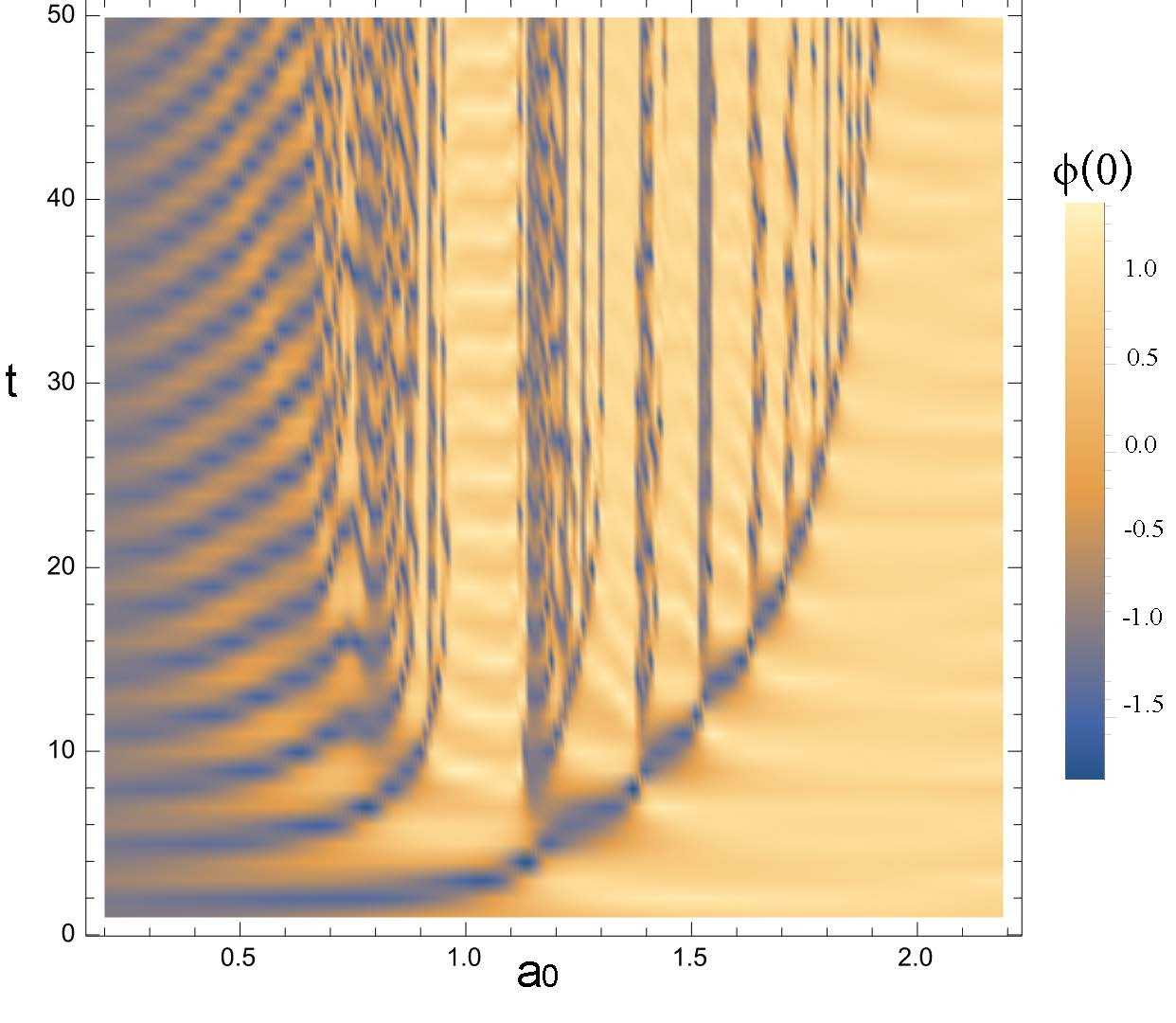}   
        }
        \subfloat[Effective model: $R=4$, $r=2$ and $n=8$.]{
        \includegraphics[scale=0.7]{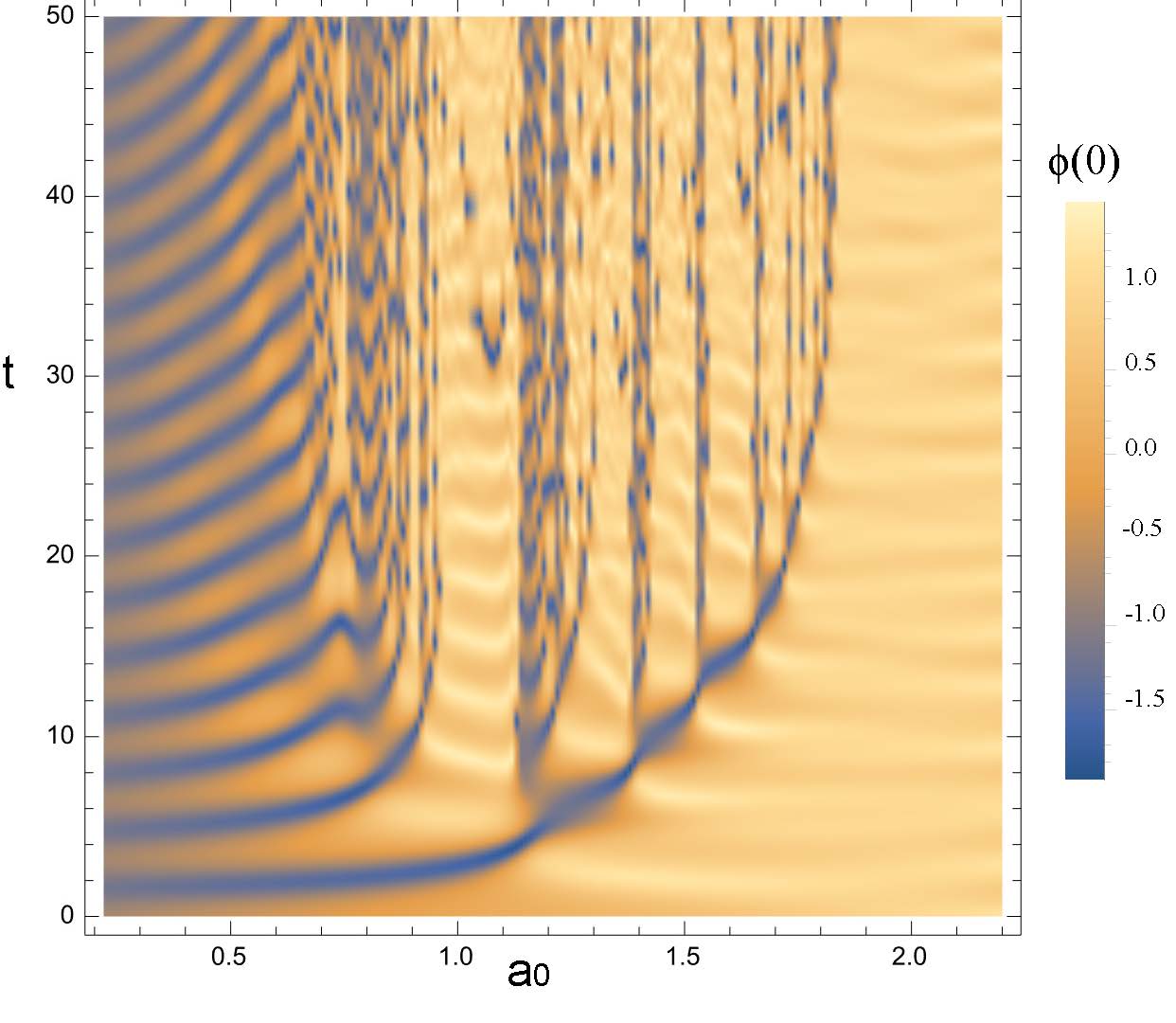}   
        }
        \caption{Comparison between field theory and the effective model (\ref{e:Lag_os_final}) for the initial data (\ref{e:KAK_ansazt_size_3}). The colour palette indicates the value of the field $\phi$ at the origin, $\phi(0,t)$.}
    \label{f:oscillon_internal_full_1}
    \end{figure}

    A remarkable feature of the effective model (\ref{e:Lag_os_final}) is its ability to describe the formation of a kink–antikink pair originating from the oscillon profile. In order to understand this phenomenon, it is enough to analyse the effective action (\ref{e:Lag_effec_a}) for $a(t)$. The corresponding potential for $a(t)$ is depicted in \autoref{f:effective_pot} for a particular value of $R$. Interestingly, the effective potential develops a new local minimum around $a\approx 2$ for $R\gtrsim 2.6$. As a result, for sufficiently large values of the initial amplitude, the system is able to climb the potential barrier and settle on the upper minimum. If internal modes are absent and $a(t)$ possesses sufficient energy to overcome the potential barrier, energy conservation imposes that it must ultimately descend to the global minimum located at zero. However, when internal modes are included, they can absorb and temporarily store the excess energy, allowing $a(t)$ to oscillate around the upper minimum. This dynamical behaviour is interpreted as the formation of a kink–antikink pair.
    \vspace{-0.4cm}
    In \autoref{f:oscillon_internal_full_1} we show the value of the field at the origin $\phi(t,0)$ for different initial amplitudes $a_0$ and $R=4$. The regions where the field takes a value close to $1$ correspond to the kink-antikink creation. In order to produce a pair, the system needs to have an energy $E > 2\,M_K$, with $M_K$ the mass of a single kink. However, even for energies above this value the pair is not necessarily produced, and the initial configurations are dissipated into radiation. These possible scenarios resemble the characteristic fractal pattern in the kink-antikink scattering processes. In fact, in the intermediate stages right after the scattering, the field profile looks like a bump above the $\phi_v = -1$ vacuum, thereby resembling the profile of an oscillon. This is the connection between oscillons and kink-antikink pair scatterings we mentioned before. Recently, a similar fractal pattern has been observed in a model where oscillons are the only localised excitations \cite{Blaschke:2024}.
    In order to reproduce the fractal pattern visible at $a_0 < 2$, one needs to add modes of higher frequencies. This suggests that the scattering modes are essential to explain this structure. Similar results are obtained for values for $2 \lesssim  R \lesssim 6$. For $R < 2.6$, the effective potential has only a local minima at $a=0$, therefore the field cannot settle on the upper vacuum that describes the creation of the kink-antikink pair. On the other hand, for very large values of $R$, our modes decrease their spatial frequency, and the model based on the ansatz (\ref{e:KAK_ansazt_size_2}) captures the kink–antikink pair creation only qualitatively. However, this issue can be solved easily by adding higher frequency modes. This can be achieved by increasing the number of modes, $n$. However, this approach increases notably the computation time. An alternative consists in assuming small values of the scale $r$, which effectively increases the frequency and allows us to consider fewer modes.
    
\begin{figure}[htb]
    \begin{center}
        \includegraphics[scale=0.62]{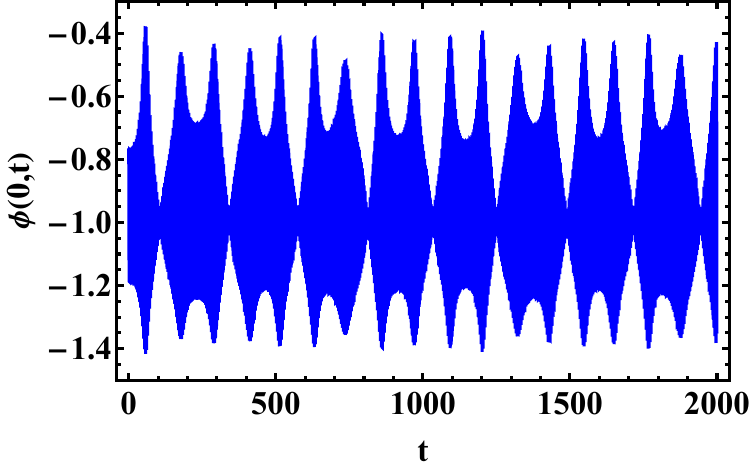}    
        \caption{Field value at the origin, $\phi(0,t)$, obtained from numerical evolution of the full field theory with initial conditions specified by (\ref{e:KAK_ansazt_1}), using as parameters $R=10$ and $a_0=0.23$.}
        \label{f:two_modes}
    \end{center}
    \end{figure}

    To conclude this section, we illustrate in \autoref{f:two_modes} the rich structure provided by the internal structure of oscillons. 
    Apart from the internal mode described by the Derrick mode illustrated in  \autoref{f:oscillon_shape} and \autoref{f:oscillon_internal_full:b}, it is possible to have oscillations with more than one internal mode excited. This gives rise to complicated long-lived oscillatory patterns which are very sensitive to the initial conditions. For example, the dynamics in \autoref{f:two_modes} is characterised by the excitation of two internal modes. From the point of view of the effective model, the amplitudes of these modes are large enough to decrease their frequencies below the mass threshold. As a consequence, they remain excited for very long times. Indeed, we have observed numerically that the oscillon only ceases to exists after $t\approx 7000$ in our time units. Although our effective model does not capture quantitatively this complex behaviour, it is able to describe oscillon configurations with more than one internal mode excited.
    
    The results of this section show the relevance of the radiative modes in oscillon dynamics. Although for certain initial data the oscillon evolution is well-approximated by the amplitude degree of freedom plus one internal mode, in general, one needs higher frequency modes to describe correctly the dynamics. In our approach, we have used modes confined to the oscillon core that, remarkably, describe the main features of oscillon dynamics through reasonably simple equations of motions, corresponding to a set of coupled anharmonics oscillators. In particular, they are able to describe the decay of the oscillon into radiation or the kink-antikink creation.

\section{Summary and conclusions}\label{s:conclusions_Radiation}

    In this chapter, we have introduced radiation degrees of freedom in the moduli space approximation of the $\phi^4$ model. We have computed, within this approach, the radiation emitted at infinity by a wobbling kink at the lowest order in the shape mode amplitude. Our results are in complete agreement with the well-known calculations from \cite{Manton:1997,Oxtoby:2009}. 

    We begin our investigation by studying in detail the interaction of radiation with the vibrational mode. In terms of the effective model, there are two leading mechanisms that explain the energy transfer between them: a Mathieu instability and a resonance. These mechanisms show that the strongest coupling between radiation and the shape mode occurs for $\omega_q = 2\omega_s$. We contemplate two different experiments: in the first one we analyse a kink with its shape mode initially excited and discuss the main frequency of the radiation emitted. In the second one, we irradiate a kink with linear radiation and study how the shape mode is triggered. Notably, we found an analytic expression for the excitation of the shape mode for frequencies away from the unstable region.
 
    We have also studied the role of the translational mode. Despite the fact that the standard collective coordinate method is non-relativistic, when considered up to second order in $\dot{a}^2(t)$, it is able to reproduce at second order the Lorentz contraction of the kink. The vibrational degree of freedom is not enough to reproduce correctly this relativistic effect, however, the inclusion of scattering modes allows for an exact Lorentz contraction at second order. This suggests that an effective model able to describe dissipative effects should be a nice candidate to describe detailed features of non-linear processes such as kink scattering. Of course, the model loses its usefulness if all radiation modes are included, as this essentially results in recovering field theory. However, a judicious choice of modes describing effectively the scattering modes could shed more light on the understanding of many non-linear processes.

    We devoted the last section to the derivation of an effective model for the $\phi^4$ oscillon. Although its natural frequency is above the mass threshold limit, the non-linear terms decrease the frequency below such a limit, avoiding the direct coupling with radiation. The numerical simulations indicate that the oscillon can host a discrete mode responsible for modifications of the width. We implemented this behaviour through the inclusion of the Derrick mode associated to the change of size of the oscillon. This new proposal gives a good agreement with the full numerical simulations. We have added higher frequency modes confined to the oscillon core. They represent scattering modes which may store energy for a certain time, acting effectively as radiation degrees of freedom. The effective equations are a system of coupled anharmonic oscillators with a trivial (i.e. constant) moduli space metric, and the relevant dynamics is completely encoded in the specific coupling between modes given by the potential. Interestingly, once these degrees of freedom are added, this simple effective model is able to describe the kink-antikink creation from initial oscillon data. A study of the internal structure of oscillons within this approach in different models deserves further research, therefore is left for a future investigation. 
 
    Our results suggest that the radiation modes play a crucial role in the study of solitons dynamics. They are of course necessary to explain the decay of non-topological solutions such as the oscillon, or long-lived internal modes such as the shape mode. But they also seem to be fundamental to disentangle the complicated patterns of soliton scattering processes. In addition, the results presented here can be easily generalised to other models. A natural extension of this approach is its application to one-dimensional models involving two scalar fields. In particular, both the Montonen–Sarker–Trullinger–Bishop (MSTB) model \cite{Montonen:1976, Sarker:1976} and the coupled two-component $\phi^4$ model \cite{Halavanau:2012} admit analytical solutions for the shape and scattering modes. As shown in \cite{Miguelez:2023, Miguelez:2024}, kink configurations in these models exhibit distinct spectra depending on the direction of fluctuation. By exciting the kink with its shape mode in one direction, it is possible to trigger the shape mode in the perpendicular direction, leading to radiation emission along both channels. Depending on the values of the coupling constants in the model, the emitted radiation can exhibit multiple frequencies simultaneously. Moreover, it has been verified that there are resonances between shape modes in different directions. All these phenomena are relevant to understand the kink-antikink collisions in these models \cite{Caballero:2025}. As a result, it would be insightful to capture this dynamics within our approach.

        \chapter{Moduli space of the excited Abelian-Higgs vortex}\label{c:Vortex}

This chapter is adapted from \cite{Miguelez:2025}:
    \hspace{-2.5cm}
    \begin{figure}[H]
        \hspace{-2.5cm}
        \centering{\includegraphics[width=1.1\linewidth]{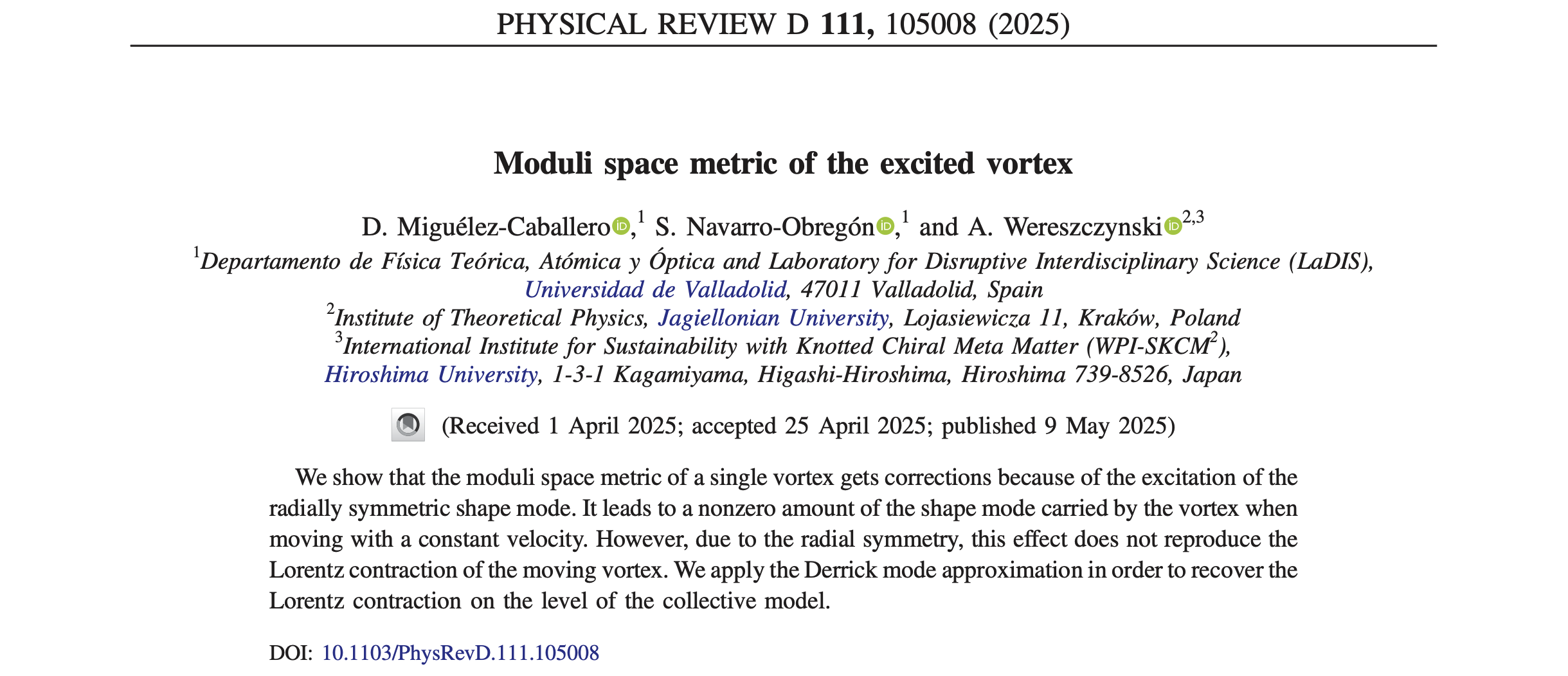}
        }
        \label{f:Thumbail_Vortex}
        \hspace{-2.5cm}
    \end{figure}

\section{Introduction}

    In \autoref{c:Topological}, we introduced the prototypical examples of topological solitons with local gauge invariance in $(2+1)$ dimensions: the vortices in the Abelian-Higgs model. These planar solutions are of particular interest because they are governed by non-linear equations analogous to those describing vortices in type II superconductors \cite{Abrikosov:1957}. Moreover, when extended into the third dimension, they find numerous applications in cosmology and astrophysics \cite{Vilenkin:2000,Manton:2002}. Remarkably, the Abelian-Higgs model in $(2+1)$ dimensions exhibits a BPS limit when the self-coupling constant of the scalar field is $\lambda = 1$. For that critical value, there is no static vortex-vortex force, and vortices can be found as solutions of the relevant Bogomolny equations, which describe energetically degenerated configurations in a topologically fixed sector. These solutions form the moduli space, where the dynamics of a multi-vortex configuration is a force-free motion governed by the moduli space metric, as explained in \autoref{s:CCM}. The moduli space metric corresponding to $N$ vortices was determined by Samols \cite{Samols:1992}.
    
    Samols's approach consists in restricting the kinetic energy $T$ given by (\ref{e:T_AH}) to vectors $(\dot{\phi}, \dot{A}_1, \dot{A}_2)^T$ satisfying both the Gauss's law and Taubes equation (\ref{e:Taubes}). The first condition is essential because, as discussed in \autoref{s:2D}, Gauss's law corresponds to the condition that guarantees that the vector $(\dot{\phi}, \dot{A}_1, \dot{A}_2)^T$ is orthogonal to the gauge orbits through $(\phi, A_1, A_2)^T$. Regarding the potential energy $V$ defined by (\ref{e:LagrangianDensity_AH}), it is gauge invariant by construction. Thus, it does not require further conditions. Moreover, the potential takes a constant value proportional to the topological charge in the BPS limit; see (\ref{e:BPS_bound_AH}). The proportionality constant is just the mass of the $N$-vortex. This approach provides semi-analytical insight not only into the static properties of BPS vortices \cite{Taubes1:1980}, but more importantly, into their dynamical behaviour \cite{Stuart:1996}. Probably, the most well-known result is the $90^\circ$ scattering of two unit charge BPS vortices \cite{Samols:1992, Ruback:1988, Ruback1:1988, Rebbi:1992, Thatcher:1997}. Samols's  method has been extended to compute the moduli space metric for vortices in the Abelian-Higgs model with magnetic impurities \cite{Cockburn:2017}, and for hyperbolic vortices \cite{Strachan:1992}.
    
    Outside of the BPS regime, the concept of moduli space is still useful \cite{Stuart:1994}, but the dynamics is modified by the appearance of a static force between the vortices, which is attractive ($\lambda > 1$) or repulsive ($\lambda < 1$) \cite{Rebbi:1979,Bettencourt:1995,Speight:1997}. Furthermore, even the dynamics of vortices when $\lambda = 1$ is much more complex and interesting than that of the standard geodesic flow. The reason is that vortices host massive normal modes \cite{Kojo:2007,Izquierdo:2016,Izquierdo1:2016}. Importantly, even though the energy of the BPS solutions does not depend on the value of the position on the moduli space, the structure of the massive normal modes does \cite{Fuertes:2024,Rees:2025}. As a consequence, there exists potential energy due to the excitation of the modes, which in the harmonic oscillator approximation, depends quadratically on the frequency of the corresponding mode \cite{Manton:2024}. Since the frequency changes on the moduli space, a mode-generated force appears. This has a strong impact on vortex scattering \cite{Krusch:2024, Rees:2024, Rees:2025}. 

    These modes are expected to generate not only a non-trivial effective potential, but also to modify the moduli space metric. While the effect of the modified metric on the dynamics is generally weaker than the force induced by the potential, it may still significantly influence the final state of vortex dynamics. The aim of the present chapter is to understand how the moduli space metric of the single vortex changes by including the normal modes.

    This chapter is organised as follows. In \autoref{s:Warm_UP}, we summarise the construction of the collective coordinate method for a $\phi^4$ kink based on its zero mode and shape mode. Building on this intuition, we generalise the procedure to the case of Abelian–Higgs vortices in \autoref{s:CCM_Shape_Vortex}. As will be shown, this collective coordinate model presents certain limitations when describing vortices moving at constant speed. To address this, we adopt a similar approach, now incorporating Derrick modes, in \autoref{s:CCM_Derrick_Vortex}. Finally, we summarise our findings in \autoref{s:Summary_Vortex}.
    
\section{Moduli space for the wobbling kink}\label{s:Warm_UP}

    In order to have a slight intuition, let us begin our discussion by analysing a simpler example. In particular, we will partly rely on the effective models that we derived in \autoref{c:Radiation} for the kink in the $\phi^4$ model. Let us assume the general ansatz (\ref{e:ansatz:trans}) with $c_s(t) = c_q(t) = 0$. Then, the resulting collective coordinate model (\ref{e:Lag_sqt}) is reduced to
    \begin{equation}\label{e:effec_lag_phi4_a}
        L[a]=\frac{1}{2} M \dot{a}^2 - M\,.
    \end{equation}
    This is the effective Lagrangian for a single moving kink with the position of the kink as the only collective coordinate. In this case, moduli space metric is constant and coincides with the kink mass $g_{aa} = M = 4/3$. Moreover, the effective potential also has a constant value $V = M$, since the modulus $a$ parametrise a set of configurations with the same static energy (see \autoref{ss:phi4}). This simple collective coordinate model has associated the equation of motion $\Ddot{a} = 0$, whose solution describes a constant velocity non-relativistic motion of the kink $a(t) = x_0 + v\,t$. 

    In \autoref{c:Radiation}, we showed that the first-order correction to a Lorentz boost requires the excitation of scattering modes. However, the inability of expression (\ref{e:effec_lag_phi4_a}) to capture relativistic motion can be partially cured by incorporating the shape mode. Let us assume then the general set of configurations (\ref{e:ansatz:trans}) but now we only omit the $c_q(t)$ amplitudes.
    The restricted set of configurations gives rise to the following effective model
    \vspace{-0.2cm}
    \begin{equation}\label{e:vib-kink-CCM}
        L[a,c_s] = \frac{1}{2}g_{aa}\dot{a}^2+\frac{1}{2}\dot{c}_s^2-V(c_s)\,,
    \end{equation}
    which corresponds to a two dimensional vibrational moduli space with a diagonal metric
    \begin{eqnarray}\label{e:metric_aC}
        g_{aa} = M + \frac{\pi}{2}\sqrt{\dfrac{3}{2}}c_s+\frac{7}{5}c_s^2 \,,  
    \end{eqnarray}
    and an effective potential
    \vspace{-0.35cm}
    \begin{equation}\label{e:Pot_kink_2dim}
        V(c_s) = M + \dfrac{1}{2}\omega_s^2\,c_s^2+ \dfrac{3\pi}{16}\sqrt{\dfrac{3}{2}} c_s^3 +\frac{9}{70}c_s^4\,.
    \end{equation}
    The off-diagonal component of the metric vanishes due to the orthogonality of the linear modes  guaranteed by \autoref{t:Sturm}. Moreover, it is important to remark that the previous constant metric component $g_{aa} = M$ receives now a non-trivial modification due to the vibrational mode, as seen in (\ref{e:metric_aC}). Such a coupling between the zero and massive mode is usually referred to as the Coriolis effect \cite{Rawlinson:2019}. 

    Naively, one could think that the modification of the moduli space metric $g_{aa}$ by the linear and quadratic term in $c_s$ is a small effect without any importance on the kink dynamics. However, this is a crucial modification for the correct description of the kink-antikink collisions within the corresponding collective coordinate model \cite{Manton:2021}. This originates from the observation that the effective model (\ref{e:vib-kink-CCM}) has a stationary solution $\dot{a} = v = \mbox{const.}$ and $c = c_s(v^2) = \mbox{const.}$, where $c$ is the constant value that satisfies 
    \vspace{-0.25cm}\begin{equation}\label{e:Stationary_Solution_Kink_C}
        \frac{v^2}{2} \frac{d g_{aa}}{d c_s}\bigg|_{c_s=c}  = \frac{dV}{dc_s}\bigg|_{c_s=c}.
    \end{equation}
    Thus, from the point of view of the collective coordinate method approach, the kink with a constant velocity has a non-zero amount of the shape mode excited. This is of great importance for the proper specification of the initial conditions for the kink-antikink collision in the collective coordinate method framework \cite{Manton:2021}. Otherwise, the effective motion does not correctly reproduce the famous chaotic pattern in the final-state formation \cite{Sugiyama:1979, Campbell:1983}. This effect is intimately related with the perturbative restoration of the Lorentz covariance of the theory. Indeed, the metric modification due to the inclusion of the shape mode allows for an approximation of the Lorentz contraction and brings the effective model closer to the original Lorentz invariant model \cite{Manton:2021}. Then, the shape mode is an alternative to the perturbative relativistic moduli space approach explained in \autoref{c:Radiation} \cite{Manton:2022}.

    An alternative to compute the moduli space metric relies on noticing that the moduli space metric and the effective potential do not depend on the trivial modulus $a$ in the single kink sector. Therefore, these terms can also be computed from the infinitesimal action of the zero mode. Indeed, let us consider an infinitesimal translation of the vibrating kink from $a=0$ to $a=\epsilon$  
    \vspace{-0.25cm}
    \begin{equation}\label{e:kink-eA}
    \scalebox{0.95}{$
    \Phi(x;\epsilon, c_s) = \tanh x - \epsilon(t)\sech^2 x 
    + c_s(t)\sqrt{\dfrac{3}{2}} \sinh x\sech^2 x 
    + \epsilon(t)c_s(t)\sqrt{\dfrac{3}{2}} \sech x \left(1 - 2\sech^2 x \right)\,,
    $}
    \end{equation}
    where the amplitude of the zero mode $\epsilon$ and the shape mode $c_s$ have been taken as moduli and promoted to time-dependent coordinates. The resulting effective model reads
    \begin{equation}
        L[\epsilon,c_s] = \dfrac{1}{2}g_{\epsilon\epsilon}\dot{\epsilon}^2 + \dfrac{1}{2}g_{\epsilon c}\dot{\epsilon}\dot{c}_s + \dfrac{1}{2}g_{cc}\dot{c}_s^2 - V(\epsilon,c_s)\,.
    \end{equation}
    The metric functions are now
    \begin{eqnarray}
        g_{\epsilon \epsilon} = M + \dfrac{\pi}{2}\sqrt{\dfrac{3}{2}}c_s + \frac{7}{5} c_s^2\,, \quad \quad
        g_{cc}= 1 + \frac{7}{5} \epsilon^2\,, \quad \quad 
        g_{\epsilon c} = \frac{3\epsilon}{40}\left( 5\pi + 28\,c_s \right)\,,
    \end{eqnarray}
    but the potential $V(\epsilon,C)$ is not shown for legibility reasons. We remark that it is crucial to include the last term in (\ref{e:kink-eA}), since it takes into account the mutual interaction of the shape and zero mode. Specifically, the linear and quadratic mode amplitude modifications of the metric component $g_{\epsilon \epsilon}$ come from this term. The new metric terms agree with those presented in (\ref{e:metric_aC}) for the limit $\epsilon\to 0$. The same happens for the effective potential (\ref{e:Pot_kink_2dim}). Hence, we reproduce the model (\ref{e:vib-kink-CCM}), which confirms the fact that the zero mode generates the motion along the $x$ direction.
    
\section{Moduli space of a single vortex with a shape mode}\label{s:CCM_Shape_Vortex}
  
    In the subsequent analysis, we want to verify whether there is a similar modification for the moduli space of the one-vortex. Samols's approach cannot be directly applied, as it is only valid in the BPS limit. Although we work in the $\lambda = 1$ regime, the inclusion of the shape mode takes us beyond the BPS case. Furthermore, our aim is to investigate also the generalisation of Samols's metric beyond the BPS limit, allowing for arbitrary values of the self-coupling $\lambda$.

    A procedure to determine the moduli space outside the BPS limit can be achieved applying the same procedure as for the kink. Let us assume the action of an infinitesimal spatial translation along one direction. Without loss of generality, let us consider that the motion is along the $x_1$ axis. Naively, one may consider that the effect of the spatial translation on the vortex configuration is
    \begin{eqnarray}\label{e:Translation_Vortex_Naive}
        \Phi^v(x_1-\epsilon, x_2) \hspace{-0.2cm}&=&\hspace{-0.2cm} \Phi^v (\vec{x}) - \epsilon\, \partial_1\Phi^v(\vec{x})\,, \label{e:Translation_Vortex_Naive_Scalar}\\ 
        A_1^v(x_1-\epsilon, x_2) \hspace{-0.2cm}&=&\hspace{-0.2cm} A_1^v(\vec{x}) -\epsilon\, \partial_1 A^v_1 (\vec{x})\,, \label{e:Translation_Vortex_Naive_A1}   \\
        A_2^v(x_1-\epsilon, x_2) \hspace{-0.2cm}&=&\hspace{-0.2cm} A_2^v(\vec{x}) - \epsilon\, \partial_1 A^v_2 (\vec{x}) \label{e:Translation_Vortex_Naive_A2}\,,
    \end{eqnarray}
    up to first order in $\epsilon$. However, this transformation does not have an associated gauge-invariant energy-momentum tensor. To solve this failure, the infinitesimal translation (\ref{e:Translation_Vortex_Naive_Scalar})-(\ref{e:Translation_Vortex_Naive_A2}) has to be replaced by 
    \begin{eqnarray}
        T_{\epsilon} \Phi^v(x_1, x_2) \hspace{-0.2cm}&=&\hspace{-0.2cm}  \Phi^v (\vec{x}) - \epsilon\, D_1 \Phi^v(\vec{x})\,,\label{e:Translation_Vortex_Scalar}\\
        T_{\epsilon} A_1^v(x_1, x_2) \hspace{-0.2cm}&=&\hspace{-0.2cm} A^v_1(\vec{x})\,, \label{e:Translation_Vortex_A1}  \\
        T_{\epsilon} A_2^v(x_1, x_2) \hspace{-0.2cm}&=&\hspace{-0.2cm} A^v_2(\vec{x}) - \epsilon\, B^v(\vec{x})\,. \label{e:Translation_Vortex_A2}
    \end{eqnarray}
    The terms proportional to $\epsilon$ can be put together in the vector $\xi_0 :=\left( D_1\Phi^v, 0, B^v \right)^T$, forming the zero mode previously found in \cite{Weinberg:1979,Tong:2014}. Promoting the infinitesimal translation parameter to a time-dependent modulus $\epsilon \rightarrow \epsilon(t)$, the time dependence is absorbed into the coordinate evolution. Differentiating with respect to time, we obtain 
    \begin{eqnarray}
        \dot{\Phi} (\vec{x}) \hspace{-0.2cm}&=&\hspace{-0.2cm} - \dot{\epsilon} D_1 \Phi^v(\vec{x})\,,\label{e:Translation_Vortex_Scalar_Der} \\
        \dot{A}_1(\vec{x}) \hspace{-0.2cm}&=&\hspace{-0.2cm} 0\,,\label{e:Translation_Vortex_A1_Der} \\
        \dot{A}_2(\vec{x}) \hspace{-0.2cm}&=&\hspace{-0.2cm} - \dot{\epsilon} B^v(\vec{x}) \label{e:Translation_Vortex_A2_Der}\,.  
    \end{eqnarray}
    With the definitions of the translated fields (\ref{e:Translation_Vortex_Scalar})-(\ref{e:Translation_Vortex_A2}) and their time derivatives (\ref{e:Translation_Vortex_Scalar_Der})-(\ref{e:Translation_Vortex_A2_Der}), Gauss's law is satisfied, thereby keeping the perturbation in the direction perpendicular to the gauge orbit. Inserting the translated field and their time derivatives into the Lagrangian \eqref{e:LagrangianDensity_AH}, we arrive at the following collective coordinate model
    \begin{equation}
        L[\epsilon] = \frac{1}{2}g_{\epsilon \epsilon} \dot{\epsilon}^2 - V(\epsilon)\,,
    \end{equation}
    where the moduli space metric reads
    \begin{eqnarray}
        g_{\epsilon \epsilon} \hspace{-0.2cm}&=&\hspace{-0.2cm} \int_{\mathbbm{R}^2} \left( |D_1\Phi^v|^2 +  (B^{v})^2 \right)\,dx_1dx_2  \nonumber \\
        &=&\hspace{-0.2cm} \int_{\mathbbm{R}^2}  \left( \frac{1}{2} |D_1\Phi^v|^2+\frac{1}{2} |D_2\Phi^v|^{2} + (B^{v})^2 \right)\,dx_1dx_2  \nonumber\\
        &=&\hspace{-0.2cm} M + \frac{1}{2} \int_{\mathbbm{R}^2} \left( (B^{v})^2 - \frac{\lambda}{4} (1-|\Phi^v|^2)^2 \right)\,dx_1dx_2 = M\,, 
    \end{eqnarray}
    and the potential is $V(\epsilon) = M + \hat{V}(\epsilon)$. Here, $\hat{V}(\epsilon)$ is a polynomial in $\epsilon$ with no constant term that we do not show for legibility reasons and $M$ is the mass of the vortex. To identify the moduli space metric $g_{\epsilon\epsilon}$ with the mass $M$, we have used that $|D_1\Phi^v|^2 = |D_2\Phi^v|^2$ due to the axial symmetry of the background vortex, as well as the following integral identity 
    \begin{equation}
        I = \displaystyle\int_{\mathbbm{R}^2} \left( (B^{v})^2 - \frac{\lambda}{4} (1-|\Phi^v|^2)^2 \right)\,dx_1dx_2 = 0\,.
    \end{equation}
    In the critical case, the vanishing of the integrand follows from the second Bogomolny equation (\ref{e:BPS_AH}). On the other hand, the integral does not contribute outside the BPS limit due to the vanishing of the spatial integral of the energy-momentum components $T_{11}$ and $T_{22}$. Physically, this vanishing occurs because the one-vortex solution is static, and there is no net flux of linear momentum. Moreover, it can be verified that it corresponds to a virial relation derived from the Derrick theorem. Therefore, we conclude that the norm of the zero mode is exactly the vortex mass. This identification is natural, as the kink exhibits the same relation.

    Finally, as nothing should depend on the specific value of $\epsilon$, we impose that $\epsilon \to 0$. By doing that, the effective potential $V(\epsilon)$ reduces to $M$. Therefore, substituting $\dot{\epsilon}$ by a continuous parameter $a$, the resulting collective coordinate model, $L[a] = \frac{1}{2}M \dot{a}^2 - M$, describes a non-relativistic constant motion of the vortex, as expected. This effective Lagrangian corresponds to the expression derived by Samols when only a single vortex is considered \cite{Samols:1992}.

    Having constructed the simplest effective model of a one-vortex including only the zero mode, and having examined the subtle aspects inherent to gauge theories, we are now in a position to extend our analysis. In what follows, we consider the effect of introducing an additional excitation, allowing us to explore how the moduli space is modified by those new collective coordinates. As previously mentioned, we are interested in the inclusion of the unique shape mode $\xi := (\varphi, a_1, a_2)^T$ supported by a one-vortex. In terms of the polar components presented in (\ref{e:SpectralProblem_AH_radial}), the Cartesian components of the vector fields are  $a_1 = - a_{\theta}\sin \theta$ and $a_2 = a_{\theta}\cos \theta$. A representation of $\varphi$ and $a_{\theta}$ is shown in \autoref{f:BoundMode_AH}.
     
\begin{figure}[htb]
        \centering
        \includegraphics[width=0.45\linewidth]{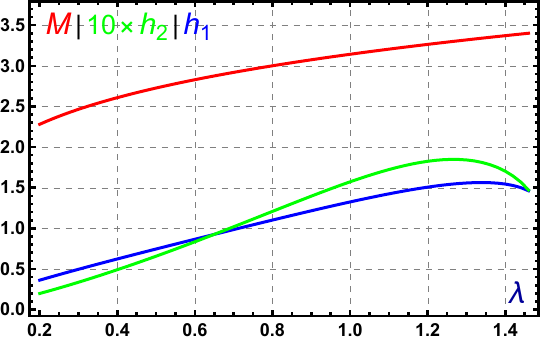}
        \hspace{0.9cm}
        \includegraphics[width=0.45\linewidth]{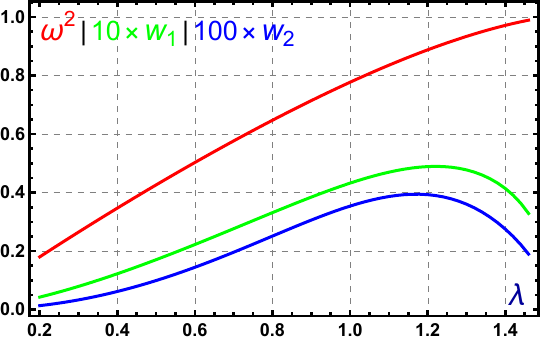}
        \caption{Coefficients of the metric (left panel) and potential (right panel) terms in the one-vortex collective coordinate model based on the shape mode (\ref{e:L_CCM_Vortex_A_C}).}
        \label{f:CCM_Vortex_A_C}
    \end{figure}

    If we consider the corrected infinitesimal translation along $x_1$-axis for a one-vortex with its shape mode $\xi$ excited, we arrive at the following set of configurations, which define the vibrational moduli space of a single vortex
    \begin{eqnarray}
        \Phi \hspace{-0.2cm}&=&\hspace{-0.2cm} \Phi^v (\vec{x}) - \epsilon\, D_1 \Phi^v(\vec{x}) + C\, \varphi (\vec{x})  - \epsilon\, C D_1 \varphi(\vec{x})\,,\label{e:Vortex_Zero_Shape_Scalar} \\
        A_1 \hspace{-0.2cm}&=&\hspace{-0.2cm} A_1^v(\vec{x}) + C\, a_1 (\vec{x})\,, \label{e:Vortex_Zero_Shape_A1}  \\
        A_2 \hspace{-0.2cm}&=&\hspace{-0.2cm} A_2^v(\vec{x}) - \epsilon\, B^v(\vec{x}) + C\, a_2 (\vec{x})- \epsilon\, C B^s (\vec{x})\,. \label{e:Vortex_Zero_Shape_A2}
    \end{eqnarray}
    Here, $C$ is the amplitude of the shape mode and $B^s(\vec{x}) = \partial_1 a_2-\partial_2a_1$ is the magnetic field associated with the vector components of the shape mode. Promoting $\epsilon$ and $C$ to time dependent moduli and taking the time derivatives 
    \begin{eqnarray}
        \dot{\Phi} \hspace{-0.2cm}&=&\hspace{-0.2cm} - \dot{\epsilon} D_1 \Phi^v(\vec{x}) + \dot{C} \varphi (\vec{x}) - \dot{\epsilon}C  D_1 \varphi (\vec{x})\,, \label{e:Vortex_Zero_Shape_Scalar_Der} \\
        \dot{A}_1 \hspace{-0.2cm}&=&\hspace{-0.2cm}  \dot{C} a_1 (\vec{x})\,, \label{e:Vortex_Zero_Shape_A1_Der}\\
        \dot{A}_2 \hspace{-0.2cm}&=&\hspace{-0.2cm} -\dot{\epsilon} B^v(\vec{x}) + \dot{C} a_2 (\vec{x}) - \dot{\epsilon} C B^s (\vec{x}) - \epsilon \dot{C} B^s (\vec{x})\,, \label{e:Vortex_Zero_Shape_A2_Der}
    \end{eqnarray}
    it is easy to check that the Gauss's law is obeyed up to first order. Now, we plug the expressions (\ref{e:Vortex_Zero_Shape_Scalar})-(\ref{e:Vortex_Zero_Shape_A2}) and (\ref{e:Vortex_Zero_Shape_Scalar_Der})-(\ref{e:Vortex_Zero_Shape_A2_Der}) into the Lagrangian density (\ref{e:LagrangianDensity_AH}). Then, we integrate over the spatial coordinates and take the limit $\epsilon \to 0$. In the end, we replace the infinitesimal variable $\epsilon$ by the finite modulus $a$, which determines the position of the vortex along the $x_1$-axis. The resulting collective coordinate model is
    \begin{equation}\label{e:L_CCM_Vortex_A_C}
        L_s[a,C]=\frac{1}{2}g_{aa}\dot{a}^2 + \frac{1}{2}\dot{C}^2 - V(C)\,.
    \end{equation}  
    The two-dimensional moduli space spanned by the coordinates $a$ and $C$ has the following diagonal metric 
    \begin{eqnarray} 
        g_{aa} = M + h_1 C + h_2 C^2\,, \qquad \qquad
        g_{CC} = 1\,.
    \end{eqnarray}
    The constants $h_1, h_2$ give the modification of the flat one-vortex moduli space due to the normal mode. We have computed them as functions of the coupling constant $\lambda$. The results are presented in \autoref{f:CCM_Vortex_A_C} (left panel), where we also show $M(\lambda)$. We clearly see that $M \gg h_1 \gg h_2$. Moreover, according to the plot, the functions $h_{1}$ and $h_2$ reach their maximum at around $\lambda \approx 1.3$. Quite surprisingly, the self-dual limit is not distinguished. A similar behaviour has been observed in the context of radiation emitted by a single vortex, where the decay rate of the excited shape mode reaches its maximum at $\lambda \approx 1.1$ \cite{Pillado:2024}. Finally, note that $g_{CC}$ is equal to $1$ for any value of $\lambda$. This reflects the normalisation of the shape mode $\xi$ to unity.
    The expression for the effective potential $V(C)$ is
    \begin{equation}
        V(C) = M + \frac{1}{2}\omega^2 C^2 + w_1C^3+  w_2C^4\,,
    \end{equation}
    where the numerical coupling constants $w_1, w_2$ and the frequency of the shape mode $\omega$ are again functions of $\lambda$. The coefficients $w_1$ and $w_2$ are plotted in \autoref{f:CCM_Vortex_A_C} (right panel). According to the scale of the plot, we deduce that $\omega^2 \gg w_1 \gg w_2$. This means that terms of higher order in the amplitude of the shape mode are not only subleading due to the assumption that the amplitude $C$ is small, but they are also multiplied by smaller coupling constants. 

    Remarkably, the two-dimensional collective coordinate model (\ref{e:L_CCM_Vortex_A_C}) has a stationary solution which describes a constant velocity motion of the vortex, $\dot{a} = v$, as observed in the kink. Due to the dependence of the metric on the amplitude of the mode, a non-zero velocity induces a non-zero amplitude of the mode $C_0$, which satisfies the following algebraic equation
    \begin{equation}
        \frac{v^2}{2}\left(h_1+2h_2C_0\right)=\omega^2 C_0+3w_1C^2_0+4w_2C^3_0\,.
    \end{equation}
    Using that $w_1$ and $w_2$ are small numbers and assuming that the amplitude of the mode is not too large, we can neglect their contributions and find an approximate formula for the excitation of the shape mode
    \begin{equation}\label{e:C0_value_Shape}
        C_0 \approx \frac{1}{2} \frac{h_1v^2}{\omega^2-h_2v^2} \approx  \frac{1}{2} \frac{h_1v^2}{\omega^2}\,.
    \end{equation}
    In the last approximation, it has been assumed that $h_2$ is also subleading with respect to $\omega^2$.
    
    We conclude that the moduli space metric of the single vortex is modified by the excitation of the internal mode, and this effect naturally extends to vortices with higher topological charge. Consequently, the effective model proposed in \cite{Guilarte:2024} should also be refined to incorporate a similar modification of the moduli space metric. In this framework, the stationary solution given by $a = x_1^{(0)} + v t$ and $C_0 = h_1v^2/(2\omega^2)$ provides an initial condition for describing the scattering of unexcited vortices.

\begin{figure}[htb]
        \includegraphics[width=0.33\linewidth]{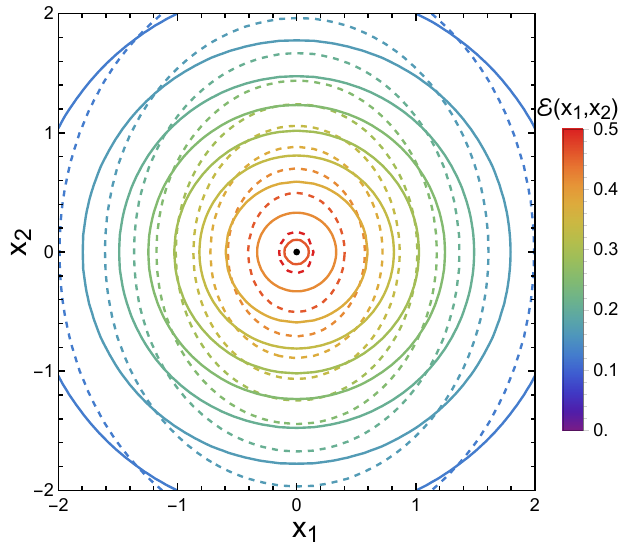}
        \includegraphics[width=0.33\linewidth]{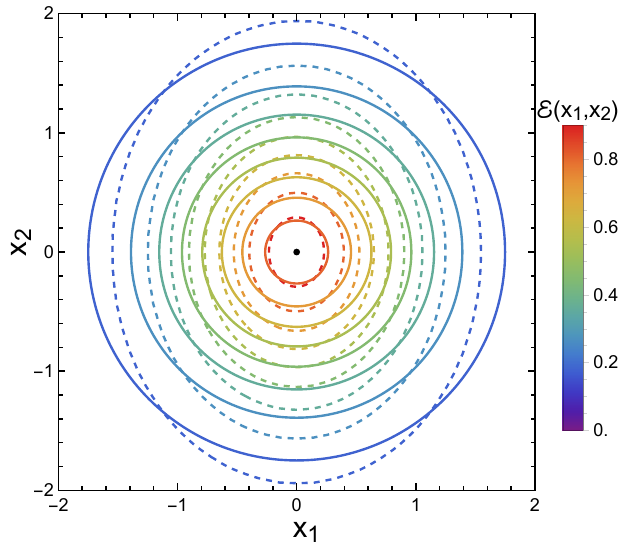}
        \includegraphics[width=0.33\linewidth]{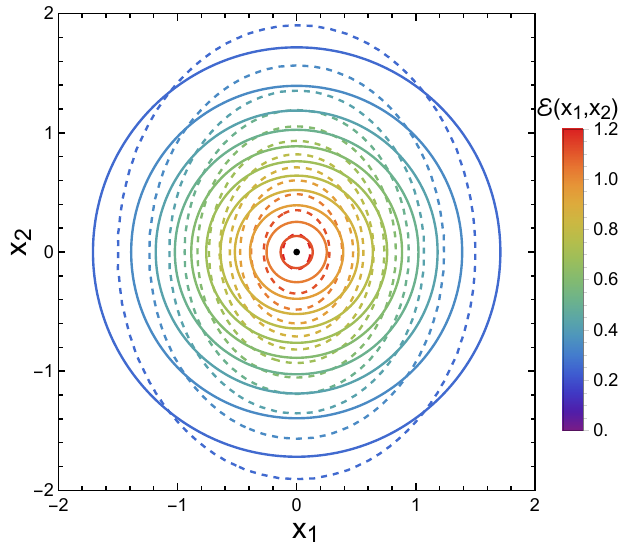}
        \caption{Comparison between the energy density obtained from the collective coordinate method with the internal mode (solid lines) given by (\ref{e:L_CCM_Vortex_A_C}) and the numerical simulation (dashed lines) for a vortex with $n=1$ and velocity $v=0.6$. Here $\lambda=0.4,1,1.4$.}
        \label{f:Energy_den_CCM_A_C}
    \end{figure}

    However, let us emphasise a crucial detail. For a one-dimensional kink, the stationary solution (\ref{e:Stationary_Solution_Kink_C}) approximately restores the Lorentz contraction of the moving soliton. In contrast, the situation here is more subtle. The inclusion of the axially symmetric shape mode modifies the vortex profile in an axially symmetric manner. This is not what occurs in a boosted vortex, since the axially symmetric vortex is instead squeezed along the direction of motion, resulting in an elliptical shape. To test the accordance between full numerics and our effective description, let us prepare the following experiment. On the one hand, we boost a static vortex solution within the field theory framework. To this end, we consider as initial condition a Lorentz-boosted scalar field  
    \begin{eqnarray}
        \Phi^v_{\gamma}(\vec{x}) = \Phi^v(\Lambda^{-1}\, \vec{x})\,,
    \end{eqnarray}
    whilst the gauge vector field transforms as 
    \begin{eqnarray}
        A_{\gamma\, 0}^{v}(\vec{x}) \hspace{-0.2cm}&=&\hspace{-0.2cm} - \gamma\,v A_{1}^v(\Lambda^{-1}\, \vec{x})\,,\\
        A_{\gamma\, 1}^{v}(\vec{x}) \hspace{-0.2cm}&=&\hspace{-0.2cm} \gamma\, A_1^v(\Lambda^{-1}\, \vec{x})\,,\\
        A_{\gamma\, 2}^{v}(\vec{x}) \hspace{-0.2cm}&=&\hspace{-0.2cm} A_2^v(\Lambda^{-1}\, \vec{x})\,.
    \end{eqnarray}
    Here, $\Lambda$ is the Lorentz boost matrix
    \begin{equation} 
        \Lambda = \left( \begin{array}{ccc}
        \gamma & -\gamma\, v & 0  \\
        -\gamma\, v & \gamma & 0 \\
        0 & 0 & 1 
        \end{array} \right) \,.
    \end{equation}
    Note that we are assuming that the motion is along the $x_1$ direction. On the other hand, in the collective coordinate model based on the shape mode, the field configuration is constructed by employing the ansatz given in (\ref{e:Vortex_Zero_Shape_Scalar})-(\ref{e:Vortex_Zero_Shape_A2}) and (\ref{e:Vortex_Zero_Shape_Scalar_Der})-(\ref{e:Vortex_Zero_Shape_A2_Der}), with the shape mode amplitude determined by (\ref{e:C0_value_Shape}). Finally, we compare the energy densities of both configurations for different values of $\lambda$, while keeping the same boost velocity in all cases. As can be clearly seen in \autoref{f:Energy_den_CCM_A_C}, the addition of the radial shape mode fails to adequately reproduce the partial Lorentz invariance within the collective coordinates framework.

\section{Moduli space of a single vortex with Derrick modes}\label{s:CCM_Derrick_Vortex}

    To address this shortcoming, we may attempt to introduce a Derrick mode. We have already discussed the origin of Derrick modes as arising from a spatial scaling transformation, $x^i \to \delta^i x^i$, and their utility in capturing the ability of the soliton to be compressed or expanded. In the present case, introducing a Derrick mode may prove useful due to the presence of two spatial directions. We can compute the Derrick mode associated with scaling along a single spatial direction, which should be aligned with the direction of motion. The resulting mode is inherently non-axially symmetric and could therefore better capture the Lorentz contraction at high velocities.  
    
    Another advantage of the Derrick mode is that it exists for any value of self-coupling $\lambda$. This is not true for the shape mode, which crosses the lowest continuum for $\lambda > 1.5$, as seen in \autoref{f:Spectrum_AH_n1}. Conversely, for values above that limit, the shape mode does not completely disappear. In reality, such a mode turns into a Feschbach resonance\footnote{For a detailed discussion of these resonances, see \cite{Feshbach:1958,Mittleman:1966}}, which, at least for not too large $\lambda$, is a rather stable excitation. Hence, it still makes sense to approximate it -concretely, its real oscillating part- in terms of a Derrick mode. This type of resonance has revealed a notable impact on soliton dynamics in one and more spatial dimensions \cite{Forgacs:2004,Russell:2011,Queiruga:2025}.

    When applied to the one-vortex, the naive effect of an infinitesimal scaling transformation acting only in the $x_1$ direction, $x_1 \to (1+\delta)x_1$, gives the following expressions up to linear order in $\delta$ 
    \begin{eqnarray}
        \Phi^v((1+\delta)x_1,x_2) \hspace{-0.2cm}&=&\hspace{-0.2cm} \Phi^v (\vec{x}) + \delta\, x_1 \partial_1\Phi^v (\vec{x})\,, \\
        A^v_1((1+\delta)x_1,x_2) \hspace{-0.2cm}&=&\hspace{-0.2cm} A^v_1(\vec{x}) + \delta\left[A^v_1(\vec{x}) + x_1 \partial_1 A_1^v(\vec{x})\right]\,,\\
        A^v_2((1+\delta)x_1,x_2) \hspace{-0.2cm}&=&\hspace{-0.2cm} A^v_2(\vec{x}) + \delta\, x_1 \partial_1 A_2^v(\vec{x})\,.
    \end{eqnarray}
    As in the case of infinitesimal translations, these expressions must be modified to ensure consistency with Gauss's law. As a result, the correct expressions for an infinitesimal scaling transformation are 
    \vspace{-0.35cm}
    \begin{eqnarray}
        T_\delta \Phi^v(x_1,x_2) \hspace{-0.2cm}&=&\hspace{-0.2cm} \Phi^v (\vec{x}) + \delta\, x_1 D_1\Phi^v (\vec{x})\,, \\
        T_\delta  A^v_1(x_1,x_2) \hspace{-0.2cm}&=&\hspace{-0.2cm} A^v_1(\vec{x})\,,\\
        T_\delta  A^v_2(x_1,x_2) \hspace{-0.2cm}&=&\hspace{-0.2cm} A^v_2(\vec{x}) + \delta\,\, x_1B^v(\vec{x})\,.
    \end{eqnarray}
    Therefore, the $x_1$-Derrick mode can be expressed in terms of the vector $\xi^D := (x_1 D_1\Phi^v, 0,x_1B^v)^T$. Combining these deformations with the zero mode translation along the $x_1$-axis we arrive at the following set of configurations
    \begin{eqnarray}\label{e:Ansatz_CCM_Vortex_A_D}
            \Phi \hspace{-0.2cm}&=&\hspace{-0.2cm} \Phi^v(\vec{x}) + \epsilon \,D_1 \Phi^v(\vec{x}) + C \,x_1 D_1 \Phi^v(\vec{x}) + \epsilon\, C D_1(x_1 D_1\Phi^v(\vec{x}))\,  ,\\
            A_1 \hspace{-0.2cm}&=&\hspace{-0.2cm} A^v_1(\vec{x})\, ,\\
            A_2 \hspace{-0.2cm}&=&\hspace{-0.2cm} A^v_2(\vec{x}) + \epsilon \, B^v(\vec{x}) + C\, x_1 B^v(\vec{x}) + \epsilon \,C\, \partial_1(x_1 B^v(\vec{x}))\, .
    \end{eqnarray}
    Here, $C$ denotes the amplitude of the $x_1$-Derrick mode. As in the kink example, we promote $\epsilon$ and $C$ to time-dependent moduli and compute the time derivatives of the restricted set of field configurations. Then, we compute the corresponding effective Lagrangian and take the limit $\epsilon \rightarrow 0$. Finally, we change the infinitesimal variable $\epsilon$ for a finite continuous variable $a$. 
    
\begin{figure}[htb]
        \hspace{-0.1cm}
        \includegraphics[width=0.46\linewidth]{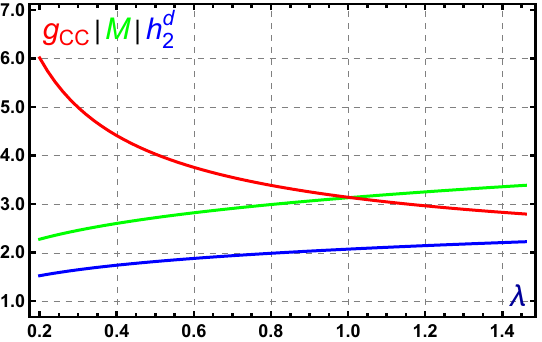}
        \hspace{0.9cm}
        \includegraphics[width=0.46\linewidth]{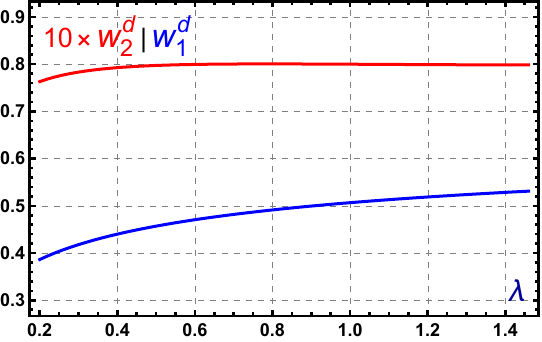}
        \caption{ Coefficients of the metric (left panel) and potential (right panel) terms in the $x_1$-Derrick mode based effective model for the one-vortex (\ref{e:L_CCM_Vortex_A_D}).}
         \label{f:CCM_Vortex_A_D}
    \end{figure}

    After these steps, we are left with the following $x_1$-Derrick mode based collective coordinate model
    \begin{equation}\label{e:L_CCM_Vortex_A_D}
        L_d[a,C] = \frac{1}{2} g_{aa} \dot{a}^2 + \frac{1}{2}g_{CC}\dot{C}^2 - V(C)\,.
    \end{equation}
    The metric functions have the same formal structure as before
    \begin{equation}
        g_{aa} = M + h_1^d C + h_2^d C^2\,. 
    \end{equation}
   One can verify that, in this case, $h_1^d = M$. This result is natural, as also occurs in the kink case. In that context, it is more straightforward to confirm this by identifying the corresponding coefficient with the norm of the zero mode. In \autoref{f:CCM_Vortex_A_D} (left panel) we represent the metric coefficients $h_1^d,h_2^d$ and $g_{CC}$. The non-trivial dependence of $g_{CC}$ on $\lambda$ arises from the fact that the $x_1$-Derrick mode is not normalised to unity.
    \vspace{0.2cm}
    
    The potential $V(C)$ is again a fourth-order function
    \begin{equation}
        V(C) = M + \frac{1}{2} w_0^d C^2 + w_1^dC^3 + w_2^dC^4\,.
    \end{equation}
    One can also verify that $w_0^d = M$. Once again, this result is analogous to the kink case. Specifically, the corresponding coefficient can be simplified by using the spectral problem for the zero mode and, after some manipulations, identifying its norm. In \autoref{f:CCM_Vortex_A_D} (right panel) we plot the couplings $w_1^d$ and $w_2^d$ as functions of $\lambda$. 
    
\begin{figure}[htb]
         \includegraphics[width=0.332\linewidth]{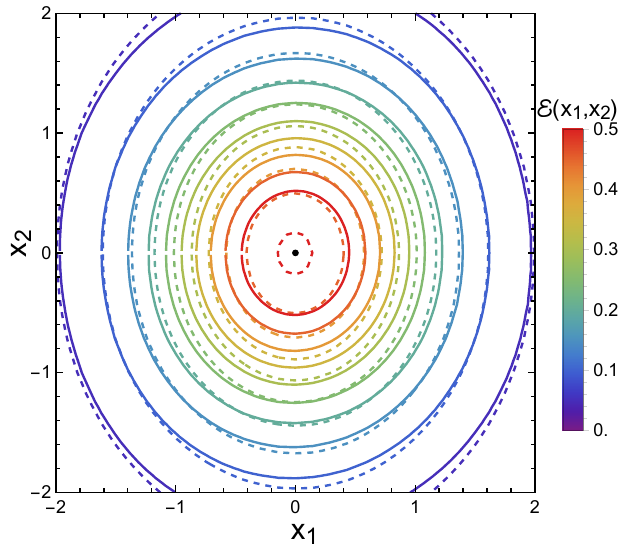}
          \includegraphics[width=0.332\linewidth]{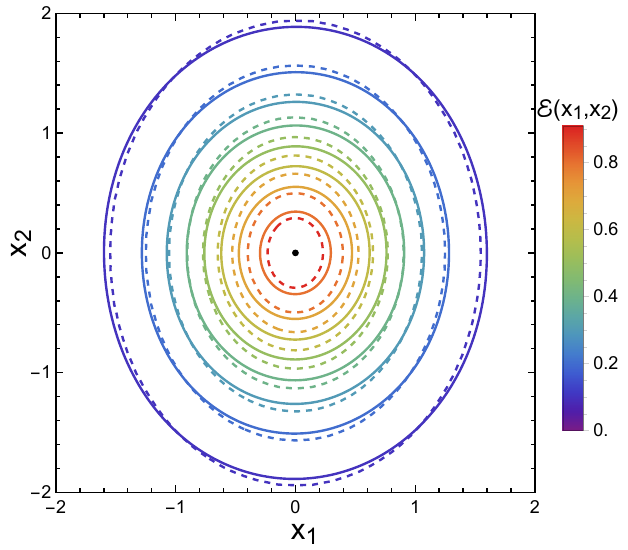}
                \includegraphics[width=0.332\linewidth]{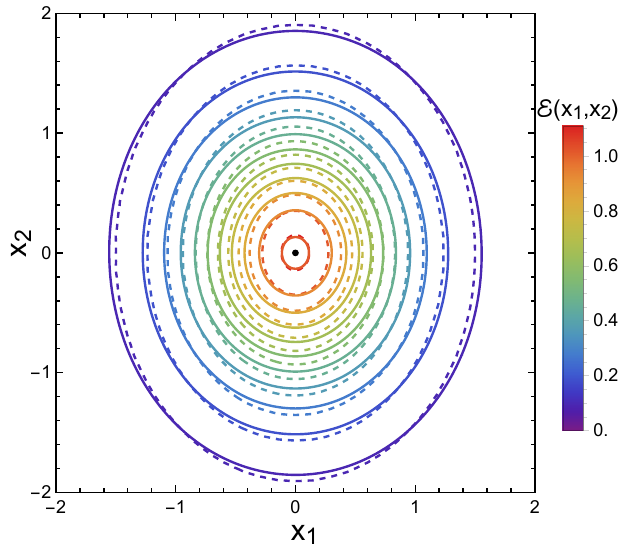}
        \caption{Comparison between  the energy density obtained from the $x$-Derrick mode collective coordinate model (solid lines) given by (\ref{e:L_CCM_Vortex_A_D}) and the numerical simulation (dashed lines) for a vortex with $n=1$ and velocity $v=0.6$. Here $\lambda=0.4,1,1.4.$}
             \label{f:Energy_den_CCM_A_D}
    \end{figure}

    This collective coordinate model also possesses a stationary solution $\dot{a}=v$ and $C=C_0$, where $C_0$ obeys the following equation
    \begin{equation}
        \frac{v^2}{2}\left(M+2h_2^dC_0\right)=M C_0+3w_1^dC^2_0+4w_2^dC^3_0\,,
    \end{equation}
    that is formally identical to (\ref{e:C0_value_Shape}). At quadratic order in velocity, the solution reduces to $C_0 = v^2/2$. An identical result is found in the collective coordinate model of a moving kink based on the Derrick mode \cite{Manton:2021}. This value is closely related to the small-velocity expansion of the Lorentz factor $\gamma$, for which the expansion up to second-order is $\gamma \approx 1 + v^2/2$. Thus, constant-velocity motion of the single vortex requires a non-zero excitation of the Derrick mode, which plays the role of effectively encoding the Lorentz contraction in the collective coordinate description. Using such a stationary solution and plotting the energy density of the associated field configuration, we confirm that it reproduces the Lorentz contraction of the moving one-vortex with very good precision, as shown in \autoref{f:Energy_den_CCM_A_D}.
    \vspace{-0.32cm}
    
\begin{figure}[htb]
        \centering
        \includegraphics[width=0.52\linewidth]{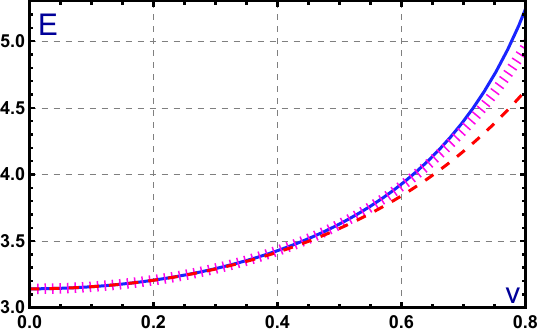}
        \caption{Comparison between the energy of the moving 1-vortex (solid blue line) with the energy of solution obtained in the collective coordinate model based on the $x$-Derrick mode (dotted line) and the shape mode (dashed line). Here $\lambda=1$.}
        \label{f:Energy_Num_Sh_Der}
    \end{figure}

    Remarkably, the agreement is significantly better than in the case of the shape-mode-based collective coordinate model. Moreover, the collective coordinate model constructed from the $x_1$-Derrick mode provides a much more accurate approximation to the relativistic energy of the single vortex. In \autoref{f:Energy_Num_Sh_Der}, we present a comparison of the velocity dependence of the energies for $\lambda = 1$. The figure contrasts the full field theory results with the predictions obtained from the two collective coordinate descriptions. These results provide strong evidence that the Derrick mode captures the essential kinematical features of boosted vortices within the collective coordinate framework.

    The following result also deserves attention. From the equations of motion for $C$, derived from (\ref{e:L_CCM_Vortex_A_D}), we deduce that the frequency of the $x_1$-Derrick mode, in the quadratic approximation, is given by
    \begin{equation}
        \omega_d^2=\frac{M}{g_{CC}}\,.
    \end{equation}
    
\begin{figure}[htb]
        \centering
        \includegraphics[width=0.5\linewidth]{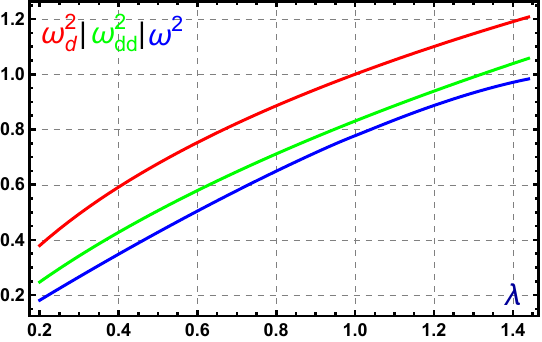}
         \caption{Comparison of the frequency of the shape mode $\omega$ and the frequencies of the radial-Derrick mode $\omega_{dd}$ and the $x_1-$Derrick mode $\omega_d$.}
        \label{f:Comparison_Freq}
    \end{figure}

    In \autoref{f:Comparison_Freq} we compare the frequency of the $x_1$-Derrick mode $\omega_d$ and the frequency of the shape mode $\omega$. It is clearly visible that the agreement is not perfect. This deviation is not surprising, as the shape mode is a radially symmetric excitation, while the $x_1$-Derrick mode explicitly breaks this symmetry. In particular, we have found an approximately constant difference $\omega^2_d - \omega^2 \approx 0.2$. 

    In order to correctly approximate the Lorentz contraction and, at the same time, be able to model the axially symmetric shape mode, we adopt the following strategy. Let us introduce a second Derrick deformation, now along the $x_2$-axis. With this further contribution, the set of configurations (\ref{e:Ansatz_CCM_Vortex_A_D}) is extended to
    \begin{eqnarray}
        \Phi \hspace{-0.2cm}&=&\hspace{-0.2cm} \Phi^v(\vec{x}) + \epsilon \,D_1 \Phi^v(\vec{x}) + C_1 x_1 D_1 \Phi^v(\vec{x}) + \epsilon\, C_1D_1(x_1 D_1\Phi^v(\vec{x}))\nonumber \\ 
        &+&\hspace{-0.2cm} C_2 x_2 D_2 \Phi^v(\vec{x})+ \epsilon\, C_2 D_1(x_2 D_2\Phi^v(\vec{x}))\,,\\
        A_1 \hspace{-0.2cm}&=&\hspace{-0.2cm} A^v_1(\vec{x}) - C_2 x_2 B^v(\vec{x})\,,\\
        A_2 \hspace{-0.2cm}&=&\hspace{-0.2cm} A^v_2(\vec{x}) + \epsilon \, B^v(\vec{x}) + C_1 x_1 B^v(\vec{x}) + \epsilon\, C_1 \partial_1( x_1 B^v(\vec{x})) + \epsilon\, C_2 \partial_2(x_2 B^v(\vec{x}))\,.  
    \end{eqnarray}
    
    Proceeding in the same manner as in the previous examples, we obtain a new collective coordinate model, now based on the zero mode and both Derrick modes 
    \begin{equation} \label{e:CCM_2Derrick}
        \begin{split}
            L_{dd}[a, C_1,C_2] &= \dfrac{1}{2}g_{aa}\dot{a}^2+ \dfrac{1}{2}g_{C_1 C_1}\dot{C}_1^2 + \dfrac{1}{2}g_{C_2 C_2}\dot{C}_2^2 + \dfrac{1}{2}g_{C_1 C_2}\dot{C}_1\dot{C}_2 - V(C_1,C_2)\,.
        \end{split}
    \end{equation}
    The metric on the three-dimensional moduli space has now a more complex form
    \begin{equation}
        \begin{split} \label{e:Metric2Derrick_AC1C2}
            g_{aa} &= M + h_{1,1}^{dd} C_1 + h_{1,2}^{dd} C_2 + h_{2,1}^{dd} C_1^2 + h_{2,2}^{dd} C_2^2 + h_3^{dd} C_1C_2\,.     
        \end{split}
    \end{equation}
    The main difference from the previous effective models is that the metric exhibits an off-diagonal term, reflecting the fact that the $x_1$ and $x_2$ scaling deformations do not give rise to orthogonal modes. The metric terms $g_{C_1C_1}$ and $g_{C_2C_2}$ coincide for all values of $\lambda$. However, the coefficients $h_{1,1}^{dd}$ and $h_{2,1}^{dd}$ differ from $h_{1,2}^{dd}$ and $h_{2,2}^{dd}$, respectively. This asymmetry arises due to their coupling to the translational degree of freedom, implying that the $x_1$ and $x_2$ directions are not equivalent. For our purpose of reproducing the frequency of the shape mode, we do not display the values of the coefficients in the metric in full generality. The same applies to the potential $V(C_1,C_2)$, which is, once again, a quartic polynomial in the mode amplitudes.  

\begin{figure}[htb]
        \hspace{0.2cm}
        \includegraphics[width=0.46\linewidth]{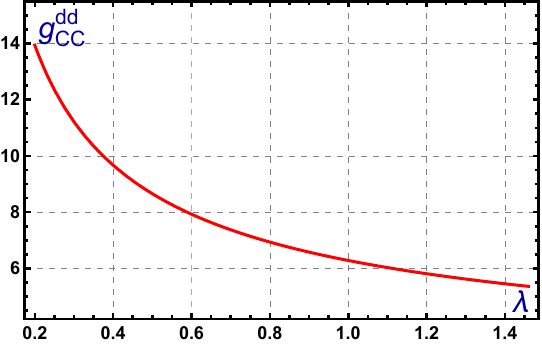}
        \hspace{0.9cm}
        \includegraphics[width=0.45\linewidth]{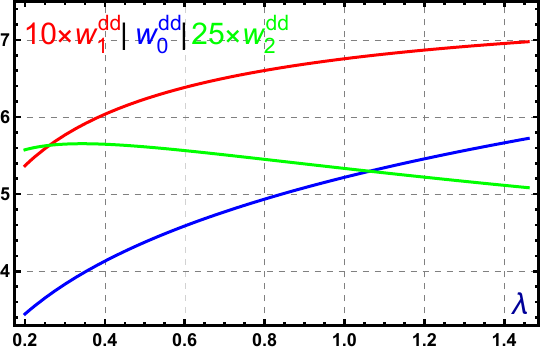}
        \caption{Coefficients of the metric (left panel) and potential (right panel) terms in the radial-Derrick mode based effective model for the one-vortex (\ref{e:CCM_RadialDerrick}).}
         \label{f:CCM_Vortex_A_DD}
    \end{figure}

    To corroborate whether the inclusion of two Derrick modes reproduces the frequency of the shape mode, we only need to assume an axially symmetric situation where $C_1 = C_2 = C$. This is a consistent assumption for a one-vortex whose centre of mass does not move, that is, $\dot{a} = 0$. The general collective coordinate model (\ref{e:CCM_2Derrick}) is then reduced to 
    \begin{equation} \label{e:CCM_RadialDerrick}
        \begin{split}
            L_{dd}[C] &= \dfrac{1}{2}g_{CC}^{dd}\dot{C}^2 - V(C)\,,
        \end{split}
    \end{equation}
    with
    \begin{equation}
        V(C) = M + \dfrac{1}{2}w_{0}^{dd}C^2 + w_1^{dd} C^3 + w_2^{dd} C^4\,.
    \end{equation}
    In \autoref{f:CCM_Vortex_A_DD}, we plot the metric coefficient $g_{CC}^{dd}$ (left panel) and the potential coefficients $w_{0}^{dd},w_{1}^{dd}$ and $w_{2}^{dd}$ (right panel). 
    Through the equation of motion for $C$, the radial collective coordinate model yields the following frequency for the radial Derrick mode at quadratic level
    \begin{equation}
        \omega_{dd}^2 = \dfrac{w_0^{dd}}{g_{CC}^{dd}}\,.
    \end{equation}
    This frequency is significantly closer to that of the shape mode than in the case of only one Derrick mode, see \autoref{f:Comparison_Freq}. Now, the difference between these frequencies is reduced to an approximately constant value $\omega_{dd}^2 - \omega^2 \approx 0.06$. 

    Therefore, we conclude that the three-dimensional collective coordinate model (\ref{e:CCM_2Derrick}) possesses sufficient degrees of freedom to incorporate the advantages of both single-mode effective models.
    
    \section{Summary and conclusions}\label{s:Summary_Vortex}

    In this chapter, we have analysed the influence of massive vibrational modes on the moduli space metric in the one-vortex case. These modes not only induce a mode-dependent effective potential interpretable as a mode-generated force \cite{Manton:2024,Rees:2025}, but also deform the metric component associated with the translational zero mode, which governs the position of the vortex. This deformation significantly impacts the vortex dynamics when described using a collective coordinate model. In particular, within this framework, a constant motion of the one-vortex requires a non-zero excitation of the vibrational mode.

    We have verified this in two single-mode models: one based on the shape mode, and the other on the $x_1$-Derrick mode, which is associated with spatial scaling along the $x_1$ direction. Although the shape mode is the proper bound mode of the one-vortex and thus a natural candidate for a collective coordinate approach, it has the drawback of failing to reproduce Lorentz contraction. On the other hand, the effective description based on the $x_1$-Derrick mode also presents limitations, as it only provides a crude approximation to the frequency of the shape mode.
    
    We have solved these shortcomings by introducing a new collective coordinate model that contains two Derrick modes apart from the zero mode. These two Derrick modes are related to spatial scalings in two orthogonal directions. Now, we can describe both the Lorentz contraction and approximate the shape mode frequency. As a consequence, this last model defines the appropriate initial conditions for the (excited) vortex-vortex and vortex-antivortex scattering in an effective theory, especially at relativistic velocities. This may also shed light on why, even in the BPS limit, highly relativistic scattering of initially unexcited one-vortices leads to vibrating vortices in the final state. A detailed investigation of this phenomenon is left for future work. A natural extension of these results could be to study the role of the internal structure of global vortices in scattering processes. They have an infinite number of internal modes capable of exchanging energy. In this context, global vortices interact strongly, and it would be interesting to examine how the internal modes affect this interaction.

    Moreover, our approach can also be applied to the collective description of the scattering of BPS monopoles \cite{Bachmaier:2024,Bachmaier:2025}. We expect a similar modification of the metric functions, even though the normal vibrational modes should be replaced by quasi-bound modes \cite{Forgacs:2004, Russell:2011}.

    Although we have not contemplated that idea in this chapter, an interesting approach arises when higher-charge vortices are considered. When $n \geq 2$, vortices support bound modes that are not axially-symmetric \cite{Hindmarsh:2008, Izquierdo:2016, Izquierdo1:2016, Izquierdo:2024}. Therefore, it would be interesting to confirm whether a non-axially-symmetric mode is able to approximate the Lorentz contraction without using Derrick modes.

        \part{On the Role of Internal Modes in the Dynamics of Sphalerons}

        \chapter{Introduction to sphalerons}\label{c:Sphalerons}

    \epigraph{"Mathematical intuition is the ability to uncover hidden truths."}{-- Emmy Noether}

    So far, we have drawn our attention to the study of topological solitons. These particle-like configurations are stable thanks to the conservation of certain topological charges. We saw that these charges are determined due the asymptotic behaviour of the fields involved, which must satisfy the requirement of finite energy. Although not explored in this thesis, there also exist field configurations that are not topologically protected but are nonetheless stable due to the conservation of other types of charges. Nevertheless, there exists another important class of solitons in physics: the so-called \textit{sphalerons}. Unlike the previously discussed cases, sphalerons are inherently unstable. This instability can be traced back to the presence of an unstable mode in the linear spectrum of perturbations around the solution. In fact, sphalerons appear as critical points of the energy functional, corresponding to saddle points. This chapter is devoted to provide a description of these objects. First, \autoref{s:Sphalerons} addresses the topological realisation behind the existence of these unstable field configurations. Then, we briefly summarise a relevant example of sphaleron from a physical point of view in \autoref{s:ElectroWeak}: the \textit{electroweak sphaleron}. Finally, we describe examples of sphalerons in $1+1$ dimensions in \autoref{s:Examples_Sph}. In particular, we will show the general framework of models with a false vacuum structure in their potential function and models defined on the circle and how sphalerons emerge in these theories. For clarity, we will provide an example in each case. Finally, \autoref{s:FurtherComments_Sph} addresses further comments on the description of sphalerons.
    
\section{Topology and Sphalerons}\label{s:Sphalerons}

    So far, we have discussed the topological reasons behind the emergence of topological solitons. We have seen that topological solitons appear when the configuration space $\mathcal{C}$ is disconnected. Mathematically, this condition translates into $\pi_0(\mathcal{C}) \neq 0$. Although it could seem that if that condition is not fulfilled there are no more static bound states, that is not the case. There is one further possibility that is related to higher homotopy groups.
    
    To investigate that possibility, let $M$ be the space of maps from the $d$-dimensional sphere $\mathbbm{S}^{d}$ to the topological space $Y$, that is,
    \begin{equation}
        M := \text{Maps}(\mathbbm{S}^{d}, Y)\,,
    \end{equation}
    and let us assume that all of them map a fixed point $\rm{x}_0 \in \mathbbm{S}^d$ to a point $\rm{y}_0 \in Y$. It can be verified that the $l$-th homotopy group of $Y$ is given by
    \begin{equation}
        \pi_{l}(M) \cong \pi_{l+d}(Y)\,.
    \end{equation}
    When $\pi_{1}(M) \neq 0$, the space $M$ is not simply-connected, what implies that there are non-contractible loops. It is under these situations where sphalerons can arise \cite{Taubes:1980,Taubes:1982,Forgacs:1984}.

    The usual technique to find these solutions is based on a \textit{min-max scheme}, usually referred to as \textit{Ljusternik-Snirelman procedure}. The mathematical statement is collected in the following theorem.
    \begin{theorem}[Ljusternik-Snirelman procedure \cite{Ljusternik:1996}] Let $E$ be a $C^2$ function defined on a compact manifold $\mathcal{N}$ of dimension $n$. Suppose that $y^{min} \in \mathcal{N}$ is an isolated minimum of $E$. Let $f_0: \mathbbm{S}^k \rightarrow \mathcal{N}$ be a generator of the homotopy group $\pi_k(\mathcal{N})$ with $f_0(x) = y^{min}$ for a given $x \in \mathbbm{S}^{k}$. 
    
    Consider the set $\Lambda$ of based maps $f: \mathbbm{S}^k \rightarrow \mathcal{N}$ homotopic to $f_0$. For each $f \in \Lambda$, choose $x^{max} \in \mathbbm{S}^k$ such that \[E\big|_{f(x^{max})} = \sup_{x \in \mathbbm{S}^k} \{E|_{f(x)}\}\,.\] As $\mathcal{N}$ is compact and the maps belonging to $\Lambda$ are not homotopic to a constant map, there is a sequence $\{f_i\} \in \Lambda$ satisfying \[\lim_{i \rightarrow \infty}E\big|_{f_{i}(x_{i}^{max})} \longrightarrow E^{*} = \inf_{\{f \in \Lambda\}}\{E|_{f(x^{max})}\} > E|_{y^{min}}\,,\] with the property that \[\lim_{i \rightarrow \infty}f_{i}(x_{i}^{max}) \longrightarrow \, y^* \in \mathcal{N}\,; \quad y^* \neq y^{min},\] where $y^*$ is a critical point of $E$.
    \end{theorem}

\begin{figure}[htb]
        \centering{\includegraphics[width=0.55\linewidth]{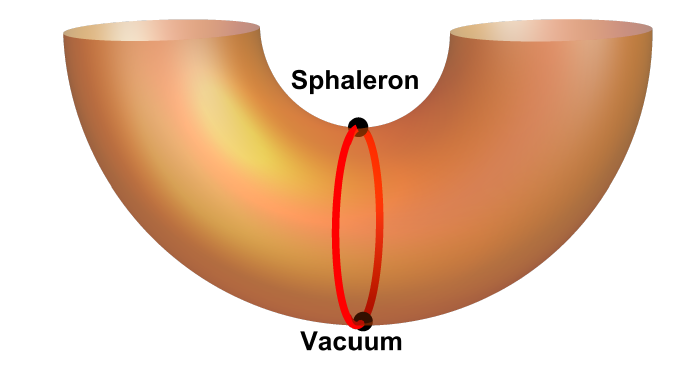}
        }
        \caption{The torus represents the value of the energy functional $E$ for the fields in the configuration space $\mathcal{C}$. The energy increases along the vertical direction. The depicted non-contractible loop corresponds to the one whose maximum configuration is the lowest possible among all non-contractible loops homotopic to the former, i.e., it contains the sphaleron configuration.}
        \label{f:min_max}
    \end{figure}

    For our purposes, the function $E$ would be the energy of the field theory and the manifold $\mathcal{N}$ would be the configuration space $\mathcal{C}$.
    The pictorial scheme behind the previous theorem is the following: assume a non-contractible continuous path consisting of finite energy field configurations, which starts and ends at the same vacuum. Then, find the field configuration with maximal energy on each of these paths and find the path where this energy is minimal. The configuration corresponding to the minimal maximum is a saddle-point solution of the theory. A representative example is illustrated in \autoref{f:min_max}.

     The difficulty here is that the configuration space $\mathcal{C}$ is not compact, thereby not complying with the claims of the previous theorem. Although the technical details are far from the scope of this thesis, we want to remark that Taubes succeeded in generalising this procedure to an infinite-dimensional configuration space in the context of the $SU(2)$ Yang-Mills-Higgs theory with the Higgs field in the adjoint representation  \cite{Taubes:1982,TaubesI:1982}. The configuration space in that theory is homotopically the space of based maps $M:= \text{Maps}(\mathbbm{S}^2,\mathbbm{S}^2$), which can be identified with the space of Higgs fields at infinity. This theory admits topological solitons, as the configuration space is disconnected $\pi_0(\mathcal{C}) \cong \pi_0(M) \cong \pi_2(\mathbbm{S}^2) \cong \mathbbm{Z}$. In this context, the topological solitons are referred to as \textit{monopoles}. Namely, the solution in the component $\mathcal{C}_1$ is the monopole, and the solution in $\mathcal{C}_{-1}$ is the antimonopole. In his seminal paper, Taubes proved the existence of at least one additional static solution, besides the vacuum, in the sector of vanishing topological charge $\mathcal{C}_0$, since $\pi_1(\mathcal{C}_0) \cong \pi_3(\mathbbm{S}^2) \cong \mathbbm{Z}$. The solution that Taubes found represented a monopole-antimonopole pair in unstable equilibrium. 
     
     Since the pioneering work of Taubes, various sphalerons have been discovered by constructing non-contractible loops in one \cite{Bochkarev:1987,Mottola:1989}, two \cite{Shurgaia:1993}, and three \cite{Manton:1983,Brien:1992,Shurgaia:1995,Kleihau:1999} spatial dimensions. 
     
\section{Electroweak sphaleron}\label{s:ElectroWeak}

    The most appealing example of a sphaleron arises in particle physics, where these configurations are believed to have mediated processes contributing to baryogenesis. To motivate this application, we provide a brief overview that will lead to the introduction of the so-called \textit{electroweak sphaleron}. For our pedagogical purposes, only the main concepts necessary to follow the reasoning will be presented, while certain technical details and intermediate results are omitted or summarised.
    
    The pure $SU(2)$ Yang-Mills theory in $3+1$ Minkowski space-time is defined by the Lagrangian density
    \begin{equation}
        \mathcal{L} = - \dfrac{1}{4}\text{Tr}\left(F^{\mu\nu}F_{\mu\nu}\right)\,,
    \end{equation}
    where $F_{\mu\nu} = \partial_{\mu}A_{\nu} - \partial_{\nu}A_{\mu} + g\,[A_{\mu},A_{\nu}]$\,, where $A_{\mu}$ is a $\mathfrak{su}(2)$-valued gauge potential. The requirement of finite energy imposes that $F_{\mu\nu}$ must vanish at infinity, thus $A_{\mu}$ must be asymptotically a  \textit{pure gauge transformation} of $A_{\mu} = 0$, that is, 
    \begin{equation}
        A_{\mu}  \xrightarrow{\|x\| \to \infty} g^{-1}U^{\infty}\partial_{\mu}(U^{\infty})^{-1}\,,
    \end{equation}
    with $U^{\infty} \in SU(2)$. Imposing that all gauge field configurations approach such a pure gauge at infinity allows us to compactify $\mathbbm{R}^3$ to $\mathbbm{S}^3$. Topologically, this means that we can classify these asymptotic configurations as $U^{\infty}: \mathbbm{S}^3 \rightarrow SU(2) \cong \mathbbm{S}^3$. Therefore, the space of gauge-inequivalent vacua corresponds to homotopy classes of such maps $\pi_0 (\text{Maps}(\mathbbm{S}^3,\mathbbm{S}^3)) \cong \pi_3(\mathbbm{S}^3) \cong \mathbbm{Z}$. Despite the topologically non-trivial structure, Derrick's theorem rules out the existence of a stationary point in this scenario. However, there is no restriction when a Wick rotation is performed, that is, when $t \rightarrow i \tau$. Upon that transformation, the model is now defined on the 4-dimensional Euclidean space. In this new framework, the theory enjoys a static solution that minimises the Euclidean action, which is referred to as the \textit{instanton} \cite{Belavin:1975}. Once again, it can be verified that the configuration space consists of $\pi_0 (\text{Maps}(\mathbbm{S}^3,\mathbbm{S}^3)) \cong \pi_3(\mathbbm{S}^3) \cong \mathbbm{Z}$ disconnected components. The relevance of the instanton solution is that the vacua are separated by a barrier energy, and the instanton describes the tunnelling between adjacent vacua. In particular, the height of the energy barrier depends inversely on the instanton scale. The subtle aspect is that the pure $SU(2)$ Yang-Mills theory in the $4$-dimensional Euclidean space is scale invariant, what implies that the instanton does not have a characteristic size. Therefore, the energy barrier is not well-defined. In addition, it has been verified that the tunnelling of instantons through the barrier is strongly suppressed \cite{Hooft:1976,Gross:1981}.

    The height of the energy barrier is fixed for the $SU(2)$ Yang-Mills-Higgs theory, where a doublet Higgs field in the fundamental representation is minimally coupled to the pure $SU(2)$ Yang-Mills theory
    \begin{equation}
        \mathcal{L} = - \dfrac{1}{4}\text{Tr}\left(F^{\mu\nu}F_{\mu\nu}\right) + \overline{D^{\mu}\Phi} D_{\mu}\Phi - \lambda\left(\dfrac{v^2}{2} - \overline{\Phi}\Phi\right)^2\,.
    \end{equation}
    Here, the covariant derivative is defined as $D_{\mu}\Phi = \partial_\mu \Phi - \dfrac{1}{2}ig\sigma_a W^a_{\mu}\Phi$, and $\sigma_a$ with $a \in \{1,2,3\}$ denotes the Pauli matrices, which are the generators of the $SU(2)$ group. Through the inclusion of the Higgs field $\Phi$, new spatial scalings are introduced, and Derrick's theorem does not rule out now the existence of a non-trivial static configuration in $3+1$ dimensions. Hence, the non-trivial static solutions have a characteristic scale. Nevertheless, no static finite-energy instantons exist now in the Euclidean version of the model. However, the topology of the field configurations space is determined completely by the asymptotics of the Higgs and gauge fields at infinity. Namely, the field configuration space is topologically equivalent to the space of maps $M := \text{Maps}(\mathbb{S}^2,\mathbb{S}^3)$. Since $\pi_1(M) \cong \pi_3(\mathbb{S}^3) \cong \mathbbm{Z}$, there are non-contractible loops in the configuration space. Manton succeeded in computing a non-contractible loop in this theory \cite{Manton:1983}, suggesting the existence of a static, unstable solution of the field equations. Such a solution was first founded numerically by Dashen, Hasslacher and Neveu \cite{Dashen:1974} almost a decade before, although they did not verify the classical stability of that configuration against small oscillations due to the complexity in performing the numerical computations. A rigorous proof of the existence of this solution was achieved by Burzlaff, who also succeeded in verifying that it was unstable \cite{Burzlaff:1984}. Moreover, Forgacs and Horvath proved that the configuration of maximum energy along the non-contractible loop found by Manton corresponded to the Dashen, Hasslacher and Neveu solution \cite{Forgacs:1984}. However, the static unstable solution was given in terms of two unknown radial functions. An appropriate ansatz for them was given by Klinkhamer and Manton as well as an estimation of the energy in terms of the self-coupling constant $\lambda$ \cite{Manton:1984}. They determined that the energy varied from $7.9$ TeV for $\lambda = 0$ to $13.7$ TeV for $\lambda \rightarrow \infty$. In that seminal paper, the authors introduced the term "sphaleron"\footnote{This word is based on the classical Greek adjective $\sigma \varphi \alpha \gamma \epsilon \rho o s$ (sphaleros), meaning "ready to fall".} to refer to this inherently unstable soliton-like solution. Regarding the internal structure of the sphaleron, it has been extensively studied in the literature \cite{Yaffe:1989,Brihayev:1990,Carson:1990}.
    
    A further extension of the $SU(2)$ Yang-Mills-Higgs model appears when a $U(1)$ gauge field is introduced in such a way that the new theory is invariant under the gauge group $SU(2)_W\times U(1)_Y$. The new theory is referred to as the \textit{Weinberg-Salam model} in the literature
    \begin{equation}\label{e:EW_Lag}
        \mathcal{L} = \dfrac{1}{4}\text{Tr}\left(F^{\mu\nu}F_{\mu\nu}\right) + \dfrac{1}{4}f^{\mu\nu}f_{\mu\nu} + \overline{D^{\mu}\Phi} D_{\mu}\Phi - \lambda\left(\dfrac{v^2}{2} - \overline{\Phi}\Phi\right)^2\,,
    \end{equation}    
    where $f_{\mu\nu} = \partial_{\mu}A_{\nu} - \partial_{\nu}A_{\mu}$ and now $D_{\mu}\Phi = \partial_\mu \Phi - \dfrac{1}{2}ig\sigma_a W^a_{\mu}\Phi - \dfrac{1}{2}ig'A_{\mu}\Phi$\,. Again, the index $a$ takes the values $a \in \{1,2,3\}$. This gauge theory is a unified model of weak and electromagnetic interactions. Concretely, it describes the electroweak sector of the standard model of particles. Through the Higgs mechanism, the gauge symmetry is spontaneously broken to $U(1)_\gamma$, which produces the charged massive bosons $W^{\pm}$ as well as the neutral massive $Z_{\mu}^0$ boson and the massless photon $B_{\mu}$. These fields are defined by
    \begin{equation}
        W^{\pm}_{\mu} = \dfrac{1}{\sqrt{2}}\left(W_{\mu}^1 \mp W_{\mu}^2 \right)\,,
    \end{equation}
    and
    \begin{equation}
        \begin{pmatrix}
            B_{\mu} \\
            Z_{\mu}^0
        \end{pmatrix}
        =
        \begin{pmatrix}
            \cos \theta_W & \sin \theta_W \\
            -\sin \theta_W & \cos \theta_W
        \end{pmatrix}
        \begin{pmatrix}
            A_{\mu} \\
            W_{\mu}^3
        \end{pmatrix},
    \end{equation}
    respectively, where $\theta_W$ is the mixing angle, defined via $\tan(\theta_W) = g'/g$, which characterises the relative strength of the gauge couplings $g$ and $g'$. As seen by 't Hooft \cite{Hooft:1976}, the electroweak theory does not conserve the baryon (resp. lepton) number due to an "anomaly" in the baryon and lepton currents. In order to verify that, we should include the fermionic part to the Lagrangian density (\ref{e:EW_Lag}), but we will omit these details here.

    In the electroweak theory, the asymptotic Higgs field defines a map $\Phi^{\infty}: \mathbbm{S}^2 \rightarrow \mathbbm{S}^3$. Therefore, the configuration space is connected $\pi_0(\mathcal{C}) \cong \pi_0(\text{Maps}(\mathbbm{S}^2,\mathbbm{S}^3)) \cong \pi_2(\mathbbm{S}^3) \cong 0$. As a consequence, it does not support topological solitons. However, the first homotopy group is not trivial, since $\pi_1(\mathcal{C}) \cong \pi_1(\text{Maps}(\mathbbm{S}^2,\mathbbm{S}^3)) \cong \pi_3(\mathbbm{S}^3) \cong \mathbbm{Z}$. As a result, sphalerons can emerge in this theory.

    Remarkably, Klinkhamer and Manton deduced a sphaleron configuration in the bosonic part of electroweak theory for small values of the mixing angle $\theta_W$  \cite{Manton:1984}. They showed the influence of a non-zero value of the mixing angle $\theta_W$ on the properties of the sphaleron in the $SU(2)$ Yang-Mills-Higgs theory. Specifically, they revealed that the energy and the symmetries of the profiles are not significantly altered at leading order, and that the sphaleron is purely magnetic, carrying a non-zero magnetic dipole moment. Moreover, they determined that the associated baryon $B$ (resp. lepton $L$) number was $B(L) = 1/2$, what motivates the picture of the sphaleron as the configuration located at the maximum of the potential separating two adjacent vacua with different baryon (resp. lepton) number. 
    
    This is where the splendour of the electroweak sphaleron truly comes into play. Identifying the height of the barrier between adjacent vacua with the energy of the sphaleron suggests that, in situations where sufficient energy is available, the system could cross the barrier rather than tunnel through it. For instance, in the early universe -- when temperatures were extremely high -- thermal fluctuations could have triggered such a process. As a result, the baryon and the lepton numbers would change $\Delta B = \Delta L$ whilst $B - L$ is conserved. This possibility has been studied extensively in the literature \cite{Carson:1990,Kuzmin:1985,Arnold:1987,Ellis:1987,Kripfganz:1989}, and implies that these processes mediated by sphalerons could have washed out any baryon number created in the early universe or even contributed to generate the asymmetry between matter and antimatter that we observe today. It is important to mention that the non-conservation of baryon number is not sufficient to fully explain the matter-antimatter asymmetry of the universe. In addition, $C$ (charge) and $CP$ (charge-parity) violation, as well as a departure from thermal equilibrium, are also required, as originally proposed by Sakharov \cite{Sakharov:1991}.
    
    \section{Sphalerons in \texorpdfstring{$1+1$}{1+1} dimensions}\label{s:Examples_Sph}
    
    The complexity of the electroweak theory makes the dynamical properties of sphalerons difficult to predict. For this reason, it is useful to begin by analysing lower-dimensional models, which can then serve as a basis for extrapolating results to more physically realistic scenarios \cite{Bochkarev:1987,Mottola:1989,Shurgaia:1993,Grigoriev:1989,Grigoriev1:1989,Carson1:1990,Tchrakian:1992}.
    
    This section is devoted to reviewing the simplest theories in which sphalerons can arise. In particular, we present sphaleron solutions in $1+1$ dimensional theories defined on the real line with a false vacuum, as well as in theories defined on the circle. We begin by outlining the general framework and then proceed to discuss illustrative examples to develop an intuitive understanding of the key features.

\subsection{Sphalerons in false vacuum models}\label{ss:Sph_FalseVacuum}

    A set of theories that allow the appearance of sphaleron solutions are the $1+1$ dimensional scalar field theories with a false vacuum structure. The Lagrangian density in these models is defined by
    \begin{equation}
        \mathcal{L} = \dfrac{1}{2}\partial_{\mu}\phi\partial^{\mu}\phi - U(\phi)\,,
    \end{equation}
    with $\mu = 0,1$, where the potential function $U(\phi)$ is assumed to have, for simplicity, two minima that will be denoted by $\phi_f$ and $\phi_v$. Let us assume that $U(\phi_f) > U(\phi_v)$. Without loss of generality we can impose that $U(\phi_f) = 0$. Thus, $\phi_v$ is the true vacuum of the theory, and $\phi_f$ is usually referred to as the false vacuum of the model. Note that the continuity of the potential guarantees the existence of an additional intermediate field value $\phi^0$, where $U(\phi^0) = 0$.

    Upon following the Bogomolny arrangement to the energy functional associated to the previous Lagrangian density, we are left with
    \begin{equation}
        \phi_x = \pm \sqrt{2\,U(\phi)}\,.
    \end{equation}
    The solutions of the individual Bogomolny equations would lead to monotonic solutions. However, no static kink or antikink solution exists for the false vacuum model under consideration because the vacua are not degenerate. In fact, the false vacuum structure produces a force that favours energetically the true vacuum. Therefore, a kink configuration would undergo a force in the opposite direction than the antikink \cite{Gonzalez:1987,Gonzalez:1989,Shnir:1998}, which is called \textit{vacuum pressure} in the literature. However, if the kink is followed by an antikink, this configuration interpolating asymptotically the same false vacuum suffers an attractive scalar force. As a consequence, the competition between these opposite forces opens up the possibility that there exists a critical separation for which the kink-antikink pair exhibits an unstable equilibrium. Intuitively, this configuration will be the sphaleron solution in this type of model\footnote{This unstable configuration is called \textit{bounce} in the literature \cite{Coleman:1975}; however, we will continue to refer to it as a sphaleron throughout this discussion.}.  

    In order to find such a sphaleron configuration, we have to note that the profile is necessarily non-monotonic, what demands the existence of a turning point. Naturally, this turning point is given by the zero of the potential $\phi_0$. Note that the turning points is attained linearly, that is,
    \begin{equation}\label{e:approx_cero_false}
        U(\phi) \approx \alpha (\phi - \phi_0) + O(\phi - \phi_0)^2 \;\; \mbox{as} \;\; \phi \to \phi_0\,.
    \end{equation}
    This allows the zero to be attained in a finite spatial extension (unlike a higher-order zero, which is attained exponentially). Therefore, the Bogomolny equation for that configuration is a first-order, piecewise differential equation
    \vspace{-0.1cm}
    \begin{equation}\label{e:Bogomolny_falseSph}
        \phi_x = \left\{
        \begin{array}{cc}
           +\sqrt{2\,U(\phi)}  & x\leq 0\,, \\
            -\sqrt{2\,U(\phi)} & x\geq 0\,.
        \end{array} 
        \right.
    \end{equation}
    The non-monotony of the profile implies that the first spatial derivative of $\phi(x)$ has one spatial node. Moreover, the Lorentz invariance of the theory guarantees that there is a zero mode, that can be written as $\eta_0 \propto \phi'(x)$ as seen in \autoref{s:Modes}. Therefore, we can claim that a solution of (\ref{e:Bogomolny_falseSph}) must hold a mode with zero spatial nodes and negative eigenfrequency in its linear spectrum of perturbations by virtue of the Sturm-Liouville theorem (see \autoref{t:Sturm}). As a consequence, the static solutions of (\ref{e:Bogomolny_falseSph}) are unstable and describe sphaleron-like configurations.
    
\begin{figure}[htb]
      \centering
      \noindent
      \begin{minipage}[t]{0.45\textwidth}
        \includegraphics[width=\linewidth]{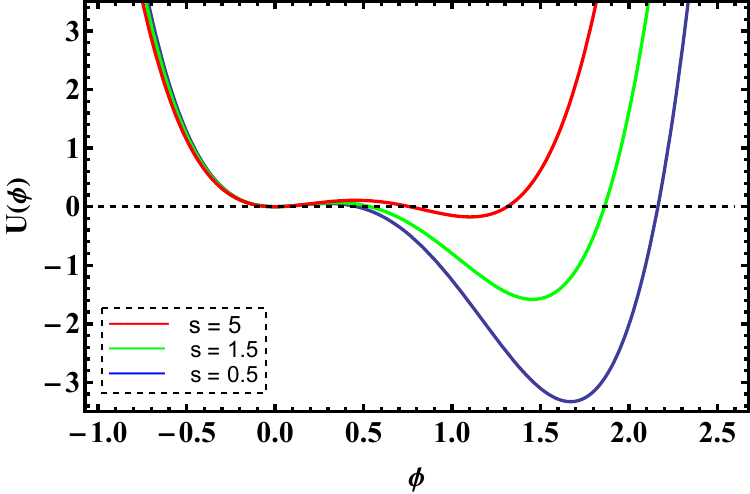}
      \end{minipage}%
      \hspace{0.8cm}
      \begin{minipage}[t]{0.45\textwidth}
        \includegraphics[width=\linewidth]{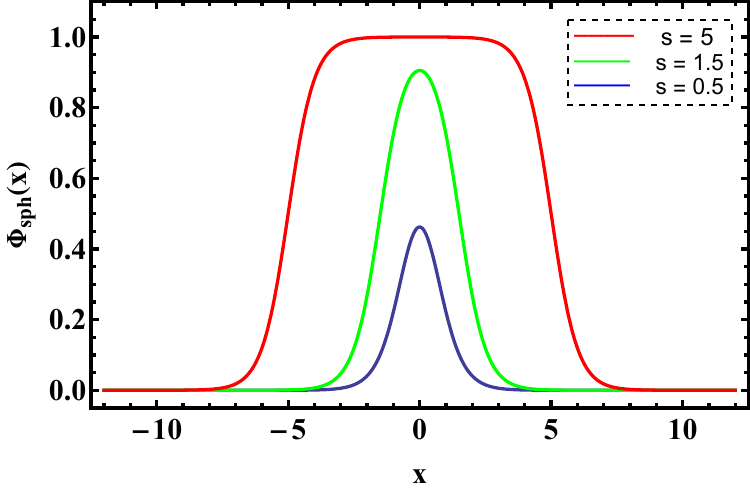}
      \end{minipage}
    
      \caption{Illustration of the false vacuum potential (\ref{e:pot_falseSph}) on the left and the sphalerons solutions (\ref{e:sph_false}) on the right for different values of the parameter $s$ in the modified $\phi^4$ model. The cut of the dashed line with the potential function represents its zeros.}
      \label{f:Phi4_Sph_False}
    \end{figure}

    The simplest example of a sphaleron in a model with false vacuum is the $\phi^3$ theory \cite{Bazeia:2008}. However, this potential is not bounded from below. A physically more appealing model is the modified $\phi^4$ theory, defined through the following potential function \cite{Bazeia:2008}
    \begin{equation}\label{e:pot_falseSph}
        U(\phi) = 2\phi^2(\phi - \tanh s)(\phi - \coth s)\,, \quad s > 0\,,
    \end{equation}
    where the constant $s$ accounts for the shift of the true vacuum. The profile of the potential function $U(\phi)$ is depicted in \autoref{f:Phi4_Sph_False} for different values of the parameter $s$. 
    Note that the original $\phi^4$ model is recovered as $s \rightarrow \infty$. Specifically, the false vacuum is identified with the configuration $\phi_f = 0$, whereas the true vacuum is given by
    \begin{equation}\label{e:true_vacuum}
        \phi_v=\frac{3}{4} \coth 2s + \frac{1}{4} \sqrt{9 \coth^22s-8}\,.
    \end{equation}
    Moreover, the zero of the potential between both vacua is $\phi_0 = \tanh s$. 

\begin{figure}[htb]
        \centering{\includegraphics[width=0.52\linewidth]{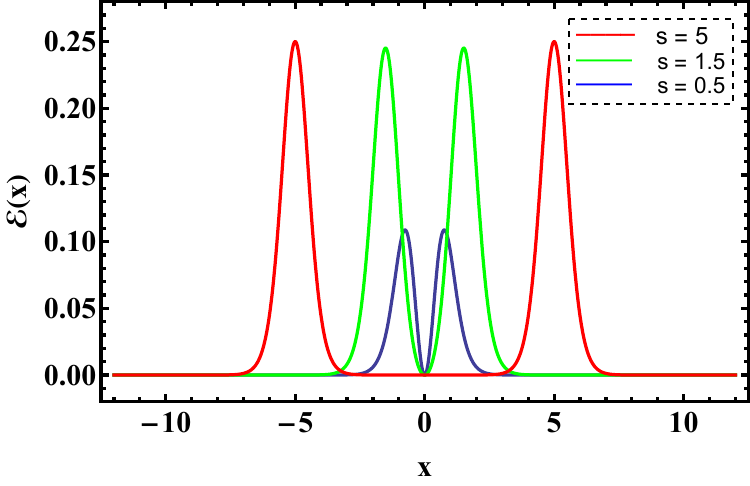}
        }
        \caption{Energy density (\ref{e:dens_energy_false}) of the sphaleron in the modified $\phi^4$ model.}
        \label{f:density_sph_false}
    \end{figure}

    Upon solving (\ref{e:Bogomolny_falseSph}) with the potential function (\ref{e:pot_falseSph}) we are left with the sphaleron-like solutions
    \begin{equation}\label{e:sph_false}
        \Phi_{sph}(x;s) = \dfrac{1}{2}(\tanh(x + s) - \tanh(x - s))\,.
    \end{equation}
    Here, the parameter $s$ effectively changes the size of the sphaleron. For small values of $s$, the sphaleron has a lump-like profile, whilst as $s \rightarrow \infty$ the sphaleron profile resembles a kink-antikink pair with increasing separation (see \autoref{f:Phi4_Sph_False}). Indeed, the energy density associated with the sphaleron configuration (\ref{e:sph_false}) is
    \begin{equation}\label{e:dens_energy_false}
        \mathcal{E} = \dfrac{1}{4}\left(\tanh(x + s)^2 - \tanh(x - s)^2\right)^2\,,
    \end{equation}
    which shows two peaks of increasing amplitude and separation (see \autoref{f:density_sph_false}), and the corresponding energy 
    \begin{equation}\label{e:energy_false}
        E[\Phi_{sph}(x;s)] = \dfrac{2}{3} - 2\cosech^2 2s (2s\coth{2s} -1)\,,
    \end{equation}
    approaches the energy of a kink-antikink pair\footnote{Note that the asymptotic energy is $E_{KAK} = \dfrac{2}{3}$. This discrepancy with the value $E_{KAK} = \dfrac{8}{3}$ arises from the different definitions of the $\phi^4$ model employed in (\ref{e:pot_falseSph}) and (\ref{e:potential_phi4}).} as $s \rightarrow \infty$.

\begin{figure}[htb]
        \centering{\includegraphics[width=0.52\linewidth]{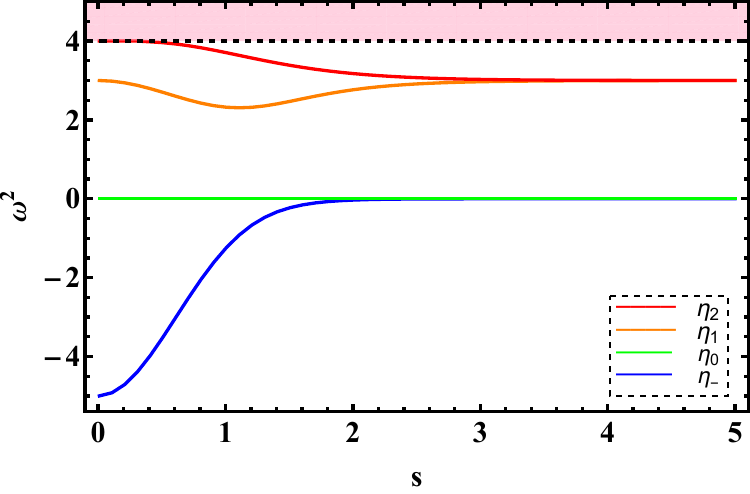}
        }
        \caption{Linear spectrum of perturbations around the sphaleron in the modified $\phi^4$ model given by (\ref{e:pot_falseSph}). We only represent the bound modes, consisting in an unstable mode $\eta_u$, a zero mode $\eta_0$, and two positive bound modes $\eta_1,\eta_2$. The shaded area represents the continuum spectrum.}
        \label{f:spectrum_sph_false}
    \end{figure}

    Now we can analyse the spectral structure of the sphaleron. In order to do that, we consider a perturbation around the sphaleron profile of the form $\phi(x,t) = \Phi_{sph}(x;s) + \eta(x;s)e^{i\omega t}$. Introducing this field configuration into the equations of motion and expanding up to first order in the perturbation, we obtain the following Schrödinger-like equation 
    \begin{equation}\label{e:Schrodinger_false}
        - \eta''(x;s) + V_{eff}(x;s) = \omega^2 \eta(x;s)\,,
    \end{equation}
    where
    \begin{equation}\label{e:Effec_Pot_False}
        V_{eff}(x;s) = 4 + 24\,\Phi_{sph}^2(x;s) - 12\,(\tanh s + \coth s)\,\Phi_{sph}(x;s)\,.
    \end{equation}
    This spectral problem must be solved numerically. The discrete modes are represented in \autoref{f:spectrum_sph_false} for the range of values $s \in [0,5]$. We can observe the presence of a negative mode accounting for the instability of the sphaleron as we anticipated, a zero mode related to the translational invariance of the theory, and two positive bound modes. Note that in the limit $s \rightarrow \infty$, the linear spectrum is doubly degenerated, coinciding with the spectrum of perturbations around each single kink.

    To conclude, we want to emphasise that the arising of the sphaleron in these type of models is not related to topological considerations. Indeed, the configuration space is topologically trivial, and the loops are contractible. The appearance of these unstable solutions to the equations of motion can be attributed to the existence of a continuous family of finite-energy field configurations, which can be parametrised, for example, by the maximum value of the field $\phi(x_{max}) = \phi_{max}$. One can verify that the sphaleron configuration -- which is a solution to the piecewise Bogomolny equation (\ref{e:Bogomolny_falseSph}) --  represents the configuration of highest energy within the continuous family. Any other configuration along the path must have strictly lower energy. 

\subsection{Sphalerons on the circle}\label{ss:Sph_Circle}

    Let us assume now a general $1+1$ dimensional scalar field model defined on the circle. The Lagrangian density is given by 
    \begin{equation} \label{e:lag_sph_circle}
    \mathcal{L}=\frac{1}{2}\partial_\mu\phi\partial^\mu\phi - U(\phi)\,,
    \end{equation}
    where the field $\phi$ is a map that satisfies the following periodicity condition
    \begin{equation}\label{e:condition_period}
    \phi: \mathbb{S}^1\rightarrow \mathbb{R}\,,\quad \phi(t,x)=\phi(t,x + L)\,.
    \end{equation}
    Here, the spatial period $L$ is regarded as the circumference of the circle. We assume that $U(\phi)$ is a positive semidefinite potential with $n$ vacua, denoted by $\phi_i$ and, for simplicity, we also assume that the vacua are zeros of the potential.  
    
    Assuming a static field configuration, the equation of motion associated to the Lagrangian density (\ref{e:lag_sph_circle}) can be integrated once
    \begin{equation}\label{e:Bogomolny_circle}
        \frac{1}{2}\phi_x^2 - U(\phi) = - \frac{1}{2}C^2\,,
    \end{equation}
    providing us with the Bogomolny equation, where $C$ is a real constant. Unlike the models defined on the real line, we do not have to impose $C = 0$ for finite energy configurations to exist. This difference is due to the compactness of the base space. Let us call $\phi_i^{(m)}$ the position of the greatest maxima of the potential $U$. Remarkably, the periodicity condition imposes that $0\leq C^2 \leq 2 \,U(\phi_i^{(m)})$. Otherwise, the Bogomolny equation (\ref{e:Bogomolny_circle}) is not fulfilled.

    In the family of periodic solutions of (\ref{e:Bogomolny_circle}) we find constant and non-trivial configurations. Specifically, the trivial solutions arise for all circumferences $L$, and are identified with the vacua of the theory
    \begin{equation}
        \phi(x)=\phi_i\,, \,\, \text{with $C=0$}\,,
    \end{equation}
    where $E(\phi_i)=0$, and with the highest maxima 
    \begin{equation}
        \phi(x)=\phi_i^{(m)}\,, \,\, \text{with $C=\sqrt{2 U(\phi_i^{(m)})}$}\,,
    \end{equation}
    whose energy is $E(\phi_i^{(m)}) = L\, U(\phi_i^{(m)})$. Regarding the non-trivial solutions, they only appear above a critical value $L_{min}$ of the circumference due to the periodicity requirement, and are identified with non-trivial sphalerons. Such non-trivial sphalerons reduce to the constant solution $\phi_i^{(m)}$ in the limit $C \rightarrow \sqrt{2 U(\phi_i^{(m)})}$, where the circle of circumference $L$ takes its minimum possible value $L = L_{min}$. On the other hand, the circumference of the circle becomes infinite, $L \rightarrow \infty$, as $C \rightarrow 0$, and the non-trivial sphaleron solution approaches a kink-antikink pair configuration with increasing separation. In fact, configurations consisting of $n$ sphalerons appear when $L > L_{min}^{(n)} = n\, L_{min}$, and they resemble $n$ kink-antikink pairs when $L \gg L_{min}^{(n)} = n\, L_{min}$.
 
    Regarding the spectral structure around the static solutions $\phi(x)$ of (\ref{e:Bogomolny_circle}), the normal modes $\eta(x)$ are obtained as usual from the associated Schrödinger-like problem
    \begin{equation}\label{e:modes_circle}
        - \eta''(x) + U''(\phi(x))\eta(x) = \omega^2 \eta(x)\,,
    \end{equation}
    although they must comply now with the periodicity condition $\eta(x) = \eta(x + L)$. When the perturbations are above the trivial solutions $\varphi = \{\phi_i,\phi_i^{(m)}\}$, the equation gives rise to a harmonic oscillator equation. When imposing the periodicity condition, one obtains the allowed eigenfunctions 
    \begin{equation}
        \eta_n^s(x) =
        \sin \left(\dfrac{2\pi n x}{L}\right)\,,  \quad \quad \eta_n^c(x) = 
        \cos \left(\dfrac{2\pi n x}{L}\right)\,,
    \end{equation}
    and eigenvalues
    \begin{equation}
        \omega_n^2 = U''(\varphi) + \dfrac{4\pi^2 n^2}{L^2}\,.
    \end{equation}
    Since the vacuum solutions $\phi_i$ are minima, we can say in advance that all the frequencies $\omega_n^2$ are positive, whereas the maximum configurations $\phi_i^{(m)}$ are inherently unstable and exhibit a negative mode. When non-trivial sphalerons $\Phi_{sph}(x)$ are assumed, the spectral problem $(\ref{e:modes_circle})$ is in general difficult to be solved analytically, and sometimes only a quasi-exactly solvable problem is obtained. However, we can remark something in advance. Unlike the real line case, the spectrum of perturbations on the circle consists of an infinite tower of discrete square-integrable modes due to the compactness of the base space, which implies that even the non-localised scattering modes have a finite norm. 

\begin{figure}[htb]
        \centering
        \noindent
        \begin{minipage}[t]{0.46\textwidth}
            \includegraphics[width=\linewidth]{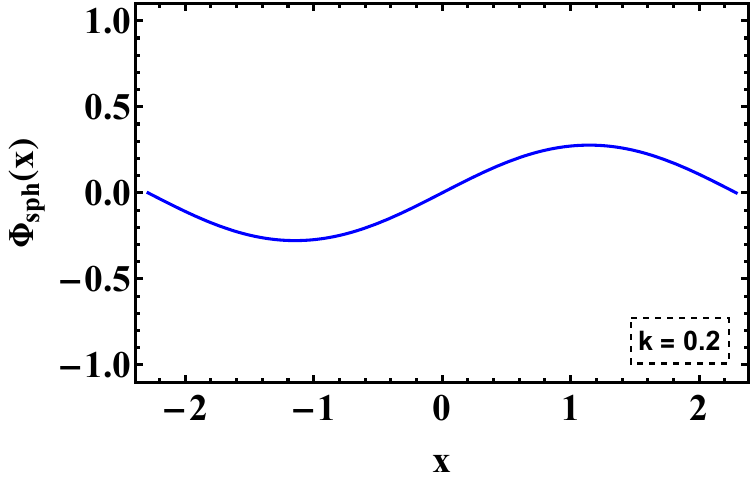}
        \end{minipage}%
        \hspace{0.8cm}
        \begin{minipage}[t]{0.46\textwidth}
            \includegraphics[width=\linewidth]{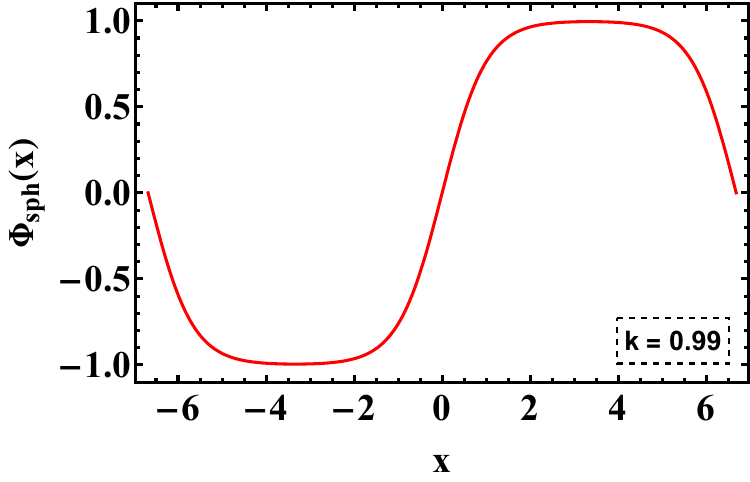}
        \end{minipage}
    \caption{Profiles of the sphaleron $\Phi_{sph}(x,k)$ given by (\ref{e:spha_phi4_circle}), plotted for representative small and large values of the parameter $k$.}
    \label{f:Phi4_Sph_Cirlce_Profiles}
    \end{figure}

    As a particular example of model defined on the circle, let us consider the $\phi^4$ theory with $U(\phi) = \dfrac{1}{2}(1 - \phi^2)^2$. Here, the first order differential equation reads
    \begin{equation}\label{e:Bogo_phi4_circle}
        \phi_x^2 = (1 - \phi^2)^2 - C^2\,.
    \end{equation}
    It can be verified that the non-trivial sphaleron solution of (\ref{e:Bogo_phi4_circle}) reads
    \begin{equation}\label{e:spha_C_phi4_circle}
        \Phi_{sph}(x;C) = \pm \sqrt{1 - C}\, \text{sn}\left( \sqrt{1 + C}\,x, \dfrac{1 - C}{1 + C}\right), \quad C \in \lbrack 0,1\rbrack\,,
    \end{equation}    
    where $\text{sn}(x,m)$ denotes the Jacobi sine function of argument $x$ and parameter $m$. Assuming the parametrisation
    \begin{equation}
        C = \dfrac{1 - k^2}{1 + k^2}\,,
    \end{equation}
    the sphaleron solution is simplified to \cite{Manton1:1988}
    \begin{equation}\label{e:spha_phi4_circle}
        \Phi_{sph}(x;k) = \pm k\,a(k)\, \text{sn}( a(k)x, k^2)\,,\quad a(k)=\sqrt{\frac{2}{1+k^2}}\,, \quad k\in \lbrack 0,1\rbrack\,.
    \end{equation} 
    We illustrate representative profiles of these sphalerons in \autoref{f:Phi4_Sph_Cirlce_Profiles}.
    
    The Jacobi sine function is a periodic function of period $4\mathrm{K}(m)$, with $\mathrm{K}(m)$ the complete elliptic integral of the first kind. Hence, we deduce the following relation between the circumference of the circle on which the sphaleron resides and the parameter $k$ of the model 
    \begin{equation}\label{e:L_k_phi4}
        L = \dfrac{4 \mathrm{K}(k^2)}{a(k)}\,.
    \end{equation}
    As aforementioned, the non-trivial sphaleron cannot arise for all the values of the circumference $L$, but there is a smallest circle where the sphaleron holds. This regime appears for the maximum value of $C$, in other words, when $k = 0$. There, the non-trivial sphaleron reduces to $\phi^{(m)} = 0$, and the smallest circle has a circumference $L_{min} = L|_{k = 0} = \sqrt{2} \pi$. On the other hand, the circumference of the circle becomes infinite when $k \rightarrow 1$, and $\Phi_{sph}(x;k)$ resembles an infinitely separated kink-antikink pair. Indeed, the energy of the sphaleron $\Phi_{sph}(x;k)$, that reads as
    \begin{equation}
        E[\Phi_{sph}] =    \dfrac{\sqrt{2}}{3(1+k^2)^{3/2}}\left(8(1+k^2)\mathrm{E}(k^2) - (1 - k^2)(5 + 3k^2)\mathrm{K}(k^2) \right)\,, 
    \end{equation}
    where $\mathrm{E}(m)$ is the complete elliptic integral of the second kind. Taking the asymptotic limit when $L \gg L_{min}$, we obtain 
    \begin{equation}
        E[\Phi_{sph}] = \dfrac{8}{3} - 32\, e^{-L}\,, 
    \end{equation}
    which implies that the energy of the single sphaleron approaches exponentially the energy of the kink-antikink pair. The exponential term in the asymptotic limit accounts for the interaction energy of the kink-antikink pair. It makes evident that the separation between the constituent subkinks is $L/2$ \cite{Manton1:1988}.

    By introducing the perturbation of the non-trivial sphaleron (\ref{e:spha_phi4_circle}) by a normal mode of the form $\phi(x,t) = \Phi_{sph}(x;k) + \eta(x;k)\,e^{i \omega t}$ into the field equation, and expanding up to first order in the perturbation, we are left with the following spectral problem
    \begin{equation}
        - a^2(k)\, \eta''(z) - 2\left[1 - 3\, a^2(k)\,k^2 sn(z, k^2)^2\right]\eta(z) = \omega^2 \eta(z)\,,
    \end{equation}
    with $z = a(k)\,x$. Remarkably, this expression is a Lamé-like differential equation. These equations have the general form
    \begin{equation}\label{e:Lame_general}
        - \eta''(x) - \left[ \lambda - N(N + 1) k^2 sn(x, k)^2\right]\eta(x) = \omega^2 \eta(x)\,.
    \end{equation}
    The relevance of this equation lies in the fact that, when $N$ is a positive integer, the first $2N + 1$ eigenfunctions are polynomials; the so-called \textit{Lamé polynomials}. In our case, $N = 2$, what guarantees us the existence of five explicit expressions. We identify these eigenfunctions as the zero mode $\eta_0$ responsible for the rigid translations of the sphaleron, a negative mode $\eta_{-}$ related to the instability of the solution, and the first three positive modes that we denote by $\eta_1,\eta_2,\eta_+$. 
    
\begin{figure}[htb]
        \centering{\includegraphics[width=0.55\linewidth]{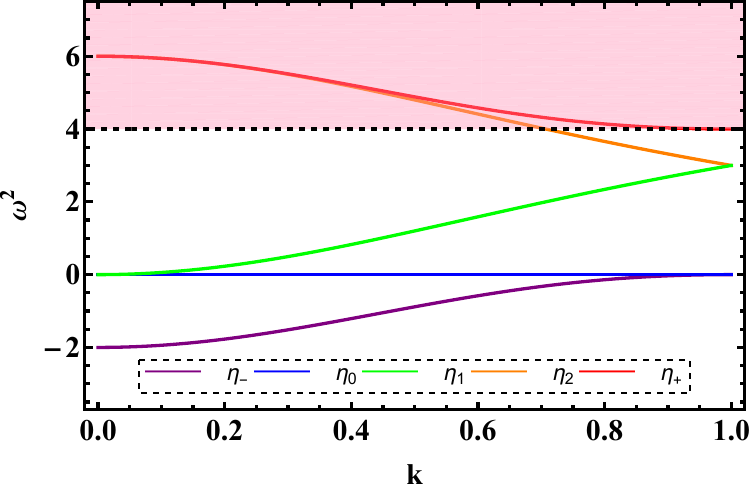}
        }
        \caption{Linear spectrum of perturbations around the sphaleron in the $\phi^4$ model on the circle. We represent the unstable mode $\eta_-$, the zero mode $\eta_0$, and three positive bound modes $\eta_1,\eta_2,\eta_+$. The modes that can appear in the shaded area are above the mass threshold of the theory.}
        \label{f:spectrum_sph4_circle}
    \end{figure}

    The corresponding expressions and their eigenfrequencies are listed below \cite{Liang:1992}
    \begin{align}\label{eq:internal_phi4}
            \eta_{\mp} &= \text{sn}^2(z,k^2) - \dfrac{1}{3k^2}\left(1 + k^2 \pm \sqrt{1 - k^2(1 - k^2)} \right)\,, \quad \omega_{\mp}^2 = 2\left(1 \mp 2\dfrac{\sqrt{1 - k^2(1 - k^2)}}{1 + k^2} \right)\, ,\\
            \eta_0 &= \text{cn}(z,k^2)\text{dn}(z,k^2)\,, \quad \omega_0^2 = 0\,,\\
            \eta_1 &= \text{sn}(z,k^2)\text{cn}(z,k^2)\,, \quad \omega_1^2 = \dfrac{6}{1 + k^2}\,,\\
            \eta_2 &= \text{sn}(z,k^2)\text{dn}(z,k^2)\,, \quad \omega_2^2 = \dfrac{6k^2}{1 + k^2}\,,
    \end{align}
    where the functions $\text{cn}(x)$ and $\text{dn}(x)$ are the Jacobi elliptic cosine and the Jacobi delta amplitude respectively. A representation of the spectral flow of these modes is shown in \autoref{f:spectrum_sph4_circle}. 
    It is important to remark that, for $k$ close to $1$ (or equivalently large $L$), the modes of the sphaleron can be interpreted in terms of the individual modes of a kink-antikink pair. Here, the zero mode and the unstable mode can be interpreted respectively as an antisymmetric or a symmetric combination of the zero modes of the individual subkinks. On the other hand, the positive modes $\eta_1$ and $\eta_2$ are symmetric and antisymmetric combinations of the shape mode of the individual kinks respectively. Finally, $\eta_+$ is above the mass threshold in the kink-antikink picture, therefore this is a mode genuinely related to the periodicity of $\mathbb{S}^1$. 

\section{Further comments}\label{s:FurtherComments_Sph}
    
    Despite the importance of the electroweak sphaleron, its understanding is far from being complete. Among the recent developments, we find for example that the behaviour of the electroweak sphaleron in presence of a strong external magnetic field has been considered \cite{Rajantie:2020,Annala:2023}. There, it has been verified that, for a critical value of the magnetic field, the energy barrier vanishes, what would result in an enhanced rate of $B + L$ violation. Additionally, the sphaleron transition rate in the Standard Model has been analysed as a function of temperature \cite{Onofrio:2012,Onofrio:2014}. Moreover, a generalisation of the electroweak sphaleron admitting an electric field and supporting a non-zero angular momentum has been found \cite{Radu:2009,Kleihaus:2009,Ibadov:2010}. These solutions could enlarge the number of transition channels to sectors with non-zero angular momentum.

    However, the complexity of the theory and the inherent instability of the solution means that its dynamical properties are not yet fully explored. Specifically, the influence of the internal structure of this solution on its dynamics has not been analysed in detail. The aim of the following chapters is to shed more light into this question. For simplicity reasons, we will work with one-dimensional models. In particular, we will study novel false vacua models and theories defined on the circle. The reason is that these models are the simplest examples where sphalerons can arise. Moreover, it has been revealed that the one-dimensional theories defined on the circle share some similarities with the $SU(2)$ Yang-Mills-Higgs theory \cite{Brihayev:1990}. With these studies, we intend to extract valuable information that can be extrapolated to higher-dimensional cases of physical interest.

        \chapter{Semi-BPS sphalerons}\label{c:BPS_Sph}

This chapter is adapted from \cite{Queiruga:2023}:
    \hspace{-2.5cm}
    \vspace{-0.2cm}
    \begin{figure}[H]
        \hspace{-2.5cm}
        \vspace{-0.2cm}
        \centering{\includegraphics[width=1.1\linewidth]{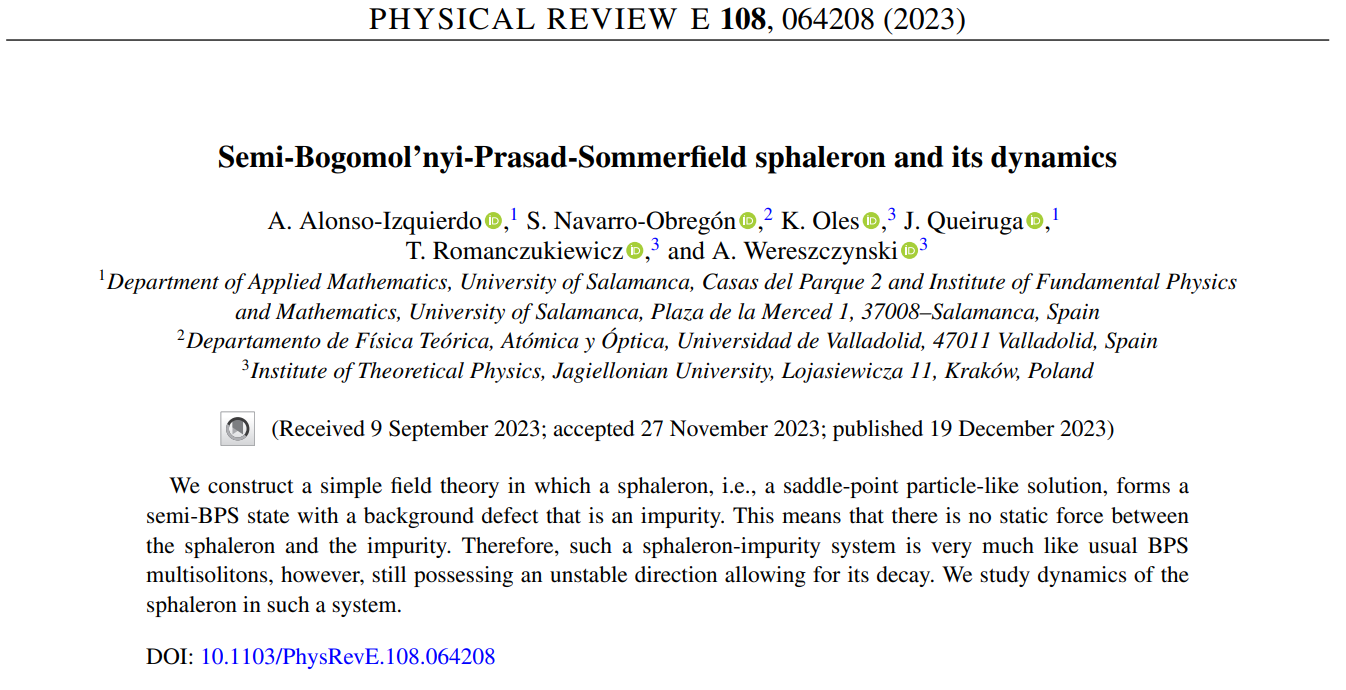}
        }
        \label{f:Thumbail_BPS}
        \hspace{-2.5cm}
    \end{figure}

\section{Introduction}

    Although the analysis of sphaleron dynamics in $1+1$ dimensions is simpler than in higher dimensions, its dynamical properties are still not fully understood. This complexity arises from the multiple interaction channels. A sphaleron, similarly to stable solitons, can be either attracted to or repelled by other sphalerons, solitons, or localised excitations of matter fields, implying the presence of static forces. Moreover, a sphaleron may possess internal degrees of freedom that can be excited during scattering processes and temporarily store part of the available energy. In the case of topological solitons, we have already discussed how this leads to the emergence of fractal structures in the final-state formation \cite{Campbell:1983, Manton:2021}. In fact, it has recently been shown that sphalerons can provide internal degrees of freedom that trigger resonant energy transfer in kink–antikink collisions \cite{Romanczukiewicz:2021}. Finally, the interaction of a sphaleron with radiation can also alter its dynamics. All of these effects must be considered along with its unstable nature.

    A more tractable situation occurs when the model under study is BPS. As we have discussed previously, in this framework a soliton does not interact statically with other localised objects, allowing them to be positioned at arbitrary distances from each other without affecting the total energy. In order to have multi-soliton BPS solutions in $1+1$ dimensions, it is usually necessary to introduce at least one additional scalar field with very specific couplings \cite{Ferreira:2019, Izquierdo:2019}. Naturally, the inclusion of a second dynamical field increases the complexity of the analysis, both analytically and numerically. Nevertheless, it has been shown recently that the scalar field can be coupled to a non-dynamical background field in such a way that the BPS property is preserved \cite{Adam1:2019,Romanczukiewicz:2019}. This results in a simpler BPS-type theory that involves only a single dynamical field.     A physically relevant example of a non-dynamical background field is an impurity. 

    In this chapter, we aim to propose a framework in which dynamics of a sphaleron does enjoy a sort of BPS limit. We will refer to such objects as \textit{semi-BPS sphalerons}. With that denomination, we intend to underline the similarities as well as differences from the usual BPS stable solitons. The reason behind these differences is that, although the BPS construction guarantees that there is no static force between such a sphaleron and other localised object (e.g. an impurity or other solitons) or that the solution possesses a zero mode, a semi-BPS sphaleron does not saturate any topological bound. This is related to the presence of a negative mode, which can lower the energy of this semi-BPS solution by decaying along the unstable direction. Due to that, even in its BPS-like limit, the sphaleron reveals a much more involved pattern of interactions than its stable BPS counterparts.

    Specifically, our aim is to introduce the simplest family of theories in which a sphaleron remains to be a BPS-like state. In order to do that, we include a background non-dynamical defect, i.e., an impurity. The importance of these defects is that, although providing a non-trivial background, they do not introduce new kinetic degrees of freedom. This feature, along with the BPS-like property, may simplify the complexity of the sphaleron dynamics and allow for new insight. In fact, a related problem is how properties of sphalerons are affected by external conditions. This includes, for example, the appearance of external magnetic and electric fields, whose impact on the sphaleron properties is of great physical importance \cite{Rajantie:2020}. 

    Finally, we would like to remark that the notion of a semi-BPS solution has recently appeared in the literature \cite{Manton:2023}. In that context, the term "semi-BPS" refers to solutions of the pertinent Bogomolny equation defined on a double cover of the complex $\phi$-plane. This construction arises due to the presence of a branch point when taking the square root of the potential function, $\sqrt{U}$. Such a branch point occurs when $U$ has a zero of odd multiplicity. As shown in (\ref{e:approx_cero_false}), a simple zero is required in order to obtain a sphaleron configuration. Therefore, our notion of semi-BPS is fully consistent with the interpretation given in \cite{Manton:2023}. 

    This chapter is organised as follows. In \autoref{s:Non_chiral}, we construct a scalar field theory with an impurity that allows for a BPS system between an impurity and the sphaleron of the free-impurity model. The internal structure of this solution is analysed in \autoref{s:Modes_Non_chiral}. We then investigate the dynamical properties of the sphaleron, taking into account its internal structure, in \autoref{s:Geodesic_Non_chiral} and \autoref{s:SW_Non_chiral}. Subsequently, we discuss the possible decay processes when the sphaleron is either located on the impurity or collides with it, in \autoref{s:Decay_Impurity} and \autoref{s:Collision_Impurity}, respectively. Finally, we present our conclusions in \autoref{s:Summary_Semi_BPS}.

\section{Non-chiral BPS-impurity models}\label{s:Non_chiral}

    The literature contains a large number of works on soliton dynamics in impurity-scalar field models; see, for example,  \cite{Gredeskul:1992, Malomed:1992, Piette:2007, Ashcroft:2020}. These models exhibit a variety of interesting phenomena, including the emergence of chaotic behaviour in kink–impurity collisions \cite{Fei:1992,Fei2:1992}. However, it is important to note that in all such models, the impurity is coupled in a non-BPS manner, typically through a modification of the potential term. Therefore, such systems exhibit static attractive or repulsive forces between the soliton and the impurity.

    A BPS-impurity model can be achieved if the original field theory is deformed as \cite{Adam1:2019} 
    \begin{equation}
        \mathcal{L} = \frac{1}{2}\phi_t^2 -\frac{1}{2}\left( \phi_x - \sigma(x) W(\phi) \right)^2\,.
    \end{equation}
    From the associated energy functional it is easy to verify that the Bogomolny equations in this model reads \cite{Adam1:2019, Oles:2020}
    \begin{equation}
        \phi_x = \sigma(x) W(\phi)\,, 
    \end{equation}
    where $W$ is a well-behaved function of $\phi$ and $W^2=2U$. Note that for $\sigma(x) = 1$ we recover the usual scalar field theory up to a boundary term. The existence of a first-order Bogomolny equation implies the presence of a BPS sector. Unlike the no-impurity case, such a BPS sector is non-trivial in this framework. For example, the linear mode structure of the soliton-impurity BPS solution depends on the distance between the soliton and impurity. This renders the BPS-impurity model an ideal, simplified laboratory for studying the role of the internal modes in kink dynamics in the limit of vanishing static force. Furthermore, the BPS-impurity framework provides a scenario in which one-loop corrections to kink–antikink processes can be computed; see \cite{Takyi:2023}.
  
    Importantly, there is only a single Bogomolny equation in the previous deformed model. As a result, the corresponding soliton can be interpreted as a chiral kink. However, this theory only allows for kink-impurity or even BPS kink-antikink-impurity solutions. As discussed in \autoref{ss:Sph_FalseVacuum}, the existence of a sphaleron-like solution requires two Bogomolny equations. Therefore, the BPS-impurity Lagrangian must be modified in such a way that two Bogomolny equations emerge. This enables the construction of a model that admits a BPS sphaleron–impurity solution. 
    
    For that reason let us consider the following background field deformed Lagrangian
    \begin{equation}\label{e:Lag_BPS-imp}
        \mathcal{L} = \frac{1}{2}\phi_t^2 -\frac{1}{2\sigma}\phi_x^2 - \sigma U(\phi)\,, 
    \end{equation}
    Here again, $\sigma$ denotes a prescribed impurity. However, in this case, the background field also modifies the gradient term. In fact, such a modification can be interpreted as arising from the non-trivial geometry of the medium, e.g., curved Josephson junction \cite{Dobrowolski:2002, Gatlik:2023}. To derive the Bogomolny equations, we analyse the static energy functional associated to the Lagrangian density (\ref{e:Lag_BPS-imp}) and perform a Bogomolny arrangement
    \begin{eqnarray}
        E&=&\int_{-\infty}^{\infty} \left( 
        \frac{1}{2\sigma}\phi_x^2 + \sigma U(\phi) \right)dx = \int_{-\infty}^{\infty} \left(\frac{1}{\sqrt{2\sigma}}\phi_x \mp \sqrt{\sigma U} \right)^2dx \pm  \int_{-\infty}^{\infty}  \sqrt{2U} \,\phi_x\, dx \nonumber \\
        &\geq&  \left| \int_{\phi(-\infty)}^{\phi(\infty)}  \sqrt{2U} \,d\phi \right| = |Q|\int_{\phi_-}^{\phi_+}  \sqrt{2U}\, d\phi \,. 
    \end{eqnarray}
    The inequality is saturated if and only if
    \begin{equation}\label{e:BPS-imp}
        \frac{1}{\sqrt{2\sigma}}\phi_x = \pm \sqrt{\sigma U}\,.  
    \end{equation}
    This computation proves that such a simple impurity model has associated two Bogomolny equations and therefore does not treat kink or antikink differently. As a result, the model consist in a non-chiral BPS-impurity theory. The Bogomolny equation implies that the impurity function must be non-negative. In all other aspects, it is not subject to any fine-tuned restrictions. However, we underline that the impurity deforms the original impurity-free field equations in the region where $\sigma(x) \neq 1$. Typically, in the literature, the impurity is assumed to be a localised deformation that vanishes sufficiently rapidly at spatial infinity. From this perspective, one may introduce an impurity function $\Sigma$ such that $\sigma = 1 + \Sigma$, where $\Sigma$ represents a deviation from the impurity-free theory. To maintain consistency with previous works on the BPS impurity, we prefer to use the formulation based on the $\sigma$ function. 

    The impurity-dependent Bogomolny equations \eqref{e:BPS-imp} can be recast into their impurity-free form by introducing a new base space coordinate $y$
    \begin{equation}
        \frac{dy}{dx} = |\sigma(x)| \; \Rightarrow \; y= \int |\sigma(x)|\, dx + a\,,
    \end{equation}
    where $a$ is an integration constant. This change of variable leads us to the usual impurity-free Bogomolny equations
    \begin{equation}
        \phi_y = \pm \sqrt{2 U}\,. 
    \end{equation}
    Solving these equations and expressing the coordinate $y$ in terms of the original coordinate $x$ leads to a family of BPS sphaleron–impurity solutions. Notably, this family is parametrised by a modulus $a$, which reflects the BPS nature of the solution.
    
\begin{figure}[htb]
        \centering
        \includegraphics[width=0.9\linewidth]{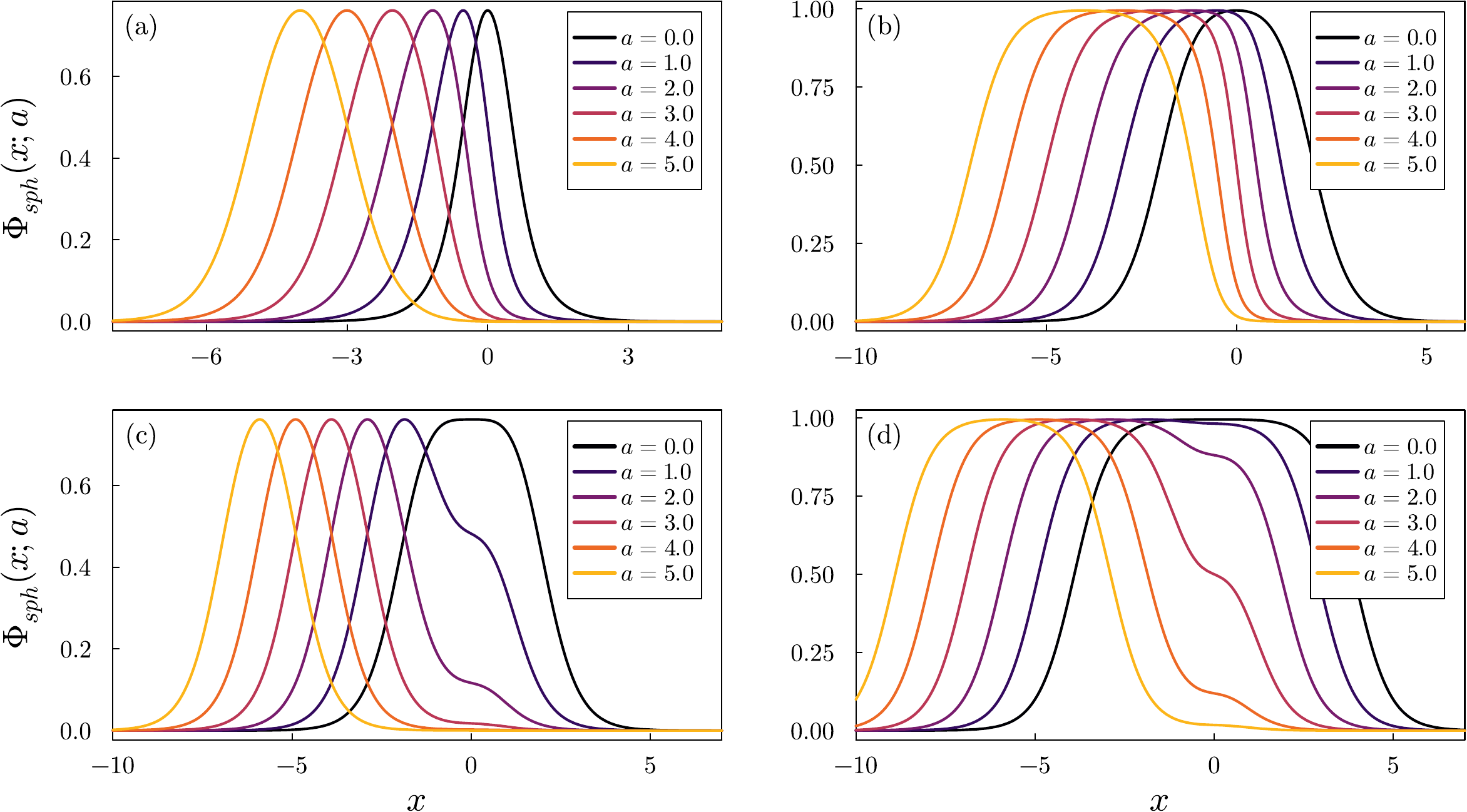}
        \caption{The BPS sphaleron-impurity solution (\ref{e:sph-imp}) for different values of the modulus $a$. Upper panels: $\alpha=1$, lower panels $\alpha=-0.9$. Left panels: $s=1$, right panels: $s=3$.} 
        \label{f:sph-imp-plot}
     \end{figure}

    As mentioned earlier in \autoref{ss:Sph_FalseVacuum}, a potential with a false vacuum structure is required in order to obtain a sphaleron configuration that satisfies the BPS equations piece-wisely. For concreteness, let us assume the potential function (\ref{e:pot_falseSph}) and the following form for the background field
    \begin{equation}
        \sigma(x) = 1 + \alpha\sech^2 x\,,
    \end{equation}
    which is an exponentially-like localised impurity centred at the origin. Here, the constant $\alpha$ is a free parameter measuring the strength of the impurity that satisfied $\alpha > -1$. For $\alpha=0$ the impurity vanishes and we recover the usual scalar field theory. With this choice of impurity, the sphaleron profile given in 
    (\ref{e:sph_false}) is modified into the following sphaleron-impurity solution
    \begin{eqnarray}\label{e:sph-imp}
        \Phi_{sph}(x; a) = \frac{1}{2} \left( \tanh(x + \alpha\tanh x + a + s) - \tanh(x + \alpha \tanh x + a - s)\right)\,, 
    \end{eqnarray}
    where the modulus $a$ can be interpreted as the distance between the sphaleron --identified with the position of the field maximum-- and the impurity, which is located at the origin. When $|a| \gg 1$, the sphaleron is located far from the impurity, and the solution corresponds to the original sphaleron. On the other hand, as $a \to 0$, the sphaleron approaches the impurity and undergoes a deformation. Moreover, for positive values of $\alpha$, the original sphaleron becomes increasingly squeezed as it nears the impurity, whereas for $\alpha \in (-1, 0)$, it becomes broader. This behaviour is illustrated in \autoref{f:sph-imp-plot}.
    The sphaleron solutions given by (\ref{e:sph-imp}) have the same energy independently of the value of $a$  
    \begin{equation}\label{e:Energy_Sph_imp}
        E[\Phi_{sph}(x;a)] = \frac{9 \sinh 2s + \sinh 6s - 24 s \cosh 2s}{6 \sinh^3 2s}\,. 
    \end{equation}
    Thus, there is no static force between the sphaleron and impurity, and form a semi-BPS-type family of sphalerons in the sense explained before. As $s \to \infty$, the static energy of the sphaleron tends to the energy of a well-separated $\phi^4$ kink-antikink pair $E_{KAK} = 2/3$. Recall that the potential $U$ given by (\ref{e:pot_falseSph}) employs a different convention than the standard $\phi^4$ potential (\ref{e:potential_phi4}), where $E_{KAK} = 4/3$.
    
\section{The mode structure}\label{s:Modes_Non_chiral}

    It is important to note that although the energy of the sphaleron solutions remains degenerate under variations of the modulus $a$, this degeneracy does not extend to the linear mode structure. This is due to the fact that the impurity does not correspond to a uniform vacuum and, as shown in \autoref{f:sph-imp-plot}, it non-trivially affects the shape of the sphaleron. To illustrate this, we consider the static BPS solution perturbed by a small fluctuation $\eta(x,t)$.
    \begin{equation}
    \phi(x,t) = \Phi_{sph}(x;a) + \eta(x,t)\,,
    \end{equation}
    and insert it into the full time-dependent field equation 
    \begin{equation}
        \phi_{tt}-\frac{1}{\sigma}\phi_{xx}+\frac{\sigma_x}{\sigma^2} \phi_x + \sigma\, U_\phi = 0\,.
    \end{equation}
    At linear order we get
    \begin{eqnarray}
    \left[- \frac{d}{dx} \left( \frac{1}{\sigma} \frac{d}{dx}  \right) + \sigma\, U''(\Phi_{sph}(x;a))\right] \eta(x; a) = \omega^2(a)\eta (x;a)\,,
    \end{eqnarray}
    where we have assumed that $\eta(x,t;a) = \eta(x;a)e^{i\omega(a) t}$. Here, $\omega(a)$ is the frequency of the pertinent mode $\eta(x;a)$.  

    Let us now examine this problem in greater detail. Asymptotically, for $|a| \to \infty$, the BPS sphaleron $\Phi_{sph}(x; a)$ corresponds to a configuration where the sphaleron in the impurity-free model is infinitely separated from the impurity. In this limit, the mode structure is simply a superposition of the modes associated with the impurity-free sphaleron and the modes introduced by the impurity itself. The former are localised around the sphaleron, while the latter are localised at the impurity.

    In our example, for not too small values of the parameter $s$, the impurity-free sphaleron can be interpreted as a well-separated kink–antikink pair. Consequently, it possesses two positive discrete modes, $\eta_{1,2}(x;a)$, originating from the shape modes of each constituent soliton. They can be approximately identified as the symmetric and antisymmetric combinations of the shape mode of the individual kinks. As $s$ increases, their frequencies approach $\omega^2_{1,2} = 3$. In addition, there is a translational zero mode, $\eta_0(x;a)$. Finally, there exists a negative mode $\eta_{-}(x;a)$, whose frequency tends to zero as $s \to \infty$. This indicates that, in the large $s$ limit, the sphaleron becomes quasi-stable.

\begin{figure}[htb]
        \centering
        \includegraphics[width=0.59\linewidth]{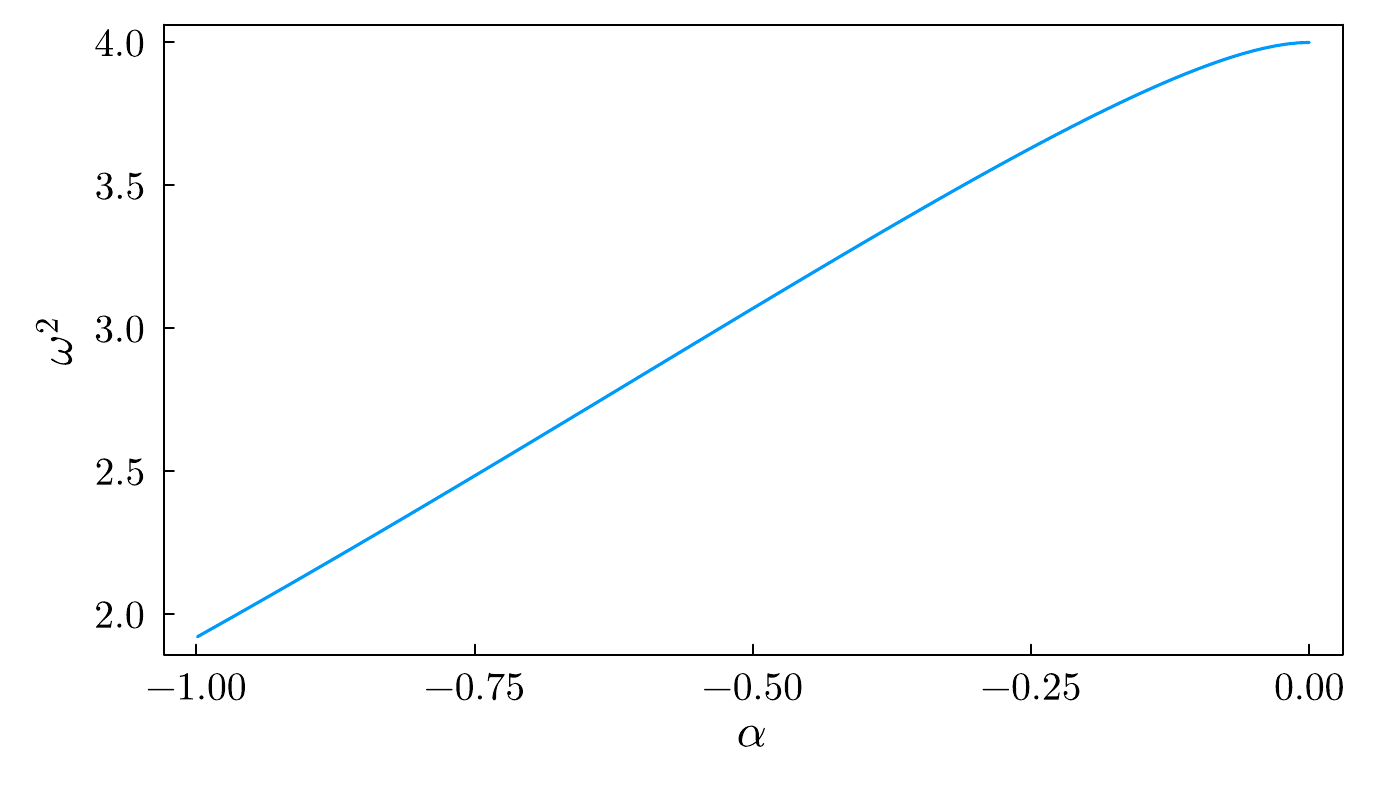}
        \caption{The frequency of the bound mode of the impurity.}
        \label{f:imp_flow}
    \end{figure}

    Regarding the impurity, it may introduce additional localised modes. This can be revealed by performing a linear perturbation analysis around the trivial solution $\phi = 0$. In our case, there exists only one such mode, which is localised at the origin and appears for $\alpha < 0$; see \autoref{f:imp_flow}. 
    Notably, this plot is independent of the model parameter $s$, what stems from the fact that $U''(\phi = 0) = 4$ is constant. As we will discuss later, this impurity-induced mode plays an active role in the dynamics of the sphaleron.

\begin{figure}[htb]
        \centering
        \includegraphics[width=0.6\linewidth]{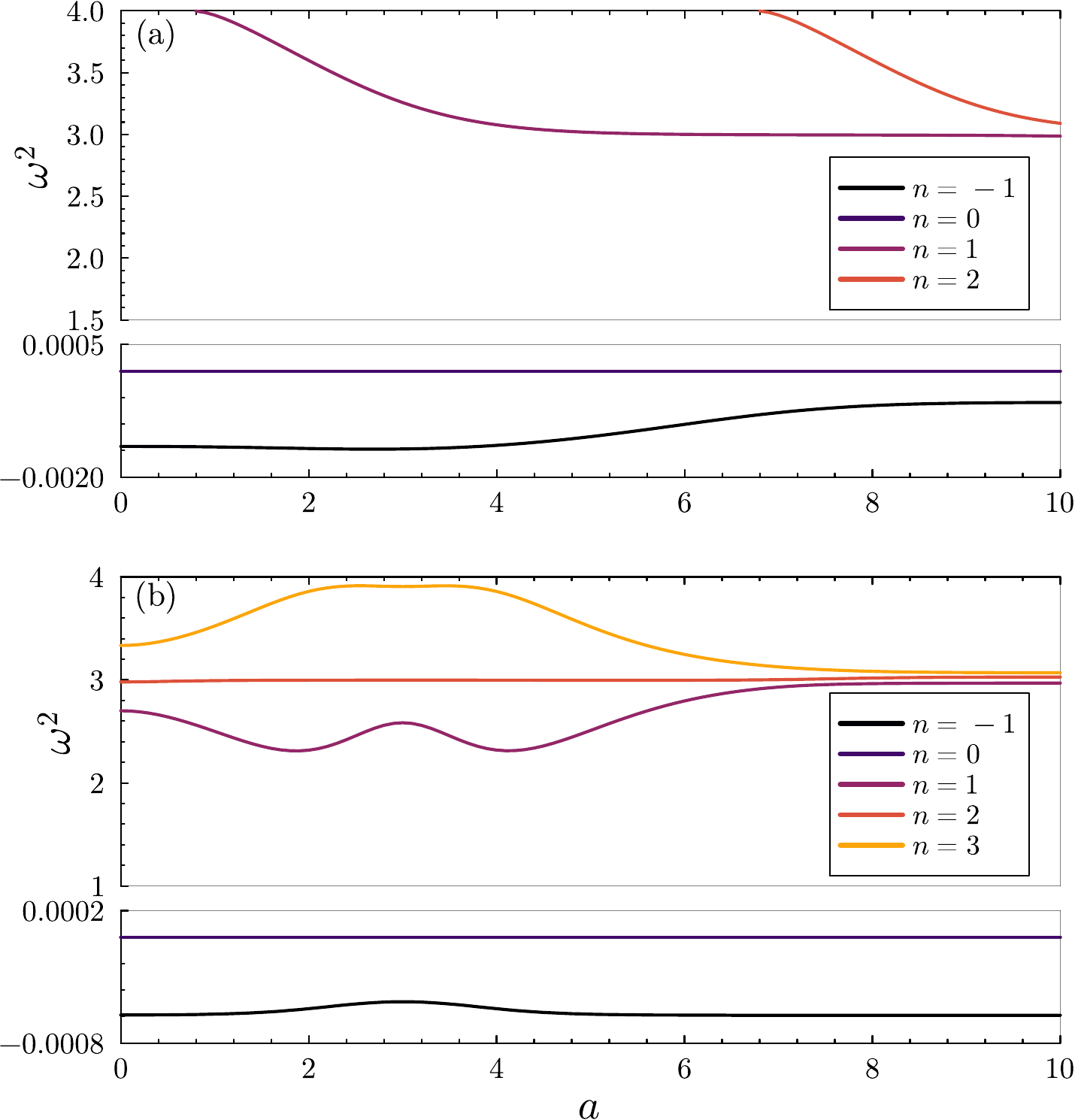}
        \caption{The dependence of the mode structure of the sphaleron-impurity solution on the modulus $a$. Upper panel: $s=3$, $\alpha=3$. Lower  panel: $s=3$, $\alpha=-0.5$. }
        \label{f:spectral_flow}
     \end{figure}

    As the sphaleron approaches the impurity, it becomes increasingly deformed, and its spectral structure undergoes significant changes; see \autoref{f:spectral_flow}. 
    Due to the BPS nature of the solution, a zero mode persists, even though translational symmetry is explicitly broken
    \begin{equation}
        \eta_0(x;a) \propto \dfrac{\partial\Phi_{sph}(x;a)}{\partial a}\,.
    \end{equation}
    Importantly, the zero mode has one node, which implies that the solution is a genuine sphaleron with one unstable mode by virtue of \autoref{t:Sturm}. The impurity can influence significantly the unstable mode. Specifically, as $a \to 0$, the frequency of the unstable mode increases for negative impurity strength ($\alpha < 0$), and decreases for positive impurity strength ($\alpha > 0$). Consequently, as the sphaleron approaches the origin, it becomes more or less stable depending on the nature of the impurity. This behaviour reveals an interesting possibility: by appropriately choosing the impurity, a genuinely unstable, rapidly decaying sphaleron could be transformed into a quasi-stable configuration, in which the negative mode frequency would approach to zero. This stabilisation would occur at a specific distance between the sphaleron and the impurity, as a consequence of their mutual interaction. Finally, the spectrum of the massive bound modes is also affected by the impurity. For sufficiently large positive impurity strength, the bound modes of the impurity-free sphaleron may cross the mass threshold at a certain critical value of the modulus $a$. This phenomenon is known to induce the so-called \textit{spectral wall}, which has a significant impact on kink dynamics \cite{Adam:2019}. 

    In the following sections, we shall analyse the dynamics of the semi-BPS sphalerons, attending to their internal structure. We will show that some of the phenomena observed in the soliton dynamics are also observed in the context of sphalerons. 

\section{Geodesic motion on an unstable moduli space}\label{s:Geodesic_Non_chiral} 

    Our analysis begins with the lowest-energy dynamics. In the usual BPS limit for topological solitons, BPS solutions saturate the pertinent topological bound. As we have previously seen, this means that in a given topological sector, there are no solutions with lower energy. A natural consequence of that is that the lowest energy dynamics occurs via transitions between energetically equivalent BPS solutions, which reflects an excitation of the zero mode. This transition finds an elegant formulation as a geodesic motion on the corresponding moduli space of the BPS states. In the case of the semi-BPS sphalerons, we do have a family of energetically degenerate solutions given by $\mathcal{M} = \{ \Phi_{sph}(x;a)\}$, which suggests that the concept of geodesic motion remains applicable. However, such a family of sphaleron solutions is not the global energy minimiser in its topological sector. This feature is related to the existence of an unstable direction in the space of field configurations along the negative mode $\eta_{-}(x;a)$. As a result, the family of semi-BPS sphalerons seems to span an unstable moduli space \cite{Manton:1988}.

\begin{figure}[H]
    \centering
    \includegraphics[width=0.6\linewidth]{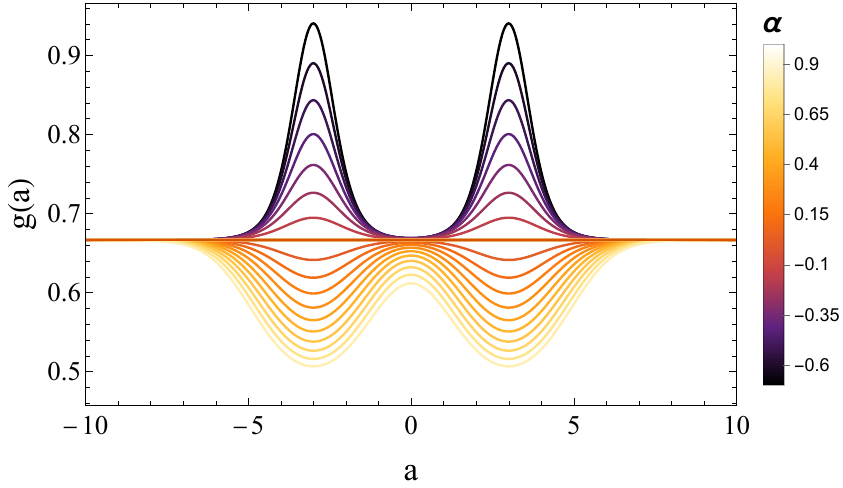} 
    \caption{Unstable moduli space metric for different values of the impurity strengths $\alpha$ and $s = 3$.}
    \label{f:metric_imp}
    \end{figure}

    By promoting the modulus $a$ to a time-dependent collective coordinate $a(t)$ and substituting the BPS sphaleron-impurity solutions (\ref{e:sph-imp}) into the Lagrangian density (\ref{e:Lag_BPS-imp}), we arrive, after integrating over the spatial coordinate, at the following one-dimensional collective coordinate model
    \begin{equation}\label{e:CCM_Sph_Imp}
        L[a]= \frac{1}{2} g(a) \dot{a}^2 - M\,. 
    \end{equation}
    The metric on the unstable moduli space is  
    \begin{equation}
        g(a) = \bigintss_{\mathbbm{R}} \left( \frac{\partial\Phi_{sph}(x;a)}{\partial a} \right)^2 dx\,,
    \end{equation}
    and the effective potential is simply $M$, the mass of the BPS solutions (\ref{e:Energy_Sph_imp}). The corresponding equation of motion describes geodesic flow on the unstable moduli space $\mathcal{M}[a]$. A representation of the unstable moduli space metric $g(a)$ is depicted in \autoref{f:metric_imp} for $s = 3$. Note that this metric is non-trivial, and it only coincides with the mass of the semi-BPS sphaleron in the limit $|a| \gg 1$. 

\begin{figure}[htb]
        \centering
        \includegraphics[width=0.55\linewidth]{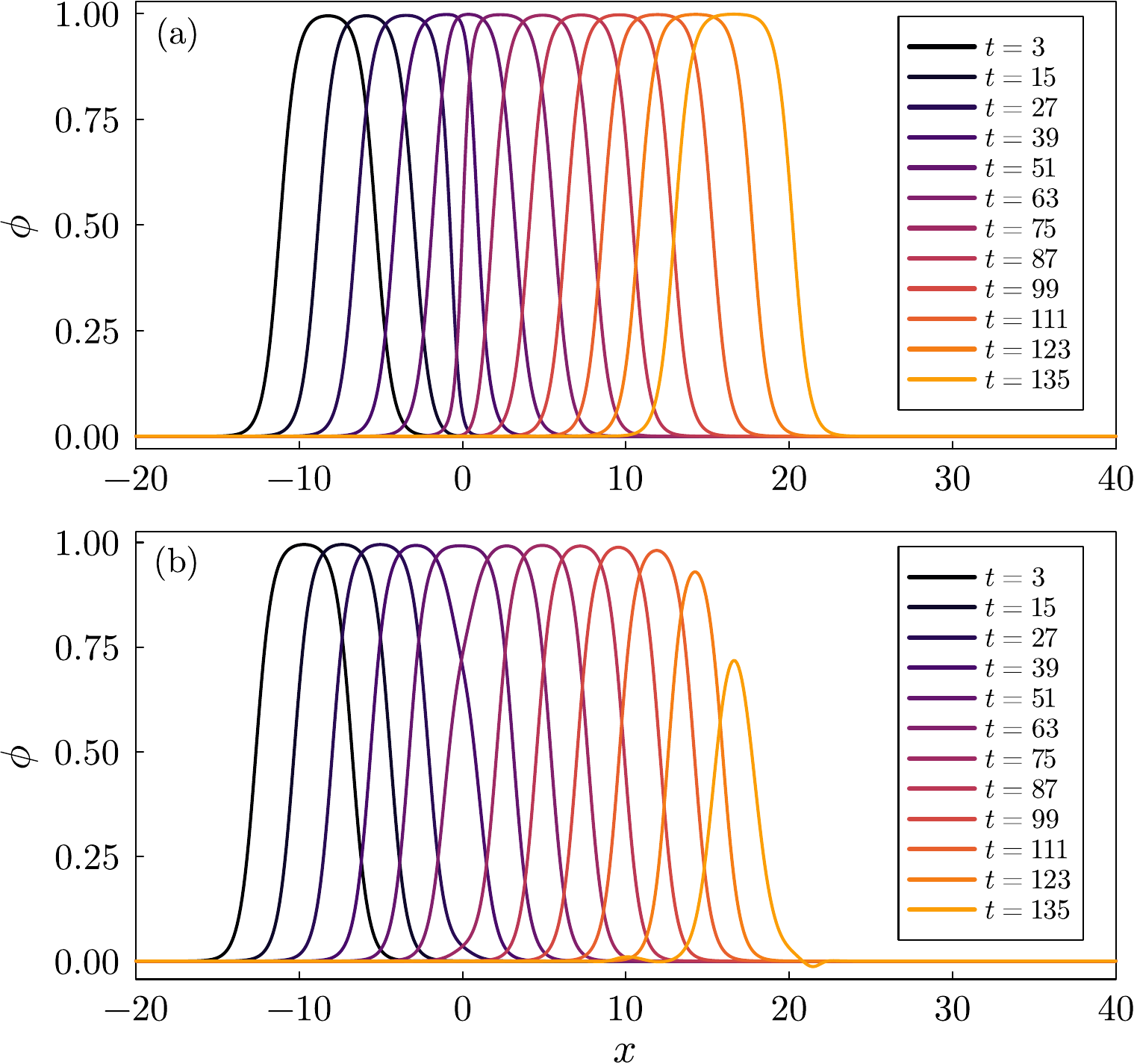}
        \caption{Collision of the sphaleron ($s=3$) with the impurity for repulsive (upper panel: $\alpha=1$) and attractive (lower panel: $\alpha=-0.5$) strength. Here, we have chosen $v_{in}=0.195$.} 
        \label{f:sph-geo-dyn-plot}
     \end{figure}

\begin{figure}[htb]
    \centering
    \includegraphics[width=0.56\linewidth]{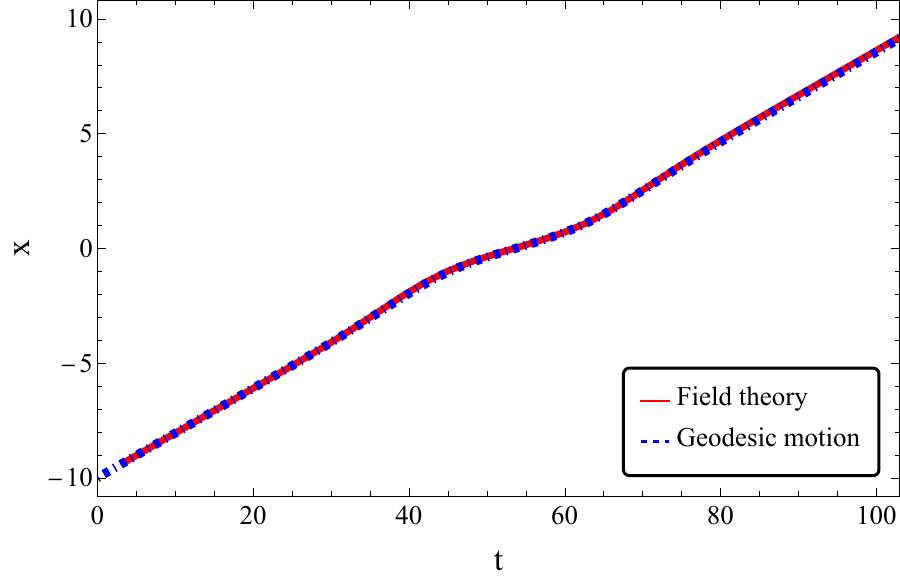} 
    \caption{Comparison between field theory and the moduli space flow for a non-decaying boosted sphaleron. The parameter of the model are $s = 3$ and $\alpha = 2$. The initial velocity is $v = 0.15$.}
    \label{fig:geodesic_flow}
    \end{figure}

    Under certain conditions, the unstable mode may remain unexcited even as the sphaleron traverses the impurity. This behaviour is commonly observed for sufficiently large values of the parameter $s$, where the sphaleron becomes a quasi-stable configuration. In such cases, one can expect that, for a long time, the evolution closely follows the geodesic flow on the unstable moduli space $\mathcal{M}[a]$. This is precisely what we saw in our numerical analysis. Specifically, in \autoref{f:sph-geo-dyn-plot} we present field profiles during the evolution of the sphaleron through the impurity. 
    In the upper panel, corresponding to $s = 3$ and $\alpha = 2$, the sphaleron undergoes a deformation due to the impurity that precisely follows the predictions of the moduli space analysis. The initial velocity is set to $v_{in} = 0.15$. The field configurations at fixed times $t$ match the static BPS solutions for corresponding values of the modulus $a$. Moreover, the full field-theoretic evolution is very accurately reproduced by the geodesic flow, as illustrated in \autoref{fig:geodesic_flow}.

    However, by definition, the sphaleron possesses an unstable mode, which may become excited as the sphaleron traverses the impurity due to higher-order non-linear effects. In the lower panel of \autoref{f:sph-geo-dyn-plot}, we present an example of such a process: after passing through the impurity, the sphaleron becomes destabilised and begins to decay. Once the unstable mode is excited, the geodesic flow on the moduli space $\mathcal{M}[a]$ ceases to provide an accurate description of the dynamics, and the sphaleron falls apart. 

    As a result, we conclude that the geodesic motion derived from the collective coordinate model (\ref{e:CCM_Sph_Imp}) can still provide a reasonably good description of the simplest dynamical behaviour of a semi-BPS sphaleron. However, as expected, it is a less powerful tool when compared to the case of genuinely stable BPS solitons.
    
\section{Spectral wall in sphaleron dynamics} \label{s:SW_Non_chiral} 

    Now we will go beyond the simplest geodesic dynamics and investigate how excitation of massive bound modes changes the interaction between the semi-BPS sphaleron and impurity. In particular, we are interested in the case where a bound mode of the static sphaleron-impurity solution crosses the mass threshold at some $a$. It is known that such a property gives rise to the so-called spectral wall phenomenon. In fact, the spectral wall phenomenon was originally discovered in the context of a BPS antikink-impurity system \cite{Adam:2019}, and it has only recently been observed in various models without impurities \cite{Oles:2021, Oles:2022}.
    
    Speaking precisely, the spectral wall phenomenon represents an obstacle in soliton dynamics, arising from the transition of a bound mode into the continuum spectrum. This transition occurs at a specific value of a moduli space coordinate, $a=a_{sw}$, which can be translated into a corresponding distance $x_{sw}$ between the soliton and its collision partner (either another soliton or an impurity). The effect is highly sensitive to the amount of mode excitation present on the soliton. If the amplitude $A$ of the excited mode is below a critical value $A < A_{crit}$, the soliton passes through $x_{sw}$, although with increasing distortion as $A$ increases. In this case, we say the soliton passes through the spectral wall. Conversely, if $A>A_{crit}$ the soliton is reflected by the spectral wall, with the reflection point occurring earlier as $A$ increases. Finally, for the critical case, $A=A_{crit}$, the soliton becomes trapped at $x=x_{sw}$, forming a long-lived quasi-stationary state. Interestingly, the spectral wall phenomenon exhibits a highly selective character. Specifically, each mode that crosses the mass threshold has its own distinct spectral wall.  
    
\begin{figure}[htb]
        \centering
        \includegraphics[width=0.6\linewidth]{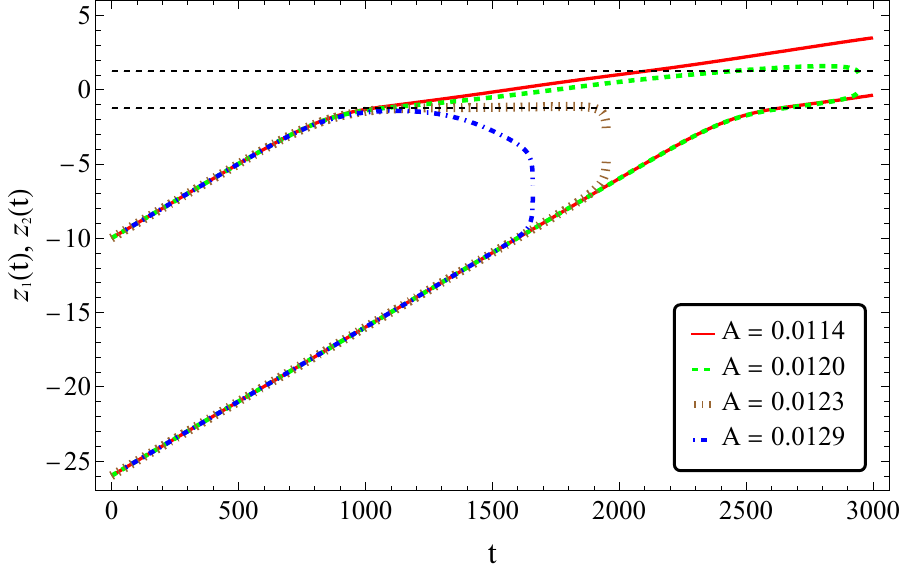} 
        \caption{Collision of the sphaleron with a spectral wall with $s = 8, \alpha = 3$ and $x_0 = 18$. The lines represent the positions $z_1$ and $z_2$ of the composite kink and antikink components of the sphaleron. Different colours correspond to varying amplitudes of the initially excited mode. The dashed horizontal lines indicate the location of the spectral wall. The initial velocity is $v_{in} = 0.01$.}
        \label{f:SW_imp}
    \end{figure}

    To investigate this issue in the context of sphaleron dynamics, we consider as the initial configuration a boosted BPS sphaleron with one of its massive bound modes $\eta(x;a)$ excited. We further assume that, at time $t=0$, the sphaleron is located at a large distance $x_0$ from the impurity. Under this assumption, the sphaleron profile can be well approximated by the free-impurity solution given in (\ref{e:sph_false}). Consequently, the field configuration at $t=0$ is taken to be 
    \begin{eqnarray}
        \phi(x,t) = \Phi_{sph}(\gamma(x - vt + x_0)) + A\, \eta(\gamma(x - vt + x_0)) \cos (\omega\,\gamma (t - vx))\,,
    \end{eqnarray}
    which provides Cauchy data for the numerical analysis. Here, $\gamma$ denotes the Lorentz factor. Naturally, the excited mode must reach the mass threshold for some value of the modulus $a$, that is, at a specific distance between the sphaleron and the origin. This condition is satisfied for sufficiently large positive values of the impurity strength $\alpha$. For simplicity, we focus on the quasi-stable sphaleron regime, corresponding to large values of the parameter $s$. In particular, in our analysis, we choose $s = 8$, $\alpha = 3$ and $x_0 = 18$. Under these conditions, the free sphaleron resembles a kink–antikink molecule in the $\phi^4$ theory, where the constituent solitons are widely separated. Hence, its massive modes are very well approximated by a symmetric and antisymmetric superposition of the shape modes of the $\phi^4$ kinks
    \begin{equation}
    \eta_{1,2} (x) \approx N_{1,2} \left( \eta_{s} (x+s) \pm  \eta_{s} (x-s)  \right)\,,
    \end{equation}
    where
    \begin{equation}
        \eta_{s}(x) = \sinh x \sech^2 x\,,
    \end{equation}
    is the usual shape mode of the $\phi^4$ kink and $N_{1,2}$ are normalisation constants. The initial velocity of the sphaleron is $v_{in}=0.01$. In \autoref{f:SW_imp} we present the dynamics of the sphaleron with the $\eta_2$ mode excited, corresponding to the highest positive bound mode. 
    From the perspective of the composite antikink, the corresponding spectral wall is located at $x_{sw}=1.25$. At this specific distance from the origin, the bound mode reaches the mass threshold. This effect is clearly observed in our numerical simulations. In particular, at a critical value of the mode amplitude, the composite antikink becomes trapped and forms a stationary state precisely at $x=x_{sw}$. 

    A similar pattern is observed for smaller values of the parameter $s$. However, for $s < 4$, the excitation of the shape mode responsible for the spectral wall must be accompanied by the excitation of the unstable mode in order to prevent the collapse of the sphaleron \cite{Manton:2023}. Finally, remark that the spectral wall also acts as another destabilizing factor which contributes to the decay of the sphaleron.

\section{Decay of the semi-BPS sphaleron}\label{s:Decay_Impurity}  

    As evidenced in the analysis above, the inherent instability of the sphaleron implies that its unstable mode can be excited by arbitrarily small perturbations and will eventually dominate the evolution after a sufficiently long time. It is thus worthwhile to study the dynamics that follow the decay.

    In our case, the sphaleron possesses two decay channels: the first is trivial, and describes the decay into the true vacuum. The sphaleron splits into kink and antikink which continually accelerate in opposite directions. The second possibility is much more interesting. It corresponds to the decay into the false vacuum and creation of an oscillon. However, this oscillon is not a BPS-type object. Furthermore, it is not immersed in the trivial vacuum but in a rather complicated background provided by the impurity. Hence, its patterns of interactions are much more involved than in the usual non-impurity model. 

    For simplicity, we will analyse the decay of the sphaleron when it is located at the origin. As a result, the problem maintains the reflection symmetry $\phi(x) \to -\phi(x)$. However, we remark that the decay of the sphaleron is highly dependent on its position with respect to the impurity. We have not explored that dependence here.
    
\begin{figure}[htb]
        \centering
        \includegraphics[width=0.65\linewidth]{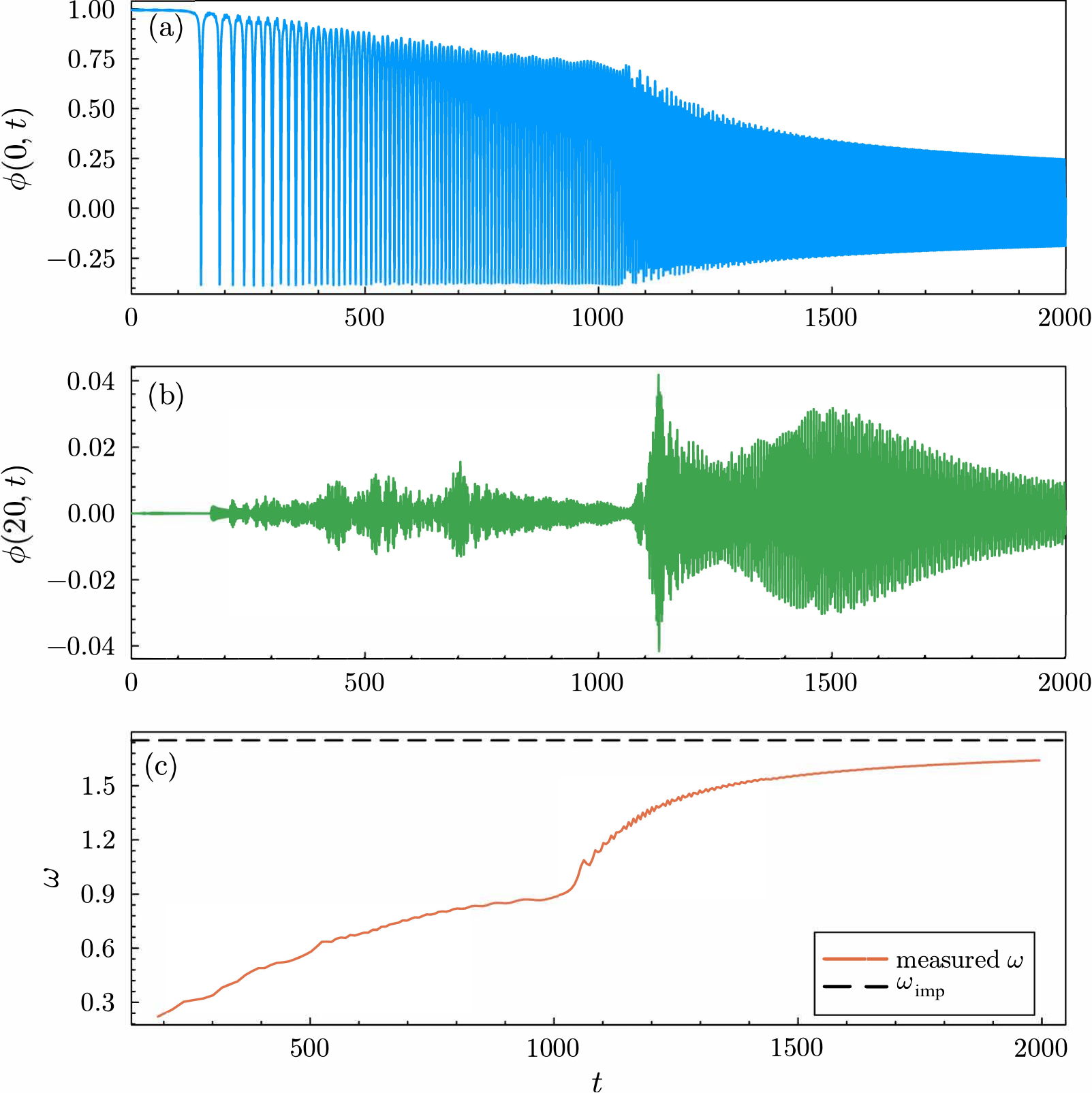}
        \caption{Decay of the sphaleron for $s=3$ and $\alpha=-0.5$. } \label{f:StoO-1}
    \end{figure}

    One interesting scenario occurs when the impurity possesses its own internal mode, that is, when $-1 < \alpha < 0$. In \autoref{f:StoO-1} we present an example of such a behaviour for $s=3$ and $\alpha=-0.5$.
    \vspace{0.15cm}
    The sequence of events is as follows: initially, the sphaleron decays into a large oscillon, which oscillates within the background field generated by the impurity. During the initial stage of the evolution, the sphaleron re-emerges in an excited state at the turning points of the oscillon oscillations. This behaviour indicates the activation of internal modes associated with the sphaleron internal structure. This behaviour is clearly manifested as the small wiggles when the value of the field at the origin, $\phi(0)$, is close to $\phi=1$. For $t < 1050$, the oscillon gradually loses energy through the emission of small-amplitude radiation, resulting in a decrease in amplitude and an increase in oscillation frequency. At approximately $t \approx 1050$, a sudden and pronounced increase in both the amplitude of the radiation and the frequency of the oscillon is observed, as shown in the upper and lower panels. This abrupt transition is attributed to a radiation-burst phenomenon \cite{Dorey:2020,Nagy:2021}. The underlying cause is that the frequency of the oscillon approaches half the mass threshold ($\omega_c^2 = 4$), and therefore the double harmonics may freely propagate and thus facilitate efficient energy emission through radiation. This increase in radiation is clearly visible in the central panel, which displays the field measured at $x = 20$. Following this radiation burst, the system evolves toward a mode associated with the impurity of frequency $\omega_{imp}$, which slowly decays over time due to the continued emission of radiation. In summary, this analysis demonstrates that, in this scenario, the oscillon is effectively trapped by the impurity and transforms into a long-lived bound state. In other words, the existence of a mode associated with the impurity effectively manifests as an attractive interaction between the oscillon and the impurity.
    
\begin{figure*}[htb]
    \centering
    \begin{subfigure}[t]{0.496\textwidth}
        \centering
        \includegraphics[width=\linewidth]{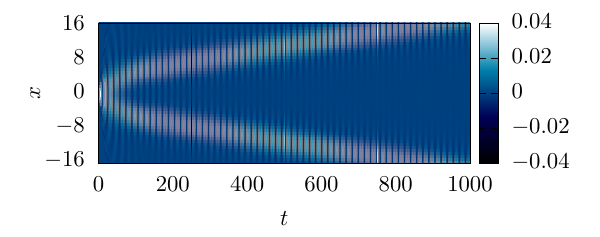}
        \caption{\small $\alpha=0.0595$: Ejection of the oscillons.}
        \label{f:scater}
    \end{subfigure}
    \hfill
    \begin{subfigure}[t]{0.496\textwidth}
        \centering
        \includegraphics[width=\linewidth]{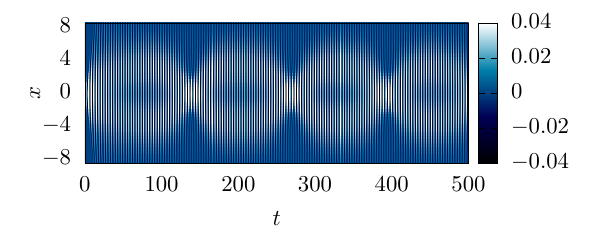}
        \caption{\small $\alpha=0.0575$: Bouncing of oscillons.}
        \label{f:bounce}
    \end{subfigure}
    \vspace{1em}
    \begin{subfigure}[t]{0.496\textwidth}
        \centering
        \includegraphics[width=\linewidth]{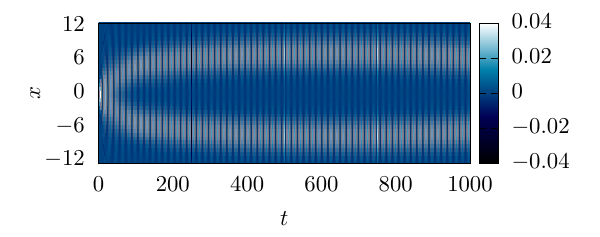}
        \caption{\small $\alpha=0.05925$: Quasi-stationary state.} 
        \label{f:qs}
    \end{subfigure}
    \caption{\small Decay of the sphaleron and interaction of the emerging oscillon with the impurity. We represent $\phi(x,t)$ for $s=0.2$.}
    \label{f:StoO-2}
    \end{figure*}

    A significantly different scenario emerges when the impurity does not support any bound mode, i.e., for $\alpha > 0$. In this case, the sphaleron once again decays into an oscillon, which evolves in the background field of the impurity. However, after a short time, this oscillon becomes destabilised by the impurity and subsequently splits into two smaller oscillons that are ejected in opposite directions as shown in \autoref{f:StoO-2}, upper panel, with parameters $s=0.2$ and $\alpha = 0.0595$. This behaviour indicates the presence of a repulsive interaction between the oscillon and the impurity. Nevertheless, a more detailed analysis reveals that the interaction is more complex. In particular, it also contains an attractive channel. An example of such dynamics is shown in \autoref{f:StoO-2}, central panel, where the two ejected oscillons eventually reverse their motion and oscillate around the origin. This occurs for $s = 0.2$ and $\alpha = 0.0575$. Interestingly, we have identified a limiting case in which the oscillons form a long-lived, quasi-stationary configuration. In this regime, their positions remain approximately fixed at a certain distance from the impurity for an extended period of time. This phenomenon is illustrated in \autoref{f:StoO-2}, lower panel, corresponding to $s=0.2$ and $\alpha = 0.05925$.

\section{Sphaleron-impurity collisions}\label{s:Collision_Impurity} 

    Another scenario that reveals the non-trivial pattern of interaction of the oscillon with the impurity is in the collision of the semi-BPS sphaleron with the impurity.

\begin{figure}[htb]
    \centering
    \includegraphics[width=0.98\linewidth]{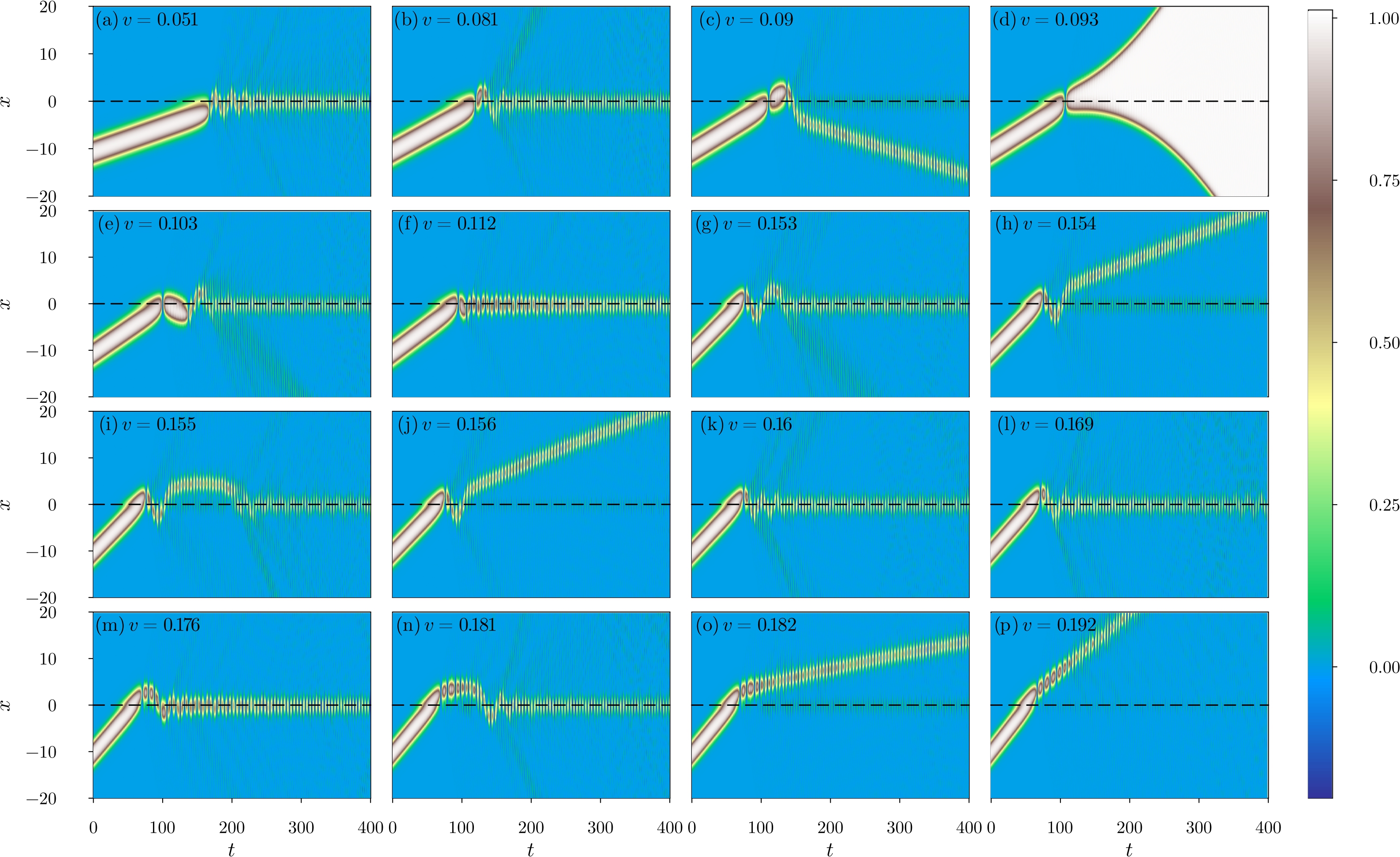}
     \caption{Examples of sphaleron–impurity collision scenarios for $s=2.5$ and $\alpha=-0.5$, shown for increasing initial velocities in the range $v_{in} \in [0.051,01.92].$} \label{f:S-I-scan-1}
     \end{figure}

    In \autoref{f:S-I-scan-1}, we present an example of the collision between a quasi-stable sphaleron with $s=2.5$ and an attractive impurity characterised by $\alpha = - 0.5$. 
    Our analysis shows that for initial velocities below $v_{in} < v_1 \approx 0.051$, the sphaleron is destabilised by the impurity and decays into an oscillon in the same manner as we explained in the previous section. Then, the oscillon is trapped by the attractive impurity and transforms into a bound mode of the impurity. Conversely, for initial velocities exceeding $v_{in} > v_2 \approx 0.182$, the sphaleron is able to pass through the impurity. However, this transit also leads to the destabilisation of the sphaleron, and at some distance beyond the impurity, the sphaleron decays into a boosted oscillon.
    
    Nevertheless, between these two cases there is a variety of different processes. We define the following qualitatively distinct classes of behaviour:
    \begin{enumerate}
        \item {\it Bounces of the oscillon.} 
        \\
        The initial sphaleron decays into an oscillon which is then attracted by the impurity. As a consequence, the oscillon performs oscillations around the impurity (e.g., a), b), g), k) and l) panels).  
        \item {\it Oscillon trapping or ejection.} 
        \\
        These oscillations release energy, and after a few bounces, the system either results in the trapping of the oscillon, eventually leading to an excited impurity (e.g., a), b), f), g), k), l), and m) panels), or in the ejection of the oscillon, either in the forward direction (e.g., h) and j) panels) or in the backward direction (e.g., c) panel).
        \item {\it Oscillon stationary state.}
        \\
        The oscillon can also form a stationary state in which it remains at a fixed distance from the impurity (e.g., i) panel). This behaviour closely resembles the stationary configuration observed in the decay of a semi-BPS sphaleron in the presence of a repulsive impurity. 
        \item {Reappearance of sphaleron.} 
        \\
        Sometimes, the sphaleron collapses and re-emerges (e.g., c) and e) panels). After its creation, it can decay into an oscillon or into a pair of kink and antikink which escape from each other (e.g., d) panel). 
    \end{enumerate}
    
\begin{figure}[H]
        \centering
        \includegraphics[width=0.7\linewidth]{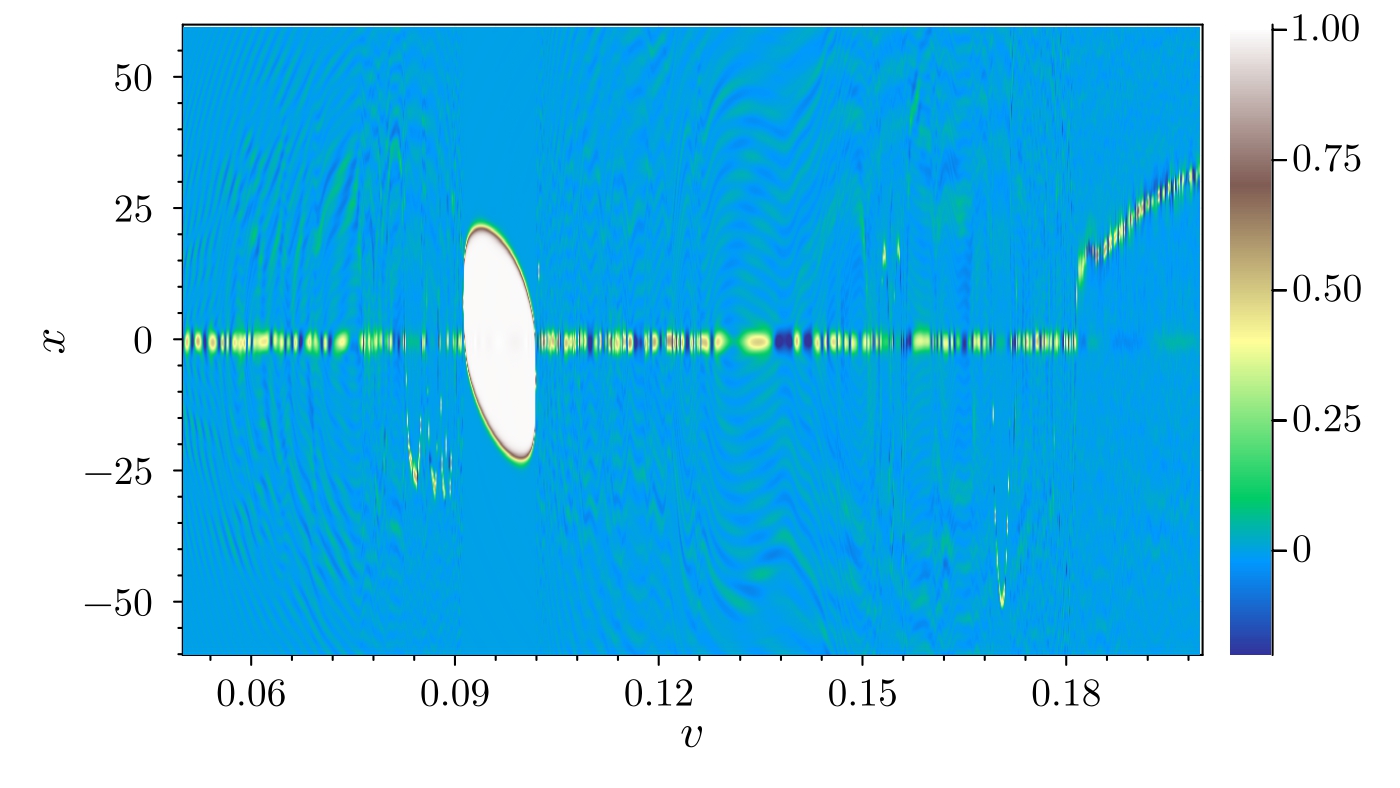}
        \caption{Spatial dependence of the field $\phi(x)$ at large time for different initial velocities $v_{in}$.} \label{f:SI-final}
    \end{figure}

    By varying the initial velocity of the sphaleron, we found evidence that the formation of the final state exhibits a structure likely chaotic. Indeed, distinct scenarios follow one another in an apparently irregular manner, see \autoref{f:SI-final}. Among these, the trapping of the oscillon by the impurity appears as the dominant behaviour ($x = 0$ axis). However, other phenomena such as backward and forward ejection of the oscillon (green areas excluding the $x = 0$ axis), as well as kink–antikink pair creation (white region), are also clearly observed.

\section{Summary and conclusions}\label{s:Summary_Semi_BPS} 

    In this chapter, we have established the existence of a type of sphaleron-impurity solutions. We have referred to them as semi-BPS sphalerons, due to the fact that they combine two distinct features. On the one hand, they resemble the usual BPS systems known from the theory of topological solitons. Specifically, there exists a family of energetically equivalent static solutions which obey a lower-order differential equation. This suggests the existence of a zero mode which, in our example, reflects that the sphaleron can be located at any distance from the impurity without altering the energy of the system. As a consequence, there are no static forces between the sphaleron and the impurity. On the other hand, there is a fundamental difference in comparison with stable BPS solitons. Namely, the semi-BPS sphaleron does not saturate any topological energy bound, and therefore there is an unstable direction along which such a static solution can decay. In terms of the linear spectrum of perturbations, the sphaleron holds a negative mode.

    After proving the existence of semi-BPS sphaleron-impurity solutions, we have investigated their dynamical properties. The absence of static forces between the sphaleron and the impurity provides a simplified environment and allows for a deeper insight into dynamical properties of sphalerons. We have found that the semi-BPS sphaleron-impurity system reveals many properties known from the usual BPS multi-soliton systems. For instance, the semi-BPS sphaleron evolves along a geodesic on the corresponding moduli space when only the translational (kinetic) degree of freedom is excited. In this case, the field configuration evolves through the continuum of available BPS states. Moreover, when the positive bound mode is excited, the sphaleron may be subject to the  spectral wall phenomenon. As usual, a necessary condition is that the excited mode crosses the mass threshold at a certain value of the modulus. Such a spectral wall appears to always destabilise sphalerons, leading to their collapse into an oscillon. This behaviour is clearly observed in cases where the sphaleron can be regarded as an unstable bound state of a kink and an antikink. Notably, it resembles the effect of spectral walls in certain supersymmetric BPS-impurity models, where they destroy the bound state between bosonic and fermionic degrees of freedom \cite{Campos:2023}.

   This characteristic behaviour of BPS soliton solutions is strongly modified once the unstable mode is excited. In such a situation, the field explores a direction in configuration space along which the sphaleron can decay, leading, for instance, to the formation of an oscillon. Naturally, due to the saddle-point nature of the sphaleron, its unstable mode can be excited by an arbitrarily small perturbation and, consequently, it will inevitably dominate the evolution after a sufficiently long time. Importantly, the resulting oscillon is not a BPS object and thus interacts with the impurity in a more intricate way. Indeed, we have observed the appearance of an intriguing oscillon-impurity stationary state. However, the origin of this phenomenon, and more generally the understanding of the oscillon in the non-trivial environment induced by the impurity, are beyond the scope of this chapter and require further investigation to determine their properties. 

    The current work can be naturally extended to models in which the background field is replaced by a dynamical one. The resulting field theories then involve at least two coupled scalar fields. Some of these models indeed support families of energetically degenerate unstable solutions. For instance, the MSTB model \cite{Montonen:1976, Sarker:1976} contains a one-parameter family of sphalerons that, in this case, consist in non-topological solutions that connect and return to the same point in a two-dimensional internal space \cite{Izquierdo:2019}. These sphalerons may decay into one of the vacua, possibly accompanied by radiation and oscillon formation, or into two less energetic kinks. It would be interesting to investigate whether the dynamical features of sphalerons and oscillons described in the present chapter persist in this model. Sphalerons also appear in the Bazeia–Nascimento–Ribeiro–Toledo (BNRT) model \cite{Bazeia:1997} for specific values of the coupling constant. In that case, they correspond to unstable topological kinks that can decay into other lower-energy kinks \cite{Izquierdo:2002}. From our perspective, studying these decay channels is equally relevant. Additional examples of field theories with sphalerons can be found along the lines discussed in \cite{Halcrow1:2023}. 

    Interestingly, another class of BPS sphalerons exists in higher-dimensional theories. These configurations are again energetically degenerate, unstable static solutions whose moduli space is parametrised by a continuous parameter, thus admitting at least one zero mode. In contrast to the semi-BPS sphalerons discussed in this chapter, they do not satisfy the original Bogomolny equations. The most extensively studied examples arise in non-linear $(2+1)$-dimensional sigma models with target spaces such as $\mathbb{CP}^n$ and $F_2$ \cite{Din:1980, Amari:2018}. It would be desirable to understand a possible connection between these sphalerons and the semi-BPS configurations considered here. It is also worth noting that, due to the scale invariance of these models, such sphalerons cannot support any bound modes, and therefore their dynamical properties may differ significantly.

    A more general question is whether any sphaleron configuration can be interpreted as an unstable multi-soliton state. In the present case, this interpretation indeed holds. Specifically, for all values of the parameter $s$, the sphaleron can be viewed as a superposition of a kink and an antikink. Remarkably, this composite picture successfully reproduces both the unstable mode and the zero mode of the sphaleron. Namely, the zero mode corresponds to a simultaneous shift of the positions of the constituent kink and antikink in the same direction, $z_{1,2} \to z_{1,2} + b$, while the unstable mode corresponds to an antisymmetric shift, $z_{1} \to z_{1} + b$, $z_{2} \to z_{2} - b$. It is clear that the second transformation modifies the separation between the constituent kinks and, as such, its effect may be interpreted as a kink-antikink scattering process. This interpretation will be studied in more detail in \autoref{c:Sph_Circle}.

        \chapter{Sphaleron decay in false-vacuum theories}\label{c:Sph_False}

This chapter is adapted from \cite{Navarro:2024}:
    \hspace{-2.5cm}
    \vspace{-0.3cm}
    \begin{figure}[H]
        \hspace{-2.5cm}
        \vspace{-0.3cm}
        \centering{\includegraphics[width=1.1\linewidth]{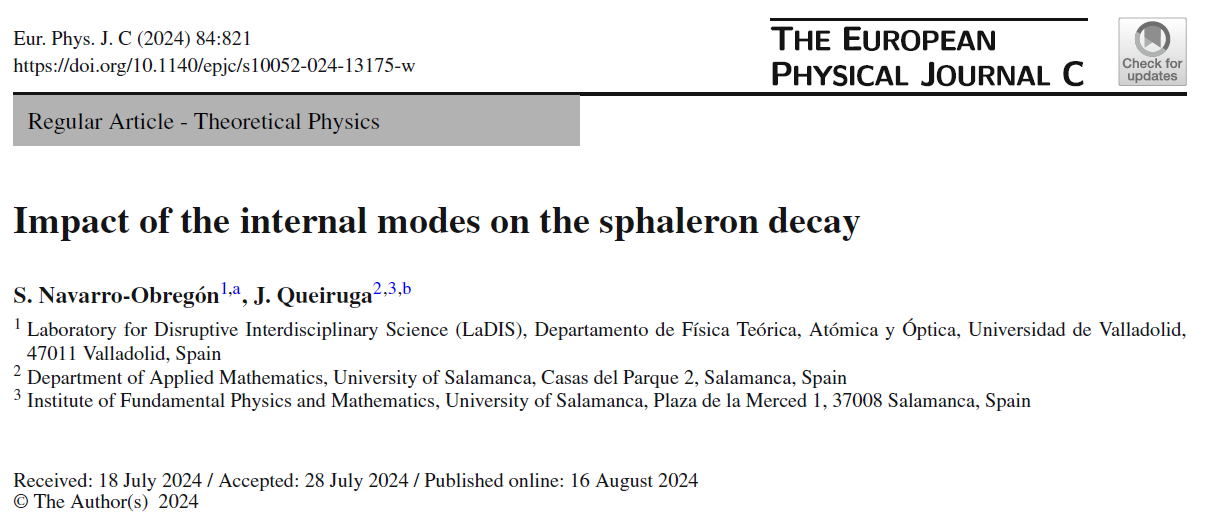}
        }
        \label{f:Thumbail_Deformation}
        \hspace{-2.5cm}
    \end{figure}

\section{Introduction}

    The inherent instability of sphalerons plays a central role in their dynamics. As shown in \autoref{c:BPS_Sph}, even small perturbations or external influences can trigger their decay. Since the most energy-efficient path between two topologically distinct vacua in a gauge theory must, at some point, approximate a sphaleron configuration, the formation and decay of sphalerons are of significant physical relevance, particularly for understanding their possible influence on the baryon asymmetry. Nevertheless, the real-time dynamics of sphaleron decay toward the vacuum state remain poorly understood and have been explored only to a limited extent in the literature \cite{Heilmund:1991,Zadrozny:1992,Cottingham:1993,Copi:2008,Chu:2011}. In particular, the influence of the internal structure on sphaleron decay has only recently begun to be investigated \cite{Romanczukiewicz:2021, Manton:2023,Wereszczynski:2023}.  

    In this chapter, we aim to analyse how the internal degrees of freedom of the sphaleron influence its subsequent decay. To this end, it is particularly interesting to construct models that admit sphalerons with markedly distinct internal structures. As shown in \autoref{s:Examples_Sph}, the internal structure of a sphaleron can be modified by deforming the underlying potential. Notably, there exists a regime of the deformation parameter in which the internal structure  of the sphaleron reproduces the spectrum of the individual kinks that constitute it, as seen in \autoref{f:spectrum_sph_false}. A natural and coherent strategy, therefore, is to start with a model whose internal structure is known and advantageous for our analysis, and then deform the theory in such a way that it allows for sphaleron solutions. In doing so, these sphalerons are expected to inherit the desired internal structure. 

     The structure of this chapter is as follows. In \autoref{s:Deformation_Models}, two deformations of the $\phi^6$ model are introduced and the sphalerons solutions are analysed, along with the linear fluctuation spectra of excitations around the corresponding sphalerons. In \autoref{s:decay_Deformed}, the oscillon formation from the sphaleron decay is discussed. Then, the role of internal modes in the possible decay channels of both models is examined in \autoref{s:CCM_Deformed}. Finally, \autoref{s:Summary_Deformed} presents our conclusions. 

\section{The \texorpdfstring{$\phi^6$}{phi6} family of models}\label{s:Deformation_Models}

\begin{figure}[H]
        \centering
        \includegraphics[width=1.02\linewidth]{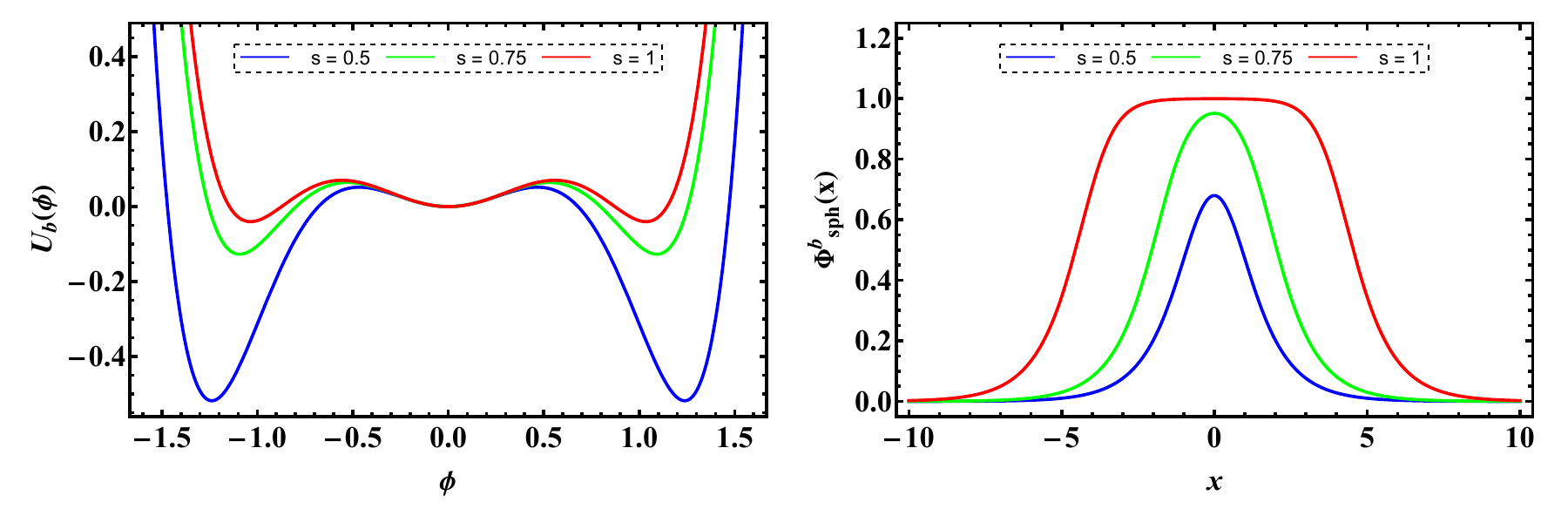}
        \caption{Plot of the potential function (\ref{e:potential_Barrier}) and its sphaleron configurations (\ref{e:sphaleron_barrier}) for various values of the parameter $s$.} \label{f:barrier_deformation}
    \end{figure}

    Let us assume a family of $\phi^6$ models with the following Lagrangian density 
    \begin{equation}\label{e:Action_Phi6_Deformed}
        \mathcal{L} = \dfrac{1}{2}\partial_{\mu}\phi\partial^{\mu}\phi - U_{6}(\phi;s)\,,
    \end{equation}
    where $s$ controls the deformation of the usual $\phi^6$ theory (\ref{e:lagrangian_phi6}). A possible deformation is given by
    \begin{equation}\label{e:potential_Barrier}
        U_{b}(\phi;s) = \dfrac{1}{2}\phi^2(\tanh s - \phi^2)(\coth s - \phi^2)\,,\quad s > 0\,.
    \end{equation}
    The potential is shown for representative values of $s$ in \autoref{f:barrier_deformation}. As observed, for any finite $s$, the potential develops a false vacuum at $\phi^{\footnotesize f}_v=0$ and two symmetric true vacua at 
    \begin{equation}
        \phi_v^t = \pm \dfrac{\sqrt{\coth s + \tanh s + \sqrt{\coth^2s + \tanh^2 s - 1}}}{\sqrt{3}}\,.    
    \end{equation}
    Note that in the limit $s \to \infty$, one recovers the standard $\phi^6$ theory \eqref{e:lagrangian_phi6}, characterised by three degenerate vacua located at $\phi_v = \{-1, 0, 1\}$.
    
    The equation of motion associated with the potential function given by (\ref{e:potential_Barrier}) is the following
    \begin{equation}\label{e:barrier_EOM}
        \partial_\mu\partial^{\mu} \phi + \phi\,(1 - 4\coth 2s \, \phi^2 + 3\,\phi^4) = 0\,.
    \end{equation} 
    The presence of the false vacuum structure permits the existence of sphaleron configurations. To obtain these saddle-point solutions, we solve the piecewise BPS equation
    \begin{equation}
        \phi_x = \left\{
        \begin{array}{cc}
           +\sqrt{2\,U_b(\phi;s)}  & x\leq 0\,, \\
            -\sqrt{2\,U_b(\phi;s)} & x\geq 0\,.
        \end{array} 
        \right.
    \end{equation}
    A direct integration of this equation leads to the following sphaleron solution
    \begin{equation}\label{e:sphaleron_barrier}
        \Phi^b_{sph}(x;s) = \pm \sqrt{\dfrac{\sinh 2s }{\cosh 2s + \cosh 2x}}\,.
    \end{equation}
    As shown in \autoref{f:barrier_deformation}, the sphaleron profile exhibits a lump-like shape for small values of the deformation parameter $s$, becoming wider as $s$ increases, and eventually resembling a kink–antikink pair. The energy associated with the sphaleron configuration given in (\ref{e:sphaleron_barrier}) can be computed analytically as
    \begin{equation}\label{eq:ener_bar}
        E(s) = \frac{1}{4}\csch^2 2s\left(\sinh 4s - 4s\right)\,,
    \end{equation}
    which increases monotonically with $s$, from $E = 0$ up to $E = 1/2$, the energy corresponding to a widely separated kink–antikink pair in the standard $\phi^6$ theory.

\begin{figure}[H]
        \centering
        \includegraphics[width=1.02\linewidth]{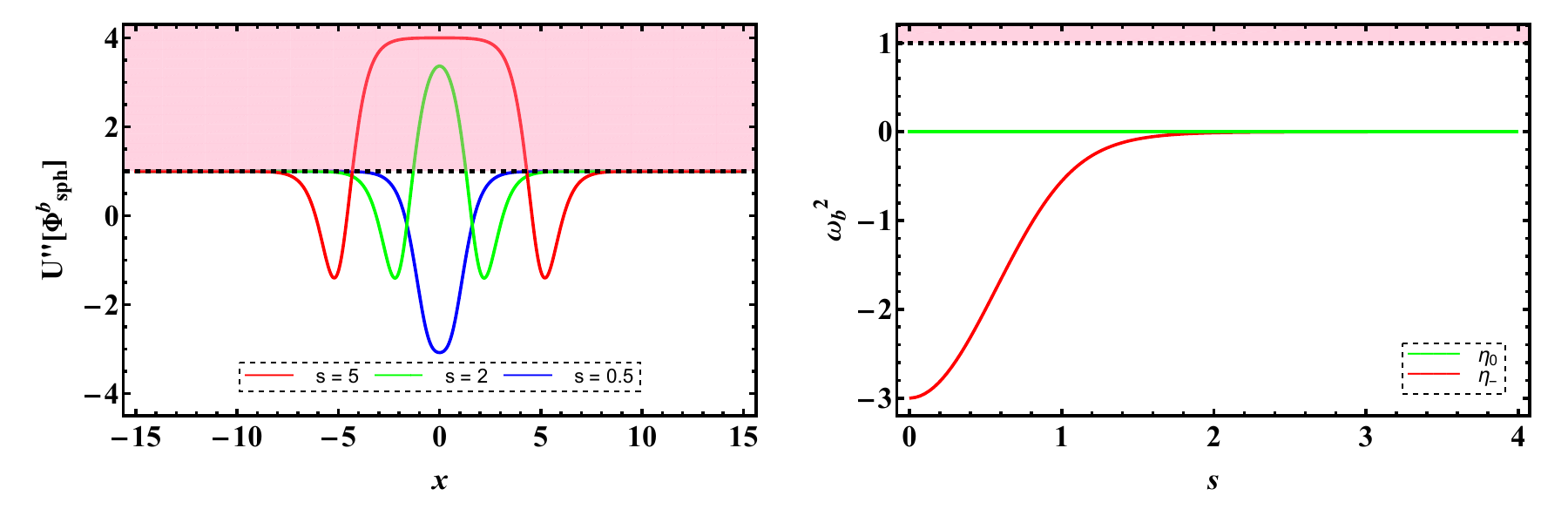}
        \caption{Effective potential of the Schrödinger-like equation (\ref{e:Linearised_Barrier}) and the linear perturbation spectrum around the sphaleron configuration (\ref{e:sphaleron_barrier}). The shaded area represents the continuum spectrum.} \label{f:barrier_SP}
    \end{figure}

    Now, we can compute the linear spectrum of perturbations around the sphaleron (\ref{e:sphaleron_barrier}) by assuming the standard linear ansatz
    \begin{equation}\label{e:perturbation_SP_barrier}
        \phi(x,t;s) = \Phi^b_{sph}(x;s) + \eta^b(x;s)e^{i \omega t}\,.
    \end{equation}
    Inserting the perturbed sphaleron (\ref{e:perturbation_SP_barrier}) into the field equation (\ref{e:barrier_EOM}) and expanding up to linear order in $\eta^b(x;s)$, we obtain
    \begin{equation}\label{e:Linearised_Barrier}
        \left[ - \dfrac{d^2}{dx^2} + 1 - 12\coth 2s\,(\Phi_{sph}^b)^2 + 15(\Phi_{sph}^b)^4 \right]\eta^b(x;s) = \omega^2(s)\,\eta^b(x;s)\,.
    \end{equation}
    The normal modes around the sphaleron are shown in \autoref{f:barrier_SP}. As observed, the discrete part of the linear spectrum contains only an unstable mode and a zero mode. Notably, as $s \rightarrow \infty$, the unstable mode approaches the zero mode. Consequently, the spectrum of the individual kinks is recovered.   
    It is noteworthy that the spectrum lacks any bound states with positive squared frequency. This absence can be qualitatively understood by observing that, for large values of $s$, the sphaleron approximates a kink-antikink pair, $(0,1)+(1,0)$, in the standard $\phi^6$ model. In this context, the inner vacuum at $\phi_v = 1$ possesses a higher mass threshold ($m_v = 2$) compared to the outer vacuum at $\phi_v = 0$ ($m_v = 1$). As a result, the effective potential in the spectral problem features a central barrier between the constituent kinks, as illustrated in \autoref{f:barrier_SP}. For this reason, we refer to the scalar theory with potential (\ref{e:potential_Barrier}) as the \textit{barrier model}.

\begin{figure}[H]
        \centering
        \includegraphics[width=1.02\linewidth]{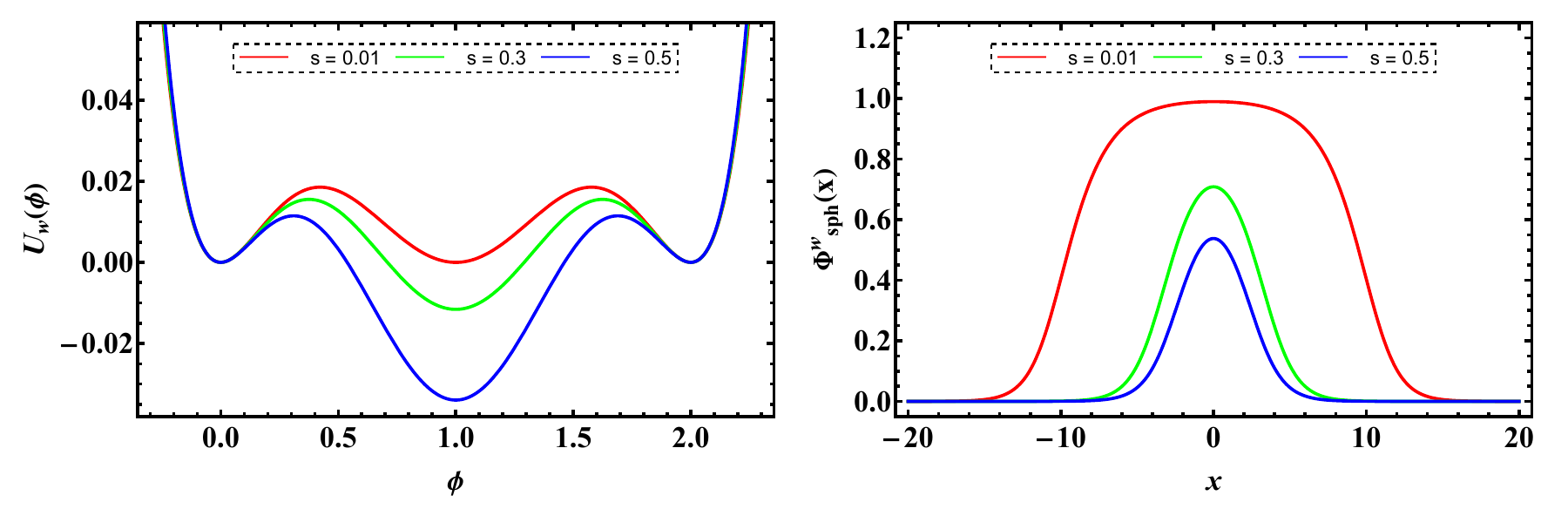}
        \caption{Plot of the potential function (\ref{e:potential_Well}) and its sphaleron configurations (\ref{e:sphaleron_well}).} \label{f:well_deformation}
    \end{figure}

    Following this reasoning, we introduce a new deformation by shifting the central vacuum so that it becomes the new true vacuum\footnote{This model shares similarities with the Christ-Lee potential \cite{Christ:1975,Dorey:2023}.}. Specifically, we consider the potential
    \begin{equation}\label{e:potential_Well}
        U_{w}(\phi;s) = \dfrac{(\phi - 1)^2 - \tanh^2 s}{8(1 -\tanh^2 s)}\left((\phi - 1)^2 - 1\right)^2\,.
    \end{equation}
    A plot of this potential for various values of the parameter $s$ is shown in \autoref{f:well_deformation}. In this case, the false vacua are located at $\phi_v^f = \{0,2\}$, while the true vacuum is located at $\phi_v^t = 1$. Note that we have shifted the field as $\phi \to \phi - 1$ in order to ensure that the sphaleron interpolates asymptotically between the same false vacuum as in the barrier model. Additionally, a factor of $1/4$ has been introduced relative to the standard $\phi^6$ theory, which is recovered in the limit $s \to 0$. This factor guarantees that the mass threshold at the false vacuum remains $m_v = 1$.
    
    The equation of motion associated to this new deformation is
    \begin{align}\label{e:well_EOM}
        \partial_{\mu}\partial^{\mu} \phi 
        + \phi \Bigg(&1 - \dfrac{3}{2}\hspace{0.1cm}\!\left(2 + \cosh 2s \right)\phi 
        + \left(\tfrac{7}{2} + 3 \cosh 2s\right)\phi^2 \nonumber\\
        &\quad - \dfrac{15}{4 (1 - \tanh^2 s)}\phi^3 
        + \dfrac{3}{4 (1 - \tanh^2 s)}\phi^4\Bigg) = 0\,.
    \end{align}

    Solving the corresponding piece-wise Bogomolny equation we obtain the sphaleron-type solutions 
    \begin{equation}\label{e:sphaleron_well}
        \phi^w_{sph}(x;s) = 1 \pm \dfrac{2 \sinh s \cosh x/2}{\sqrt{3 + \cosh 2s + 2\cosh x \sinh^2 s}}\,.
    \end{equation}
    Various sphaleron profiles are shown in \autoref{f:well_deformation}. As expected, in the limit $s \to 0$, the sphaleron approaches the form of a kink–antikink pair. Although the energy can be computed analytically, its explicit expression is not particularly insightful. Nevertheless, it can be verified that the energy is bounded from above by that of a well-separated kink–antikink pair, appropriately scaled.
    
    By perturbing the sphaleron (\ref{e:sphaleron_well}) in the same way as before, we obtain the corresponding Schrödinger-like equation that governs their normal modes 
    \begin{equation}\label{e:Linearised_Well}
        \left[ - \dfrac{d^2}{dx^2} +  U_w''(\Phi_{sph}^w)\right]\eta^w(x;s) = \omega^2(s)\,\eta^w(x;s)\,,
    \end{equation}
    where
    \begin{align}\label{e:Linearised_Well_Veff}
        U_w''(\Phi_{sph}^w) 
       &= \dfrac{\cosh^2 s}{4}\left( 4\,\sech^2 s 
          - 12\,(2 + \sech^2 s)\,\Phi_{sph}^w \right. \nonumber \\ 
       &\quad \left. + 6\,(12 + \sech^2 s)\,(\Phi_{sph}^w)^2
          - 60\,(\Phi_{sph}^w)^3 
          + 15\,(\Phi_{sph}^w)^4 \right)\,.
    \end{align}

\begin{figure}[H]
    \hspace{-0.15cm}
        \centering
        \includegraphics[width=1.02\linewidth]{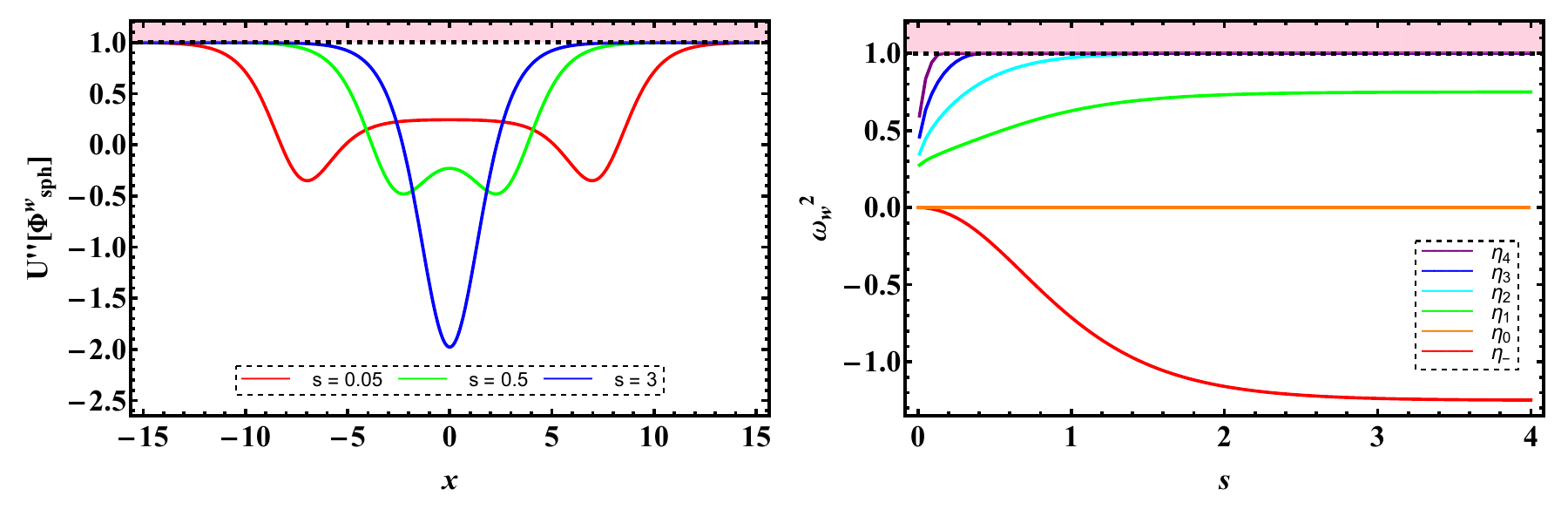}
        \caption{Effective potential of the Schrödinger-like equation (\ref{e:Linearised_Well}) and lowest modes in the linear perturbation spectrum around the sphaleron configuration (\ref{e:sphaleron_well}). The shaded area represents the continuum spectrum.} \label{f:well_SP}
    \end{figure}

    Unlike the sphaleron in the barrier model, the sphaleron in this new model possesses at least one positive bound mode, in addition to the zero mode and the unstable mode. Specifically, the number of shape modes increases as $s \to 0$; see \autoref{f:well_SP}. This behaviour can again be qualitatively understood through the kink–antikink picture. As $s \to 0$, the sphaleron increasingly resembles an antikink–kink pair $(0,1)+(1,0)$, which induces an effective potential in the linearised problem featuring a central well. This potential well, shown in \autoref{f:well_SP}, supports an increasing number of bound modes as the separation between the constituent kinks grows. For this reason, we refer to this model as the \textit{well model}.
    It is important to note that we have represented only the lowest bound modes in the previous picture. Numerically, we have seen that the number of positive bound modes increases without bound. To verify that, we can adopt the following strategy. Let us assume the spatial asymptotic limit of the effective potential (\ref{e:Linearised_Well_Veff}). The asymptotic form of the eigenstate with eigenfrequency $\omega_c^2 = 1$ is
    \begin{equation}
        \eta^w_{thres} \propto \text{I}\left[ 0,\dfrac{2\, e^{-x/2}\sqrt{-12 - 48e^{2s} - 12e^{4s}}}{\sqrt{1 - 2e^{2s} + e^{4s}}}\right]\,,
    \end{equation}
    where $I[n,x]$ denotes the modified Bessel function of the first kind. This eigenstate exhibits an unbounded number of nodes as $s$ decreases. By virtue of \autoref{t:Sturm}, this implies the existence of an unbounded number of eigenstates lying below it, in full agreement with our numerical findings.

    As a result, we have constructed two deformations of the $\phi^6$ theory that exhibit a false-vacuum structure, thereby allowing for the existence of sphaleron configurations. Remarkably, depending on the deformation, the corresponding sphalerons possess completely different internal structures. In particular, we have been able to build a model that supports a sphaleron without shape modes, and another in which we have direct control over the number of shape modes. Therefore, these deformed models may be the simplest theories where we can strictly compare the impact of the internal modes on the sphaleron decay.

\section{Sphaleron decay}\label{s:decay_Deformed}

    The sphaleron can decay along its unstable direction through various mechanisms. To analyse this decay numerically in a controlled manner, we consider the following initial condition
    \begin{align}\label{e:IC_decay}
        \phi(x,0) &= \Phi_{sph}(x;s) \pm A\,\eta_{-}(x;s)\,,\\
        \dot{\phi}(x,0) &= 0\,.
    \end{align}
    Here, $\eta_{-}(x;s)$ represents the sphaleron unstable mode, which is assumed normalised, and $A$ is its initial amplitude. Therefore, this initial condition represents a sphaleron displaced slightly along its unstable direction. The signs accompanying the amplitude account for the two different decay channels: the plus sign denotes a perturbation in the expanding direction, whereas the minus sign accounts for an excitation in the collapsing direction. In other words, one leads to the creation of an accelerating kink-antikink pair (associated with the true vacuum), while the other results in the collapse into an oscillon (linked to the false vacuum).

    The observed acceleration of the kink-antikink pair in the expanding direction can be understood through an energy conservation argument: as the maximum of the sphaleron descends towards the true minima, the field reaches negative values of the potential. During the expansion of the sphaleron, a plateau of negative potential energy forms at its centre. This negative contribution must be compensated by an increase in kinetic energy at the sphaleron walls, leading to their acceleration. 
    
\begin{figure*}[htb]
    \centering
    \begin{subfigure}[t]{0.495\textwidth}
        \centering
        \includegraphics[width=\linewidth]{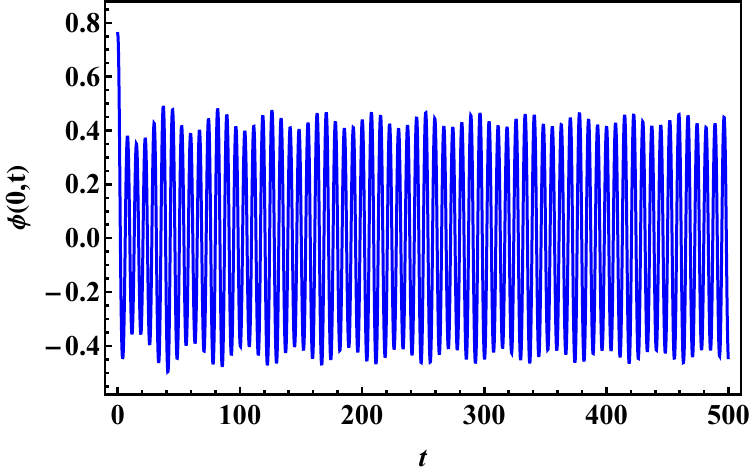}
        \caption{\small Barrier model: $s = 1.267$ and $A = 0.6$.}
        \label{f:Barrier_1267}
    \end{subfigure}
    \hfill
    \begin{subfigure}[t]{0.495\textwidth}
        \centering
        \includegraphics[width=\linewidth]{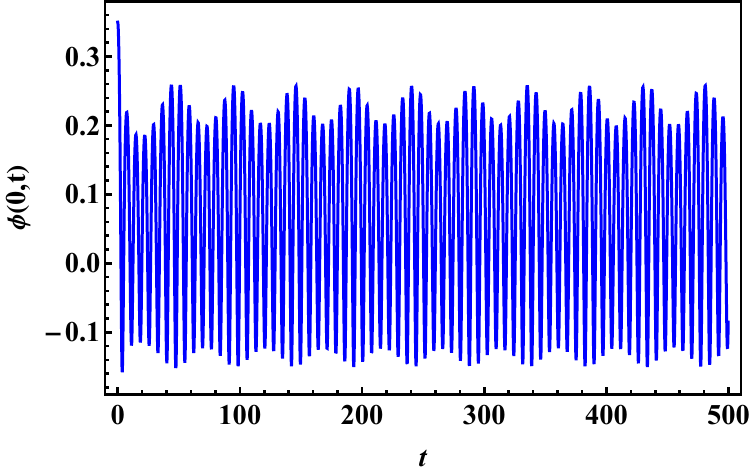}
        \caption{\small Well model: $s = 0.45$ and $A = 0.6$.}
        \label{f:Well_045}
    \end{subfigure}
    \caption{\small Decay of the sphaleron in the barrier model and in the well model. The figures show the field at the origin $\phi(0, t)$. The frequency of the unstable mode is $\omega_-^2 = - 0.2056$.}
    \label{f:decay_comparison_low}
    \end{figure*}

\begin{figure*}[htb]
    \centering
    \begin{subfigure}[t]{0.491\textwidth}
        \centering
        \includegraphics[width=\linewidth]{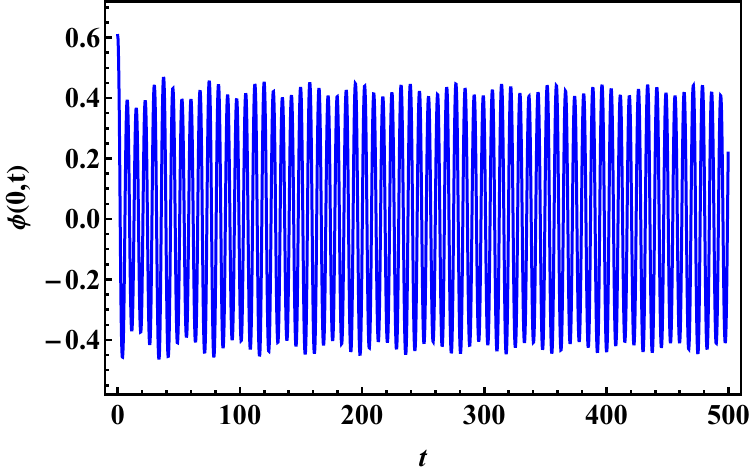}
        \caption{\small Barrier model: $s = 0.75$ and $A = 0.3$.}
        \label{f:Barrier_075}
    \end{subfigure}
    \hfill
    \begin{subfigure}[t]{0.498\textwidth}
        \centering
        \includegraphics[width=\linewidth]{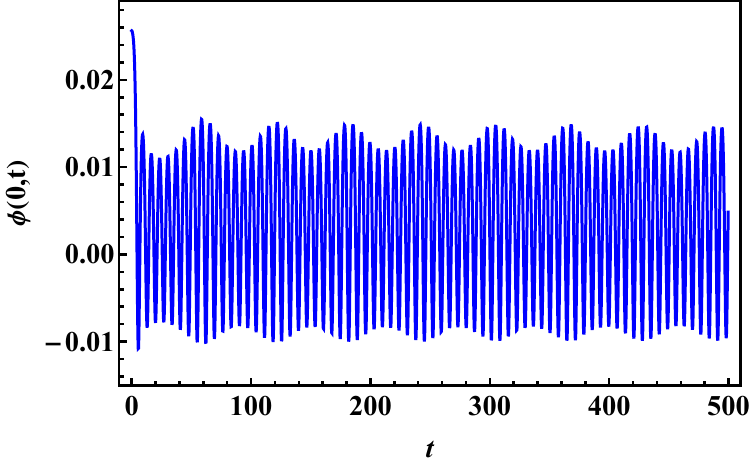}
        \caption{\small Well model: $s = 2.16$ and $A = 0.001$.}
        \label{f:Well_216}
    \end{subfigure}
    \caption{\small Decay of the sphaleron in the barrier model and in the well model. The figures show the field at the origin $\phi(0, t)$. The frequency of the unstable mode is $\omega_-^2 = - 1.183$.}
    \label{f:decay_comparison_high}
    \end{figure*}

    Regarding the analysis in the collapsing direction, we have selected appropriate values of $s$ so that the unstable frequencies $\omega^2_{-}$ are equal in the barrier and well model, enabling us to compare the decays of their respective sphalerons. When examining the decay of the sphalerons, we observe qualitatively the same behaviour in both models. For small amplitudes, a long-lived oscillon forms, initially exhibiting an imperceptible periodic modulation that becomes noticeable at later times. For moderate amplitudes, the double quasi-periodic structure is distinguishable from the beginning. Finally, for high amplitudes, the sphaleron collapses into a rapidly decaying oscillon. Naturally, the different regimes depend on the size of the sphaleron and the model under study. Plots of the decay processes in the barrier and well models for different values of $\omega^2_{-}$ are shown in \autoref{f:decay_comparison_low} (corresponding to a low negative frequency case) and in \autoref{f:decay_comparison_high} (corresponding to a high negative frequency case). 

    This behaviour may suggest that the double quasi-periodic structure is an intermediate state between stable and unstable oscillons \cite{Ghersi:2023}, although a complete understanding of the nature of this internal degree of freedom is still lacking. However, we remark that a recent non-trivial relationship between oscillons and Q-balls has been established, which has shed more light on the origin of the modulation \cite{Blaschke:2024,Blaschke1:2025,Blaschke:2025}. Another conjecture relates the original sphaleron degrees of freedom to the dynamics of the oscillon formed after the collapse \cite{Manton:2023}. However, as demonstrated in \cite{Wereszczynski:2023}, the oscillon formed after sphaleron decay exhibits a modulated amplitude even when the original sphaleron lacks any shape mode. Our findings are in full agreement with this result. In that work, the authors also argue that viewing the sphaleron as a kink–antikink pair might suggest that the oscillon evolution is influenced by the dynamics of the constituent kinks of the sphaleron. Nevertheless, we have shown here that the double quasi-periodic structure arises even when the individual kinks do not possess positive bound modes. This suggests that the influence of the constituent kinks on the resulting oscillon may be manifested in a highly non-trivial way, and needs further research.

\begin{figure*}[htb]
    \centering
    \begin{subfigure}[t]{0.495\textwidth}
        \centering
        \includegraphics[width=\linewidth]{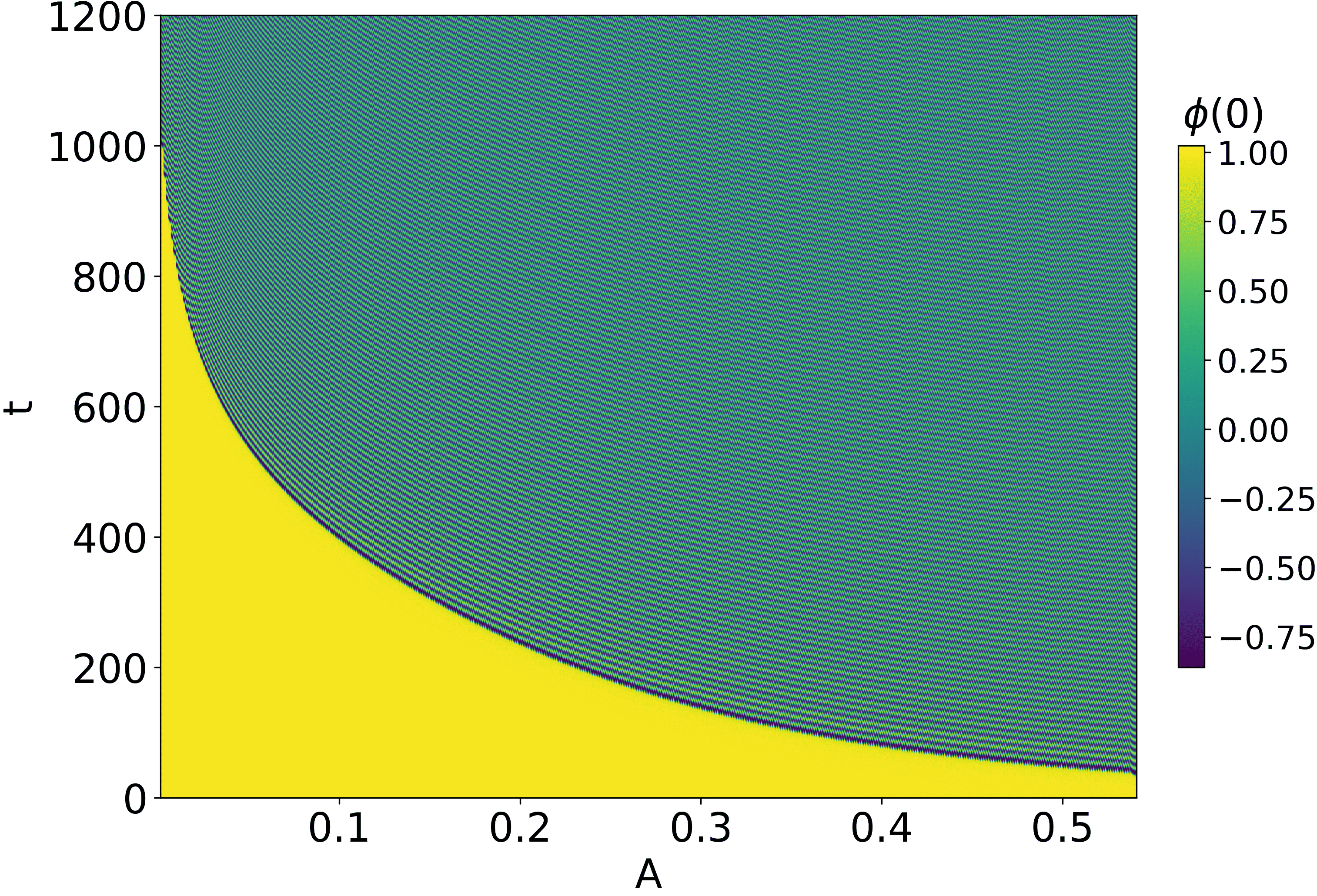}
        \caption{\small Barrier model: $s = 3.5$.}
        \label{f:Barrier_351267}
    \end{subfigure}
    \hfill
    \begin{subfigure}[t]{0.4855\textwidth}
        \centering
        \includegraphics[width=\linewidth]{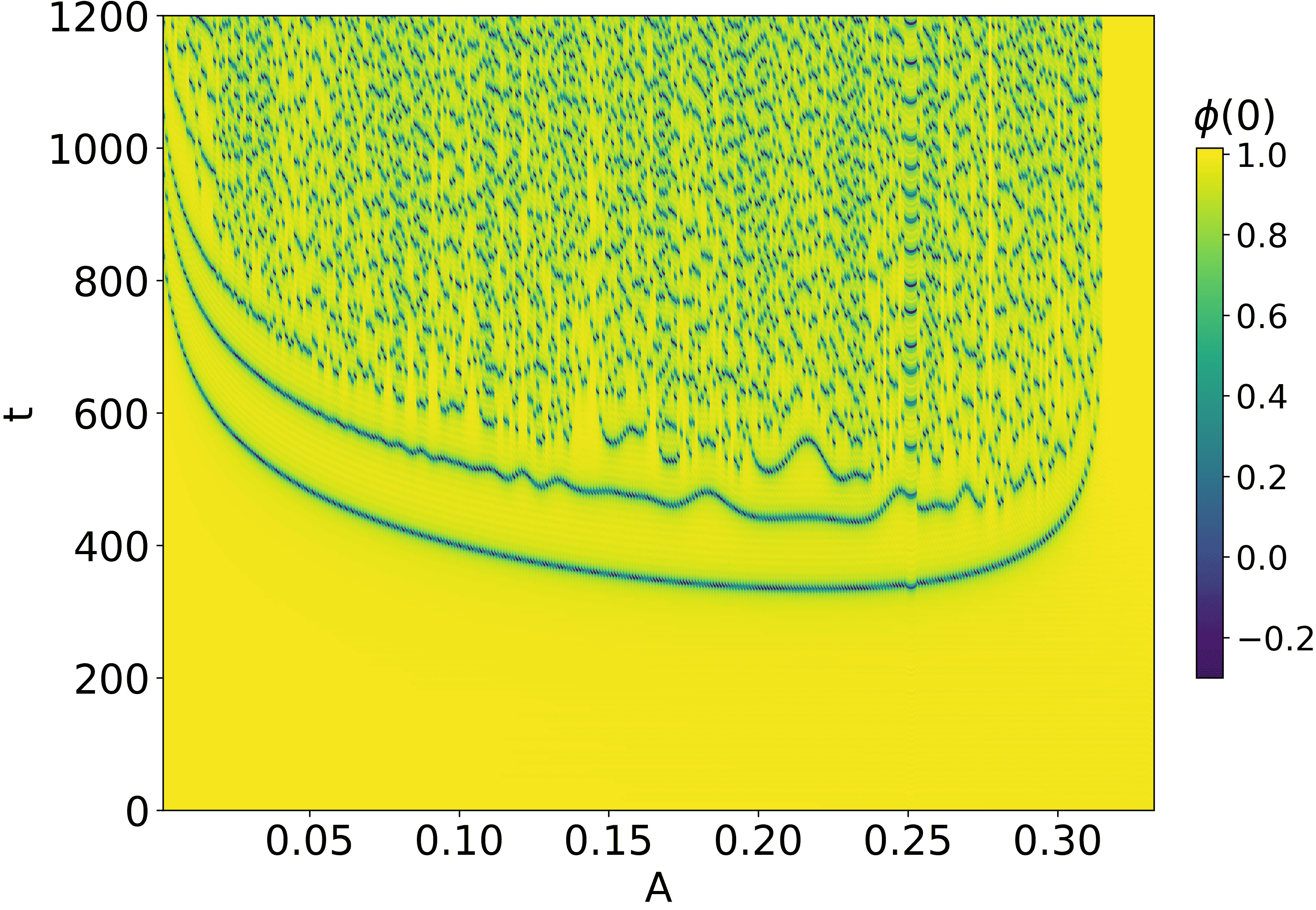}
        \caption{\small Well model: $s = 0.008$.}
        \label{f:Well_0008}
    \end{subfigure}
    \caption{\small Comparison of the decay of the sphaleron between the barrier model and the well model for a well-established  kink-antikink pair. The colour palette shows the field at the origin.}
    \label{f:decay_comparison_Density}
    \end{figure*}

\begin{figure}[htb]
        \centering
        \includegraphics[width=1.0\linewidth]{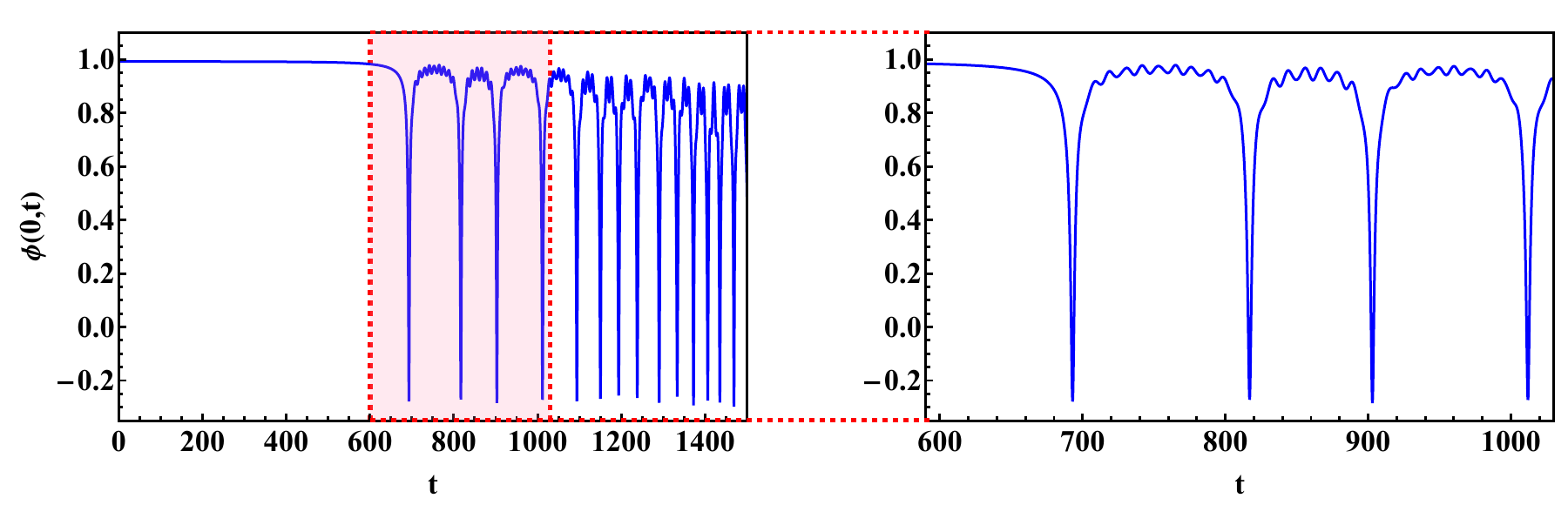}
        \caption{Intermediate state of bounces for $s = 0.008$ and $A = 0.01$ in the well model. The shaded region is magnified to highlight the wiggle more clearly.} \label{f:decay_Well_Small_S}
    \end{figure}

    An interesting regime emerges when the sphaleron can be interpreted as a well-separated kink–antikink pair. As representative values of the parameter $s$, we consider $s = 3.5$ for the barrier model and $s = 0.008$ for the well model. The time evolution of the sphaleron for various initial amplitudes $A$ of the unstable mode is shown in \autoref{f:decay_comparison_Density}, where the colour palette indicates the value of the field at the origin. 
    In the barrier model, it is clearly observed that the resulting oscillon after the decay is insensitive to the initial amplitude $A$. Similarly to smaller sphalerons, it collapses into an oscillon in this regime as well. Nevertheless, the sphaleron decay in the well model exhibits new features. Specifically, it can be distinguished a more chaotic pattern where the sphaleron, before forming an oscillon, is able to bounce a certain number of times. The evolution of the field value at the origin is shown in \autoref{f:decay_Well_Small_S} for $s = 0.008$ and $A = 0.01$. 
    In this figure, small oscillations are visible following each bounce. This observation suggests that the mechanism driving the bounces is closely related to the presence of internal modes. The qualitative picture is as follows: as the sphaleron begins to decay, part of its energy is transferred to internal modes. These modes temporarily store the energy and subsequently return it in the form of kinetic energy to the constituent kinks. As a result, the sphaleron re-emerges and collapses once again. However, since this energy exchange mechanism is not observed for large values of $s$, it appears that the energy storage mechanism is no longer effective when there are not sufficient positive bound modes. Recall that the number of shape modes decreases for increasing values of the parameter $s$.

\begin{figure}[htb]
        \centering
        \includegraphics[width=0.45\linewidth]{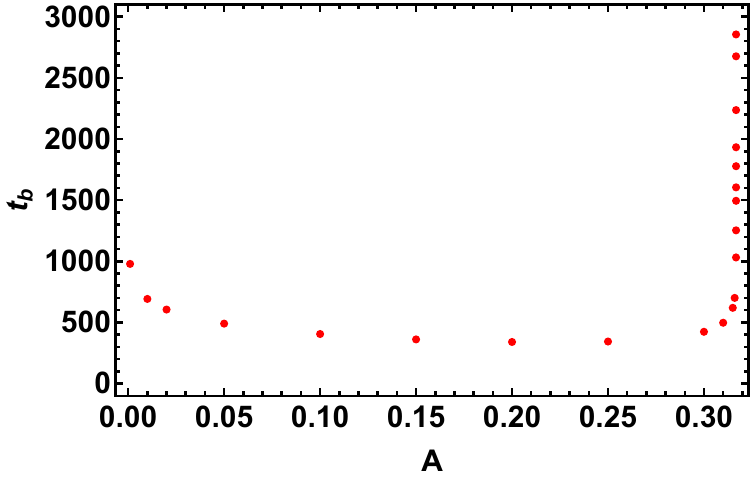}
        \caption{Time elapsed before the first bounce $t_b$ of the sphaleron during its decay in the well model for $s = 0.008$ and different values of the unstable mode amplitude $A$.} \label{f:LifeSpan}
    \end{figure}

    Remarkably, a novel phenomenon emerges for sufficiently large amplitudes and small values of $s$ in the well model. Whilst in the barrier model the sphaleron decays earlier as the amplitude $A$ of the unstable mode increases, in the well model this tendency is not monotonic. 
    Initially, the decay time decreases with increasing $A$, until a certain threshold is reached. Beyond that point, the time elapsed before the first bounce starts to increase, as illustrated in \autoref{f:LifeSpan}. This enhancement of the lifespan persists up to a critical amplitude $A_c$, above which the sphaleron no longer collapses but instead expands into two outgoing, accelerating kinks. In \autoref{f:decay_comparison_Density}, this critical amplitude is identified as $A_{c} \approx 0.31673$. Notably, as the system approaches this critical amplitude from below, the decay is significantly delayed. This behaviour suggests a form of dynamical stabilisation of the sphaleron, where it neither collapses nor expands, but it oscillates for a long time.

\section{Collective coordinate method based on sphaleron degrees of freedom}\label{s:CCM_Deformed}

    In this section, we aim to describe the dynamics of the sphaleron once its unstable mode has been triggered. We begin by analysing the evolution of the oscillon that forms following the collapse of the sphaleron. To this end, we construct an effective model based on the collective coordinate method. As shown in \cite{Manton:2023,Wereszczynski:2023}, a good approximation of the oscillon dynamics can be obtained by considering the discrete, unstable, and stable deformation modes of the sphaleron. Nevertheless, this effective description is expected to break down when the initial sphaleron is significantly larger than the resulting oscillon. In this regime, the unstable mode $\eta_{-1}(x;s)$ is equivalent to the symmetric superposition of the zero modes of the individual kinks. As a result, this mode does not overlap with the oscillon profile, implying that it does not serve as an appropriate degree of freedom to describe the oscillon. Therefore, we will instead adopt a description directly based on the oscillon itself. Later, we will introduce a new effective model formulated in terms of sphaleron degrees of freedom, but suitably adapted to the oscillon profile for arbitrary values of the parameter $s$.
    
    To conclude, we will analyse the underlying mechanism responsible for the dynamical stabilisation observed in the well model at small values of the parameter $s$, and provide a detailed explanation.

\subsection{Effective description of oscillon dynamics in the barrier model}

    A natural description of the subsequent oscillon formed from the sphaleron decay can be given by constructing an \textit{ad hoc} effective model based on the oscillon itself \cite{Navarro:2023,Fodor:2008}. As shown in \cite{Wereszczynski:2023}, a reasonable effective model contains the leading profile of the oscillon from a small-amplitude expansion \cite{Fodor:2008} and a second contribution accounting for the associated Derrick mode
    \begin{equation}\label{e:oscillon_proposal}
        \Phi(x;a,b) = \dfrac{a(t)}{\cosh(x \lambda)} + b(t)\dfrac{x\tanh(x \lambda)}{\cosh(x\lambda)}\,.
    \end{equation}
    Here, $X = \{a,b\}$ is the set of collective coordinates, and $\lambda$ is a constant parameter accounting for the size of the oscillon. 

    As usual, in order to compute the effective model by means of the collective coordinate method, we introduce the ansatz (\ref{e:oscillon_proposal}) into the Lagrangian density (\ref{e:Action_Phi6_Deformed}) and integrate over the space. By doing that, we obtain an effective model of the form
    \begin{equation}
        L[a,b] = \dfrac{1}{2}g_{ij}\dot{X}^i\dot{X}^j - V_{eff}(a,b)\,,
    \end{equation} 
    with the following metric
    \begin{equation}\label{e:metric_oscillon}
        g_{ij} = \begin{pmatrix}
        \dfrac{2}{\lambda} & \dfrac{1}{\lambda^2} \\
        \dfrac{1}{\lambda^2} & \dfrac{12 + \pi^2}{18\lambda^3} \\
        \end{pmatrix}\,.
    \end{equation} 
    The explicit expression of the effective potential $V_{eff}(a,b)$ has been omitted for the sake of clarity and readability, but we highlight that it can be obtained analytically.
    
    To evolve the effective model, it is necessary to specify the initial conditions $a(0),\, b(0)$, as well as to fix the parameter $\lambda$. To achieve this, we evolve the sphaleron in full field theory using the initial conditions given in (\ref{e:IC_decay}). At a specific instant of time $t_0$, when the oscillon has stabilised, we fit our proposed field configuration (\ref{e:oscillon_proposal}) to the actual field configuration obtained from the field theory simulation. In particular, we choose $t_0$ to correspond to a turning point in the evolution of the oscillon, allowing the kinetic energy to be neglected \cite{Wereszczynski:2023}. Consequently, it is reasonable to assume that $\dot{a}(0) = \dot{b}(0) = 0$.
     
\begin{figure}[htb]
        \centering
        \includegraphics[width=1.00\linewidth]{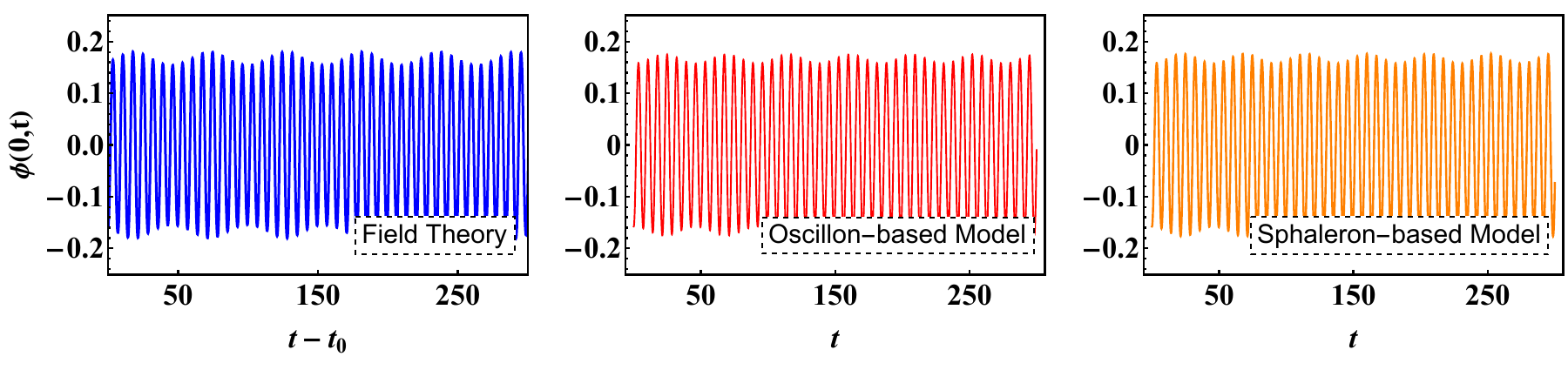}
        \caption{Comparison between the dynamics of field theory and least squares fitting in oscillon-based and sphaleron-based models for $s = 0.1$, with the fitting performed at time $t_0 = 7007.69$. The initial amplitude of the unstable mode used in the field theory simulation is $A = 0.001$.} \label{f:comparison_barrier_01}
    \end{figure}

\begin{figure}[htb]
        \centering
        \includegraphics[width=1.00\linewidth]{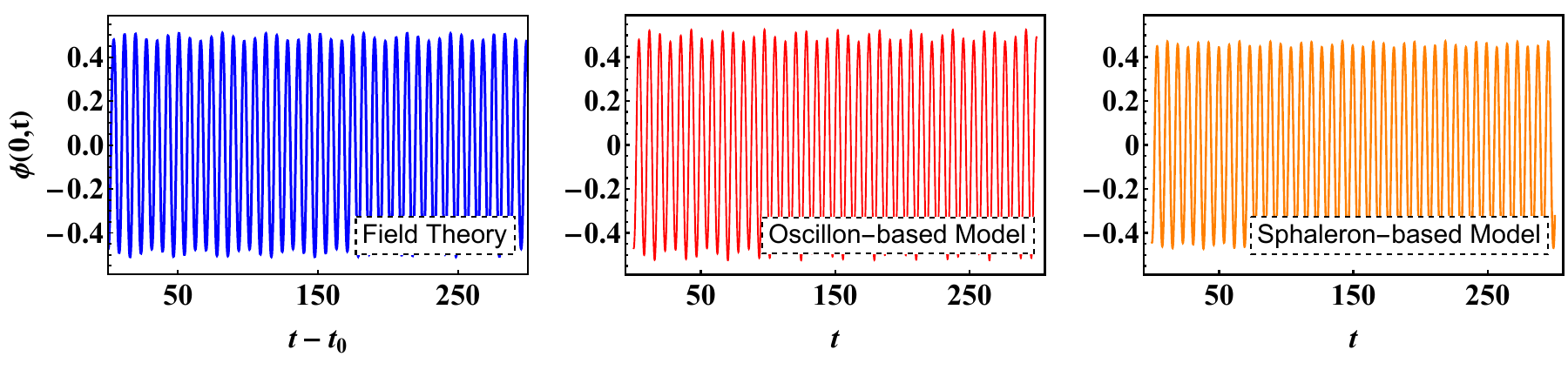}
        \caption{Comparison between the dynamics of field theory and least squares fitting in oscillon-based and sphaleron-based models for $s = 1.267$, with the fitting performed at time $t_0 = 7109.87$. The initial amplitude of the unstable mode used in the field theory simulation is $A = 0.1$.} \label{f:comparison_barrier_1267}
    \end{figure}

    We consider two representative examples corresponding to sphalerons with different unstable mode frequencies, and therefore, different spatial sizes. Specifically, we have assumed the values $s = 0.1$ and $s = 1.267$. The comparison between field theory and the effective theory based on the oscillon (\ref{e:oscillon_proposal}) is shown in \autoref{f:comparison_barrier_01} and \autoref{f:comparison_barrier_1267} by representing the values of the field configuration at the origin $\phi(0,t)$.
    
\begin{figure}[htb]
        \centering
        \includegraphics[width=1.00\linewidth]{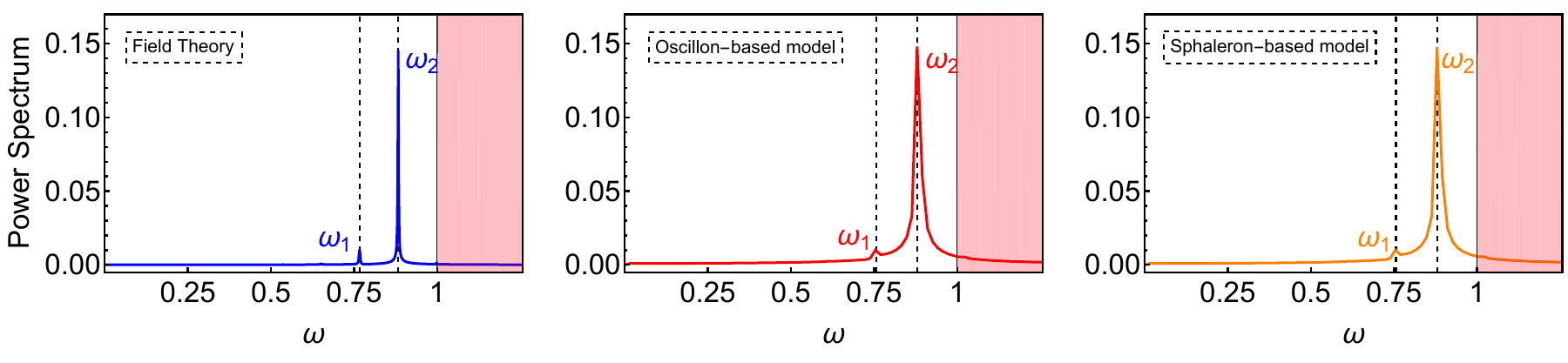}
        \caption{Fourier transform of the field values at the origin $\phi(0,t)$ for the barrier model. The shaded region represent the continuum, and the vertical dashed lines the main frequencies involved in the oscillon evolution for $s = 0.1$.} \label{f:PW_barrier_s0.1}
    \end{figure}

\begin{figure}[htb]
        \centering
        \includegraphics[width=1.00\linewidth]{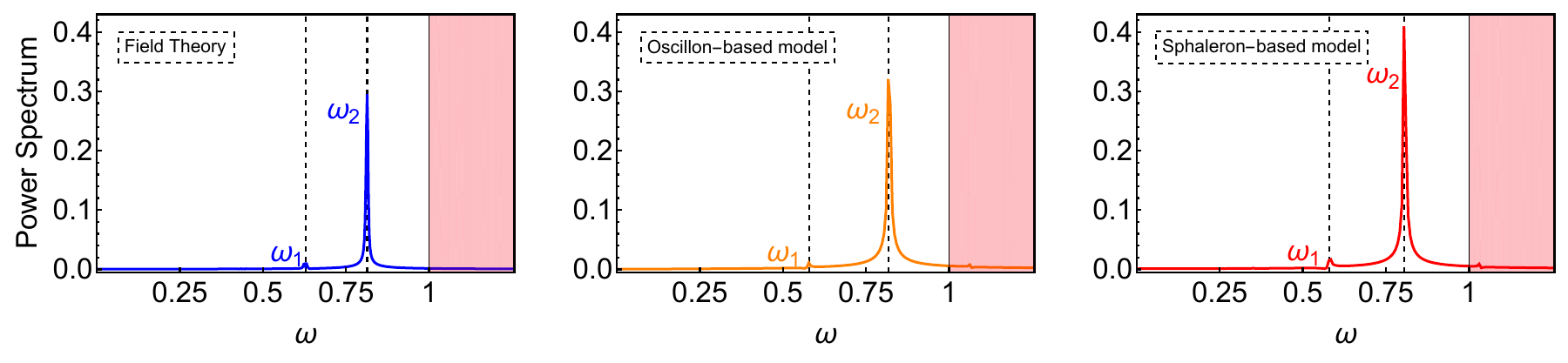}
        \caption{Fourier transform of the field values at the origin $\phi(0,t)$ for the barrier model. The shaded region represent the continuum, and the vertical dashed lines the main frequencies involved in the oscillon evolution for $s = 1.267$.} \label{f:PW_barrier_s1267}
    \end{figure}

    It is clearly visible that the effective model captures the amplitude modulation in both cases. This double quasi-periodic structure arises from the presence of two dominant frequencies: the oscillation frequency $\omega_{\text{osc}}$ and the modulation frequency $\omega_{\text{mod}}$. We have computed the Fourier transform of the field values at the origin, $\phi(0,t)$, for each model. In \autoref{f:PW_barrier_s0.1} and \autoref{f:PW_barrier_s1267}, we show the corresponding power spectra for $s = 0.1$ and $s = 1.267$, respectively. In both cases, two prominent peaks are observed at frequencies $\omega_1$ and $\omega_2$, from which we identify $\omega_{\text{osc}} = \omega_2$ and $\omega_{\text{mod}} = \omega_2 - \omega_1$. The corresponding oscillation and modulation frequencies are listed in \autoref{tab:oscilation_1} and \autoref{tab:modulation_1}. These tables show a spectacular agreement in the oscillation frequency, with only a slight deviation appearing in the modulation frequency, being more noticeable for the $s = 1.267$ case. This agreement suggests that one can describe the oscillon evolution with a degree of freedom accounting for the oscillon profile amplitude and an additional degree of freedom describing small changes in the oscillon size.

    \vspace{0.2cm}
    \begin{table}[H]
    \centering
        \begin{tabularx}{\textwidth}{YYYY}
            \toprule
            $s$ & $\omega^{ET(O)}_{\text{osc}}$ & $\omega^{ET(S)}_{\text{osc}}$ & $\omega^{FT}_{\text{osc}}$ \\
            \midrule
            0.1   & 0.881 & 0.880 & 0.881 \\
            1.267 & 0.805 & 0.817 & 0.813 \\
            \bottomrule
        \end{tabularx}
    \caption{Oscillation frequencies for $s = 0.1$ and $s = 1.267$ in field theory and in the effective models for the barrier model. The upper index O or S denotes the oscillon-based effective model or the sphaleron-based effective model. The initial amplitude of the unstable mode used in the field theory simulation is $A = 0.001$, and the fitting was performed at time $t_0 = 7007.69$.}
    \label{tab:oscilation_1}
    \end{table}
    \begin{table}[H]
    \centering
        \begin{tabularx}{\textwidth}{YYYY}
            \toprule
            $s$ & $\omega^{ET(O)}_{\text{mod}}$ & $\omega^{ET(S)}_{\text{mod}}$ & $\omega^{FT}_{\text{mod}}$ \\
            \midrule
            0.1 & 0.124 & 0.124 & 0.114 \\
            1.267 & 0.225 & 0.238 & 0.185 \\
            \bottomrule
        \end{tabularx}
    \caption{Modulation frequencies for $s = 0.1$ and $s = 1.267$ in field theory and in the effective models for the barrier model. The upper index O or S denotes the oscillon-based effective model or the sphaleron-based effective model. The initial amplitude of the unstable mode used in the field theory simulation is $A = 0.1$, and the fitting was performed at time $t_0 = 7109.87$.}
    \label{tab:modulation_1}
    \end{table}
        
    To support this hypothesis, we suggest an oscillon profile entirely based on the sphaleron itself. As mentioned at the beginning of this section, describing the oscillon in terms of the unstable mode of the sphaleron likely does not yield accurate results. However, we can adopt an alternative strategy to ensure a significant overlap between the sphaleron degrees of freedom and the oscillon. 
    
    In particular, let us now approximate the oscillon profile by
    \begin{eqnarray}
        \Phi = a\, \sqrt{\dfrac{\sinh 2s }{\cosh 2s + \cosh 2 \lambda x}}\,,
    \end{eqnarray}
    that is, the sphaleron profile $\Phi_{sph}^b(x;s)$ modified by the parameters $a$ and $\lambda$. These coefficients account for the amplitude and the size, respectively. Assuming in addition a small spatial-scale deformation $\lambda \rightarrow \lambda + \epsilon$ and expanding up to first order, we obtain
    \begin{eqnarray}
        \Phi = a\,\sqrt{\dfrac{\sinh 2s}{\cosh 2s + \cosh 2 \lambda x}} -  \epsilon\, a\,\dfrac{x\, \sqrt{\sinh 2s} \sinh 2\lambda x}{(\cosh 2s + \cosh 2\lambda x )^{3/2}}\,.
    \end{eqnarray}
     
    Finally, we promote the amplitudes to independent collective coordinates, so the previous expression reads  
    \begin{eqnarray}\label{e:sphaleron_proposal}
        \Phi(x;a,b) = a(t)\sqrt{\dfrac{\sinh 2s }{\cosh 2s + \cosh 2 \lambda x}}  + b(t)\dfrac{x\, \sqrt{\sinh 2s} \sinh 2\lambda x}{(\cosh 2s + \cosh 2\lambda x )^{3/2}}\,.
    \end{eqnarray}
    
    The new field configuration (\ref{e:sphaleron_proposal}), unlike the ansatz (\ref{e:oscillon_proposal}), is entirely based on sphaleron degrees of freedom. Remarkably, the metric coefficients and the effective potential of the associated effective model can be computed analytically. However, due to the length of some of the expressions involved, they have been omitted here for the sake of clarity. To test the validity of this new effective model, we consider the previously analysed cases $s = 0.1$ and $s = 1.267$ , using the same initial conditions for the field theory simulations (i.e., the initial value of the unstable mode amplitude $A$) and the same turning points, defined by the instant of time $t_0$. The values of the field configuration at the origin, $\phi(0,t)$, obtained from the sphaleron-based effective model (\ref{e:sphaleron_proposal}), are compared with those from full field theory and the oscillon-based effective model in \autoref{f:comparison_barrier_01} and \autoref{f:comparison_barrier_1267}. The involved frequencies are presented in \autoref{tab:oscilation_1} and in \autoref{tab:modulation_1} respectively.

    In light of the results obtained, the sphaleron-based model adapted to the oscillon (\ref{e:sphaleron_proposal}) also determines the  frequencies involved in the oscillon evolution with high accuracy. Furthermore, the agreement is comparable to that obtained with the oscillon-based model (\ref{e:oscillon_proposal}). In both cases, the Derrick mode can be interpreted as a kind of internal excitation of the oscillon that is responsible for the amplitude modulation.    

\subsection{Effective description of oscillon dynamics in the well model}

   As seen in \autoref{s:decay_Deformed}, for moderate and large values of the parameter $s$, the evolution of the oscillon formed after the sphaleron decay in the well model is analogous to the one observed in the barrier model. In particular, the oscillon exhibits the characteristic double quasi-periodic structure that we have already identified in the previous case. This close similarity suggests that the same idea underlying the effective models introduced before can also be implemented here to study the dynamics of oscillons in the well model.

    Regarding the oscillon-based model, the effective description is again obtained by substituting the field proposal (\ref{e:oscillon_proposal}) into the Lagrangian density (\ref{e:Action_Phi6_Deformed}) with the potential given by (\ref{e:potential_Well}) and subsequently integrating over space. Since the metric coefficients only depend on the time derivatives of the field ansatz, they coincide exactly with those given in (\ref{e:metric_oscillon}). The differences therefore arise solely in the effective potential. Once again, the calculation proceeds straightforwardly and can be done analytically; however, the explicit expression is rather lengthy and will not be displayed here explicitly.

    In addition, the sphaleron profile can be adapted to the oscillon profile by assuming the following deformation
    \vspace{0.3cm}
    \begin{equation}
        \Phi = a \,\left(1 - \dfrac{2 \sinh s \cosh \lambda \,x/2}{\sqrt{3 + \cosh 2s + 2\cosh \lambda \,x \sinh^2 s}}\right)\,. 
    \end{equation}
    Once again, the coefficients $a$ and $\lambda$ determine the amplitude and the spatial extent of the sphaleron, respectively. By performing a small spatial-scale deformation $\lambda \rightarrow\lambda + \epsilon$ and promoting the amplitudes to time-dependent parameters we are left with
    \begin{align}\label{e:Spaleron_proposal_well}
        \Phi(x;a,b) =& \hspace{0.1cm} a(t)\,\left(1 - \dfrac{2 \sinh s \cosh \lambda \,x/2}{\sqrt{3 + \cosh 2s + 2\cosh \lambda \,x \sinh^2 s}}\right)\nonumber\\
        &+ b(t)\,\dfrac{4\,x \sinh(x\, \lambda / 2) \sinh s}{\left( 3 + \cosh 2s + 2 \cosh \lambda\,x\sinh^2 s \right)^{3/2}}\,.
    \end{align}
    To determine the initial conditions $a(0),\, b(0)$ and fix the value of $\lambda$, we adopt the same strategy as in the barrier model. First, we evolve the sphaleron in full field theory using the initial condition given in  (\ref{e:IC_decay}) for a given amplitude $A$. Then, we adjust the field theory profile to our field proposal at an instant of time $t_0$ consisting in a turning point, which fixes $a(0),\, b(0)$ and $\lambda$ and allow us to assume that $\dot{a}(0) = \dot{b}(0) = 0$.

\begin{figure}[htb]
        \centering
        \includegraphics[width=1.02\linewidth]{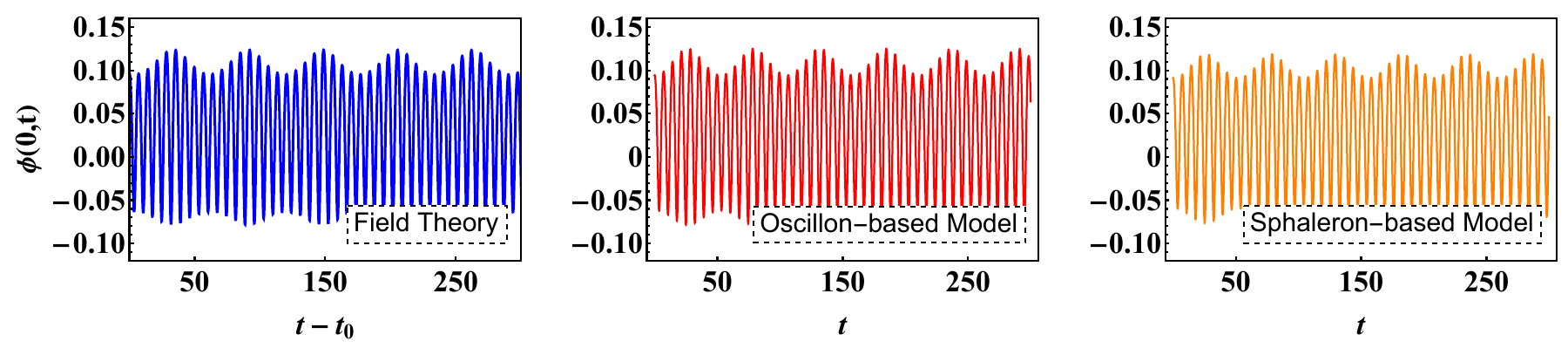}
        \caption{Comparison between the dynamics of field theory and least squares fitting in oscillon-based and sphaleron-based models for $s = 1.0$, with the fitting performed at time $t_0 = 7009.29$. The initial amplitude of the unstable mode used in the field theory simulation is $A = 0.01$.} \label{f:comparison_well_1}
    \end{figure}

\begin{figure}[htb]
        \centering
        \includegraphics[width=1.02\linewidth]{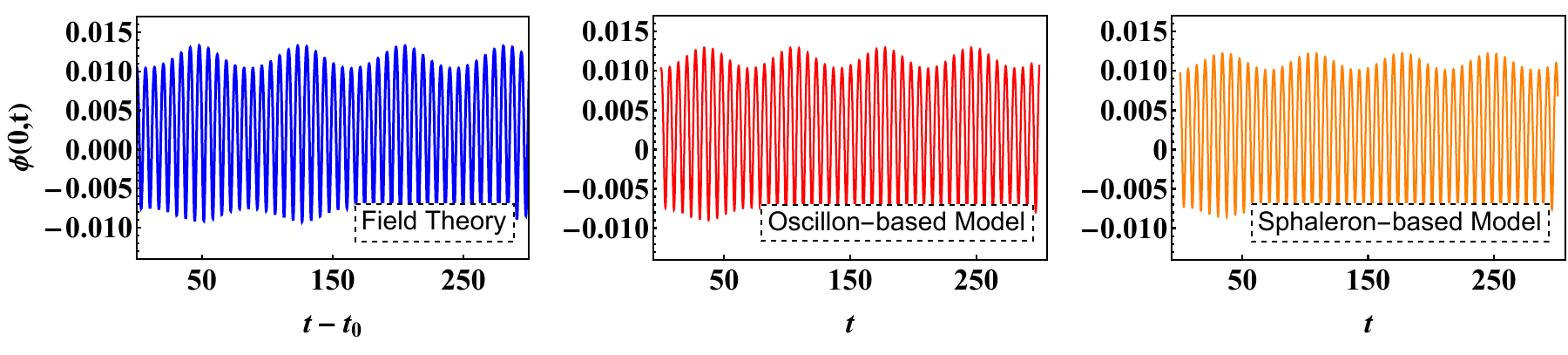}
        \caption{Comparison between the dynamics of field theory and least squares fitting in oscillon-based and sphaleron-based models for $s = 2.16$, with the fitting performed at time $t_0 = 7051.58$. The initial amplitude of the unstable mode used in the field theory simulation is $A = 0.001$.} \label{f:comparison_well_216}
    \end{figure}

    We have considered the cases $s = 1.0$ and $s = 2.16$, for which the sphalerons exhibit significantly different spatial sizes. In \autoref{f:comparison_well_1} and \autoref{f:comparison_well_216}, we illustrate the values of the field at the origin, $\phi(0,t)$, as obtained from the full field theory and from the effective models based on equations (\ref{e:oscillon_proposal}) and (\ref{e:Spaleron_proposal_well}) for $s = 1.0$ and $s = 2.16$, respectively. 
    \vspace{-0.35cm}
    
\begin{figure}[htb]
        \centering
        \includegraphics[width=1.00\linewidth]{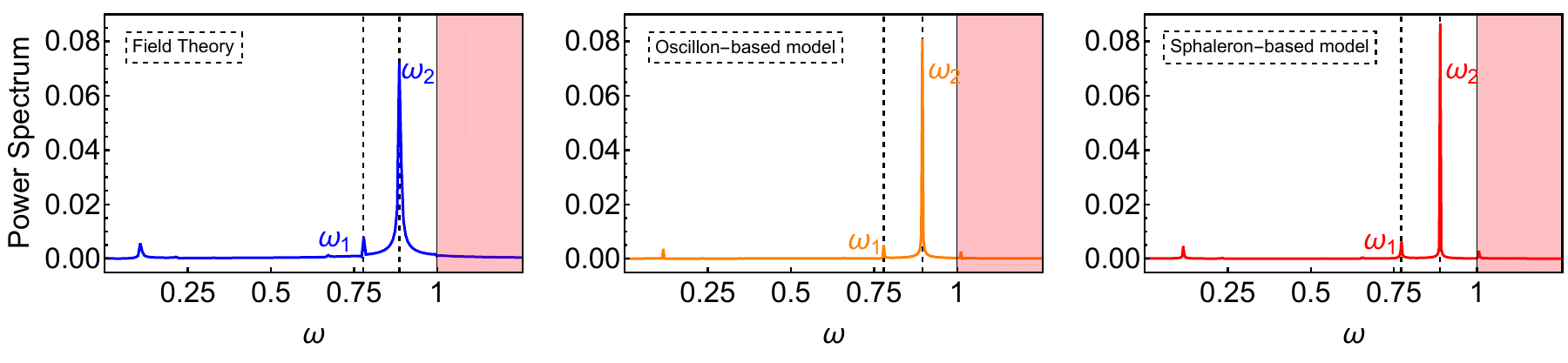}
        \caption{Fourier transform of the field values at the origin $\phi(0,t)$ for the well model. The shaded region represent the continuum, and the vertical dashed lines the main frequencies involved in the oscillon evolution for $s = 1.0$.} \label{f:PW_well_s1}
    \end{figure}

\begin{figure}[htb]
        \centering
        \includegraphics[width=1.00\linewidth]{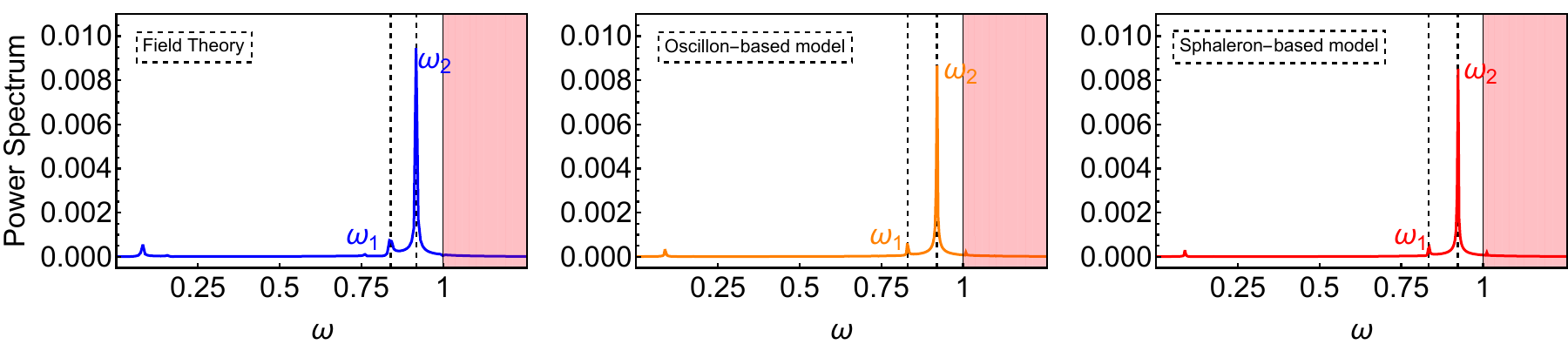}
        \caption{Fourier transform of the field values at the origin $\phi(0,t)$ for the well model. The shaded region represent the continuum, and the vertical dashed lines the main frequencies involved in the oscillon evolution for $s = 2.16$.} \label{f:PW_well_s216}
    \end{figure}

    It is evident that these models successfully reproduce the amplitude modulation characteristic of the oscillon in field theory. The relevant frequencies are extracted by performing a Fourier transform of $\phi(0,t)$, and the resulting spectra are shown in \autoref{f:PW_well_s1} and \autoref{f:PW_well_s216} for $s = 1.0$ and $s = 2.16$, respectively. From this analysis, we identify the primary oscillation frequency $\omega_{\text{osc}} = \omega_2$ and the modulation frequency $\omega_{\text{mod}} = \omega_2 - \omega_1$, which are summarised in \autoref{tab:oscilation_2} and \autoref{tab:modulation_2}. Once again, we observe that the effective models capture the oscillation frequency with high precision, while the modulation frequency shows only a slight deviation.

    \begin{table}[H]
    \centering
        \begin{tabularx}{\textwidth}{YYYY}
            \toprule
            $s$ & $\omega^{ET(O)}_{\text{osc}}$ & $\omega^{ET(S)}_{\text{osc}}$ & $\omega^{FT}_{\text{osc}}$ \\
            \midrule
            1.0 & 0.889 & 0.894 & 0.886 \\
            2.16 & 0.920 & 0.923 & 0.918 \\
            \bottomrule
        \end{tabularx}
    \caption{Oscillation frequencies for $s = 1.0$ and $s = 2.16$ in field theory and in the effective models for the well model. The upper index O or S denotes the oscillon-based effective model or the sphaleron-based effective model. The initial amplitude of the unstable mode used in the field theory simulation is $A = 0.01$, and the fitting was performed at time $t_0 = 7009.29$.}
    \label{tab:oscilation_2}
    \end{table}
    \begin{table}[H]
    \centering
        \begin{tabularx}{\textwidth}{YYYY}
            \toprule
            $s$ & $\omega^{ET(O)}_{\text{mod}}$ & $\omega^{ET(S)}_{\text{mod}}$ & $\omega^{FT}_{\text{mod}}$ \\
            \midrule
            1.0 & 0.117 & 0.115 & 0.108 \\
            2.16 & 0.09 & 0.088 & 0.078 \\
            \bottomrule
        \end{tabularx}
    \caption{Modulation frequencies for $s = 1.0$ and $s = 2.16$ in field theory and in the effective models for the well model. The upper index O or S denotes the oscillon-based effective model or the sphaleron-based effective model. The initial amplitude of the unstable mode used in the field theory simulation is $A = 0.001$, and the fitting was performed at time $t_0 = 7051.58$.}
    \label{tab:modulation_2}
    \end{table}

    Finally, we would like to mention that the presence of shape modes in this model naturally suggests the possibility of constructing an ansatz for the field configuration based on the sphaleron profile together with the positive bound modes, namely
        \begin{eqnarray}\label{e:modes_well_ansatz}
            \Phi(x;a,b) = \Phi^w_{sph}(x;s) + a(t)\,\eta^w_{-}(x;s) + \sum_{i = 1}^{N} b_i(t)\,\eta^w_{i}(x;s)\,.
        \end{eqnarray}
    Here $N$ denotes the maximum number of shape modes for a given value of $s$, and all internal modes are assumed to be properly normalised. However, contrary to the initial expectation, the tests performed with this ansatz in the cases studied so far do not exhibit a significant agreement with the oscillon evolution observed in the full field theory simulations. This mismatch seems to originate from the inability of expression (\ref{e:modes_well_ansatz}) to faithfully reproduce the actual oscillon profile. Although it may be some range of $s$-values for which the accordance could be better, there is not mathematical or physical guarantee for this to happen.

\subsection{Kink-antikink formation in the well model}

\begin{figure}[htb]
        \centering
        \includegraphics[width=0.55\linewidth]{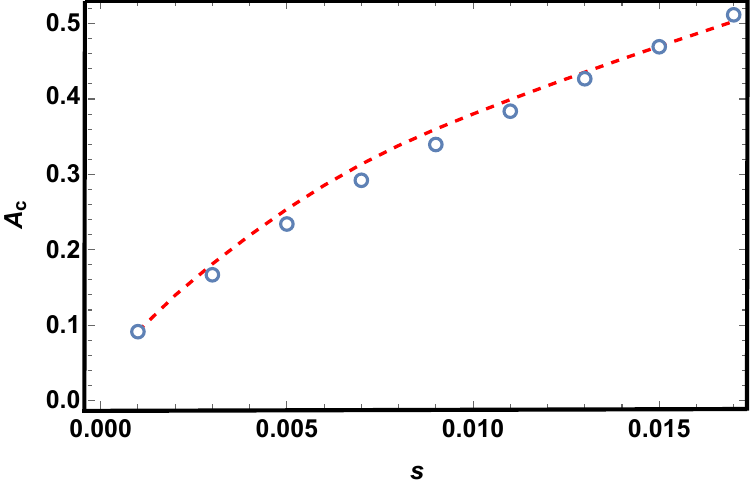}
        \caption{Dependence of the critical amplitude $A_c$ on the model parameter $s$ in the well model. The dashed line represents the theoretical prediction given by (\ref{e:Acrit}), while the dots correspond to the full numerical results.} \label{f:critical_amplitude}
    \end{figure}

    So far, we have used an effective model to describe the evolution of the oscillon following the collapse of the sphaleron. However, as briefly mentioned in \autoref{s:decay_Deformed}, a counter-intuitive phenomenon emerges for small values of the parameter $s$ in the well model. Specifically, we observed numerically that, contrary to expectations, increasing the initial amplitude of the unstable mode does not always lead to a faster decay. Instead, there exists a critical amplitude of the unstable mode, $A_c$, at which the sphaleron exhibits oscillatory behaviour with a significantly enhanced lifespan. Furthermore, for amplitudes above $A_c$, the sphaleron expands as two outgoing, accelerating kinks. This is not a singular phenomenon, but it occurs when $s \lesssim 0.017$. The dependence of the critical amplitude $A_c$ on the parameter $s$ is illustrated in \autoref{f:critical_amplitude}. Notably, this intriguing evolution can be seen for arbitrarily small values of the initial unstable amplitude $A$ as long as the parameter $s$ is small enough.
    
\begin{figure}[htb]
        \centering
        \includegraphics[width=0.83\linewidth]{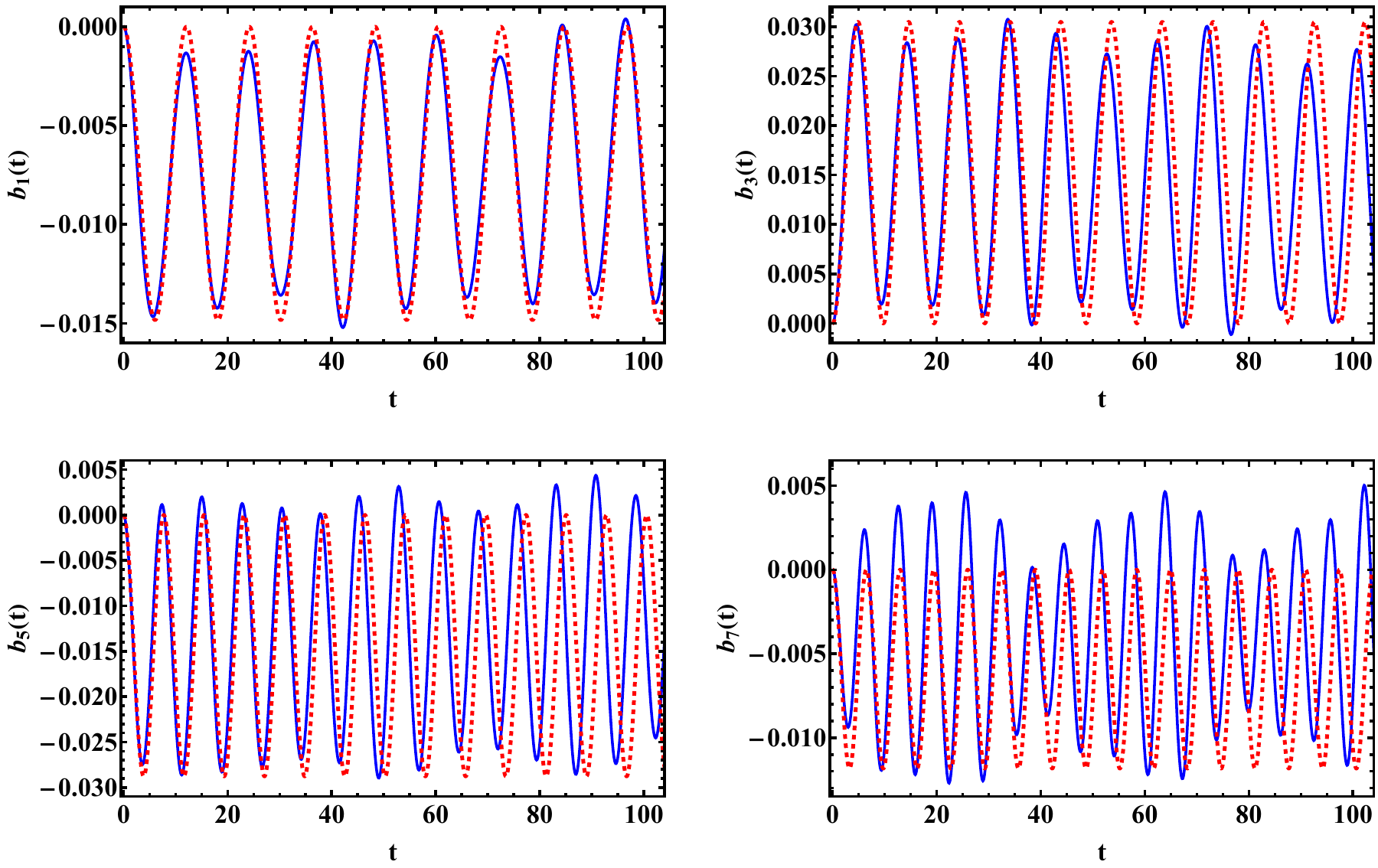}
        \caption{Excitation of the even-parity shape modes by their coupling to the unstable mode for $s = 0.005$ and $A = 0.25$. The solid line represents the field theory dynamics and the dashed line the approximate analytical expression (\ref{e:bi}).} \label{f:projections}
    \end{figure}

\begin{figure}[htb]
        \centering
        \includegraphics[width=1.00\linewidth]{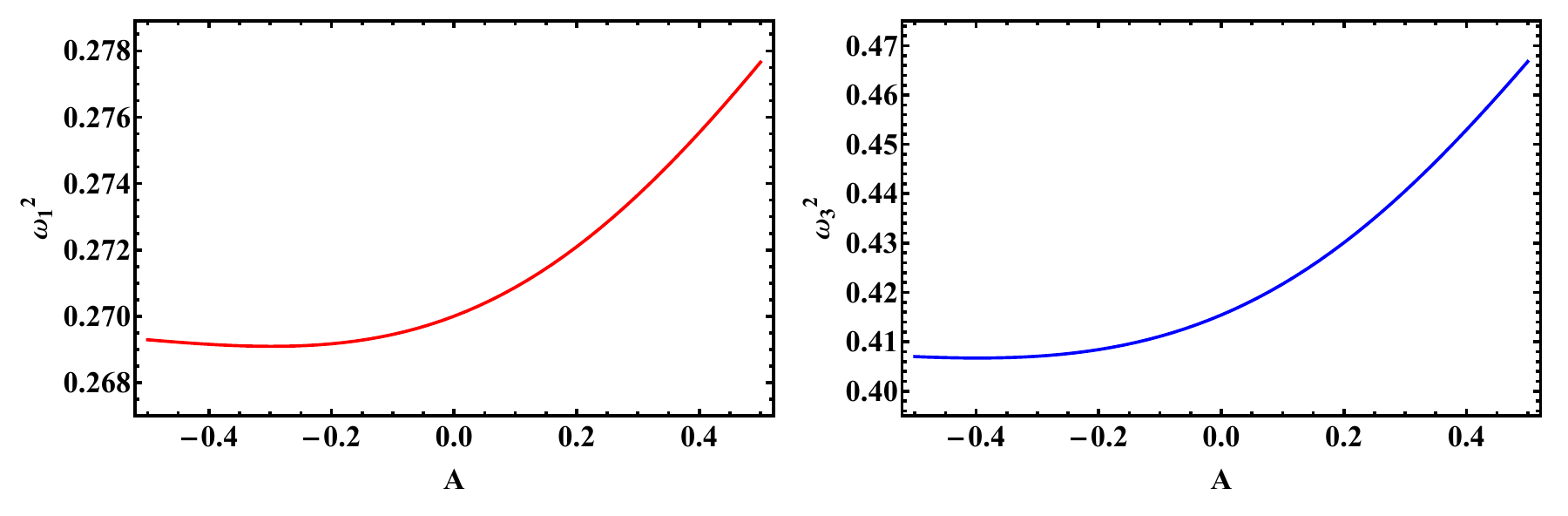}
        \caption{Spectral flow of the first two even-parity modes as a function of the amplitude $A$ of the unstable mode, for a sphaleron with $s = 0.005$.} \label{f:SpectralFlow_Modes}
    \end{figure}

    This dynamical process must be related to the influence of the internal modes. Therefore, the first step is to explore the subsequent excitation of the internal modes in the sphaleron evolution. In order to do that, we consider the effective model based on the ansatz (\ref{e:modes_well_ansatz}), as it includes the unstable mode and the positive bound modes. Assuming that $a(t) \sim \mathcal{O}(A)$ and $b_i(t) \sim \mathcal{O}(A^2)$, where $A$ is the initial amplitude of the unstable mode, the following effective Lagrangian results up to $\mathcal{O}(A^4)$
    \begin{eqnarray}
        L[a,b_i] = \dfrac{1}{2}\dot{a}^2 - \dfrac{1}{2}\omega_{-}^2\,a^2 + \dfrac{1}{2}\sum_{i = 1}^{N}\dot{b}_i^2 - \dfrac{1}{2}\sum_{i = 1}^{N}\omega_{i}^2\,b_i^2- \alpha\, a^3 - \beta\, a^4 - \sum_{i = 1}^{N}\gamma_i\, a^2\, b_i\,. \label{e:effective_Lagrangian}
    \end{eqnarray}
    The coefficients $\alpha$, $\beta$, and $\gamma_i$ are explicitly given by
    \begin{eqnarray}
        \alpha \hspace{-0.2cm}&=&\hspace{-0.2cm} \dfrac{1}{6}\int_{\mathbbm{R}} U_{w}^{(3)}[\Phi^w_{sph}(x;s)]\,(\eta^w_{-}(x;s))^3\, dx\,,\\
        \beta\hspace{-0.2cm}&=&\hspace{-0.2cm} \dfrac{1}{24}\int_{\mathbbm{R}} U_{w}^{(4)}[\Phi^w_{sph}(x;s)]\,(\eta^w_{-}(x;s))^4 \, dx\,,\\
        \gamma_i \hspace{-0.2cm}&=&\hspace{-0.2cm} \dfrac{1}{2} \int_{\mathbbm{R}} U_{w}^{(3)}[\Phi^w_{sph}(x;s)]\,(\eta^w_{-}(x;s))^2\,\eta^w_{i}(x;s)\, dx\label{e:gamma_i}\,.
    \end{eqnarray} 
    The equation of motion for the $b_i$ variables can be readily integrated, yielding the following expression
    \begin{equation}\label{e:bi}
        b_i(t) = A^2\,\dfrac{\gamma_i}{\omega_i^2}(\cos(\omega_i t) - 1)\,,
        \end{equation}
    for the initial conditions
    \begin{eqnarray}
        a(0) = A\,,\hspace{0.6cm} b_i(0) = 0\,,
    \end{eqnarray}
    corresponding to a sphaleron with its unstable mode initially excited with amplitude $A$, while all shape modes are initially unexcited. Note that only the shape modes with the same symmetry as the unstable mode are excited at lowest order, since the shape modes with opposite symmetry do not contribute to the integral (\ref{e:gamma_i}). Indeed, we have verified this numerically, and the approximate solution (\ref{e:bi}) captures with good agreement the initial excitation of the shape modes in field theory. A comparison between our approximate analytical expression and the full field theory dynamics is shown in \autoref{f:projections}. 
    As seen in the figure, there is initially very good agreement in both the amplitude and oscillation frequency. The deviations observed at later times are expected and likely stem from higher-order corrections, as well as from the squeezing of the sphaleron, which modifies its internal structure and, consequently, the frequency of the modes. Indeed, if we compute the frequency of the internal modes for the configuration $\phi(x;s) = \Phi^w_{sph}(x;s) - A\eta^w_-(x;s)$, we observe a dependence on $A$; see \autoref{f:SpectralFlow_Modes}.
    
    Therefore, we conclude that the excitation of the unstable mode in the collapsing direction has two main effects: on the one hand, the sphaleron shrinks, giving rise to a static attractive force between the constituent kinks. On the other hand, the positive bound modes are excited through non-linear couplings between the unstable mode and each of them. Moreover, the squeezing of the sphaleron causes a spectral flow of these modes. The effective Lagrangian (\ref{e:effective_Lagrangian}) suggests therefore that the system is subject to an additional force of the form
    \begin{equation}\label{e:Force}
        F_i = - \dfrac{1}{2}\dfrac{d \omega^2(A)}{dA}b_i^2\,.
    \end{equation}
    Since all these modes exhibit a minimum, the induced force is repulsive\footnote{Note that the minimum is not located at $A = 0$, which is a consequence of the lack of reflection symmetry in the theory.}. As a result, there are two competing opposite forces that may balance under certain conditions. 
    
    Remarkably, an approximate analytical expression that predicts the critical amplitude required for sphaleron stabilisation can be found. To achieve this, we adopt a perturbative approach. First, we assume the following approximate field configuration
    \begin{equation}\label{e:crit_ansatz}
        \phi=\phi_0(x,A_c) + A\,\eta^w_{-} + \sum_i A_i\,\phi_i^{(1)} + \sum_i A_i^2\,\phi_i^{(2)}\,,    
    \end{equation}
    where $\phi_0(x,A_c) = \Phi^w_{sph}(x;s) - A_c \,\eta^w_{-}(x;s)$ represents the initial sphaleron deformed along the unstable mode direction. The terms $\phi^{(1)}$ and $\phi^{(2)}$ denote the first and second-order corrections to the internal modes, respectively. Then, we have introduced the expression (\ref{e:crit_ansatz}) into the equation of motion (\ref{e:well_EOM}) and we have solved the resulting equation order by order in $A_i$, looking for the value of $A_c$ such that $A$ is purely oscillatory. This condition should lead to the stationary solution we are searching for.

    At zeroth order in $A_i$ and first order in $A$ we obtain 
    \begin{equation}\label{e:A_neg}
        \ddot{A} + \omega_{-}^2A = A_c\,\omega_{-}^2\,, 
    \end{equation}
    after projecting onto the unstable mode. At first order in $A_i$, we identify the Schrödinger equation for the internal modes, therefore
    \vspace{0.2cm}
    \begin{equation}\label{e:mod_i}
        \phi_i^{(1)} = \eta^w_i(x;s)\cos(\omega_i t)\,,
    \end{equation}
    and $A_i$ is given by (\ref{e:bi}). Expanding now at second order in $A_i$, we deduce the equation for the second order correction to the internal modes
    \begin{equation}\label{e:second_mod}
        \partial_{\mu}\partial^{\mu} \phi_i^{(2)} + U''[\Phi^w_{sph}]\,\phi_i^{(2)} = - \frac{1}{2}U^{(3)}[\Phi^w_{sph}]\,\left(\sum_i \phi_i^{(1)} \right)^2\,.
    \end{equation}
    Projecting (\ref{e:second_mod}) onto the unstable mode and taking into account (\ref{e:A_neg}), we finally obtain
    \begin{eqnarray}
        \ddot{A} + \omega_{-}^2\,A = A_c\,\omega_{-}^2 - \frac{1}{2}\int_{\mathbbm{R}} U^{(3)}[\Phi^w_{sph}]\,\eta^w_{-}\,\left(\sum_i A_i\, \phi_i^{(1)}\right)^2\,dx\,. \label{e:tot_neg}
    \end{eqnarray}
    Expanding the squared term in (\ref{e:tot_neg}) we note that there are oscillatory terms of the form $\cos(\omega_i t)\cos(\omega_j t)$ and constants. Naturally, the oscillatory terms do not lead to exponential growth of the unstable amplitude $A$. Therefore, the vanishing condition of the constant terms should account for the stabilisation at lowest order. The constant terms are
    \begin{equation}
        P(A_c) = A_c\,\omega_{-1}^2 - \frac{1}{4} A_c^4\sum_i \frac{\,\gamma_i^2\,\delta_i}{\omega_i^4}\,, 
    \end{equation}
    where 
    \begin{equation}
        \delta_i = \int_{\mathbbm{R}} U^{(3)}[\Phi^w_{sph}(x;s)]\,\eta^w_{-}(x;s)\,\eta^w_i(x;s)^2\,dx\,.
    \end{equation}
    Imposing $P(A_c) = 0$ we obtain an approximate expression for the critical amplitude 
    \begin{equation}\label{e:Acrit}
        A_c = \biggl(\frac{4\,\omega_{-}^2}{\sum_i \tfrac{\gamma_i^2 \delta_i}{\omega_i^4}}\biggr)^{1/3}\,.
    \end{equation}

    In \autoref{f:critical_amplitude} we compare the theoretical prediction (\ref{e:Acrit}) with the full numerical computation. It is clearly visible that there is an astonishing agreement between both results. Moreover, we have to emphasise that the expression (\ref{e:Acrit}) is completely general, thereby being valid in principle for any theory where sphalerons have internal modes. Although our analysis has focused on a particular model, it is expected that this mechanism can appear in any model with unstable sphaleron-like solutions that support internal modes. As we have seen, the key ingredient responsible for the stabilisation of the sphaleron is the excitation of internal modes, which are capable of compensating the sphaleron instability.

    To conclude, we note that a vibrating sphaleron has also been obtained in \cite{Manton:2023} by fine-tuning the initial amplitude of the unstable mode. However, this approach also requires the initial excitation of the internal modes. 
    
\section{Summary and conclusions}\label{s:Summary_Deformed}

    In this chapter, we have examined the decay of unstable sphaleron-type solutions within two deformations of the $\phi^6$ model. The motivation for studying these particular models is threefold. Firstly, both reproduce kink–antikink configurations of the $\phi^6$ model in a certain limit. Secondly, each supports analytical sphaleron-type solutions. Thirdly, despite their similarities, only one of the models possesses positive internal modes. This distinction leads to crucial differences in their respective decay dynamics.

    In the first case, the $\phi^6$ barrier model, the sphaleron lacks positive internal modes. Exciting the sphaleron along the contraction direction causes it to collapse, while excitation in the opposite direction leads to an accelerated expansion. Upon collapse, the sphaleron decays into an oscillon, which then radiates very slowly. We have investigated this resulting oscillon from various perspectives. In the first approach, we have considered the amplitude of the leading term in a small-amplitude expansion as an effective degree of freedom \cite{Navarro:2023,Fodor:2008}, along with its Derrick mode. In the second, we have used a profile adapted to the oscillon profile, based on the sphaleron itself. Typically, the oscillon exhibits a modulated amplitude characterised by two distinct frequencies, a feature that is also reflected in the effective models.

    In the second case, the $\phi^6$ well model, the sphaleron possesses positive internal modes, whose number increases as the deformation parameter decreases. For large values of the deformation parameter, the sphaleron collapses and decays into a modulated oscillon, as in the barrier model. However, for small values, the dynamics change significantly. Under small-amplitude excitations of the unstable mode, the sphaleron collapses and then undergoes bounces. This phenomenon is intimately linked to the presence of positive internal modes: during collapse, these modes can store energy, which is subsequently transferred back to the constituent kinks, i.e., the sphaleron walls, in the form of kinetic energy.

    Even more intriguingly, as the sphaleron becomes increasingly compressed by the action of the unstable mode, a critical value for its amplitude emerges at which a stationary configuration is formed. At this threshold, the excitation of the internal modes counterbalances the attractive force along the unstable direction, resulting in an oscillatory sphaleron. This mechanism can be interpreted as a means of stabilising unstable solutions. As we have seen, the lifespan of the sphaleron at the critical value is considerably extended with respect to the non-perturbed sphaleron. Beyond this critical value, the sphaleron expands. Notably, this mechanism is only triggered during contraction, as the internal modes always exert an outward force on the constituent kinks. A related effect, known as the \textit{thick spectral wall}, has also been observed in kink–antikink pairs within near-BPS theories \cite{Wereszczynski:2019}. Notably, we have succeeded in determining a general expression for the critical amplitude that can be applied to any model hosting sphaleron-like solutions with internal modes.

    As a result, we conclude that the direction along which the sphaleron decays depends crucially on the critical amplitude of the negative mode. Although we have assumed a particular initial perturbation, we remark that a generic perturbation of the sphaleron has an overlap with the unstable mode, in such a way that other scenarios could trigger this phenomenon. From a broader perspective, the mechanisms described here are expected to be reproducible in more general $1+1$ dimensional models \cite{Manton1:1988,Bochkarev:1987,Mottola:1989,Funakubo:1990} and in higher-dimensional theories. Naturally, this phenomenon would be especially relevant in the study of the electroweak sphaleron, which is known to host at least one internal positive mode \cite{Brihayev:1990, Brihaye:1992}. Concretely, most mechanisms of magnetic field generation in the universe produce non-helical fields. However, it is well known that electroweak baryogenesis generates helical fields. During the decay of the electroweak sphaleron, electromagnetic currents are produced, which in turn generate electromagnetic fields. The sign of the magnetic helicity is determined by the direction in which the sphaleron rolls down from the saddle point \cite{Copi:2008,Chu:2011}. As a consequence, the sign of the magnetic helicity is directly related to whether baryons or antibaryons are produced during sphaleron decay. As we have shown in this chapter, the direction along which the sphaleron decays can be modified during the evolution. Therefore, the proposed mechanism should be extended to more realistic scenarios, where it may have physical implications and could be used to probe aspects of early-universe cosmology.

        \chapter{Non-linear dynamics of sphalerons on the circle}\label{c:Sph_Circle}

This chapter is adapted from \cite{Navarro:2025}:
    \hspace{-2.5cm}
    \begin{figure}[H]
        \hspace{-2.5cm}
        \centering{\includegraphics[width=1.1\linewidth]{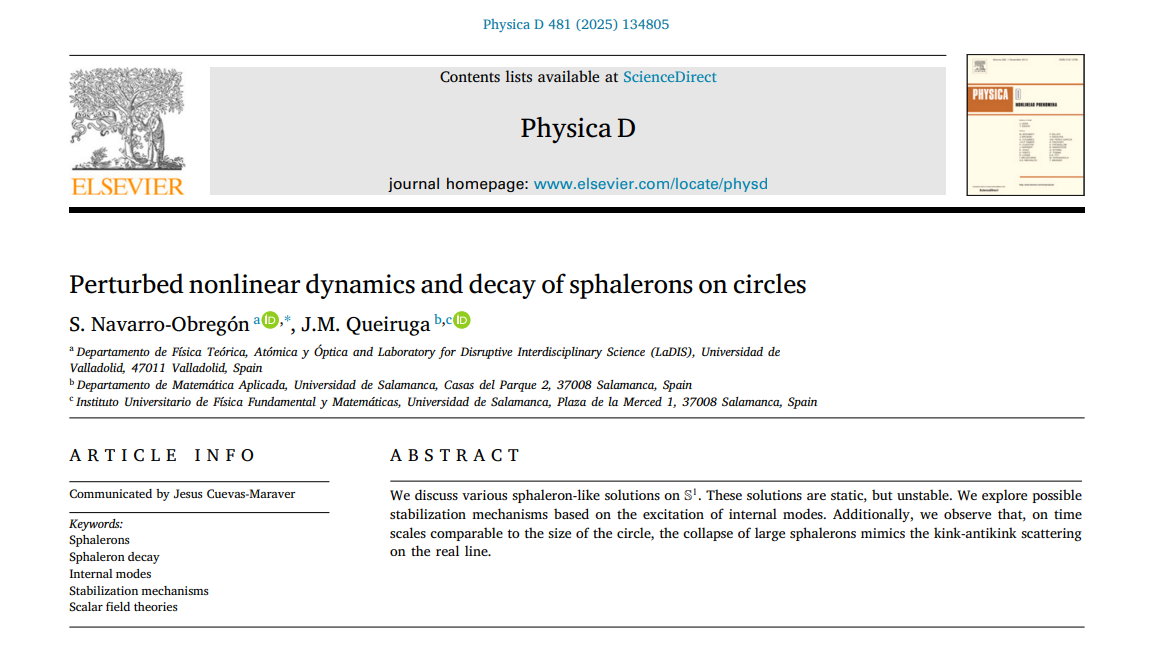}
        }
        \label{f:Thumbail_Circle}
        \hspace{-2.5cm}
    \end{figure}

\section{Introduction}

    So far, we have focused on field theories with a false vacuum structure in $1+1$ dimensions. These can be regarded as the simplest models that admit sphaleron solutions. However, a more intriguing situation in $1+1$ arises when the base space manifold is compact, e.g., a circle \cite{Manton1:1988,Funakubo:1990,Costabile:1978,Giller:1992,Park:2000}. On the one hand, quantum field theories in finite volume are closely related to those at finite temperature \cite{Pawellek:2009}. On the other hand, theories defined on a circle exhibit features that resemble sphaleron configurations in higher-dimensional models \cite{Brihayev:1990,Funakubo:1990}. Furthermore, such simplified settings often admit analytical expressions for sphaleron profiles and, in some cases, even for their internal modes. An analytical understanding of these modes is particularly valuable when computing the decay rate of the sphalerons.

    We introduced the general theory of scalar field theories defined on the circle in \autoref{c:Sphalerons}. In particular, we exposed the easiest example: the $\phi^4$ model. The existence of a sphaleron in that theory can be understood by means of the following idea: assume that a kink-antikink pair is created from one of the vacua. Such a configuration can travel around the circle in opposite directions and annihilate, leaving behind the other vacuum. Intuitively, there should exist an intermediate configuration where the attraction in one direction is exactly compensated by the attraction in the other direction when the pair is located at two opposite points on the circle. That configuration should represent the sphaleron solution of the model.

    However, the kink-antikink picture of the sphaleron is only reliable for large sphalerons. In our scenario, this regime is achieved when the size of the base space manifold is large enough. As shown in \autoref{c:BPS_Sph} and \autoref{c:Sph_False}, when the sphaleron resembles a kink-antikink profile, the unstable mode of the sphaleron corresponds to a symmetric combination of the zero mode of the individual kinks. This interpretation suggests that the sphaleron decay might mimic the kink-antikink scattering. 

    The aim of this chapter is to explore the sphaleron decay in different scalar models defined on the circle. Specifically, we will study the prototypical $\phi^4$, $\phi^6$ and the sine-Gordon models. These models differ in terms of their underlying symmetries and whether or not they admit an integrable structure. We shall therefore discuss the different phenomena arising in these models, highlighting the influence of their underlying properties. The purpose of this chapter is also to verify the expected correspondence between the decay of large sphalerons on the circle and the kink-antikink scattering on the real line. Moreover, we aim to investigate the influence of positive internal modes on the subsequent evolution of the sphaleron. Evidence that these modes can alter the sphaleron's lifetime was presented in \autoref{c:Sph_False} for a specific model. Extending this analysis to more general settings would support the idea that this mechanism is ubiquitous.

    The chapter is organised as follows. In \autoref{s:phi4-SG-phi6} we introduce the $\phi^6$ and sine-Gordon on the circle and compute the sphaleron solutions and their spectrum of perturbations. In \autoref{s:Decay_Circle}, we explore the sphaleron decay in the $\phi^4$, $\phi^6$ and the sine-Gordon models. Moreover, the influence of the positive internal modes on the lifetime of the sphaleron in \autoref{s:Stabilization_circle}. Finally, \autoref{s:Conclusions_Circle} is devoted to the conclusions.

\section{Sphalerons and their internal structure in scalar models on \texorpdfstring{$\mathbbm{S}^1$}{S1}} \label{s:phi4-SG-phi6}

    In a previous chapter, we discussed the general framework of scalar field theories defined on the circle, and we presented the $\phi^4$ model defined on this compact space. In this section, we turn our attention to other scalar field theories defined on the circle that exhibit distinct features. In particular, we focus on the $\phi^6$ and the sine-Gordon models, which will enable us to further investigate sphaleron-like configurations and their internal structure.

\subsection{\texorpdfstring{$\phi^6$}{phi6} sphaleron on the circle} 
\label{ss:phi6_circle}

\begin{figure*}[ht]
        \centering
        \includegraphics[width=1.02\textwidth]{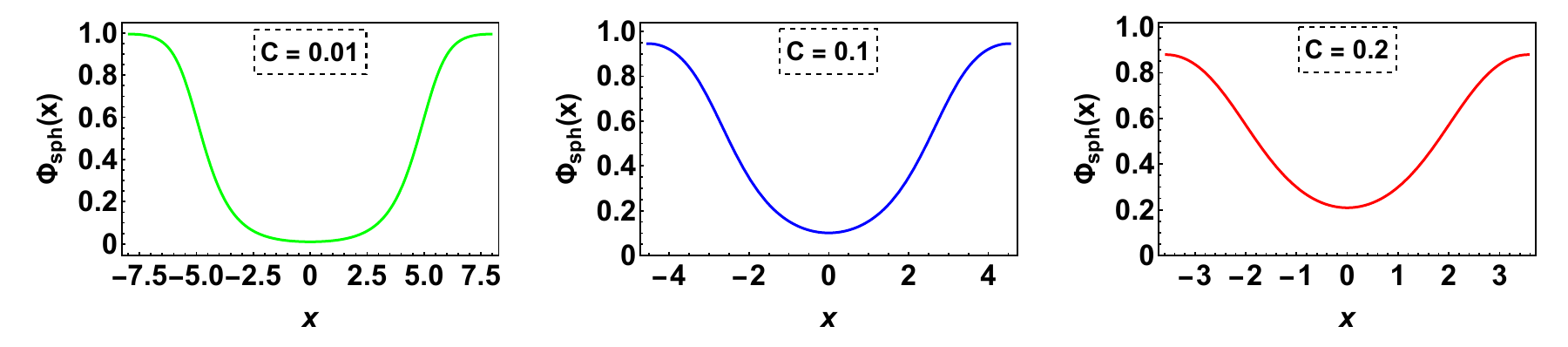}
    \caption{Profiles of the $\phi^6$ sphaleron $(\ref{e:sph_phi6})$ for different values of the parameter $C$. For $C \rightarrow 0$ the sphaleron profile resembles the antikink-kink pair in the $\phi^6$ theory on the real line.}
    \label{f:profile_phi6_circle}
    \end{figure*}

    Let us begin by studying the $\phi^6$ sphaleron, whose potential function is $U(\phi)=\frac{1}{2}\phi^2\left(\phi^2-1\right)^2$. The second-order static equation of motion associated to this model reads
    \begin{equation}\label{e:FE_phi6_circle}
        \partial_{\mu}\partial^{\mu}\phi - (1 - \phi^2)(-1+3\phi^2)\phi = 0\,.
    \end{equation}
    Assuming the static case, the resulting equation of motion can be integrated once to obtain the following first-order differential equation
    \begin{equation}\label{e:first_eq_phi6_circle}
        \phi'(x)^2-\phi^2 (\phi^2-1)^2 = -C^2\,,
    \end{equation}
    where $C$ is the integration constant. For the model at hand, the integration constant belongs to the range $0\leq C\leq 2/(3\sqrt{3})$. As explained in \autoref{ss:Sph_Circle}, values of $C$ beyond the upper bound are excluded, since in that case the periodicity condition cannot be fulfilled and no non-trivial periodic solutions exist. Since the differential equation (\ref{e:first_eq_phi6_circle}) is separable, we can integrate it straightforwardly
    \begin{equation}
        \int_{\phi(x_0)}^{\phi(x)}\frac{d\phi}{\sqrt{\phi^2(\phi^2-1)^2-C^2}} = x-x_0\,.
    \end{equation}
    This is a hyperelliptic integral that can be computed after some algebraic manipulations\footnote{Similar solutions were found in \cite{Sanati:1999} in the context of kink lattices.}
    \begin{equation} \label{e:sph_phi6}
        \Phi_{sph}(x;C) = \pm\frac{1}{\sqrt{\alpha_3 - (\alpha_3-\alpha_2)\,\text{sn}(a x, b)^2}}\,,
    \end{equation}
    where
    \begin{equation}
        a(C) = C \sqrt{\alpha_3-\alpha_1}\,,\quad b(C) =\sqrt{\frac{\alpha_3-\alpha_2}{\alpha_3-\alpha_1}}\,,
    \end{equation}
    and $\alpha_1, \alpha_2, \alpha_3$ are the roots of the following cubic polynomial
    \begin{equation}\label{e:polynomial_phi6_circle}
        C^2 \alpha_i^3-\alpha_i^2+2 \alpha_i-1=0\,,
    \end{equation} 
    satisfying $\alpha_1<\alpha_2<\alpha_3$. It can be trivially shown that the roots $\alpha_i$ of the cubic polynomial (\ref{e:polynomial_phi6_circle}) cannot degenerate except in the limiting cases $C=0$ and $C = 2/(3\sqrt{3})$. Through the spatial periodicity of the sphaleron profile (\ref{e:sph_phi6}), we can determine the relation between the circumference of the circle to which it belongs and the integration constant
    \begin{equation}\label{e:L_phi6_circle}
        L = \dfrac{2 \mathrm{K}(b)}{a(C)}\,,
    \end{equation}
    where $\mathrm{K}$ is the complete elliptic integral of the first kind introduced in \autoref{ss:Sph_Circle}. From this expression we deduce that the circumference below which the non-trivial sphaleron no longer exists is $L_{min} = L|_{C = 2/(3\sqrt{3})} = \sqrt{3} \pi$. In that limit, the sphaleron configuration reduces to the trivial solution $\phi^{(m)}= \pm 1/\sqrt{3}$ corresponding to the maxima of the potential. Conversely, the non-trivial sphaleron (\ref{e:sph_phi6}) resembles an infinitely separated antikink-kink pair when $C \to 0$. The profile of the $\phi^6$ sphaleron $(\ref{e:sph_phi6})$ is depicted in \autoref{f:profile_phi6_circle} for different values of the parameter $C$.
    
\begin{figure}[htb]
        \centering
        \includegraphics[width=0.95\linewidth]{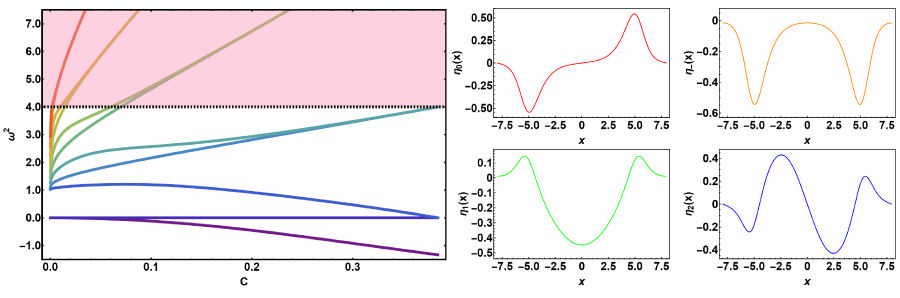}
        \caption{Left panel: Spectrum of the first eleven linear perturbations around the sphaleron (\ref{e:sph_phi6}) for different values of the parameter $C$. The modes that can appear in the shaded area are above the mass threshold of the theory. Right panel: Profiles of the zero mode $\eta_0$, the unstable mode $\eta_{-}$ and the first two internal modes $\eta_{1},\eta_{2}$ hold by the $\phi^6$ sphaleron (\ref{e:sph_phi6}) for $C = 0.01$.}
        \label{f:phi6_eigenfunctions}
    \end{figure}

    Regarding the linear spectrum of perturbation around the $\phi^6$ sphaleron, we can proceed as usual. First, we consider a perturbation of the form $\phi(x,t) = \Phi_{sph}(x;C) + \eta(x;C)e^{i\omega t}$ and introduce it into (\ref{e:FE_phi6_circle}). Then, we expand up to first order in the perturbations. After doing that, we are left with the following spectral problem 
    \begin{equation}\label{e:modes_phi6_circle}
        - \eta''(x;C) + \left[1 - 12 \Phi_{sph}^2(x;C) + 15 \Phi_{sph}^4(x;C)\right]\eta(x;C) = \omega^2 \eta(x;C)\,,
    \end{equation}
    where one must impose the constraint $\eta(x + L) = \eta(x)$. It is important to stress that, contrary to what happens in the $\phi^4$ model, the spectral problem \eqref{e:modes_phi6_circle} does not reduce to a Lam\'e-type equation; see (\ref{e:Lame_general}). This lack of analytic simplification implies that the associated Schrödinger problem must be addressed entirely by numerical means. In \autoref{f:phi6_eigenfunctions} we show the spectral flow of the modes with the model parameter $C$. We see that, except for the unstable mode, the modes degenerate in pairs in the limit of $C \rightarrow 2/(3\sqrt{3})$. Nevertheless, this degeneration breaks gradually from the lower to the higher modes as $C \rightarrow 0$. 

    As a representative example, the profiles of the first eigenfunctions corresponding to the solution $(\ref{e:sph_phi6})$ are shown in \autoref{f:phi6_eigenfunctions} for $C = 0.01$. It is easy to verify that, in the limit of $C$ close to zero, the unstable mode and the zero mode of the sphaleron can be interpreted as the antisymmetric and the symmetric combinations of the zero mode of the individual kinks, respectively. Contrary to the individual $\phi^6$ kink on the real line, the $\phi^6$ sphaleron hosts internal modes. This is in agreement with the internal mode structure of the antikink-kink pair on the real line \cite{Dorey:2011}. Therefore, we recover the delocalised modes extended across the region between the individual kinks; see \autoref{ss:phi6}. 
    
\subsection{Sine-Gordon model on the circle} \label{ss:SG_circle}

    Let us now consider the sine-Gordon model on the circle, where the theory is defined through the potential $U(\phi)= 1 + \cos \phi$. It is worth noting that there is a sign difference compared to the conventional definition given in (\ref{e:equation_SG}). Nevertheless, both formulations are related through the discrete transformation $\phi \to \phi + (2k + 1)\pi$, with $k \in \mathbbm{Z}$. As we will see, this convention will allow us to write the sphaleron solution in a more compact way.
    
\begin{figure}[H]
        \centering
        \noindent
        \begin{minipage}[t]{0.45\textwidth}
            \includegraphics[width=\linewidth]{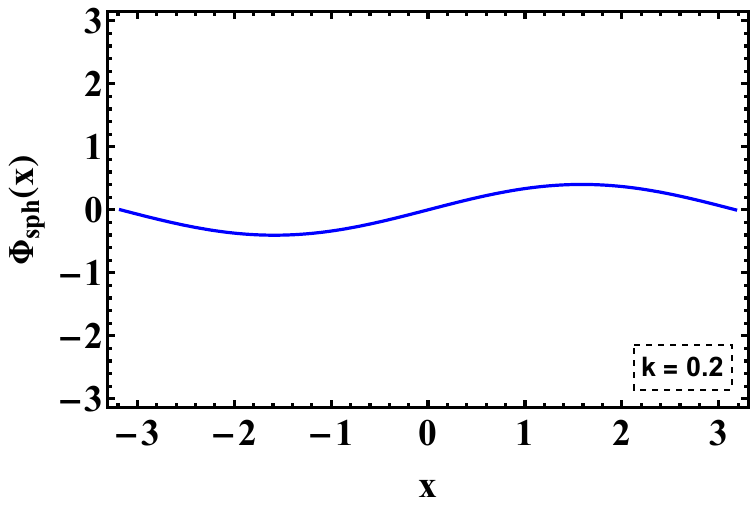}
        \end{minipage}%
        \hspace{0.8cm}
        \begin{minipage}[t]{0.45\textwidth}
            \includegraphics[width=\linewidth]{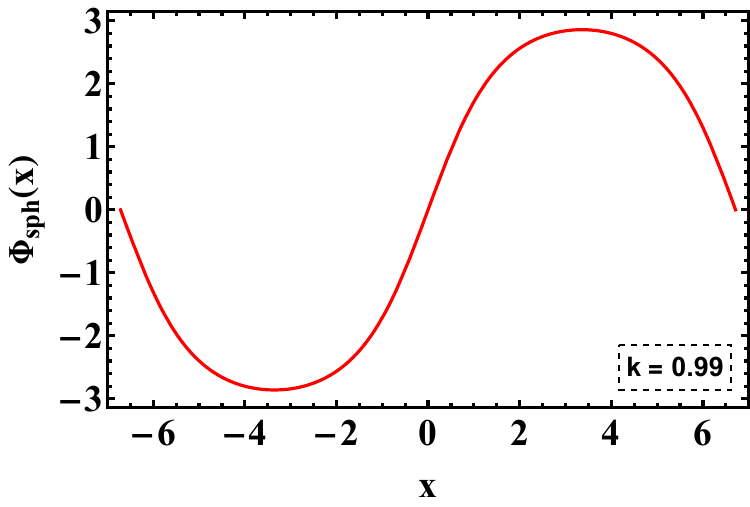}
        \end{minipage}
    \caption{Profiles of the sphaleron $\Phi_{sph}(x,k)$ given by (\ref{e:spha_sine}), plotted for representative small and large values of the parameter $k$.}
    \label{f:SG_Sph_Circle_Profiles}
    \end{figure}

    The equation of motion associated to this potential function reads
    \begin{equation}\label{e:equation_SG_Circle}
        \partial_{\mu}\partial^{\mu}\phi - \sin \phi = 0\,.
    \end{equation}
    Integrating once the static equation of motion we are left with
    \begin{equation}\label{e:Bogo_SG_circle}
        \phi_x^2 = 2(1 + \cos\phi) - C^2\,.
    \end{equation}
    This first-order differential equation holds the following sphaleron solution  
    \begin{equation}
        \Phi_{sph}(x;k) = \pm 2\, \arcsin\left(\sqrt{\dfrac{4-C^2}{4}}\, \text{sn}\left(x, \dfrac{4-C^2}{4}\right)\right)\,, \quad C\in \lbrack 0,2\rbrack\,.
    \end{equation}
    However, assuming the parametrisation
    \begin{equation}
        C = 2\sqrt{1 - k^2}\,,
    \end{equation}
    it is straightforward to verify that the sphaleron solution is simplified to the following expression \cite{Liang:1992}
    \begin{equation}\label{e:spha_sine}
        \Phi_{sph}(x;k) = \pm 2\, \arcsin\left( k\, \text{sn}(x, k^2)\right)\,, \quad k\in \lbrack 0,1\rbrack\,.
    \end{equation}
    Once more, the spatial periodicity of the solution allows us to determine the relation between the circumference of the circle and the new parameter $k$ 
    \begin{equation}
        L = 4 \mathrm{K}(k^2)\,.
    \end{equation}
    Therefore, for $k=0$, the non-trivial sphaleron reduces to the trivial one $\phi^{(m)} = 0$ when the circumference of the circle reaches the critical value $L_{min} = L|_{k = 0} = 2 \pi$. We illustrate representative profiles of these sphalerons in \autoref{f:SG_Sph_Circle_Profiles}. It is clearly visible that the sphaleron transitions from the trivial configuration to a well-separated kink–antikink pair.
    Moreover, it can be shown that the energy of the sphaleron given by $(\ref{e:spha_sine})$
    \begin{equation}
        E\left[\Phi_{sph}\right] = 8\left(\mathrm{E}(k^2) + (k^2 - 1)\mathrm{K}(k^2)\right)\,,
    \end{equation}
    can be written as
    \begin{equation}
        E\left[\Phi_{sph}\right] = 16 - 64\, e^{-L/2}\,, 
    \end{equation}
    in the limit $k \rightarrow 1$, that is, when $L \gg L_{min}$. From that expression, we deduce that the energy of the sphaleron approaches twice the sine-Gordon kink energy and that the separation between the constituent kinks is $s = L/2$ \cite{Rajamaran:1987}.

\begin{figure}[htb]
        \centering{\includegraphics[width=0.52\linewidth]{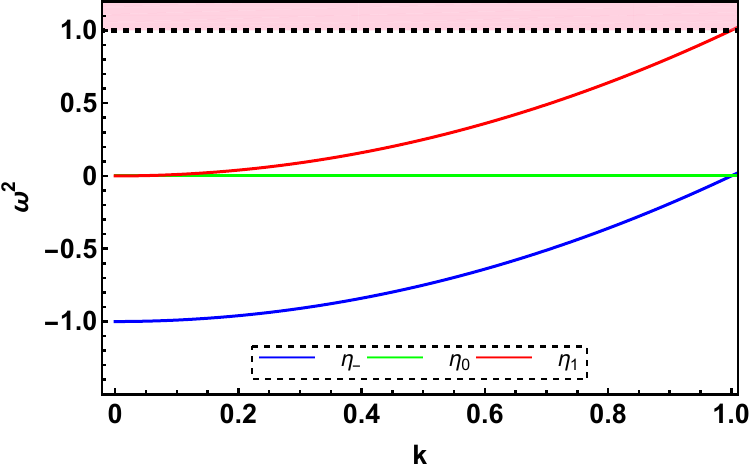}
        }
        \caption{Linear spectrum of perturbations around the sphaleron in the sine-Gordon model on the circle. We represent the unstable mode, $\eta_-$, the zero mode,  $\eta_0$ and the first positive mode $\eta_1$. The modes that can appear in the shaded area are above the mass threshold of the theory.}
        \label{f:spectrum_sph_SG_circle}
    \end{figure}

    Interestingly, as in the $\phi^4$ model, the spectral problem governing the internal structure of the non-trivial sphaleron in the sine-Gordon model can be cast in the form of a Lamé-like equation
    \begin{equation}
        - \eta''(x;k) - \left[1 - 2 k^2 sn(x, k^2)^2\right]\eta(x;k) = \omega^2 \eta(x;k)\,,
    \end{equation}
    where comparing with the general equations (\ref{e:Lame_general}), we see that $\lambda = 1$ and $N = 1$. For this particular case, the general theory of Lamé equations ensures that the first three eigenfunctions and eigenfrequencies can be obtained analytically \cite{Liang:1992}
    \begin{align}
            \eta_{-} &= \text{dn}(x,k^2)\,, \quad \omega_{-}^2 = k^2 - 1\,,\\
            \eta_0  &= \text{cn}(x,k^2)\,, \quad \omega_0^2 = 0\,,\\
            \eta_1  &= \text{sn}(x,k^2)\,, \quad \omega_1^2 =  k^2\,.
    \end{align}
    The spectrum consists of an unstable mode, $\eta_-$, a zero mode, $\eta_0$, and a massive mode $\eta_1$. Naturally, the sphaleron also supports an infinite tower of modes above the mass threshold that, in the large $L$ limit, correspond to the scattering modes. Once more, as the size of the sphaleron increases the unstable mode can be interpreted as a symmetric combination of the zero modes of the component kinks. Moreover, the unstable mode and the zero mode become degenerate, whereas the massive mode becomes the threshold mode of the sine-Gordon on the real line. The spectral flow of these mode are represented in \autoref{f:spectrum_sph_SG_circle}.

    To conclude the description of these models on the circle, we want to remark that the sine-Gordon model, like the $\phi^4$ case, exhibits reflection symmetry. On the real line this symmetry takes the form $\phi(x) = - \phi(-x)$ but, on the circle, this property translates into
    \begin{equation}
        \phi(x) = - \phi(x + L/2)\,.
    \end{equation}
   However, the $\phi^6$ theory does not exhibit reflection symmetry. We shall analyse in detail, in the next section, the consequences that the absence or presence of such a discrete symmetry has on the dynamics of the sphaleron decay. More notably, the sine-Gordon model is additionally an integrable theory on the circle. As we will show in \autoref{ss:SG_circle_decay}, this integrability strongly constrains the dynamics and has profound consequences for the study of sphaleron decay.
    \vspace{0.2cm}

\section{Sphaleron dynamics on the circle}\label{s:Decay_Circle}

    Now that we have introduced the models under consideration and discussed the sphaleron configurations they support, we proceed to the analysis of their decay. To study the decay process, we consider the following initial condition
    \begin{align} 
        \phi(x,0) &= \Phi_{sph}(x;k) + A\, \eta_-(x;k)\,, \label{e:lin_ansatz_circlePhi}\\
        \dot{\phi}(x,0) &= A\, |\omega_{-}|\eta_-(x;k)\,, \label{e:lin_ansatz_circlePhiP}
    \end{align}
    where the unstable mode $\eta_-(x;k)$ is assumed normalised, $A$ denotes its initial amplitude and $\omega^2_- < 0$ is the corresponding eigenfrequency. Note that this initial configuration differs from that used in \autoref{s:decay_Deformed}. Now, not only is the sphaleron slightly displaced along the unstable direction, but an initial velocity along that direction is given. This is a more general configuration that models a dynamical excitation of the sphaleron. Moreover, this initial condition is linearly exact. This means that for $A$ small enough, the configuration converges to an exact solution of the equations of motion.

    As the sphaleron on the circle resembles a kink-antikink pair in the limit $L \gg L_{min}$, the initial condition (\ref{e:lin_ansatz_circlePhi})-(\ref{e:lin_ansatz_circlePhiP}) effectively represents a kink-antikink pair boosted towards each other. Therefore, it is natural to relate the factor $A\,\omega_-$ to the initial velocity of the pair. However, the negative frequency $\omega_-^2$ is generally close to zero in that regime. This means that through that initial condition we cannot model an "abrupt" sphaleron decay. In other words, it does not serve to represent high-speed contractions of the sphaleron. For this reason, we will also explore the following initial configuration 
    \begin{align}
        \phi(x,0) &= \phi_{KAK}(x; x_0, v),\label{e:rel_ansatz_circlePhi}\\
        \dot{\phi}(x,0) &= v\,\gamma(v) \phi'_{KAK}(x; x_0, v) \,,\label{e:rel_ansatz_circlePhiP}
    \end{align}
    where the prime denotes the derivative with respect to the argument of the function. Here, $\phi_{KAK}(x; x_0, v)$ denotes the kink–antikink profile of the theory under consideration, with $x_0$ specifying the position of the constituent kinks. If we identify the intrinsic width of a kink or antikink with the spatial region over which the field interpolates between vacua, then the initial configuration (\ref{e:rel_ansatz_circlePhi})-(\ref{e:rel_ansatz_circlePhiP}) is no longer meaningful if the separation between the component kinks is comparable to the kink width itself. Consequently, on a circle of circumference $L$, such a configuration is only valid and becomes exact in the regime where $L \gg L_{min}$. Finally, $\gamma(v)$ represents the Lorentz factor. Note that, with this choice, we gain direct control over the initial velocities of the individual kinks. 

    In what follows, the numerical simulations will be performed in a simulation box where the spatial variable ranges in the interval  $-L/2 \leq x \leq L/2$. As we shall see, the evolution leads to a variety of final states, depending on the initial conditions and the specific model under consideration. Notably, we will show that it is possible to relate the sphaleron collapse to kink scattering processes on the real line. It is worthwhile to remark that, unlike in the real line case, the kink-antikink pair that constitutes the sphaleron on the circle is truly static.

\subsection{Dynamics of the perturbed \texorpdfstring{$\phi^4$}{phi4} sphaleron on the circle}

\begin{figure*}[htb]
        \centering
        \includegraphics[width=1.00\textwidth]{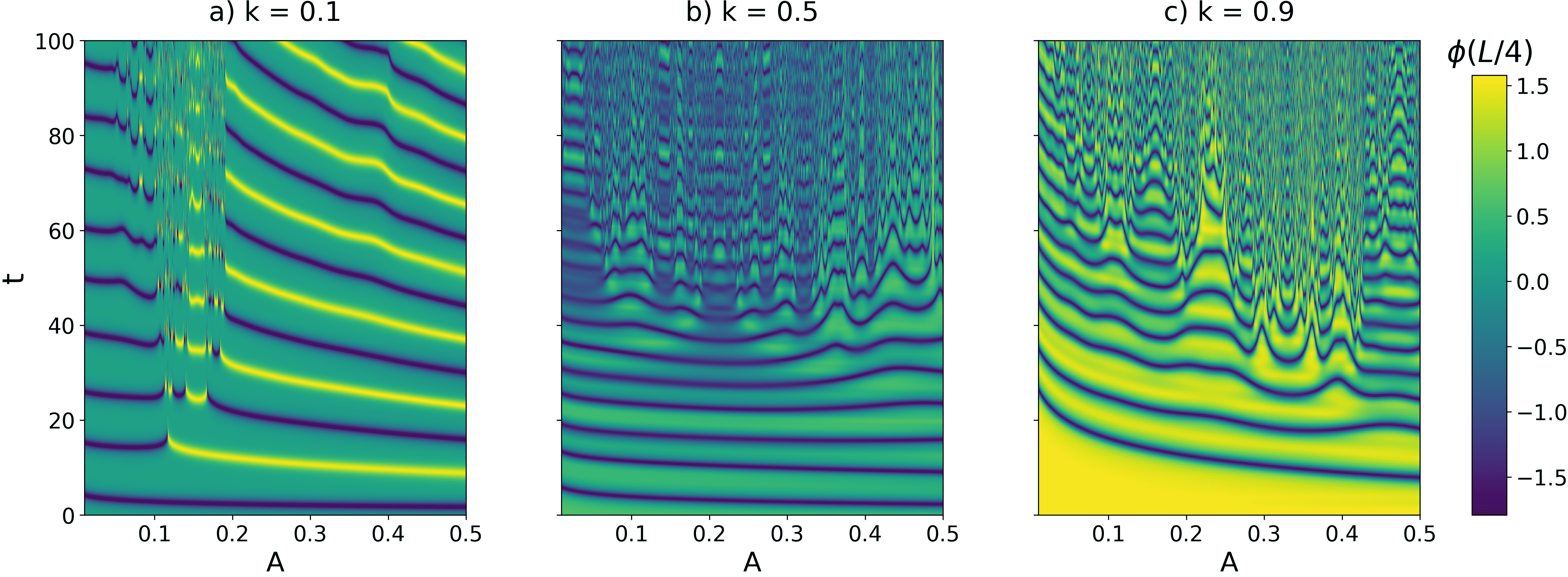}
    \caption{Decay of the $\phi^4$ sphaleron for different initial amplitudes of the unstable mode $A$ using the initial condition (\ref{e:lin_ansatz_circlePhi})-(\ref{e:lin_ansatz_circlePhiP}).}
    \label{f:decay_phi4_circle}
    \end{figure*}

\begin{figure*}[htb]
    \centering
    \begin{subfigure}[t]{0.45\textwidth}
        \centering
        \includegraphics[width=0.99\textwidth]{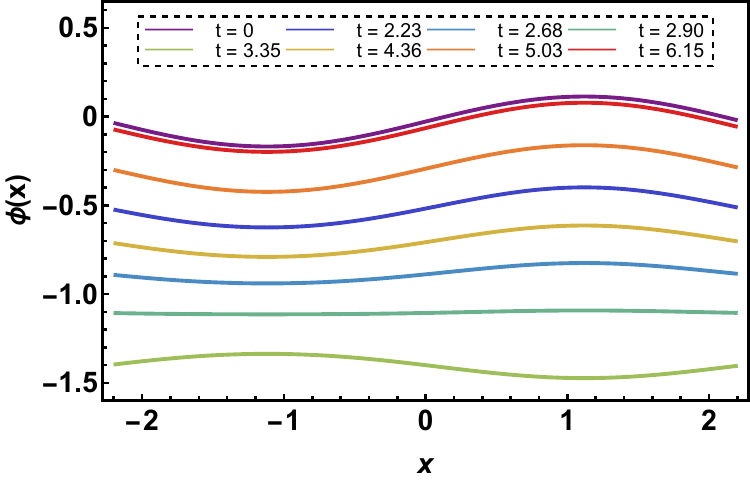}
        \caption{\small $A = 0.05$ and $k = 0.1$.}
        \label{fig:phi4_k01_A005}
    \end{subfigure}
    \begin{subfigure}[t]{0.45\textwidth}
        \centering
        \includegraphics[width=0.99\textwidth]{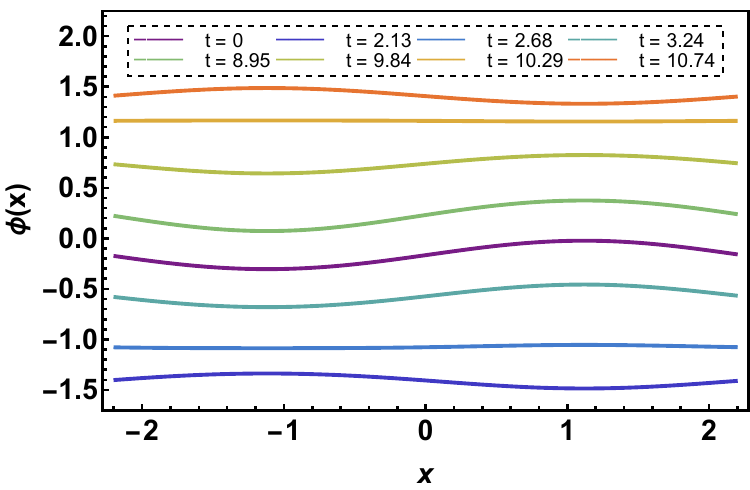}
        \caption{\small $A = 0.3$ and $k = 0.1$.}
        \label{fig:phi4_k05_A01}
    \end{subfigure}
    \begin{subfigure}[t]{0.45\textwidth}
        \centering
        \includegraphics[width=0.99\textwidth]{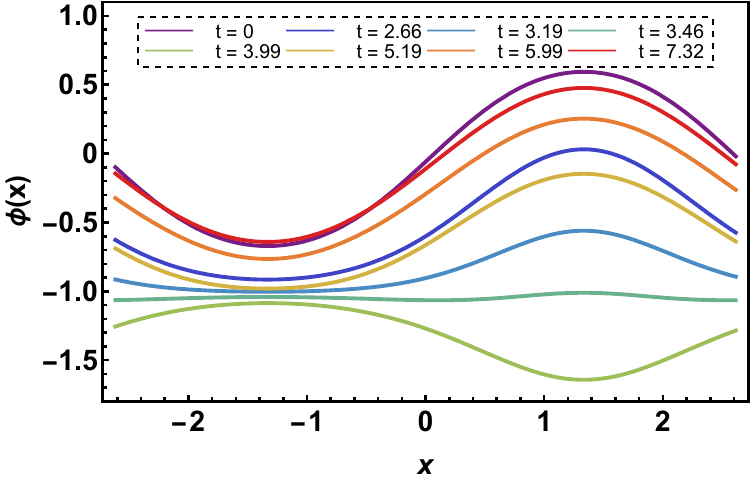}
        \caption{\small $A = 0.1$ and $k = 0.5$.}
        \label{fig:phi4_k05_A005}
    \end{subfigure}
   \begin{subfigure}[t]{0.45\textwidth}
        \centering
        \includegraphics[width=0.99\textwidth]{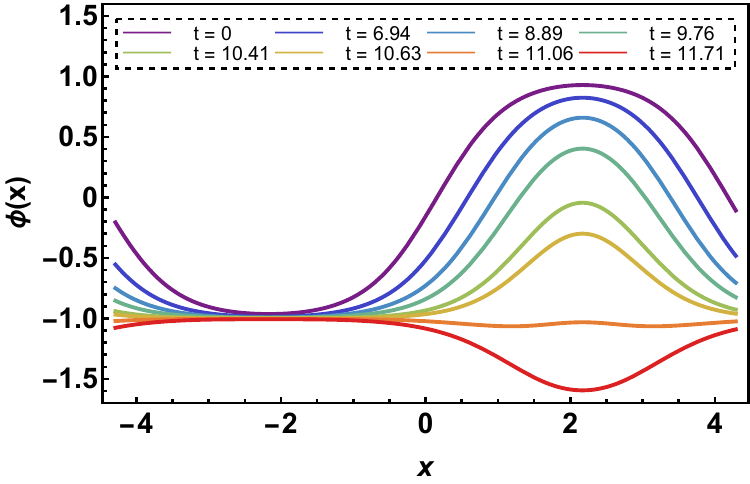}
        \caption{\small $A = 0.25$ and $k = 0.9$.}
        \label{fig:phi4_k09_A025}
    \end{subfigure}
    \caption{Time evolution of the field profile $\phi(x)$ in the $\phi^4$ model, shown for different time steps. The initial condition corresponds to the linear ansatz (\ref{e:lin_ansatz_circlePhi})-(\ref{e:lin_ansatz_circlePhiP}).}
    \label{f:phi4_k}
    \end{figure*}

    The $\phi^4$ model was introduced in \autoref{c:Sphalerons}, where its formulation on the circle and the corresponding sphaleron solution were discussed. We now turn to the study of its decay. We begin our analysis by studying the sphaleron decay with the initial condition (\ref{e:lin_ansatz_circlePhi})-(\ref{e:lin_ansatz_circlePhiP}). Specifically, we have considered three representative values for $k$ covering the whole set: $k = 0.1,\, 0.5$ and $0.9$. Recall that the parameter $k$ defines the size of the base space of the sphaleron. The corresponding evolution of the field for different initial values of the unstable amplitude $A$ is illustrated in \autoref{f:decay_phi4_circle}. The colour palette accounts for the field value at $x = L/4$, which corresponds to the position on the circle where the sphaleron collapses. Moreover, representative examples of the time evolution of the scalar field profile are shown in \autoref{f:phi4_k} for different values of the unstable amplitude $A$ and the model parameter $k$.

    For $k = 0.1$, where the circumference of the circle is small, three regions can be distinguished: in the region $A \lessapprox 0.06$, the sphaleron initially decays along the direction determined by the unstable mode. Due to our choice of sign in the initial condition, the sphaleron decays toward the negative vacuum and, at that point, the configuration bounces back and approaches the initial perturbed profile. For these small values of $k$, the sphaleron decays once more along the same direction, and this process is repeated almost periodically. The decay process changes completely for moderate values of the unstable amplitude. Specifically, in the region $0.06 \lessapprox A \lessapprox 0.2$, the system undergoes a chaotic decay after the first bounce, with the decay proceeding along either the original or the opposite unstable direction, that is, toward the negative or the positive vacuum. This behaviour suggests that the evolution is extremely sensible on the unstable amplitude. Finally, for $A \gtrapprox 0.2$, the sphaleron exhibits again a periodic decay, although now the process alternates between both vacua.
    
    For $k = 0.5$, the dynamics differs significantly. Now, the sphaleron initially evolves along the direction of the unstable mode and returns to a configuration close to the initial one during the first few periods. Subsequently, the period appears to exhibit a sensitive dependence on the initial amplitude $A$.

    Finally, for $k = 0.9$, the sphaleron collapses and forms an oscillon state. It is worth noting that, for intermediate values $0.2 < A < 0.3$, the oscillon period increases, resembling the back-scattering behaviour of a kink–antikink pair that scatters on the real line \cite{Campbell:1983}. This is not surprising since, as we have commented, the sphaleron profile approaches the profile of a kink antikink pair as $k\rightarrow 1$ and $L$ grows. Therefore, the sphaleron decay resembles the kink-antikink scattering on the real line as $k \rightarrow 1$. Nonetheless, it is well-known that, on the real line, the non-linear evolution of the kink-antikink scattering exhibits a fractal dependence on the initial velocity of the kinks. This translates into two possible outcomes: after the scattering, the pair may bounce back and forth several times, with number of bounces also depending sensitively on the initial velocity. Eventually, the pair either decays into an oscillon state or escapes to infinity. Remarkably, there exists a critical velocity above which the pair always escapes to infinity after the first bounce. Although we can see in \autoref{f:decay_phi4_circle} and \autoref{f:phi4_k} that the evolution depends on both the size of the sphaleron and the amplitude of the unstable mode, we do not recover that non-linear evolution for the sphaleron. These differences are still visible even for values of $k$ closer to $1$. 

    The reason is that as $k$ increases to $1$, the frequency associated with the unstable mode, $\omega^2_{-}$, vanishes. Consequently, the kinetic contribution to the initial configuration becomes negligible, and expression (\ref{e:lin_ansatz_circlePhi})-(\ref{e:lin_ansatz_circlePhiP}) effectively describes a kink–antikink pair separated by a large distance and interacting via the static inter-solitonic force. In this limit, the initial configuration corresponds to a kink-antikink pair with vanishing centre of mass velocity. Therefore, the expected dynamics resembles a kink–antikink scattering at very low initial velocities. 
    
\begin{figure*}[htb]
    \centering
    \begin{subfigure}[t]{0.5\textwidth}
        \centering
        \includegraphics[width=0.9\textwidth]{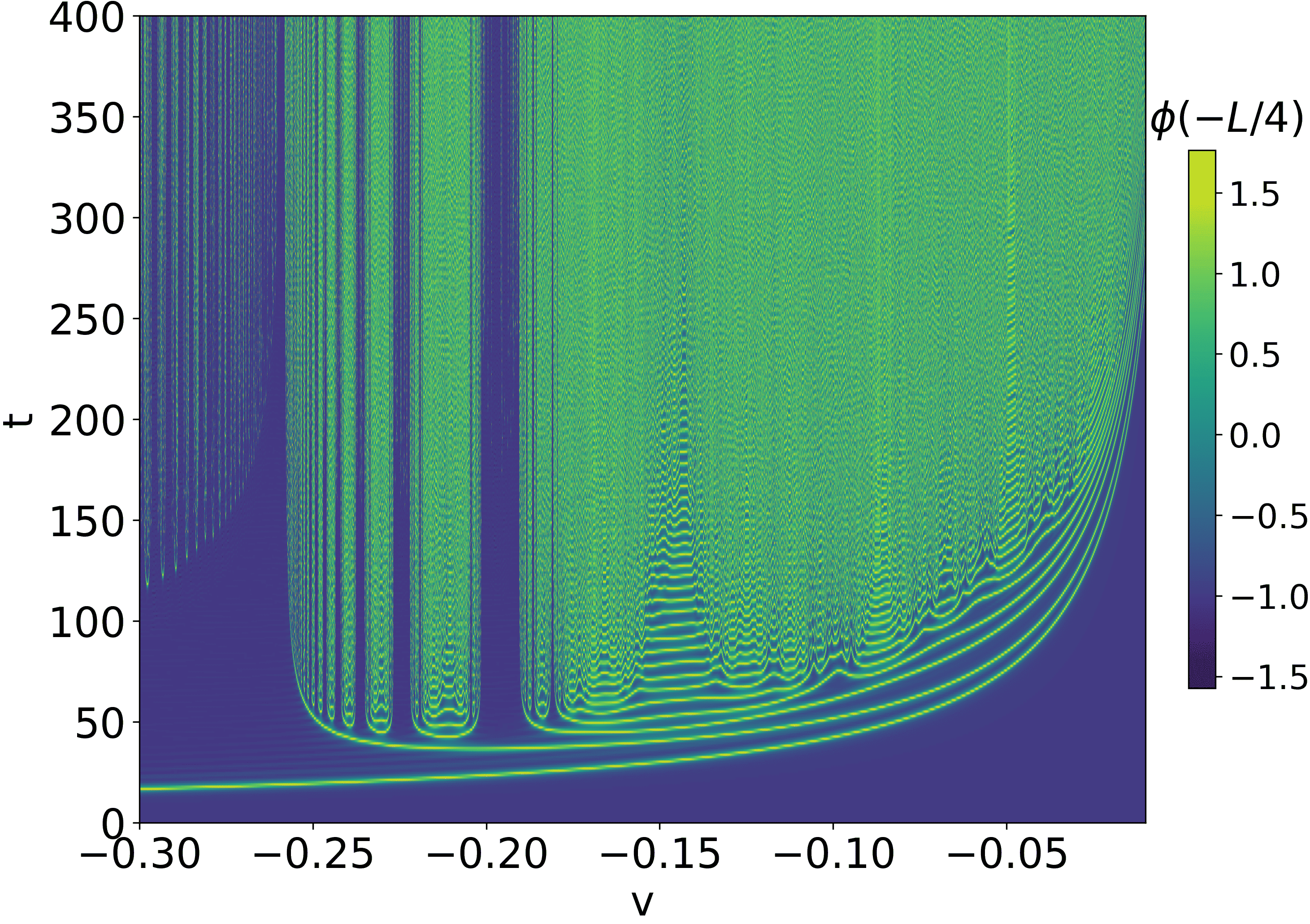}
        \caption{\small Antikink-kink scattering.}
        \label{fig:decay_phi4_positive_rel}
    \end{subfigure}%
    \begin{subfigure}[t]{0.5\textwidth}
        \centering
        \includegraphics[width=0.9\textwidth]{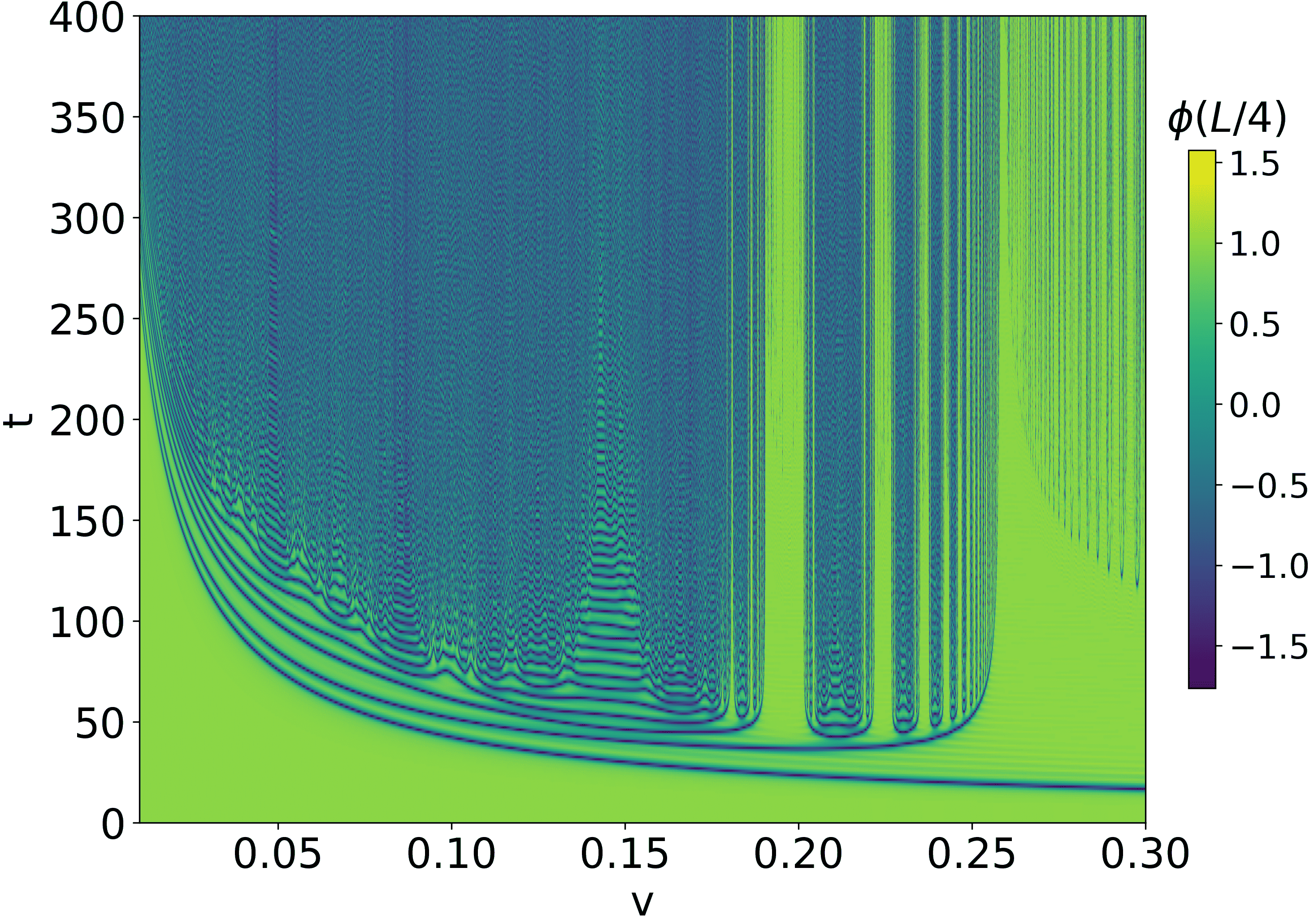}
        \caption{\small Kink-antikink scattering.}
        \label{fig:decay_phi4_negative_rel}
    \end{subfigure}

    \caption{Dynamics of the $\phi^4$ sphaleron for the initial condition given in (\ref{e:rel_phi4_circlePhi})-(\ref{e:rel_phi4_circlePhiP}), with $L$ given by (\ref{e:L_k_phi4}) and $k = 0.9999$. The bounce windows and the critical velocity of the $\phi^4$ kink-antikink and antikink-kink pair scattering are recovered.}
    \label{f:phi4_rel}
    \end{figure*}

    The parallelism between kink-antikink and perturbed sphaleron dynamics will become clear in our next numerical experiment. In order to study a rapid collapse of the sphaleron, we can consider the regime where $k$ is close to 1 and use the initial configuration (\ref{e:rel_ansatz_circlePhi})-(\ref{e:rel_ansatz_circlePhiP}). For the $\phi^4$ model, this initial configuration reads as
    \begin{align} 
            \phi(x,0) &= \tanh(\gamma(v) \,x) - \tanh(\gamma(v) \,(x - L/2)) - \tanh(\gamma(v) \,(x + L/2))\,, \label{e:rel_phi4_circlePhi}\\
            \dot{\phi}(x,0) &= v\, \gamma(v)\left(\sech^2(\gamma(v) \,x) + \sech^2(\gamma(v) \,(x - L/2)) + \sech^2(\gamma(v) \,(x + L/2))\right)\,. \label{e:rel_phi4_circlePhiP}
    \end{align}
    This configuration corresponds to a kink–antikink pair located at diametrically opposite points on the circle, boosted towards one another. Specifically, the kink is positioned at $x = 0$ and the antikink at $x = \pm L/2$, where the circumference $L$ of the circle is determined by the parameter $k$; in particular, we set $k = 0.9999$. The resulting dynamics from this initial condition are illustrated in \autoref{f:phi4_rel}.
    We observe that the behaviour of the collapsing sphaleron now precisely reproduces the fractal scattering pattern characteristic of kink–antikink scattering on the real line \cite{Campbell:1983}. Naturally, due to the periodicity of the spatial domain nothing escapes, and at times much larger than the system size, the outgoing kinks that form the sphaleron interact again. This further structure is clearly visible in the regime above the critical velocity, $v > v_{crit} \approx 0.25$.

    The fractal pattern observed in the kink-antikink scattering in the  $\phi^4$ theory is due to the existence of internal modes located around the solitons, which are able to store energy for some time and released it in form of kinetic energy \cite{Campbell:1983,Manton:2021,Manton:2022}. The similarity of the perturbed sphaleron evolution shown in \autoref{f:phi4_rel} with the kink-antikink scattering leads us to infer that the same resonant energy transfer mechanism is present in sphaleron dynamics. 
    
    Finally, for negative velocities, i.e., when the sphaleron collapses to the opposite vacuum, the fractal pattern remains exactly the same, as seen in \autoref{f:phi4_rel}. This parallelism is due to the reflection symmetry of the $\phi^4$ model, condition that, on the circle, is given by $\Phi_{sph}(x) = -\Phi_{sph}(x + L/2)$. The same conclusion applies to \autoref{f:decay_phi4_circle} when negative values of the unstable amplitude $A$ are assumed. We therefore infer that the decays to any vacuum are equivalent in the sense that they are related by the reflection symmetry.
  
\subsection{Dynamics of the perturbed \texorpdfstring{$\phi^6$}{phi6} sphaleron on the circle}\label{ss:decay_phi6_circle}

    In this section, we shall discuss the sphaleron decay in $\phi^6$ theory. As aforementioned, the $\phi^6$ theory lacks reflection symmetry. The results for the $\phi^4$ model suggest that now, the perturbed evolution must be analysed separately along each unstable direction.
    
\begin{figure*}[htb]
    \centering
    \begin{subfigure}[t]{0.45\textwidth}
        \centering
        \includegraphics[width=0.992\textwidth]{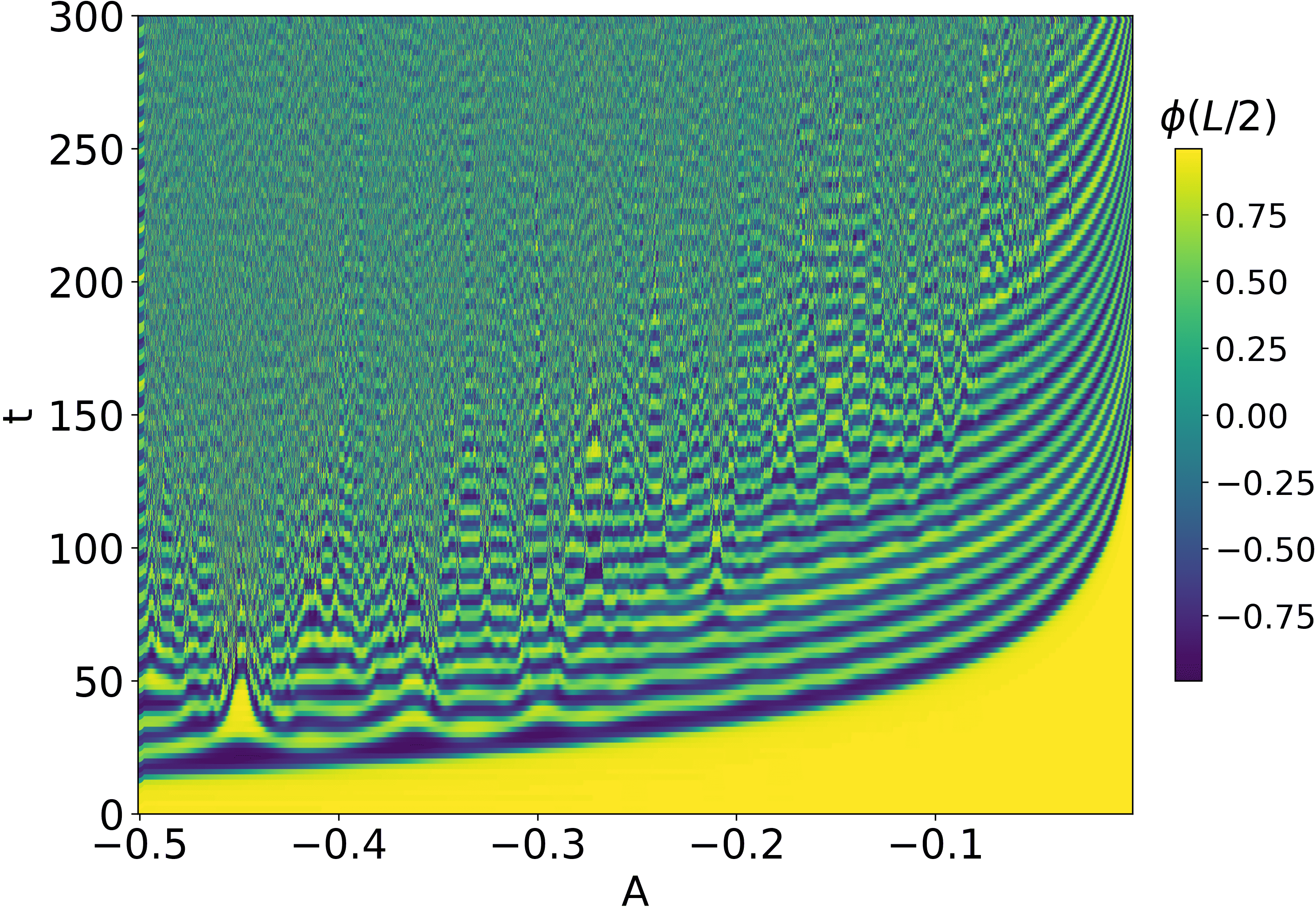}
        \caption{\small $\phi(L/2,t)$ for $C= 0.01$.}
        \label{f:decay_phi6_negative_moderateC}
    \end{subfigure}%
    \hspace{0.7cm}
    \begin{subfigure}[t]{0.45\textwidth}
        \centering
        \includegraphics[width=0.98\textwidth]{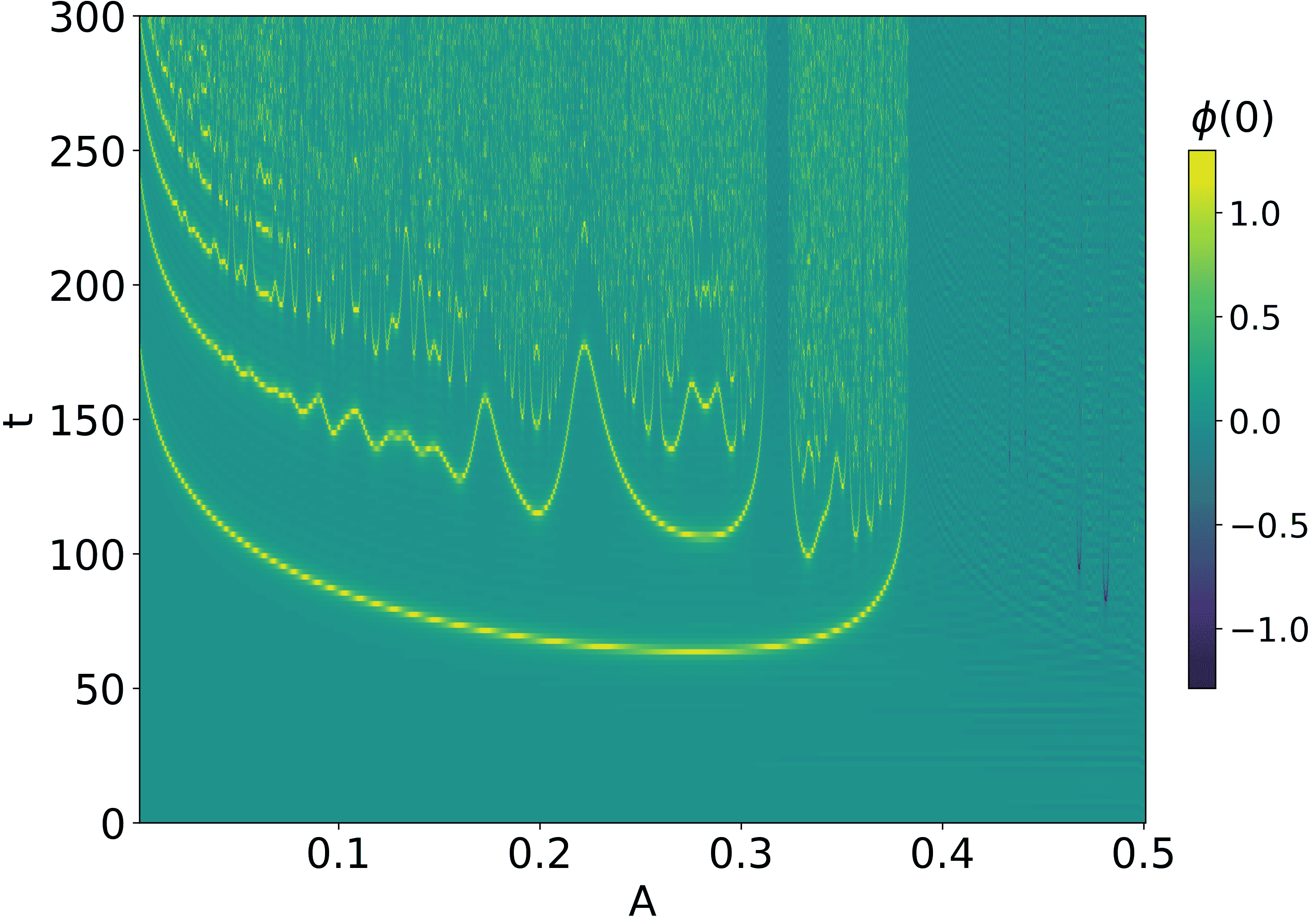}
        \caption{\small $\phi(0,t)$ for $C = 0.01$.}
        \label{f:decay_phi6_positive_smallC}
    \end{subfigure}%
    \caption{\small \justifying Decay of the $\phi^6$ sphaleron for negative values of the unstable amplitude (left panel) and for positive values of the unstable amplitude $A$ (right panel) using the initial condition (\ref{e:lin_ansatz_circlePhi})-(\ref{e:lin_ansatz_circlePhiP}).}
    \label{f:decay_phi6_linear_neg}
    \end{figure*}

\begin{figure*}[htb]
    \centering
    \begin{subfigure}[t]{0.45\textwidth}
        \centering
        \includegraphics[width=1.005\textwidth]{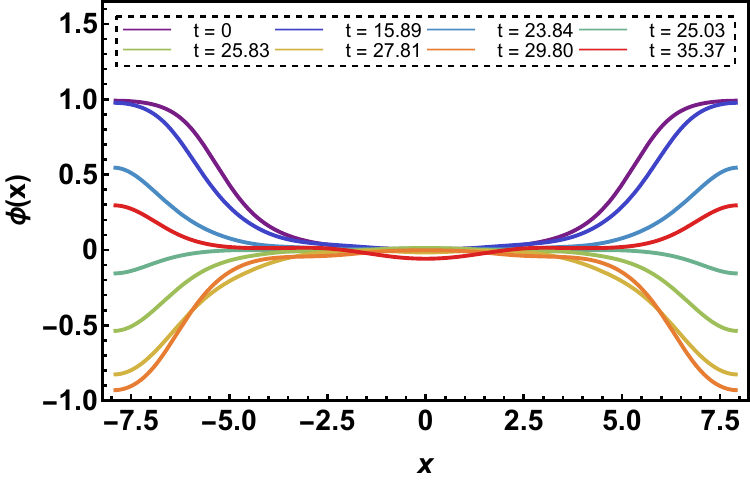}
        \caption{\small $A = - 0.3$.}
        \label{fig:phi6_C0.01_AN0.3}
    \end{subfigure}
    \begin{subfigure}[t]{0.45\textwidth}
        \centering
        \includegraphics[width=0.99\textwidth]{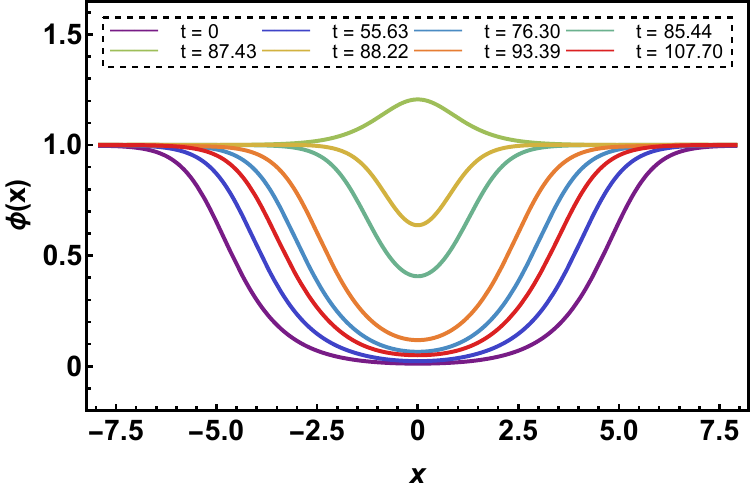}
        \caption{\small $A = 0.1$.}
        \label{fig:phi6_C0.01_A0.1}
    \end{subfigure}
    \begin{subfigure}[t]{0.45\textwidth}
        \centering
        \includegraphics[width=0.99\textwidth]{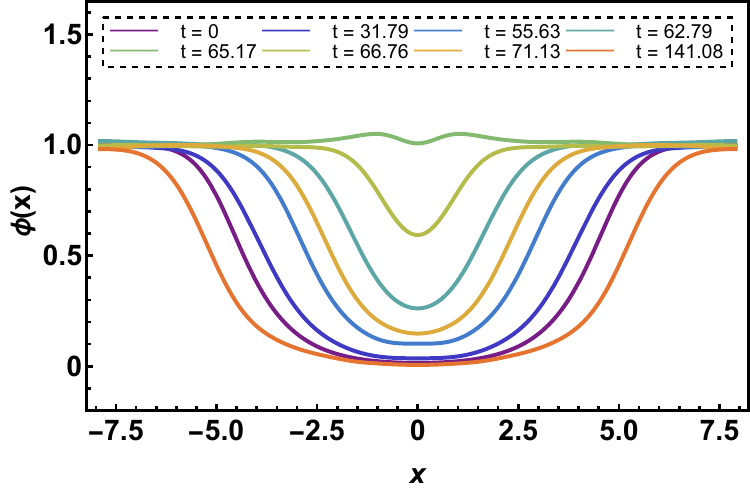}
        \caption{\small $A = 0.32$.}
        \label{fig:phi6_C0.01_A0.32}
    \end{subfigure}
    \begin{subfigure}[t]{0.45\textwidth}
        \centering
        \includegraphics[width=0.99\textwidth]{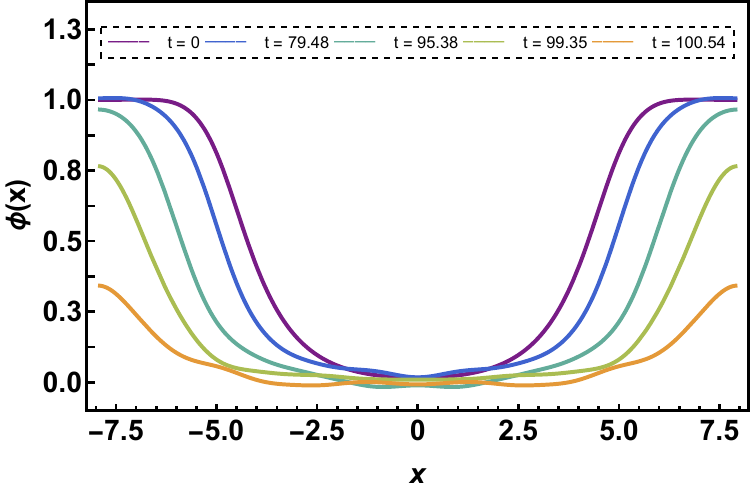}
        \caption{\small $A = 0.4$.}
        \label{fig:phi6_C0.01_A0.4}
    \end{subfigure}
    \caption{Time evolution of the field profile $\phi(x)$ in the $\phi^6$ model, shown for different time steps. The initial condition corresponds to the linear ansatz (\ref{e:lin_ansatz_circlePhi})-(\ref{e:lin_ansatz_circlePhiP}) with model parameter $C = 0.01$.}
    \label{f:phi6_C0.01}
    \end{figure*}

    We will begin the analysis using the linear ansatz (\ref{e:lin_ansatz_circlePhi})-(\ref{e:lin_ansatz_circlePhiP}) as initial configuration with the choice $C=0.01$ for the integration constant. Let us start with negative values of the unstable amplitude $A$. The numerical simulations show that for any value of the unstable amplitude, the sphaleron always collapses and produces an oscillon; see  \autoref{f:decay_phi6_negative_moderateC}. However, a much richer structure arises for positive values of the amplitude, as shown in \autoref{f:decay_phi6_positive_smallC}. Now, the sphaleron decay is more chaotic. Specifically, for a certain range of amplitudes, the sphaleron collapses and forms a bion state that undergoes several bounces before eventually settling into an oscillon configuration. Interestingly, in a small window close to $A\sim 0.32$, the sphaleron only bounces once, and the sphaleron walls move apart from $x=0$ to the opposite side of the circle, where an oscillon state is formed after they collide. More intriguing is what happens at $A \geq 0.39$. In this regime, the sphaleron begins to oscillate and expand rather than collapse.
    In \autoref{f:phi6_C0.01} we plot the time evolution of the scalar field profile in the $\phi^6$ model for different values of the unstable amplitude $A$ and $C = 0.01$.  
    These behaviours have been qualitatively observed for other values of the parameter $C$, showing only small differences.

    The different behaviours observed at negative and positive amplitudes can be traced back to the internal structure of the sphaleron. In the case of negative amplitudes, the sphaleron decay resembles a $(0, 1) + (1, 0)$ kink-antikink collision on the real line. As discussed in \autoref{c:Topological}, it is well-known that this configuration does not have internal modes \cite{Dorey:2011}. Although we have verified that the sphaleron configuration hosts positive bound modes in \autoref{ss:phi6_circle}, their support are localised in the region outside the collapse; see \autoref{f:phi6_eigenfunctions}. Effectively, this implies that the energy cannot be stored in the bound modes during the collapse, thereby only being produced oscillons after the decay. On the other hand, the situation is opposite for positive amplitudes of the unstable amplitude. Now, the internal modes do have support in the region of collapse. Consequently, they can be excited during the decay and trigger the energy transfer mechanism.

    In order to support this hypothesis, we perform a slightly different experiment. We now start with the initial configuration given by (\ref{e:rel_ansatz_circlePhi})-(\ref{e:rel_ansatz_circlePhiP}) for $C=0.0001$. More precisely, we have used the following ansatz 
    \begin{align} 
            \phi(x,0) &= \sqrt{\dfrac{1 - \tanh(\gamma(v)(x + x_0))}{2}} + \sqrt{\dfrac{1 + \tanh(\gamma(v)(x - x_0))}{2}}\,,\label{e:rel_phi6_circlePhi}\\
            \dot{\phi}(x,0) &= v\, \gamma(v)\left(\dfrac{\sech^2(\gamma(v)(x + x_0))}{2\sqrt{2}(1 - \tanh(\gamma(v)(x + x_0)))} + \dfrac{\sech^2(\gamma(v)(x - x_0))}{2\sqrt{2}(1 + \tanh(\gamma(v)(x - x_0)))}\right)\,. \label{e:rel_phi6_circlePhiP}
    \end{align}
    In order to ensure the match between the previous ansatz and the sphaleron profile, we fix the parameter $x_0 = 9.89$. 
    
\begin{figure*}[htb]
    \centering
    \begin{subfigure}[t]{0.5\textwidth}
        \centering
        \includegraphics[width=0.91\textwidth]{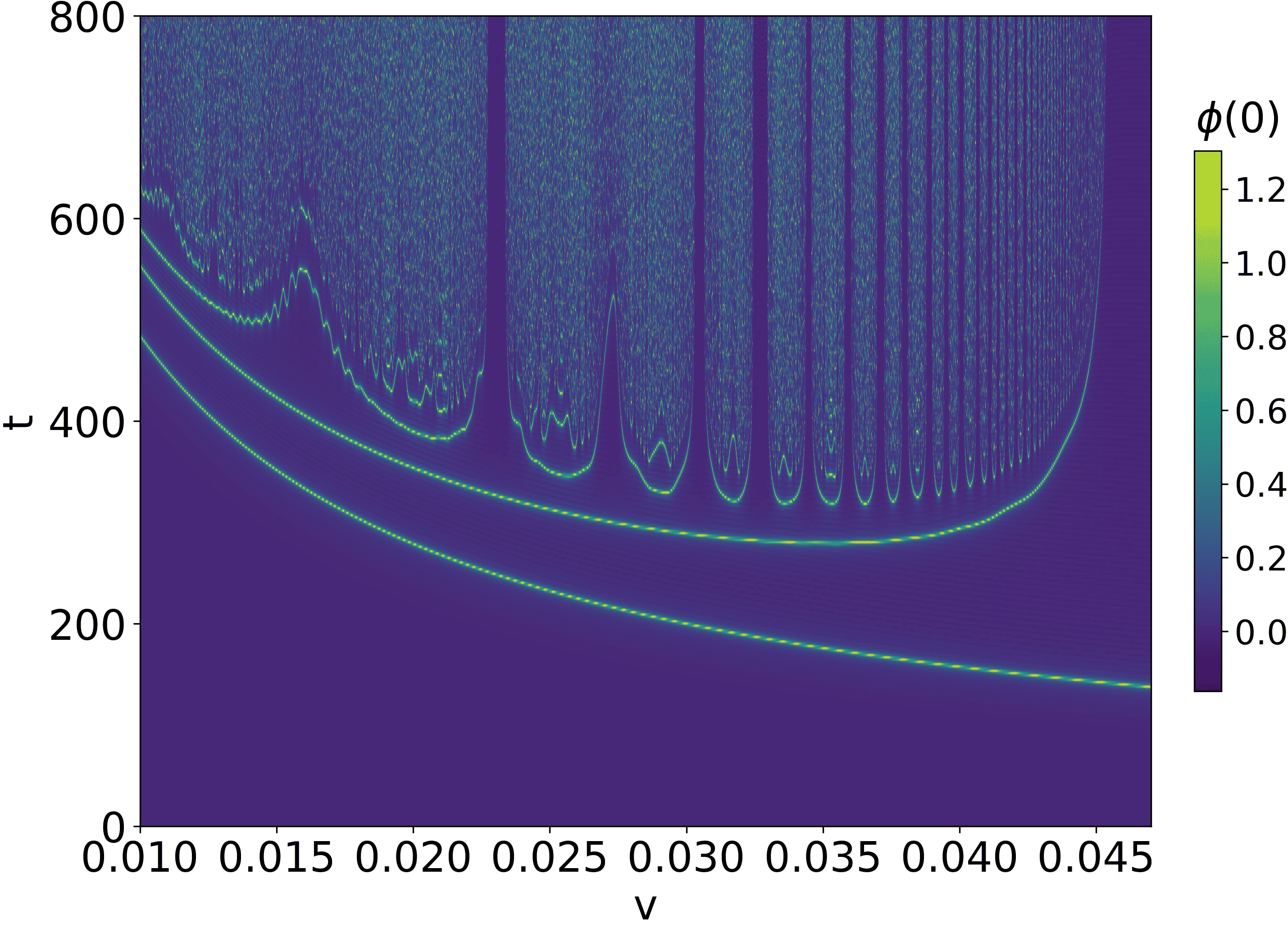}
        \caption{Antikink-kink scattering.}
        \label{fig:decay_phi6_positive_rel}
    \end{subfigure}%
    \begin{subfigure}[t]{0.5\textwidth}
        \centering
        \includegraphics[width=0.93\textwidth]{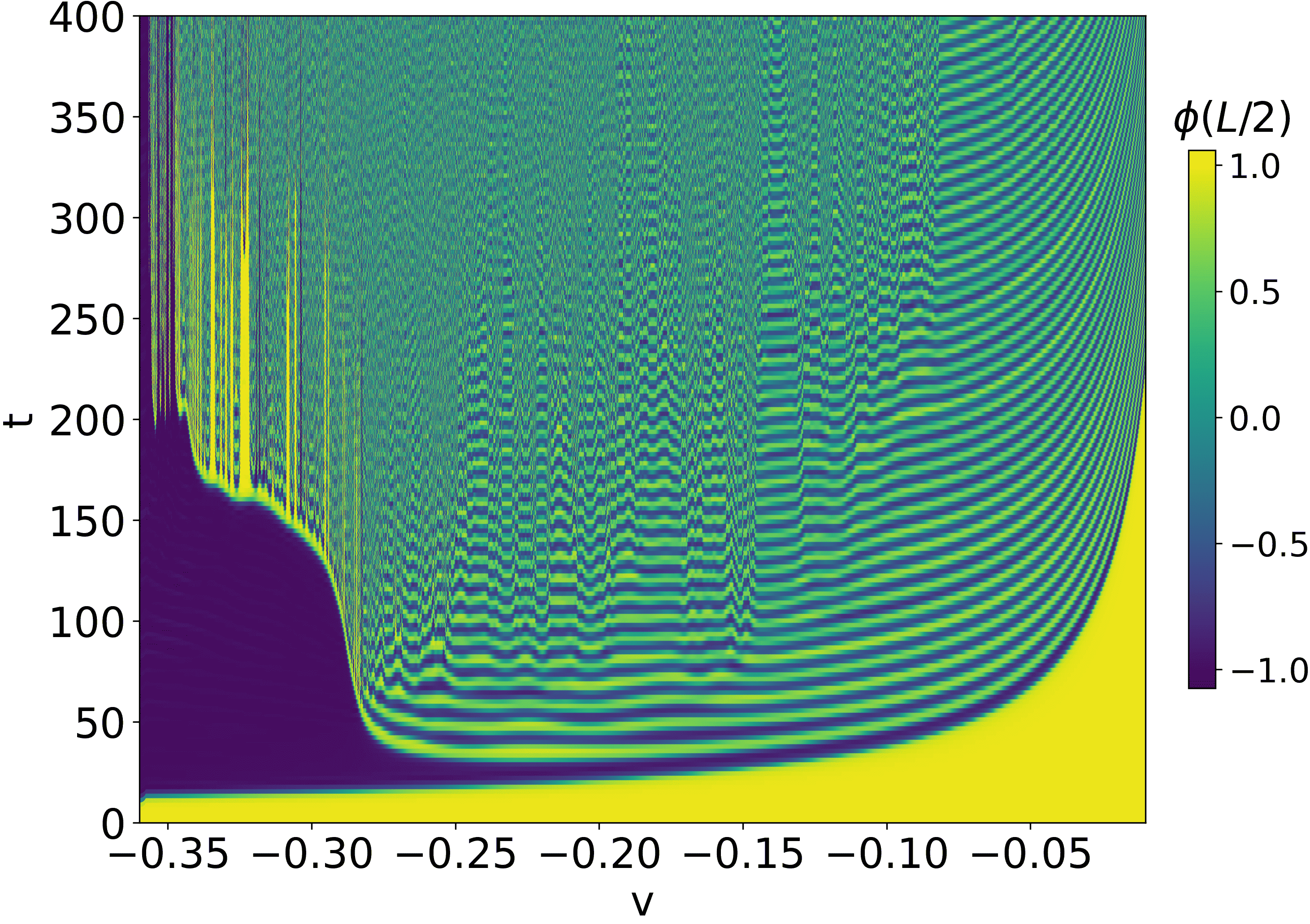}
        \caption{Kink-antikink scattering.}
        \label{fig:decay_phi6_negative_rel}
    \end{subfigure}

    \caption{Dynamics of the $\phi^6$ sphaleron for the initial condition given in (\ref{e:rel_phi6_circlePhi})-(\ref{e:rel_phi6_circlePhiP}) and $C = 0.0001$. The bounce windows and the critical velocity of the $\phi^6$ kink-antikink and antikink-kink pair scattering are recovered.}
    \label{f:phi6_rel}
    \end{figure*}

    We have performed simulations for the sphaleron decay, both for negative and positive initial velocities, corresponding respectively to $ (0, 1) + (1, 0)$ kink-antikink and $ (1, 0)+(0, 1)$ antikink-kink configurations. The time evolution for different velocities is shown in \autoref{f:phi6_rel}. The left panel shows $ (1, 0)+(0, 1)$ antikink-kink collisions, where the characteristic fractal pattern is clearly observed. The critical velocity and bounce windows are accurately determined \cite{Dorey:2011}. On the other hand, the right panel shows the $ (0, 1)+(1, 0)$ kink-antikink collisions, where the fractal pattern is absent; instead, only the decay into a bion is seen, along with the critical velocity at which the pair scatters back inelastically \cite{Dorey:2011}. 

    Therefore, we find that the lack of reflection symmetry leads to inequivalent decays along the two possible unstable directions. Such a distinction might have strong implications for more realistic scenarios. In particular, \autoref{f:decay_phi6_positive_smallC} suggests that there is an intermediate amplitude $A\sim 0.39$ between the regimes of collapse and expansion where the sphaleron freezes and its life-span is notably enhanced. We will explore this mechanism in more detail in \autoref{s:Stabilization_circle}.

\subsection{Dynamics of the perturbed sine-Gordon sphaleron on the circle }\label{ss:SG_circle_decay}

    Finally, in this section, we analyse the evolution of the sine-Gordon sphaleron on the circle after perturbing it along the unstable direction.

\begin{figure*}[htb]
    \centering     
    \begin{subfigure}[t]{0.48\textwidth}
        \centering
        \includegraphics[height=4.5cm]{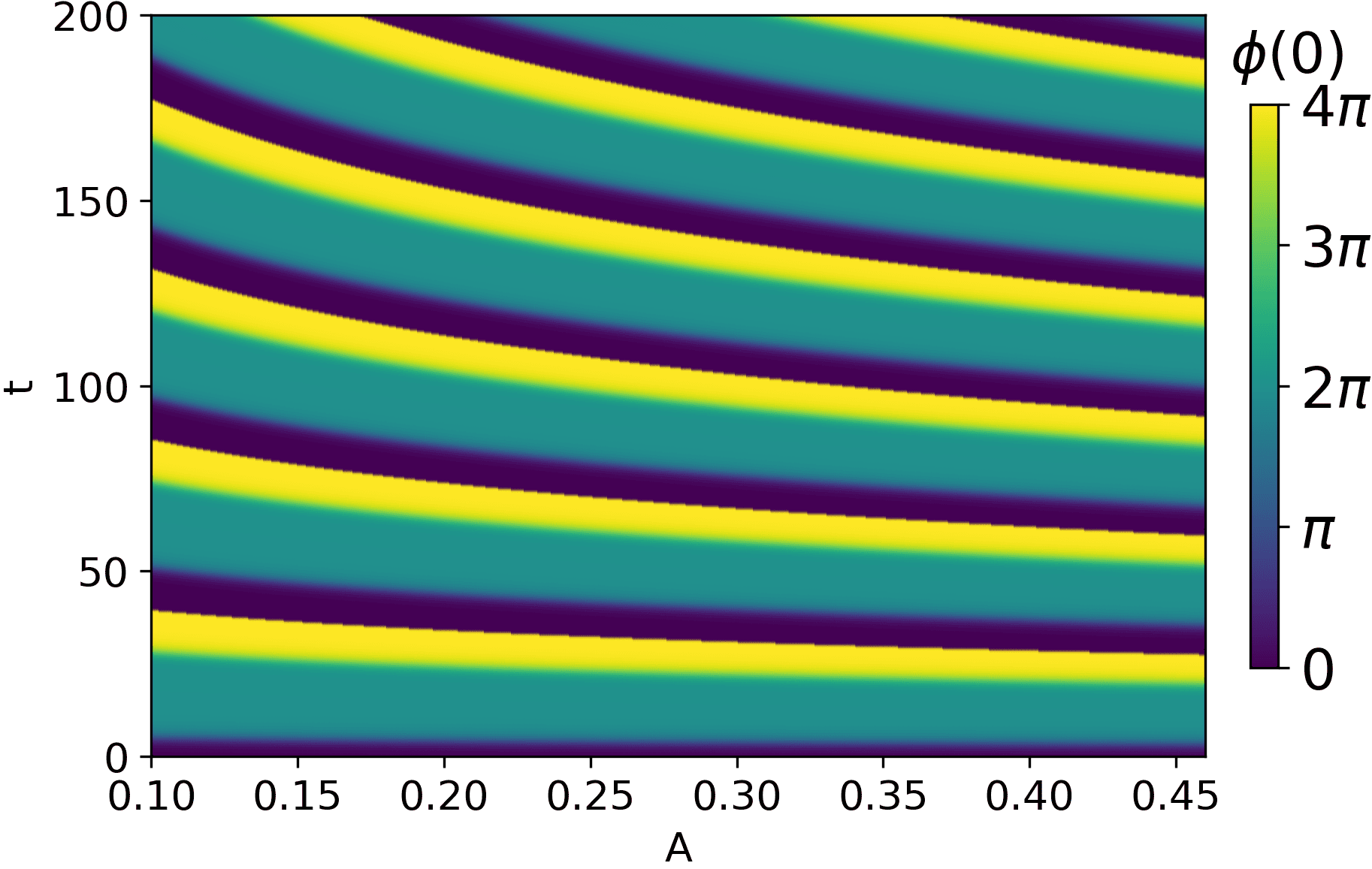}
        \caption{$\phi(0,t)$ for different unstable amplitudes $A$.}
        \label{fig:DecaySG}
    \end{subfigure}%
    \hspace{0.05\textwidth}
    \begin{subfigure}[t]{0.45\textwidth}
        \hspace{-0.48cm}
        \centering
        \includegraphics[height=4.5cm]{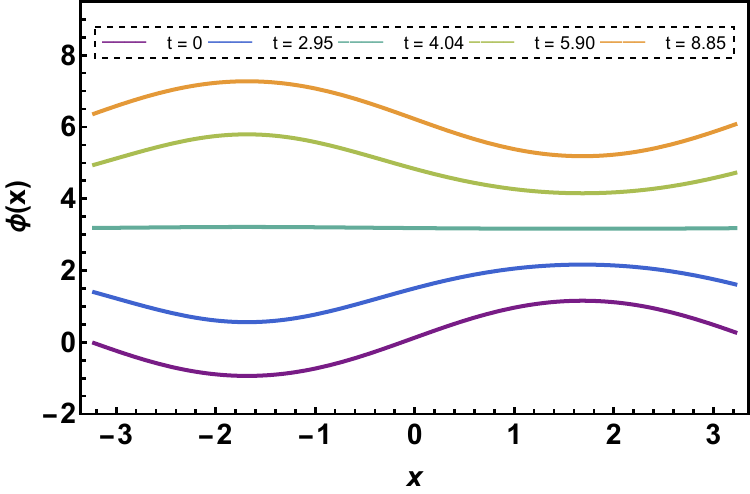}
        \caption{Time evolution of $\phi(x)$ for $A = 0.3$.}
        \label{fig:Evol_k0.5_A0.3}
    \end{subfigure}
    \vspace{1em}

    \begin{subfigure}[t]{0.45\textwidth}
        \centering
        \includegraphics[width=\textwidth]{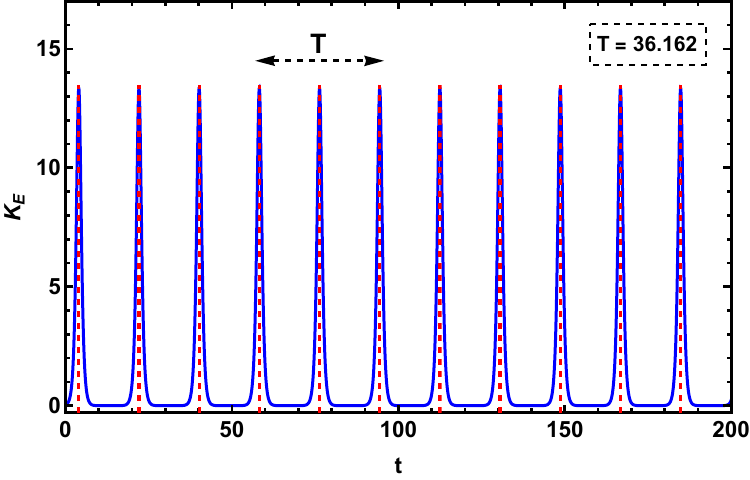}
        \caption{Kinetic energy for $A = 0.3$.}
        \label{fig:KE_k0.5_A0.3}
    \end{subfigure}%
    \hspace{0.05\textwidth}
    \begin{subfigure}[t]{0.45\textwidth}
        \centering
        \includegraphics[width=\textwidth]{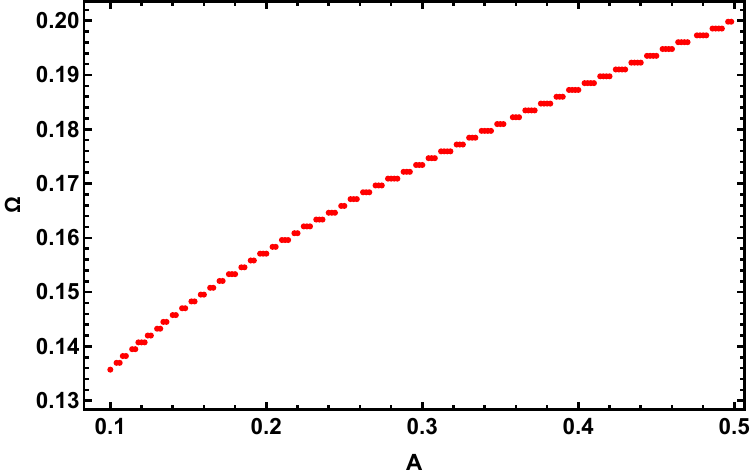}
        \caption{Dependence of angular frequency on $A$.}
        \label{fig:KE_k0.5_All}
    \end{subfigure}
    
    \caption{Upper left panel: Evolution of the sphaleron profile at $x = 0$ (mod $4\pi$). Upper right panel: Time evolution of the field profile $\phi(x)$ in the sine-Gordon model, shown for different time steps. Lower left panel: Kinetic energy as a function of time for $A = 0.3$. Here, the label $T$ stands for the period of oscillation between two sphaleron (resp. antisphalerons) configurations observed numerically. Lower right panel: Temporal angular frequency of the kinetic energy during the decay of the sine-Gordon sphaleron for different values of the initial amplitude of the unstable mode $A$. All plots assume $k = 0.5$ as illustrative example.}
    \label{f:KE_k0.5}
    \end{figure*}

    Once more, we choose the initial configuration (\ref{e:lin_ansatz_circlePhi})-(\ref{e:lin_ansatz_circlePhiP}) and allow the system to evolve. As an illustrative example, we have assumed the value $k = 0.5$. The same discussion remains valid for other choices. In \autoref{fig:DecaySG} we represent the time evolution of the field at the origin, $\phi(0)$, for a wide range of initial unstable amplitudes. The regular pattern suggests a periodic evolution. In fact, the numerical simulations reveal that, as the configuration evolves in the unstable direction, the initial sphaleron transitions to an almost antisphaleron configuration centred at the nearest maximum of potential, and stays there for a while, and evolves toward a sphaleron configuration. Importantly, the direction of decay is determined by the sign of the unstable mode amplitude, and this choice is highly sensitive to any perturbation. Therefore, the antisphaleron may evolve towards the original sphaleron, $\Phi^{(0)}_{sph}(x)$, or towards the next sphaleron configuration, $\Phi^{(0)}_{sph}(x) + 4\pi$. A representative example of the periodic evolution of the scalar field profile is shown in \autoref{fig:Evol_k0.5_A0.3}. The transition between sphaleron and antisphaleron configurations occurs periodically, with a fixed angular frequency $\Omega$. This frequency can be readily determined by monitoring the time evolution of the kinetic energy of the system. As shown in \autoref{fig:KE_k0.5_A0.3}, a series of pronounced peaks are observed. Initially, the kinetic energy is negligible, since the initial configuration of the form (\ref{e:lin_ansatz_circlePhi})-(\ref{e:lin_ansatz_circlePhiP}) contains only a small kinetic contribution, as previously mentioned. As the system evolves along the unstable direction, the kinetic energy increases significantly during the transitions between the sphaleron and antisphaleron configurations. The constant time separation between successive peaks allows us to extract the period $T$ and thus compute the angular frequency via $\Omega = 2\pi/T$. Although we have not been able to derive an analytical expression relating the amplitude of the unstable mode $A$ to the angular frequency $\Omega$, we have obtained this dependence numerically, as illustrated in \autoref{fig:KE_k0.5_All}. As expected, the frequency increases with the amplitude of the unstable mode $A$. Of course, this regular behaviour is intimately related to the integrability of the model.
    
\begin{figure}[H]
        \centering{\includegraphics[width=0.95\linewidth]{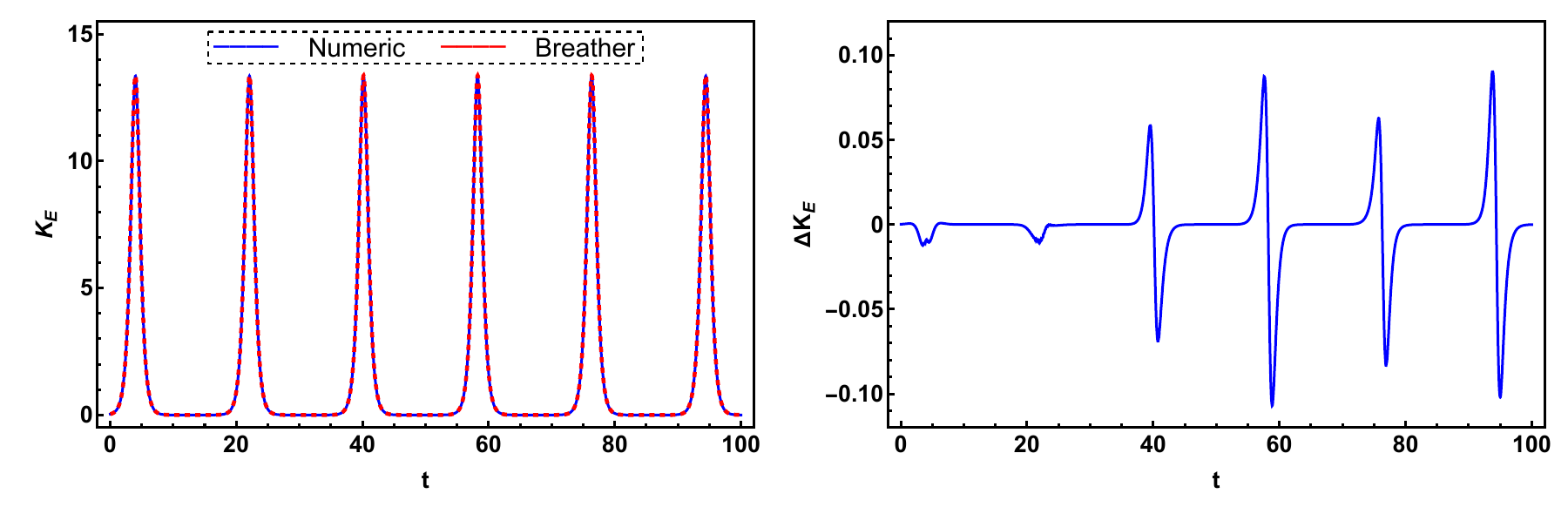}
        }
        \caption{Comparison between the perturbed sphaleron dynamics and the breather solution given by (\ref{e:spha_sine_time}). We represent the overlap between the numerical kinetic energy with that of the breather solution. Only slight deviations are observed.}
        \label{f:comparison_sph_SG_circle}
    \end{figure}

    Remarkably, the sine-Gordon model on the circle exhibits an exact time-periodic solution of the form    \begin{equation}\label{e:spha_sine_time}
        \Phi_B(x,t) = \pm \left( 4\, \arctan\left( \alpha\, \text{dn}(z, m_1)\,\text{sn}(f\, t, m_2)\right) \pm \pi\right)\,,
    \end{equation}
    where
    \begin{eqnarray}
        z\hspace{-0.2cm}&=&\hspace{-0.2cm} \beta(x - L/4)\,, \quad L = \frac{2}{\beta}\mathrm{K}(m_1)\,, \quad \beta = f \alpha\,,\\
        m_1 \hspace{-0.2cm}&=&\hspace{-0.2cm} 1 + \dfrac{1}{\alpha^2}-\dfrac{1}{\beta^2(1 + \alpha^2)}\,, \quad m_2 = \dfrac{\alpha^2}{f^2 (1 + \alpha^2)} - \alpha^2\,.
    \end{eqnarray}
    A solution with this structure is commonly referred to in the literature as a breather solution \cite{Costabile:1978,Marchesoni:1987}. This configuration corresponds to the periodic analogous to the breather solution shown in (\ref{e:Breather_SG}) for the real line case. Notably, we have succeeded in establishing a correspondence between solutions of the form (\ref{e:spha_sine_time}) and sphaleron decays, in such a way that the dynamics of the perturbed sphaleron can be related to an exact time-periodic solution. Here, by correspondence we mean that the time-periodic solution (\ref{e:spha_sine_time}) becomes numerically indistinguishable from the non-linear evolution resulting from a perturbed sphaleron, provided that specific constraints are satisfied. To verify this correspondence, we have considered a sphaleron defined on a circle of circumference $L$ and perturbed it with the linear ansatz (\ref{e:lin_ansatz_circlePhi})-(\ref{e:lin_ansatz_circlePhiP}). The angular frequency $\Omega$ of the resulting periodic motion is then extracted numerically. The condition for correspondence is that the spatial and temporal frequencies of the breather solution (\ref{e:spha_sine_time}) match the circumference of the circle and the observed angular frequency $\Omega$, respectively. These requirements lead to the following relations 
    \begin{eqnarray} \label{omega}
        L = \frac{2}{f \alpha}\mathrm{K}(m_1)\,, \qquad \Omega =  \frac{\pi f}{2\mathrm{K}(m_2)}\,.
    \end{eqnarray}
    By imposing those constraints, the parameters $\alpha$ and $f$ are determined uniquely. A comparison between the evolution obtained from field theory and the fitted breather solutions is presented in \autoref{f:comparison_sph_SG_circle}. Since the decay direction is highly sensitive to the initial conditions, a direct comparison of the field profiles is not optimal. Instead, we compare the time evolution of the kinetic energy. On the one hand, we overlap the numerical kinetic energy with that of the breather solution and, on the other hand, we measure the deviations. The agreement is appreciable, with only slight discrepancies in the transition between the sphaleron and the antisphaleron and vice versa.     

    To conclude this section, it is worth noting that the observed periodicity is a crucial distinction of the sine-Gordon model compared to the previous models due to the integrability of the theory. That property enables us to obtain analytical knowledge of the sphaleron decay at all times. For that reason, this theory may provide an interesting framework for testing sphaleron properties.
    
\section{Sphaleron stabilisation and internal modes}\label{s:Stabilization_circle}

    In \autoref{c:Sph_False} we observed that the life-span of a sphaleron can be enhanced by the effect of its internal modes, which can be understood as a sort of partial stabilisation. The general explanation of this mechanism lies on a balance between two opposite forces. Generally, any perturbation of a sphaleron can be decomposed into a linear combination of its internal modes. As a result, a generic perturbation will typically have non-zero projections onto both the unstable mode and the stable internal modes. Under certain conditions, this interplay may lead to a long-lived configuration. In the analogy where the sphaleron is viewed as an unstable solution of a mechanical system located at the maximum of a potential, the excitation of the internal modes effectively modifies the potential, creating a new local minimum above the true vacuum. The sphaleron configuration may then evolve and settle in this local minimum for a time comparable to the lifetime of the internal mode. In this section we aim to revisit that mechanism.

\begin{figure*}[htb]
        \centering
        \begin{subfigure}[t]{0.5\textwidth}
            \centering
            \includegraphics[width=0.855\textwidth]{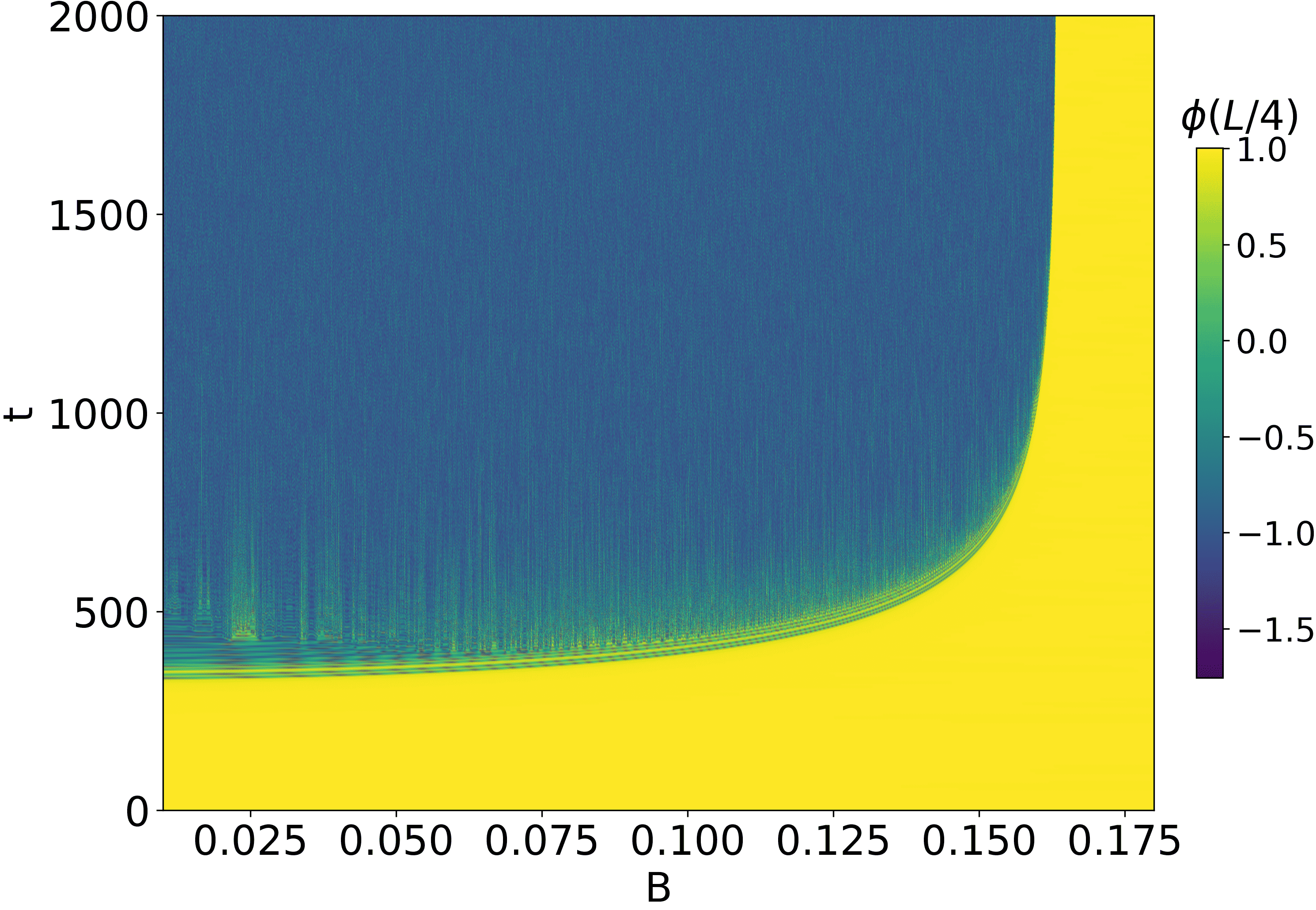}
            \caption{\small $\phi(L/4,t)$ for $A = 0.005$.}
            \label{fig:Stabilization_phi4_A0.05}
        \end{subfigure}%
        \hspace{-0.15cm}
        \begin{subfigure}[t]{0.5\textwidth}
            \centering
            \includegraphics[width=0.865\textwidth]{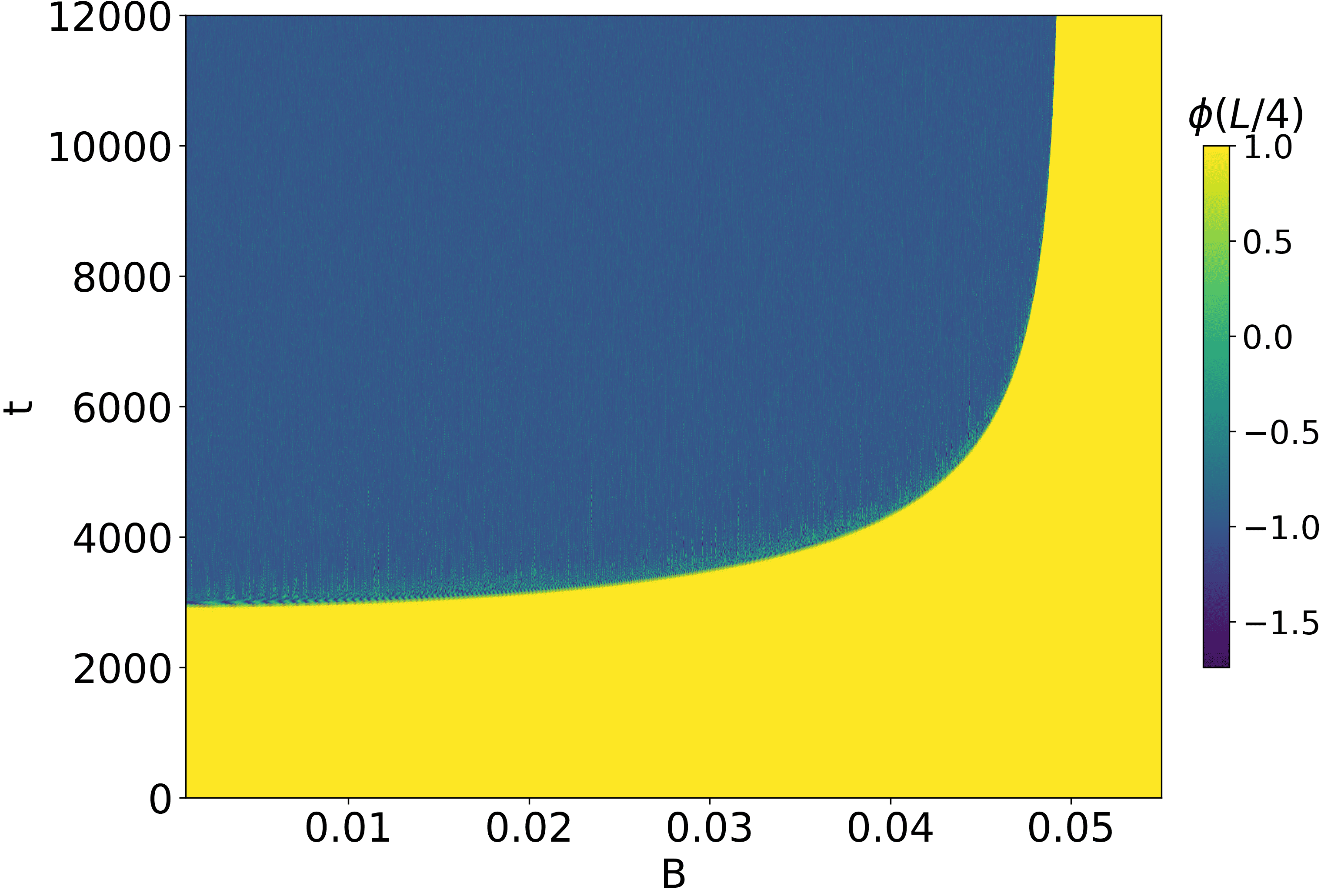}
            \caption{\small $\phi(L/4,t)$ for $A = 0.01$. }
            \label{fig:Stabilization_phi4_A0.01}
        \end{subfigure}%
    
        \caption{Dynamical stabilisation of the $\phi^4$ sphaleron with $k = 0.99$  (left panel) and $k = 0.999$ (right panel). The initial unstable amplitude is given by $A$ and the amplitude $B$ accounts for the initial excitation of the first positive internal mode $\eta_1$ around the sphaleron $(\ref{eq:internal_phi4})$.}
        \label{f:phi4_stabilization}
    \end{figure*}

    This explanation can be verified qualitatively using the following initial condition
    \begin{align} 
        \phi(x,0) &= \Phi_{sph}(x;k) + A\, \eta_-(x;k) + B \, \eta(x;k)\,, \label{e:lin_ansatz_withShape_circlePhi}\\
        \dot{\phi}(x,0) &= A\, |\omega_{-}|\eta_-(x;k)\,, \label{e:lin_ansatz_withShape_circlePhiP}
    \end{align}
    where $B$ is the amplitude of a positive mode $\eta(x;k)$ with frequency $\omega$. Specifically, we fix the amplitude of the unstable mode $A$ and vary the amplitude of the positive mode. In \autoref{f:phi4_stabilization}, we display the value of the field at the origin in the $\phi^4$ model for two sphalerons of different sizes, using the initial condition given in (\ref{e:lin_ansatz_withShape_circlePhi})-(\ref{e:lin_ansatz_withShape_circlePhiP}) with the first positive internal mode $\eta_1$ computed in (\ref{eq:internal_phi4}). We clearly observe that, above a critical value of the amplitude $B$ of the first internal mode, the sphaleron does not appear to decay within the time scale considered here. Remarkably, the effect appears to be quite efficient in the sense that within this range, the sphaleron undergoes oscillations around the new equilibrium position for an extended period of time. Numerical simulations indicate that the vibrating sphaleron configuration maintains its stability even after $10^4$ oscillation cycles.

    The stabilisation in the $\phi^6$ theory can also be achieved by imposing the condition (\ref{e:lin_ansatz_withShape_circlePhi})-(\ref{e:lin_ansatz_withShape_circlePhiP}), and qualitatively the same remarks as in the $\phi^4$ model apply. However, a counter-intuitive stabilisation mechanism occurs here. In \autoref{f:decay_phi6_positive_smallC}, we noted the existence of a critical amplitude close to $A \approx 0.39$ below which the sphaleron always undergoes at least one collapse, and above which it instead expands. Continuity considerations suggest that the sphaleron approaches a quasi-static configuration in the limiting case. Notably, this quasi-static state is reached by exciting only the unstable mode initially, i.e., with $B = 0$. While the fundamental explanation, namely a balance of forces, remains the same, the specific mechanism responsible for the stabilisation differs.
    
\begin{figure}[htb]
        \centering
        \includegraphics[width=0.75\linewidth]{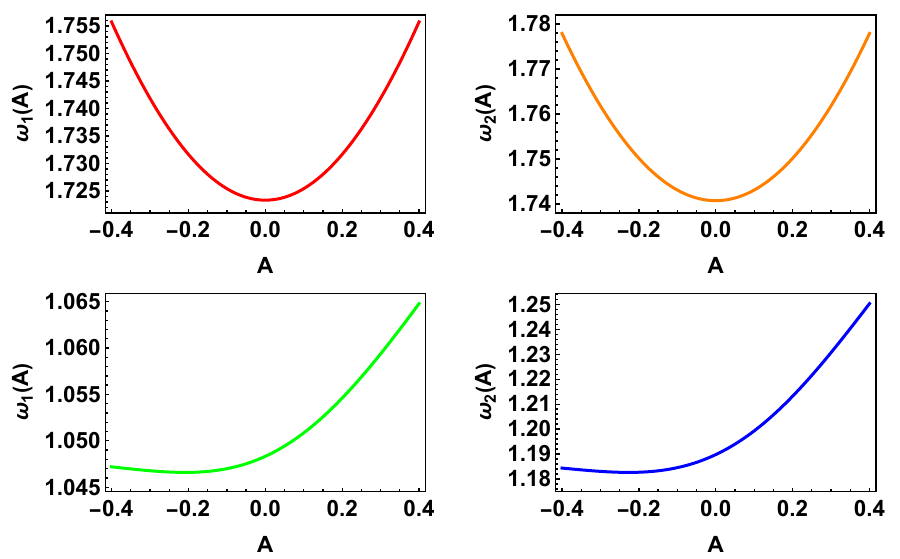}
        \caption{Upper panel: Spectral flow of the first two positive internal modes around the $\phi^4$ sphaleron for $k = 0.99$. Lower panel: Spectral flow of the first two positive internal modes around the $\phi^6$ sphaleron for $C = 0.001$. The amplitude $A$ denotes the strength of the corresponding unstable mode.}
        \label{f:spectralFlow_phi4_phi6}
    \end{figure}

    A similar behaviour was observed in the deformed $\phi^6$ well model introduced in \autoref{c:Sph_False}, where the stabilisation arises dynamically. As discussed there, when the sphaleron is perturbed by the unstable mode and forced to collapse, it departs from its unstable equilibrium and its walls approach each other,  experiencing an attractive static force. This process corresponds to the non-linear evolution of the perturbed sphaleron and, in the kink–antikink picture, can be interpreted as a force attracting the kink and the antikink to one another. Therefore, the initial excitation of the unstable mode results in an effective change in the size of the sphaleron. On the other hand, the linear modes of the sphaleron have a fixed position in the linear spectrum. As the sphaleron starts to collapse, the modes do not disappear but move through the spectrum, changing their shape and frequency. Generically, this spectral flow generates an effective force of the form (\ref{e:Force}). As illustrated in \autoref{f:spectralFlow_phi4_phi6}, the flow of the internal modes exhibits a minima with the variation of the unstable amplitude. As a consequence, the force is repulsive. The balance between these two competing forces leads to the critical value of the unstable mode amplitude $A_c$ for which the sphaleron presents the quasi-static behaviour.

     In \autoref{c:Sph_False}, we were able to determine the condition for the balance by means of a perturbative calculation. As indicated in that chapter, the derivation is completely general, and does not assume any particular model. Following the same steps, we deduce the expression for the critical amplitude 
    \begin{equation}\label{e:crit}
        A_c = \biggl(\frac{4\,\omega_{-}^2}{\sum_i \tfrac{\gamma_i^2 \delta_i}{\omega_i^4}}\biggr)^{1/3}\,.
    \end{equation}
    The only difference is that the domain of integration is the circle and not the real line. For that reason, we define $\gamma_i$ and $\delta_i$ as
    \begin{equation}
    \begin{split}
        \gamma_i = \dfrac{1}{2} \int_{\mathbbm{S}^1} U^{(3)}[\Phi_{sph}(x)]\,\eta^2_{-}(x)\,\eta_{i}(x)\,dx\,, \qquad
        \delta_i =\int_{\mathbbm{S}^1} U^{(3)}[\Phi_{sph}(x)]\,\eta_{-}(x)\,\eta^2_i(x)\,dx\,,
        \end{split}
    \end{equation}
    where $i$ denotes the positive modes.

\begin{figure*}[htb]
    \centering
    \begin{subfigure}[t]{0.5\textwidth}
        \centering
        \includegraphics[width=0.99\textwidth]{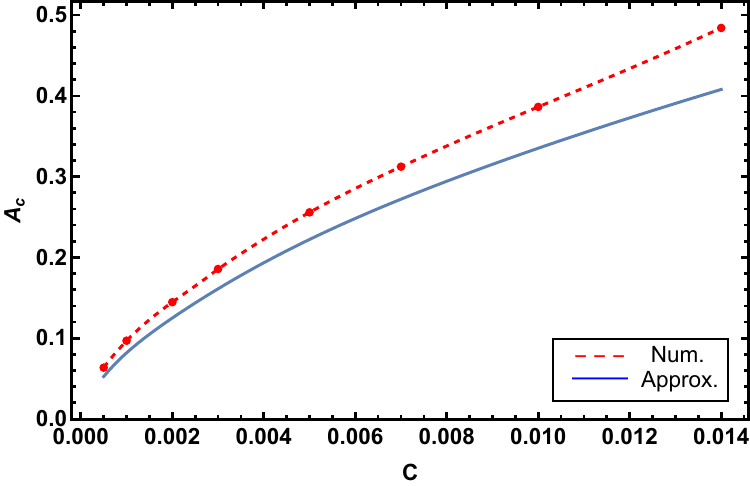}
        \caption{Critical amplitude comparison for the $\phi^6$ model}
        \label{f:ComparionAc}
    \end{subfigure} \vspace{1em}  
    \begin{subfigure}[t]{0.5\textwidth}
        \centering
        \includegraphics[width=0.93\textwidth]{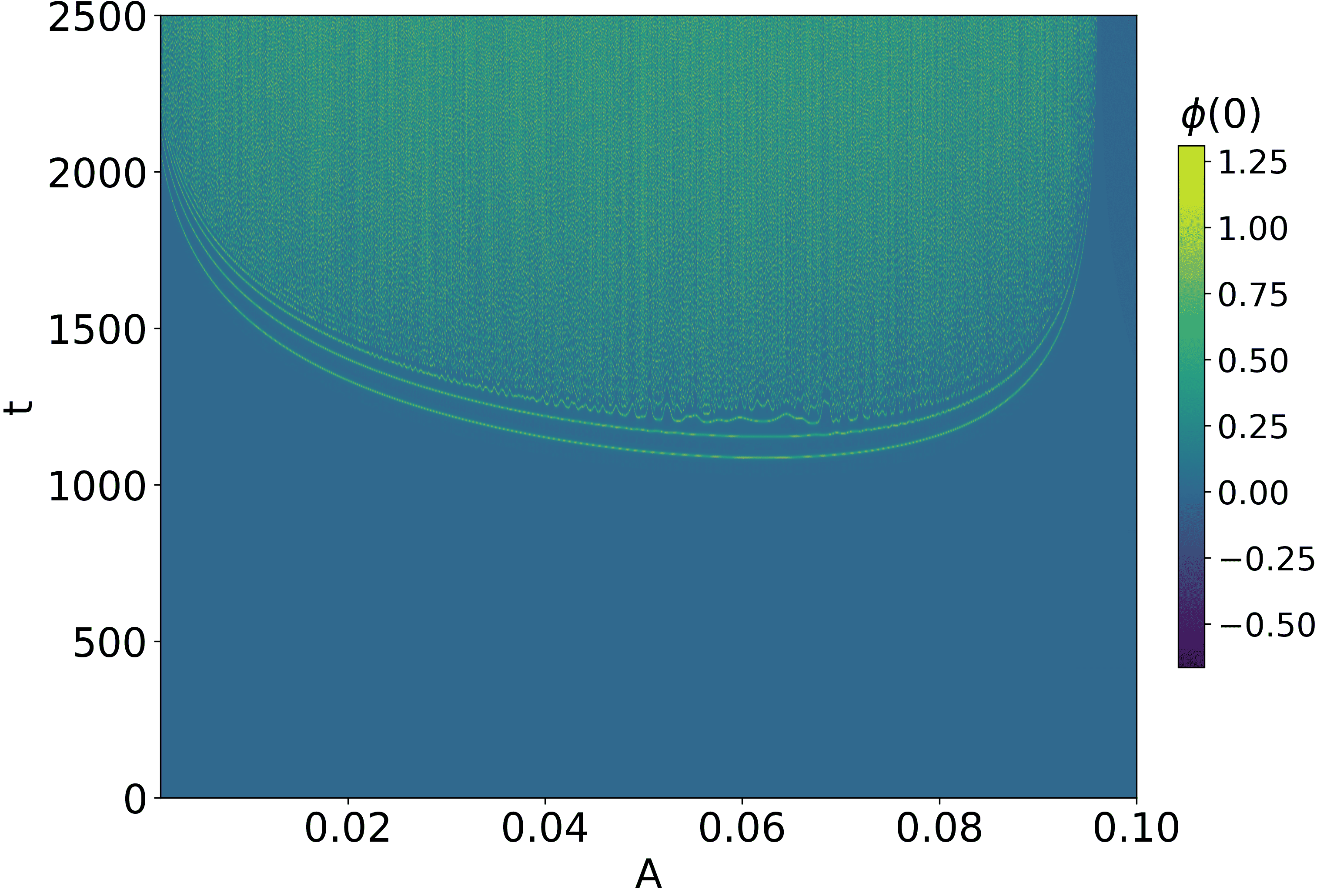}
        \caption{$\phi(0,t)$ for $C = 0.001$. }
        \label{f:Stabilization_phi6_A0.001}
    \end{subfigure}%
    \begin{subfigure}[t]{0.5\textwidth}
        \centering
        \includegraphics[width=0.895\textwidth]{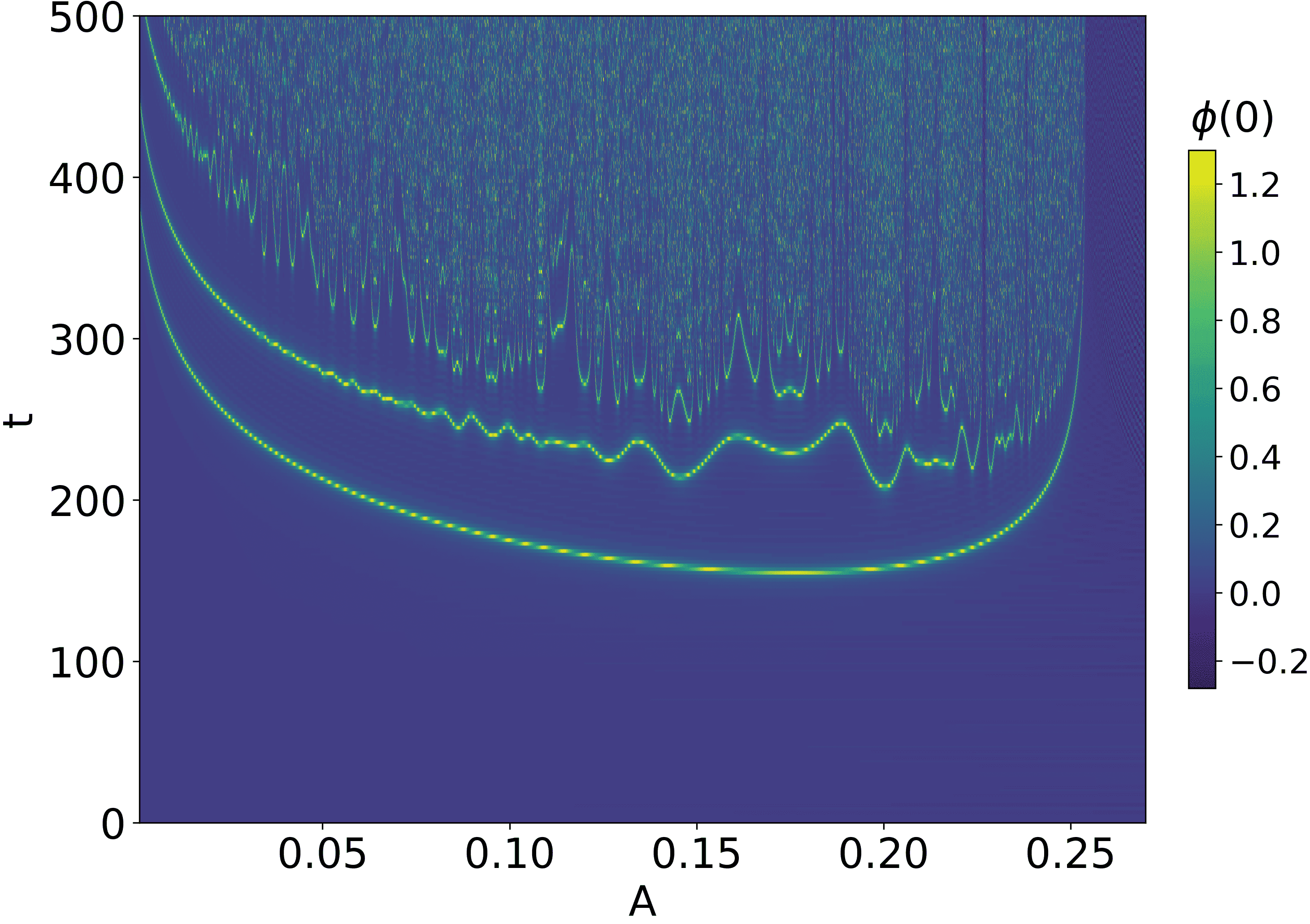}
        \caption{$\phi(0,t)$ for $C = 0.005$.}
        \label{f:Stabilization_phi6_A0.005}
    \end{subfigure}

    \caption{Upper panel: Comparison between the numerical critical amplitude and the analytical prediction given by equation $(\ref{e:crit})$ for the $\phi^6$ model for different values of the model parameter $C$. Lower panel: Stabilisation of the $\phi^6$ sphaleron with $C = 0.001$  and $C = 0.005$. The amplitude $A$ accounts for the initial excitation of the unstable mode.}
    \label{f:phi6_stabilization}
    \end{figure*}

    A comparison between (\ref{e:crit}) and full-numerical results is shown in \autoref{f:ComparionAc} for the $\phi^6$ model. Two representative examples of the stabilisation in $\phi^6$ can be found in \autoref{f:Stabilization_phi6_A0.001} and \autoref{f:Stabilization_phi6_A0.005}. The agreement between the analytical formula and full numerics is good, although at large critical amplitudes, higher-order perturbative effects have to be taken into account. This should be responsible for the small deviations observed in \autoref{f:ComparionAc} as $A_c$ grows. 

    It is natural to ask now why dynamical stabilisation is not observed in the $\phi^4$ model. In this case, the sphaleron cannot be stabilised solely by exciting the unstable mode due to the reflection symmetry inherent in the theory. This symmetry causes the $\gamma_i$ terms, which account for the excitation of the positive modes through their coupling with the unstable mode, to vanish. A similar argument applies to the $\delta_i$ terms, representing the force balance at quadratic order. Consequently, stabilisation in the $\phi^4$ model appears to be a higher-order effect that requires the independent excitation of the positive modes, as shown previously. However, a precise determination of the critical amplitude for this scenario is beyond the scope of the present chapter and is left to future work.

    To conclude, we emphasise that the results presented in this chapter, together with the analogous mechanism identified in the context of the false vacuum sphaleron in \autoref{c:Sph_False}, suggest that the stabilisation mechanism is completely general. Although we have focused on certain initial conditions, it is expected that multiple physical processes may trigger sphaleron stabilisation. The resulting dynamics depend on the relative interplay among the modes involved. For instance, the interaction of the sphaleron with radiation could excite internal modes, which in turn may facilitate the stabilisation process.

\section{Summary and conclusions}\label{s:Conclusions_Circle}

    In this chapter, we have described the perturbed non-linear dynamics of sphalerons in some relevant one-dimensional models with $\mathbbm{S}^1$ as base space. First, we have introduced the $\phi^6$ and the sine-Gordon model, which, along with the $\phi^4$ model discussed in \autoref{c:Sph_Circle}, constitute the simplest scalar field theories that support sphaleron solutions on the circle. Notably, all these models admit analytical expressions for the sphaleron configuration, although their internal structures are only partially known analytically in the cases of the $\phi^4$ and the sine-Gordon model. In the $\phi^6$ model, the equation governing the linear spectrum of perturbations does not reduce to a Lamé equation, and therefore a full numerical analysis is required. 

    These models exhibit distinct features regarding spatial reflection symmetry and the integrability of the theory. As a result, they provide valuable insight into the diverse dynamical behaviours that can arise during the decay process. A natural approach to perform this analysis is simply to excite the sphaleron along the unstable direction by perturbing it with its unstable mode. We have begun by analysing the $\phi^4$ model. We have seen that the sphaleron forms an oscillon state in most of the cases when the sphaleron is sufficiently small. On the other hand, whether the sphaleron has a considerable size and admits the interpretation of a kink-antikink composite, a very rich structure appears during the decay. Specifically, we can recover the characteristic fractal pattern of the kink-antikink scattering on the real line when suitable initial conditions are assumed. This correspondence between the sphaleron decay on the circle and the kink-antikink scattering on the real line suggests that the resonant energy transfer mechanism, which is crucial to understanding the dynamics of solitons, also plays a significant role in the dynamics of sphalerons. Moreover, the underlying reflection symmetry of the system implies that the decay along both unstable directions is equivalent.

    Conversely, the $\phi^6$ theory lacks this symmetry, and as a result, the decay processes are inequivalent, depending on the vacuum to which the configuration evolves. We have seen that, by selecting the direction of excitation of the unstable mode, one can obtain either an oscillon-like configuration or a chaotic pattern characterised by bounces. This behaviour can be traced back to the fact that the support of the internal modes becomes delocalised within the inner region of the sphaleron. Once more, when sufficiently big sphalerons are considered in $\mathbb{S}^1$, we can recover the scattering between different species of $\phi^6$ kinks on the real line. It is important to note that the analogy between the decay of the sphaleron in $\phi^4$ and $\phi^6$ with the kink-antikink/antikink-kink scattering on the real line holds at time scales of the order of the size of the base space. For longer times, the periodicity of the base space allows any mode emitted from the interaction region to interact again, transforming the configuration into a superposition of modes. This contrasts with the situation on the real line, where the possible asymptotic states resulting from the soliton scattering can be described as a superposition of kinks at certain velocities plus radiation. Here, the fact that our base space is bounded suggests that the system should have a Poincaré recurrence time. Therefore, at some possibly large but finite time, the configuration would be arbitrarily close to the initial state, leading to a sort of quasi-periodic configuration. 
    
    The evolution of sine-Gordon sphalerons is rather different. Due to the integrability of the model, the sphaleron evolves into a time-periodic solution, which we have been able to identify analytically. At all times, this evolution can be interpreted as a periodic transition between sphaleron and antisphaleron configurations.

    Finally, we have also shown that it is possible to stabilise the $\mathbb{S}^1$ sphalerons due to the internal mode structure. The excitation of the positive internal modes leads to a positive pressure that compensates the static attraction in the direction of the unstable mode. Depending on whether the model exhibits reflection symmetry, we have shown that stabilisation may or may not require the internal modes to be initially excited. This stabilisation may increase the lifespan of the sphaleron by producing an oscillatory solution that lasts as long as the internal mode remains excited and above a critical amplitude. We would like to note that the de-excitation of sphalerons is currently under investigation by P. Bizon, J. Jedrej and T. Romanczukiewicz. It should be remarked that stabilisation occurs over a wide amplitude range of positive modes; therefore, it does not appear to be a fine-tuned phenomenon. 

     This framework could be applied to Abelian-Higgs vortices with $n \geq 2$ and $\lambda > 1$, where the vortices are unstable against splitting as seen in \autoref{c:Topological}. Moreover, this mechanism may be relevant for stabilising magnetic skyrmions in the ferromagnetic background of chiral magnets \cite{Lin:2014}. Although skyrmion crystals remain stable in the low-field region, a single magnetic skyrmion is unstable at a critical magnetic field. Finally, an extension of these results to the electroweak sphaleron would be of particular interest. The role of internal modes on the decay of the electroweak sphaleron and its possible modification of the decay rate is left for future investigation.

        \chapter*{Conclusions and future research lines}
        \markboth{CONCLUSIONS AND FUTURE RESEARCH LINES}{CONCLUSIONS AND FUTURE RESEARCH LINES}
        \phantomsection 
        \addcontentsline{toc}{chapter}{\small{CONCLUSIONS AND FUTURE RESEARCH}}
        \label{c:Conclusions}

    \epigraph{"What we know is a drop, what we don't know is an ocean."}{-- Isaac Newton}

    The common thread throughout this thesis has been analysing the impact of internal modes on soliton dynamics. In order to do that, we have combined numerical and analytical techniques. Concretely, we have carried out numerical simulations within the full field theory and developed effective models using the collective coordinate method, as well as perturbative approaches, to extract approximate analytical results that account for the observed phenomenology. Now, the main conclusions regarding the results obtained in previous chapters are summarised to present an overall view of the work conducted. We will also explain the working lines that can be explored in the future.

    \begin{itemize}
        \item In this thesis, we have first introduced an effective description of radiation within the collective coordinate method in the one-dimensional $\phi^4$ model. Contrary to previous approaches, our effective model introduces genuine scattering modes appearing in the linear spectrum of perturbations around the kink. The relevance of this achievement lies in the variety of phenomena involving radiation that the corresponding effective model can capture. Through this description, we have unified well-known results from the literature within a single framework and provided a comprehensive explanation, highlighting the coupling between radiation, internal modes, and translational modes. More precisely, we have analytically derived an approximate expression, valid throughout the entire space, for the radiation field emitted by an excited kink, that reduces to the results by Manton and Merabet in the spatial asymptotic limit. Moreover, we have explained the resonant excitation of the shape mode by an incoming radiation wave whose frequency is twice that of the shape mode in terms of a Mathieu equation. On the other hand, we have obtained an expression describing the excitation of the shape mode outside the resonant regime. Interestingly, the effective model includes relativistic corrections that can be relevant for scattering processes, although it is not particularly efficient to this end unless a judicious choice of scattering modes is considered. Even though the collective coordinate model can be applied to oscillons and distinguishes between unstable and stable configurations, an alternative effective description of radiation has been proposed in this scenario, which captures more accurately both the decay rate and the double-frequency oscillation of well-established oscillons. Remarkably, the model is also able to reproduce the creation of kink-antikink pairs. 
    
        The results of this investigation suggest that radiation modes play a crucial role in the study of solitons dynamics and could be fundamental to disentangle the complicated patterns in soliton scattering processes. A promising extension of our approach consists in applying it to study kink-antikink collisions. Through the inclusion of dissipative degrees of freedom, we could describe the dissipation of energy or the energy exchange with other modes during the collision. Furthermore, we could describe the annihilation of kink-antikink pairs into radiation, either after the formation of a bion or via direct conversion. 

        \item We have then generalised Samols's metric on the moduli space of Abelian-Higgs one-vortex beyond the BPS limit. To this end, we have considered a field ansatz that includes both the zero mode and an additional mode associated with variations in the vortex size, with their amplitudes treated as collective coordinates. Specifically, we have incorporated either its only shape mode or a Derrick mode along the direction of motion. We have emphasised the need to redefine the naive notions of zero mode and Derrick mode in the context of gauge theories. In particular, the usual infinitesimal translations and spatial scalings must be reformulated to ensure gauge invariance and compliance with Gauss's law. Our results demonstrate that the presence of the additional mode modifies the moduli space metric governing vortex dynamics. Interestingly, at the level of the effective model, a constant motion of the vortex requires a non-null, constant amplitude of the mode. This solution approximates the Lorentz boost, but there are some drawbacks. On the one hand, the shape-mode-based model does not reproduce the Lorentz contraction due to the axial symmetry of this mode, and the Derrick-mode-based model only crudely approximates the frequency of the shape mode. To address this, we have incorporated two Derrick modes aligned along orthogonal spatial directions, in order to capture non-axially symmetric deformations and to recover the correct frequency of the shape mode.

        This last effective description provides the suitable initial conditions to model the collision of vortex-vortex or vortex-antivortex pairs. As further work, we could construct a collective coordinate model for two vortices using the procedure established in that chapter. This will allow us to extract semi-analytical insight into this process. A first approximation could be to assume that the vibrational modes remain frozen in the form derived in the free vortex (or antivortex) limit. This approach, commonly used in studies of kink-antikink scattering, is known as the \textit{frozen vibrational moduli space approximation}. This approach should be sufficient to explain the excitation of vortices after their collision at high velocities. However, the form of the normal modes as well as their frequencies depend non-trivially on the distance between the vortices. Therefore, phenomena such as mode-induced forces or spectral walls cannot be captured within the frozen vibrational moduli space approximation. As a consequence, studying collisions of initially excited vortices or capturing the emergence of spectral walls requires incorporating the dynamics of the internal structure, within the \textit{dynamical vibrational moduli space approximation}.

        \item Apart from studying the role of internal modes in the dynamics of topological solitons, we have also analysed their impact on the dynamics of unstable solitons, specifically, sphaleron-like solutions. To this end, we have first introduced a family of one-dimensional models with an impurity that support a new class of sphalerons, which we have termed semi-BPS sphalerons. These solutions form a BPS state with the impurity, which ensures the absence of static forces between the sphaleron and the impurity. As a result, we can guarantee that the observed dynamics arise purely from the influence of the internal modes. In particular, we have found that most of the phenomena observed in the dynamics of stable BPS configurations also appear here. For example, the simplest dynamics follows the geodesic flow, and the spectral wall phenomenon can emerge once a suitable internal mode is excited above a critical amplitude. However, the inherent unstable mode will eventually cause the sphaleron to decay. In fact, any small perturbation would have a projection onto the unstable mode. The subsequent configuration (an oscillon) is not a BPS object and therefore interacts with the impurity in a rather sophisticated manner. In this sense, the semi-BPS sphaleron is a weaker concept than the standard BPS limit.

        In the future, we plan to replace the non-dynamical impurity with a dynamical field. The idea is to identify a one-parameter family of sphalerons in these models with two coupled fields and analyse their dynamical properties. An illustrative example is the MSTB model. In this model, there exists a range of values of the coupling constant for which a family of non-topological kink solutions emerges. This family can be interpreted as a non-linear superposition of topological solitons. Due to the absence of topological protection, these configurations are unstable. They may decay into the vacuum by forming an oscillon, or alternatively, decay into two less energetic topological kinks. We should compare the dynamics with the results obtained in our simplified model.
                
        \item Since the unstable mode dominates the evolution of sphalerons, we have investigated the role of internal modes during their decay. To this end, we have introduced two deformations of the one-dimensional $\phi^6$ model that support analytical sphaleron solutions. The deformation parameter allows us to shift the vacua, thereby creating false vacua and effectively controlling the size of the sphaleron. Although the resulting configurations are qualitatively similar in appearance, they exhibit fundamentally different internal structures: in one case, the sphaleron-like solution lacks internal modes, while in the other, it possesses a non-zero and growing number of them. This contrast originates from the differing mass thresholds associated with the vacua of the theory, which leads to spectral problems with a barrier or a well in their effective potentials around the sphalerons. When studied the collapse of these configurations due to the excitation of the unstable mode, we have shown that the decay process is largely insensitive to internal modes when the sphalerons have small amplitude. The subsequent oscillons present the same modulated amplitude independently of the model. However, significant differences emerge when the sphalerons resemble a kink–antikink configuration. Specifically, we have observed that internal modes are excited during the evolution due to non-linear couplings and store energy in a very efficient way, inducing multiple bounces before the sphaleron ultimately collapses into an oscillon. We have derived an approximate analytical expression describing the excitation of these modes at the beginning of the simulation within the framework of the collective coordinate method, showing a great accordance with field theory. Remarkably, when the initial amplitude of the unstable mode is sufficiently large, we have observed that the positive pressure exerted by the internal modes can change the direction of decay during the collapse. Notably, we have identified an intermediate critical amplitude at which the lifespan of the sphaleron increases significantly, leading to a form of dynamical stabilisation. Such a critical amplitude has been determined analytically with great accordance through an approximate expression derived from a perturbative expansion.

        \item To further develop the previous work, we have finally studied the dynamics of sphalerons defined on a circle, focusing in particular on the $\phi^4$, $\phi^6$, and sine-Gordon models. Their formulation on a compact domain is particularly appealing, as it often allows for explicit expressions of the sphaleron solutions and, in some cases, a quasi-solvable internal mode structure. Each model exhibits distinct properties, and we have analysed how these influence sphaleron dynamics. Once again, we have examined the collapse of sphalerons by perturbing them with their unstable mode. We have shown that the $\phi^4$ theory displays symmetric decay along the two available unstable directions, whereas this symmetry is absent in the $\phi^6$ model. These differences arise from the presence or absence of reflection symmetry, respectively. Interestingly, we have verified that the collapse of sphalerons on the circle reproduces the typical patterns observed in kink–antikink scattering, provided that the circumference of the circle is large compared to the size of the individual kinks composing the sphaleron and a suitable initial condition is considered. Moreover, we have demonstrated that the reflection symmetry plays a central role in the stabilisation mechanisms. While the $\phi^4$ sphaleron requires the initial excitation of the internal modes for stabilisation, in the $\phi^6$ model, the configuration can be stabilised solely through the excitation of the unstable mode and subsequent excitation of the internal modes by non-linear couplings. Our results show that the stabilisation mechanism may be ubiquitous in more general models and is not a fine-tuned phenomenon. Moreover, it constitutes an efficient mechanism that significantly enhances the life-span of the configuration. On the other hand, analysing the dynamics in the sine-Gordon model, we have found that perturbing the initial sphaleron with its unstable mode causes it to evolve periodically, interpolating between the sphaleron and the antisphaleron. This behaviour is a consequence of the integrability of the theory. In fact, we have succeeded in identifying a breather-like exact analytic expression that describes this process, although a relation between the observed oscillation frequency and the initial amplitude of the unstable mode has not been determined. Therefore, the connection must be established through a quasi-empirical approach. 

        As further work, it would be enlightening to extend our results to higher dimensional theories. Naturally, the final objective is to analyse the dynamics of the electroweak sphaleron. It would be valuable to investigate whether our results are observable and relevant in that context, or whether dominant physical effects render the aforementioned phenomena subleading. Therefore, investigating the decay of the electroweak sphaleron could be engaging due to the physical implications that it could have.   
    \end{itemize}

    We would also like to emphasise that, in addition to the possible extensions of the articles that constitute this thesis, we are currently investigating two distinct topics. We present here a brief overview of the main ideas underlying each of them.

    \begin{itemize}
        \item The dynamics of $n = 1$ vortices with its only excited massive bound mode has previously been investigated \cite{Pillado:2024}. In particular, both the asymptotic behaviour of the radiation emitted by the shape mode and the decay rate of its amplitude have been derived analytically. Our objective is to extend this analysis to vortices with higher winding number. Due to the vortex instability for $\lambda > 1$, we restrict our analysis to the regime $\lambda < 1$. In this region, higher-charge vortices admit additional internal modes, including non-axially symmetric ones, which suggests that the resulting radiation emission should be anisotropic. Firstly, we shall compute the asymptotic radiation field emitted by an excited vortex and then determine the decay law governing the evolution of the amplitude of these internal modes. Furthermore, we aim to investigate the suppression of radiation in specific emission channels, a phenomenon expected to arise from the dependence of the mode frequencies on the self-coupling constant. Interestingly, the triggering of the non-axially symmetric internal mode for $n = 2$ leads to a periodic displacement of the zero with multiplicity two of the vortex: the vortex zeros oscillate alternately along the $x_1$ and $x_2$ directions, resembling a $\pi/2$ scattering. As a result, the radiation field emitted by the excited two-vortex configuration is expected to offer valuable insights into the mechanisms of energy emission during vortex–vortex scattering processes. These findings would serve as a complementary work to existing estimates of radiated energy in the scattering of slowly moving BPS monopoles \cite{Manton3:1988}.   
        
        \item While the study of kink solutions in scalar field theories involving a single field has been largely developed, significantly less is known about models with multiple degrees of freedom. In this work, we aim to investigate the Lie symmetries associated with the differential equations governing the BPS sector of a given theory. By determining these symmetries, we aim to construct new families of solutions within a given topological sector. This approach would have the potential to reveal hidden symmetries that are not manifest at the level of the Lagrangian. Consequently, this work should provide insight into the structure of the moduli space and the moduli that characterise it.
    \end{itemize}

    As a concluding remark, this thesis has highlighted the relevance of internal modes in soliton dynamics. As evidenced, these modes can drastically alter the expected behaviour of solitons. Given the potential applications of solitons and their ubiquitous presence in various areas of physics, a comprehensive understanding of their dynamics is crucial.
    There is a lot of promising work to be done regarding the extension of our results to higher dimensions. Moreover, several open questions remain that deserve further investigation and particular attention. It is hoped that the contributions presented here serve as a foundation for future research.

        \chapter*{Conclusiones y futuras líneas de investigación}
        \markboth{CONCLUSIONES Y FUTURAS LÍNEAS DE INVESTIGACIÓN}{CONCLUSIONES Y FUTURAS LÍNEAS DE INVESTIGACIÓN}
        \phantomsection 
        \addcontentsline{toc}{chapter}{\small{CONCLUSIONES E INVESTIGACIÓN FUTURA}}
        \label{c:Conclusiones}

    \epigraph{"What we know is a drop, what we don't know is an ocean."}{-- Isaac Newton}

El hilo conductor a lo largo de esta tesis ha sido el análisis del impacto de los modos internos en la dinámica de los solitones. Para ello, hemos combinado técnicas numéricas y analíticas. Concretamente, hemos llevado a cabo simulaciones numéricas dentro de la teoría de campos completa y hemos desarrollado modelos efectivos utilizando el método de coordenadas colectivas, así como enfoques perturbativos, para extraer resultados analíticos aproximados que den cuenta de la fenomenología observada. A continuación, se resumen las principales conclusiones relativas a los resultados obtenidos en los capítulos anteriores, con el fin de presentar una visión global del trabajo realizado. A mayores, explicaremos las líneas de trabajo que pueden explorarse en el futuro.

\begin{itemize}
\item En esta tesis, hemos introducido en primer lugar una descripción efectiva de la radiación dentro del método de coordenadas colectivas en el modelo $\phi^4$ unidimensional. A diferencia de enfoques anteriores, nuestro modelo efectivo introduce modos de dispersión genuinos que aparecen en el espectro lineal de perturbaciones alrededor del ''kink''. La relevancia de este logro radica en la variedad de fenómenos que involucran radiación que el modelo efectivo correspondiente es capaz de capturar. A través de esta descripción, hemos unificado resultados bien conocidos de la literatura dentro de un único marco y hemos proporcionado una explicación detallada, haciendo énfasis en el acoplamiento entre la radiación, los modos internos y los modos traslacionales. Más precisamente, hemos derivado analíticamente una expresión aproximada, válida en todo el espacio, para el campo de radiación emitido por un kink excitado, que se reduce a los resultados de Manton y Merabet en el límite asintótico espacial. Además, hemos explicado la excitación resonante del ''shape mode'' por una onda de radiación incidente cuya frecuencia es el doble de la del ''shape mode'' en términos de una ecuación de Mathieu. Por otro lado, hemos obtenido una expresión que describe la excitación del ''shape mode'' fuera del régimen resonante. Notablemente, el modelo efectivo incluye correcciones relativistas que pueden ser relevantes para procesos de dispersión, aunque no resulta particularmente eficiente con este fin a menos que se realice una elección juiciosa de los modos de dispersión. Aunque el modelo de coordenadas colectivas puede aplicarse a oscilones y distingue entre configuraciones inestables y estables, en este escenario se ha propuesto una descripción efectiva alternativa de la radiación, que captura con mayor precisión tanto la tasa de decaimiento como la oscilación de doble frecuencia de oscilones bien establecidos. Notablemente, el modelo también es capaz de reproducir la creación de pares kink–antikink.

Los resultados de esta investigación sugieren que los modos de radiación desempeñan un papel crucial en el estudio de la dinámica de solitones y podrían ser fundamentales para desentrañar los complejos patrones presentes en los procesos de dispersión de solitones. Una extensión prometedora de nuestro enfoque consiste en aplicarlo al estudio de colisiones kink–antikink. Mediante la inclusión de grados de libertad disipativos, podríamos describir la disipación de energía o el intercambio de energía con otros modos durante la colisión. A mayores, podríamos describir la aniquilación de pares ''kink–antikink'' en radiación, ya sea tras la formación de un bión o mediante una conversión directa.

\item A continuación, hemos generalizado la métrica de Samols en el espacio de módulos de un vórtice Abelian-Higgs más allá del límite BPS. Para ello, hemos considerado un ansatz para los campos que incluye tanto el modo cero como un modo adicional asociado a variaciones en el tamaño del vórtice, tratando sus amplitudes como coordenadas colectivas. En concreto, hemos incorporado bien su único ''shape mode'' o bien un modo de Derrick a lo largo de la dirección del movimiento. También, hemos enfatizado la necesidad de redefinir las nociones ingenuas de modo cero y modo de Derrick en el contexto de teorías gauge. En particular, las traslaciones infinitesimales habituales y los escalados espaciales deben reformularse para garantizar la invariancia gauge y el cumplimiento de la ley de Gauss. Nuestros resultados demuestran que la presencia del modo adicional modifica la métrica del espacio de módulos que gobierna la dinámica del vórtice. Notablemente, a nivel del modelo efectivo, un movimiento constante del vórtice requiere una amplitud no nula y constante del modo. Esta solución aproxima el boost de Lorentz, pero presenta algunas limitaciones. Por un lado, el modelo basado en el ''shape mode'' no reproduce la contracción de Lorentz debido a la simetría axial de dicho modo, y el modelo basado en el modo de Derrick solo aproxima de manera burda la frecuencia del modo de forma. Para abordar este problema, hemos incorporado dos modos de Derrick alineados a lo largo de direcciones espaciales ortogonales, con el fin de capturar deformaciones no axialmente simétricas y recuperar la frecuencia correcta del ''shape mode''.

Esta última descripción efectiva proporciona las condiciones iniciales adecuadas para modelar la colisión de pares vórtice–vórtice o vórtice–antivórtice. Como trabajo futuro, podríamos construir un modelo de coordenadas colectivas para dos vórtices utilizando el procedimiento establecido en ese capítulo. Esto nos permitirá extraer una comprensión semi-analítica de este proceso. Una primera aproximación podría consistir en suponer que los modos vibracionales permanecen congelados en la forma derivada en el límite de vórtice (o antivórtice) libre. Este enfoque, utilizado habitualmente en estudios de dispersión kink–antikink, se conoce como la \textit{aproximación del espacio de módulos vibracional congelados}. Dicho enfoque debería ser suficiente para explicar la excitación de los vórtices tras su colisión a altas velocidades. Sin embargo, la estructura de los modos normales, así como sus frecuencias, dependen de manera no trivial de la distancia entre los vórtices. Por lo tanto, fenómenos como fuerzas inducidas por modos o las ''spectral walls'' no pueden capturarse dentro de la aproximación del espacio de módulos vibracional congelados. Como consecuencia, el estudio de colisiones de vórtices inicialmente excitados o la captura de la aparición de ''spectral walls'' requiere incorporar la dinámica de la estructura interna, dentro de la \textit{aproximación dinámica del espacio de módulos vibracional}.

\item Además de estudiar el papel de los modos internos en la dinámica de solitones topológicos, también hemos analizado su impacto en la dinámica de solitones inestables, en particular, soluciones tipo esfalerón. Con este fin, hemos introducido en primer lugar una familia de modelos unidimensionales con una impureza que soportan una nueva clase de esfalerones, a los que hemos denominado esfalerones semi-BPS. Estas soluciones forman un estado BPS con la impureza, lo que garantiza la ausencia de fuerzas estáticas entre el esfalerón y la impureza. Como resultado, podemos asegurar que la dinámica observada surge puramente de la influencia de los modos internos. En particular, hemos encontrado que la mayoría de los fenómenos observados en la dinámica de configuraciones BPS estables también aparecen aquí. Por ejemplo, la dinámica más simple sigue el flujo geodésico, y el fenómeno de la ''spectral wall'' puede emerger una vez que un modo interno adecuado se excita por encima de una amplitud crítica. Sin embargo, el modo inestable inherente acabará provocando el decaimiento del esfalerón. De hecho, cualquier pequeña perturbación tendrá una proyección sobre el modo inestable. La configuración subsiguiente (un oscilón) no es un objeto BPS y, por tanto, interactúa con la impureza de una manera bastante sofisticada. En este sentido, el esfalerón semi-BPS es un concepto más débil que el límite BPS estándar.

En el futuro, planeamos sustituir la impureza no dinámica por un campo dinámico. La idea es identificar una familia uniparamétrica de esfalerones en estos modelos con dos campos acoplados y analizar sus propiedades dinámicas. Un ejemplo ilustrativo es el modelo MSTB. En este modelo, existe un rango de valores de la constante de acoplamiento para el cual emerge una familia de soluciones kink no topológicas. Esta familia puede interpretarse como una superposición no lineal de solitones topológicos. Debido a la ausencia de protección topológica, estas configuraciones son inestables. Pueden decaer al vacío formando un oscilón o, alternativamente, decaer en dos kinks topológicos menos energéticos. Deberíamos comparar la dinámica con los resultados obtenidos en nuestro modelo simplificado.
            
\item Dado que el modo inestable domina la evolución de los esfalerones, hemos investigado el papel de los modos internos durante su decaimiento. Con este fin, hemos introducido dos deformaciones del modelo unidimensional $\phi^6$ que admiten soluciones analíticas de esfalerones. El parámetro de deformación nos permite desplazar los vacíos, creando así falsos vacíos y controlando de manera efectiva el tamaño del esfalerón. Aunque las configuraciones resultantes son cualitativamente similares en apariencia, presentan estructuras internas fundamentalmente diferentes: en un caso, la solución tipo esfalerón carece de modos internos, mientras que en el otro posee un número no nulo y creciente de ellos. Este contraste se origina en los distintos umbrales de masa asociados a los vacíos de la teoría, que conducen a problemas espectrales con una barrera o un pozo en sus potenciales efectivos alrededor de los esfalerones. Al estudiar el colapso de estas configuraciones debido a la excitación del modo inestable, hemos demostrado que el proceso de decaimiento es en gran medida insensible a los modos internos cuando los esfalerones tienen pequeña amplitud. Los oscilones subsiguientes presentan la misma amplitud modulada independientemente del modelo. Sin embargo, emergen diferencias significativas cuando los esfalerones se asemejan a una configuración kink–antikink. En concreto, hemos observado que los modos internos se excitan durante la evolución debido a acoplamientos no lineales y almacenan energía de manera muy eficiente, induciendo múltiples rebotes antes de que el esfalerón colapse finalmente en un oscilón. Hemos derivado una expresión analítica aproximada que describe la excitación de estos modos al comienzo de la simulación dentro del marco del método de coordenadas colectivas, mostrando una gran concordancia con la teoría de campos. De forma notable, cuando la amplitud inicial del modo inestable es suficientemente grande, hemos observado que la presión positiva ejercida por los modos internos puede cambiar la dirección del decaimiento durante el colapso. En particular, hemos identificado una amplitud crítica intermedia en la que la vida media del esfalerón aumenta de manera significativa, dando lugar a una forma de estabilización dinámica. Dicha amplitud crítica ha sido determinada analíticamente con gran precisión mediante una expresión aproximada derivada de una expansión perturbativa.

\item Para desarrollar aún más el trabajo anterior, hemos estudiado finalmente la dinámica de esfalerones definidos sobre un círculo, centrándonos en particular en los modelos $\phi^4$, $\phi^6$ y seno-Gordon. Su formulación en un dominio compacto resulta especialmente atractiva, ya que a menudo permite expresiones explícitas de las soluciones de esfalerones y, en algunos casos, una estructura de modos internos cuasi-resoluble. Cada modelo presenta propiedades distintas, y hemos analizado cómo estas influyen en la dinámica de los esfalerones. Una vez más, hemos examinado el decaimiento de los esfalerones perturbándolos con su modo inestable. Hemos demostrado que la teoría $\phi^4$ presenta un decaimiento simétrico a lo largo de las dos direcciones inestables disponibles, mientras que esta simetría está ausente en el modelo $\phi^6$. Estas diferencias surgen de la presencia o ausencia, respectivamente, de simetría de reflexión. Notablemente, hemos verificado que el decaimiento de esfalerones en el círculo reproduce los patrones típicos observados en la dispersión kink–antikink, siempre que la circunferencia del círculo sea grande en comparación con el tamaño de los kinks individuales que componen el esfalerón y se considere una condición inicial adecuada. Además, hemos demostrado que la simetría de reflexión desempeña un papel central en los mecanismos de estabilización. Mientras que el esfalerón $\phi^4$ requiere la excitación inicial de los modos internos para su estabilización, en el modelo $\phi^6$ la configuración puede estabilizarse únicamente mediante la excitación del modo inestable y la posterior excitación de los modos internos a través de acoplamientos no lineales. Nuestros resultados muestran que el mecanismo de estabilización puede ser ubicuo en modelos más generales y no constituye un fenómeno finamente ajustado. Además, constituye un mecanismo eficiente que incrementa de manera significativa la vida media de la configuración. Por otro lado, al analizar la dinámica en el modelo seno-Gordon, hemos encontrado que perturbar el esfalerón inicial con su modo inestable provoca una evolución periódica, interpolando entre el esfalerón y el antiesfalerón. Este comportamiento es consecuencia de la integrabilidad de la teoría. De hecho, hemos logrado identificar una expresión analítica exacta de tipo breather que describe este proceso, aunque no se ha determinado una relación entre la frecuencia de oscilación observada y la amplitud inicial del modo inestable. Por lo tanto, dicha conexión debe establecerse mediante un enfoque cuasi-empírico.

Como trabajo futuro, resultaría esclarecedor extender nuestros resultados a teorías de dimensiones superiores. Naturalmente, el objetivo final es analizar la dinámica del esfalerón electrodébil. Sería valioso investigar si nuestros resultados son observables y relevantes en ese contexto, o si efectos físicos dominantes hacen que los fenómenos mencionados anteriormente sean subdominantes. Por ello, el estudio del decaimiento del esfalerón electrodébil podría resultar especialmente interesante debido a las implicaciones físicas que podría tener.   
\end{itemize}

También queremos enfatizar que, además de las posibles extensiones de los artículos que constituyen esta tesis, actualmente estamos investigando dos temas distintos. Presentamos aquí una breve visión general de las ideas principales que subyacen a cada uno de ellos.

\begin{itemize}
\item La dinámica de vórtices con $n = 1$ y su único modo ligado masivo excitado ha sido investigada previamente \cite{Pillado:2024}. En particular, se han derivado analíticamente tanto el comportamiento asintótico de la radiación emitida por el modo de forma como la tasa de decaimiento de su amplitud. Nuestro objetivo es extender este análisis a vórtices con ''winding numbers'' mayores. Debido a la inestabilidad del vórtice para $\lambda > 1$, restringimos nuestro análisis al régimen $\lambda < 1$. En esta región, los vórtices de carga superior admiten modos internos adicionales, incluidos modos no axialmente simétricos, lo que sugiere que la emisión de radiación resultante debería ser anisótropa. En primer lugar, calcularemos el campo de radiación asintótico emitido por un vórtice excitado y, a continuación, determinaremos la ley de decaimiento que rige la evolución de la amplitud de estos modos internos. Además, pretendemos investigar la supresión de la radiación en canales de emisión específicos, un fenómeno que se espera que surja de la dependencia de las frecuencias de los modos con la constante de autoacoplamiento. De manera interesante, la activación del modo interno no axialmente simétrico para $n = 2$ conduce a un desplazamiento periódico del cero de multiplicidad dos del vórtice: los ceros del vórtice oscilan alternativamente a lo largo de las direcciones $x_1$ y $x_2$, asemejándose a una dispersión de $\pi/2$. Como resultado, se espera que el campo de radiación emitido por la configuración de dos vórtices excitada ofrezca información valiosa sobre los mecanismos de emisión de energía durante los procesos de dispersión vórtice–vórtice. Estos hallazgos servirían como un trabajo complementario a las estimaciones existentes de la energía radiada en la dispersión de monopolos BPS que se mueven lentamente \cite{Manton3:1988}.

\item Mientras que el estudio de soluciones kink en teorías de campos escalares que involucran un solo campo ha sido ampliamente desarrollado, se sabe considerablemente menos sobre modelos con múltiples grados de libertad. En este trabajo, nos proponemos investigar las simetrías de Lie asociadas a las ecuaciones diferenciales que gobiernan el sector BPS de una teoría dada. Mediante la determinación de estas simetrías, pretendemos construir nuevas familias de soluciones dentro de un sector topológico dado. Este enfoque tendría el potencial de revelar simetrías ocultas que no son manifiestas a nivel del lagrangiano. Como consecuencia, este trabajo debería proporcionar información sobre la estructura del espacio de módulos y los módulos que lo caracterizan.
\end{itemize}

Como comentario final, esta tesis ha destacado la relevancia de los modos internos en la dinámica de los solitones. Tal como se ha evidenciado, estos modos pueden alterar de manera drástica el comportamiento esperado de los solitones. Dadas las posibles aplicaciones de los solitones y su presencia ubicua en diversas áreas de la física, resulta crucial una comprensión exhaustiva de su dinámica.
Existe una gran cantidad de trabajo prometedor por realizar en relación con la extensión de nuestros resultados a dimensiones superiores. Además, permanecen abiertas varias cuestiones que merecen una investigación adicional y una atención particular. Se espera que las contribuciones aquí presentadas sirvan como base para futuras investigaciones.
               
        \cleardoublepage
        \appendix
        
        \clearpage
        \thispagestyle{empty}
        \begin{center}
        \vspace*{0.35\textheight}
        {\Huge \textbf{Appendix A: First Article}}\\[1.5em]
        {\centering Authors: S. Navarro-Obreg\'on, L.M. Nieto and J.M. Queiruga.\\}
        {\centering Title: Inclusion of radiation in the collective coordinate method approach of the $\phi^4$ model.\\}
        {\centering Journal: Phys. Rev. E \textbf{108} (2023), 044216.}
        
        {\centering DOI: \href{https://doi.org/10.1103/PhysRevE.108.044216}{https://doi.org/10.1103/PhysRevE.108.044216}}
        \end{center}
        \phantomsection
        \label{app:Radiation}
        \addcontentsline{toc}{section}{Appendix A: First Article}
        \clearpage
        
        \clearpage
        \thispagestyle{empty}
        \begin{center}
        \vspace*{0.35\textheight}
        {\Huge \textbf{Appendix B: Second Article}}\\[1.5em]
        {\centering Authors: D. Migu\'elez-Caballero, S. Navarro-Obreg\'on and A. Wereszczynski.\\}
        {\centering Title: Moduli space metric of the excited vortex.\\}
        {\centering Journal: Phys. Rev. D \textbf{111} (2025), 105008.}
        
        {\centering DOI: \href{https://doi.org/10.1103/PhysRevD.111.105008}{https://doi.org/10.1103/PhysRevD.111.105008}}
        \end{center}
        \phantomsection
        \label{app:Vortex}
        \addcontentsline{toc}{section}{Appendix B: Second Article}
        \clearpage
        
        \clearpage
        \thispagestyle{empty}
        \begin{center}
        \vspace*{0.35\textheight}
        {\Huge \textbf{Appendix C: Third Article}}\\[1.5em]
        {\centering Authors: A. Alonso-Izquierdo, S. Navarro-Obreg\'on, K. Oles, J. Queiruga, T. Romanczukiewicz and A. Wereszczynski.\\}
        {\centering Title: Semi-Bogomol'nyi-Prasad-Sommerfield sphaleron and its dynamics.\\}
        {\centering Journal: Phys. Rev. E \textbf{108} (2023), 064208.}
        
        {\centering DOI: \href{https://doi.org/10.1103/PhysRevE.108.064208}{https://doi.org/10.1103/PhysRevE.108.064208}}
        \end{center}
        \phantomsection
        \label{app:SemiBPS}
        \addcontentsline{toc}{section}{Appendix C: Third Article}
        \clearpage
        
        \clearpage
        \thispagestyle{empty}
        \begin{center}
        \vspace*{0.35\textheight}
        {\Huge \textbf{Appendix D: Fourth Article}}\\[1.5em]
        {\centering Authors: S. Navarro-Obreg\'on and J. Queiruga.\\}
        {\centering Title: Impact of the internal modes on the sphaleron decay.\\}
        {\centering Journal: Eur. Phys. J. C \textbf{84} (2024), 821.}
        
        {\centering DOI: \href{https://doi.org/10.1140/epjc/s10052-024-13175-w}{https://doi.org/10.1140/epjc/s10052-024-13175-w}}
        \end{center}
        \phantomsection
        \label{app:Decay}
        \addcontentsline{toc}{section}{Appendix D: Fourth Article}
        \clearpage
        
        \clearpage
        \thispagestyle{empty}
        \begin{center}
        \vspace*{0.35\textheight}
        {\Huge \textbf{Appendix E: Fifth Article}}\\[1.5em]
        {\centering Authors: S. Navarro-Obreg\'on and J. Queiruga.\\}
        {\centering Title: Perturbed nonlinear dynamics and decay of sphalerons on circles.\\}
        {\centering Journal: Physica D \textbf{481} (2025), 134805.}
        
        {\centering DOI: \href{https://doi.org/10.1016/j.physd.2025.134805}{https://doi.org/10.1016/j.physd.2025.134805}}
        \end{center}
        \phantomsection
        \label{app:Circle}
        \addcontentsline{toc}{section}{Appendix E: Fifth Article}
        \clearpage

        \singlespacing
\label{app:Bibliography} 

\manualmark 
\markboth{\spacedlowsmallcaps{\bibname}}{\spacedlowsmallcaps{\bibname}} 
\refstepcounter{dummy}

\addtocontents{toc}{\protect\vspace{\beforebibskip}} 
\addcontentsline{toc}{chapter}{\tocEntry{\bibname}}
\printbibliography
        \cleardoublepage
      
        \pagestyle{empty}
        \onehalfspacing
        
    \end{document}